\documentclass[longauth]{aa}  
\pdfoutput=1
\usepackage{graphicx}
\usepackage{txfonts}
\usepackage{pdflscape}

\usepackage{natbib,twoopt}
\usepackage{amsmath} 
\usepackage[breaklinks=true]{hyperref} 
\bibpunct{(}{)}{;}{a}{}{,} 
\makeatletter
\newcommandtwoopt{\citeads}[3][][]{\href{http://adsabs.harvard.edu/abs/#3}%
{\def\hyper@linkstart##1##2{}%
\let\hyper@linkend\@empty\citealp[#1][#2]{#3}}}
\newcommandtwoopt{\citepads}[3][][]{\href{http://adsabs.harvard.edu/abs/#3}%
{\def\hyper@linkstart##1##2{}%
\let\hyper@linkend\@empty\citep[#1][#2]{#3}}}
\newcommandtwoopt{\citetads}[3][][]{\href{http://adsabs.harvard.edu/abs/#3}%
{\def\hyper@linkstart##1##2{}%
\let\hyper@linkend\@empty\citet[#1][#2]{#3}}}
\newcommandtwoopt{\citeyearads}[3][][]%
{\href{http://adsabs.harvard.edu/abs/#3}
{\def\hyper@linkstart##1##2{}%
\let\hyper@linkend\@empty\citeyear[#1][#2]{#3}}}
\makeatother

\usepackage{color}
\definecolor{mygreen}{RGB}{0,128,0}

\hypersetup{colorlinks=true,linkcolor=blue,citecolor=blue,urlcolor=blue}

\usepackage{etoolbox}
\makeatletter

\renewcommand{\degr}{\ensuremath{^\circ}}

\newcommand{\muas}{\ensuremath{\,{\mu}\text{as}}}
\newcommand{\muasyr}{\ensuremath{\,\mu\text{as\,yr}^{-1}}}
\newcommand{\masyr}{\ensuremath{\,\text{mas\,yr}^{-1}}}

\newcommand{\gaia}{\textit{Gaia}}

\newcommand{\gcrf}{\textit{Gaia}-CRF}

\newcommand{\gcrfthree}{\textit{Gaia}-CRF3}

\newcommand{\gdr}[1]{{\gaia~DR#1}}
\newcommand{\gedr}[1]{{\gaia~EDR#1}}

\usepackage{wasysym}

\newcommand{\orcit}[1]{\protect\href{https://orcid.org/#1}{\protect\includegraphics[width=8pt]{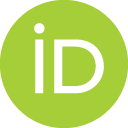}}}

\let\vec=\ve

\DeclareMathOperator{\erfc}{erfc}

\makeatletter
\ifaa@referee
\def\scalethree{0.329}
\else
\def\scalethree{0.33}
\fi
\makeatother

\makeatletter
\renewcommand*\maketitle{%
  \thispagestyle{firstpage}
\begingroup
    \if@wideboxfn
    \setlength\bibindent{1.4\parindent}
    \else
    \setlength\bibindent{\parindent}
    \fi
    \renewcommand*\thefootnote{\@fnsymbol\c@footnote}%
    \renewcommand\@makefntext[1]{%
    \ifaa@longfn\hsize\textwidth\fi
    \noindent
    \hb@xt@\bibindent{\hss\@makefnmark\enspace}##1}
  \ifaa@twocolumn
  \begingroup
    \begin{aa@strip}
          \aa@maketitle
    \end{aa@strip}
    \@thanks	  	
  \endgroup
  \else
    \begingroup
      \let\thanks\footnote
      \aa@maketitle
    \endgroup
  \fi
\endgroup
  \setcounter{footnote}{0}%
}
\makeatother

\begin{document}

   \title{\textit{Gaia} Early Data Release 3}
   \subtitle{The celestial reference frame (\textit{Gaia}-CRF3)}

   \author{
{\it Gaia} Collaboration
\and       S.A.~Klioner\orcit{0000-0003-4682-7831}\inst{\ref{inst:0001}}
\and         L.~Lindegren\orcit{0000-0002-5443-3026}\inst{\ref{inst:0002}}
\and         F.~Mignard\inst{\ref{inst:0003}}
\and         J.~Hern\'{a}ndez\orcit{0000-0002-0361-4994}\inst{\ref{inst:0004}}
\and         M.~Ramos-Lerate\inst{\ref{inst:0005}}
\and         U.~Bastian\orcit{0000-0002-8667-1715}\inst{\ref{inst:0006}}
\and         M.~Biermann\inst{\ref{inst:0006}}
\and         A.~Bombrun\inst{\ref{inst:0008}}
\and         A.~de Torres\inst{\ref{inst:0008}}
\and         E.~Gerlach\orcit{0000-0002-9533-2168}\inst{\ref{inst:0001}}
\and         R.~Geyer\orcit{0000-0001-6967-8707}\inst{\ref{inst:0001}}
\and         T.~Hilger\orcit{0000-0003-1646-0063}\inst{\ref{inst:0001}}
\and         D.~Hobbs\orcit{0000-0002-2696-1366}\inst{\ref{inst:0002}}
\and       U.L.~Lammers\orcit{0000-0001-8309-3801}\inst{\ref{inst:0004}}
\and       P.J.~McMillan\orcit{0000-0002-8861-2620}\inst{\ref{inst:0002}}
\and         H.~Steidelm\"{ u}ller\inst{\ref{inst:0001}}
\and         D.~Teyssier\orcit{0000-0002-6261-5292}\inst{\ref{inst:0005}}
\and       C.M.~Raiteri\orcit{0000-0003-1784-2784}\inst{\ref{inst:0017}}
\and         S.~Bartolom\'{e}\orcit{0000-0002-6290-6030}\inst{\ref{inst:0018}}
\and         M.~Bernet\orcit{0000-0001-7503-1010}\inst{\ref{inst:0018}}
\and         J.~Casta\~{n}eda\orcit{0000-0001-7820-946X}\inst{\ref{inst:0020}}
\and         M.~Clotet\inst{\ref{inst:0018}}
\and         M.~Davidson\inst{\ref{inst:0021}}
\and         C.~Fabricius\orcit{0000-0003-2639-1372}\inst{\ref{inst:0018}}
\and         N.~Garralda~Torres\inst{\ref{inst:0105}}
\and       J.J.~Gonz\'{a}lez-Vidal\inst{\ref{inst:0018}}
\and         J.~Portell\orcit{0000-0002-8886-8925}\inst{\ref{inst:0018}}
\and         N.~Rowell\orcit{0000-0003-3809-1895}\inst{\ref{inst:0021}}
\and         F.~Torra\orcit{0000-0002-8429-299X}\inst{\ref{inst:0020}}
\and         J.~Torra$^\dagger$\inst{\ref{inst:0018}}
\and     A.G.A.~Brown\orcit{0000-0002-7419-9679}\inst{\ref{inst:0028}}
\and         A.~Vallenari\orcit{0000-0003-0014-519X}\inst{\ref{inst:0029}}
\and         T.~Prusti\orcit{0000-0003-3120-7867}\inst{\ref{inst:0030}}
\and     J.H.J.~de Bruijne\orcit{0000-0001-6459-8599}\inst{\ref{inst:0030}}
\and         F.~Arenou\orcit{0000-0003-2837-3899}\inst{\ref{inst:0032}}
\and         C.~Babusiaux\orcit{0000-0002-7631-348X}\inst{\ref{inst:0033},\ref{inst:0032}}
\and       O.L.~Creevey\orcit{0000-0003-1853-6631}\inst{\ref{inst:0003}}
\and         C.~Ducourant\orcit{0000-0003-4843-8979}\inst{\ref{inst:0036}}
\and       D.W.~Evans\orcit{0000-0002-6685-5998}\inst{\ref{inst:0037}}
\and         L.~Eyer\orcit{0000-0002-0182-8040}\inst{\ref{inst:0038}}
\and         R.~Guerra\orcit{0000-0002-9850-8982}\inst{\ref{inst:0004}}
\and         A.~Hutton\inst{\ref{inst:0040}}
\and         C.~Jordi\orcit{0000-0001-5495-9602}\inst{\ref{inst:0018}}
\and         X.~Luri\orcit{0000-0001-5428-9397}\inst{\ref{inst:0018}}
\and         C.~Panem\inst{\ref{inst:0043}}
\and         D.~Pourbaix$^\dagger$\orcit{0000-0002-3020-1837}\inst{\ref{inst:0044},\ref{inst:0045}}
\and         S.~Randich\orcit{0000-0003-2438-0899}\inst{\ref{inst:0046}}
\and         P.~Sartoretti\inst{\ref{inst:0032}}
\and         C.~Soubiran\orcit{0000-0003-3304-8134}\inst{\ref{inst:0036}}
\and         P.~Tanga\orcit{0000-0002-2718-997X}\inst{\ref{inst:0003}}
\and       N.A.~Walton\orcit{0000-0003-3983-8778}\inst{\ref{inst:0037}}
\and     C.A.L.~Bailer-Jones\inst{\ref{inst:0051}}
\and         R.~Drimmel\orcit{0000-0002-1777-5502}\inst{\ref{inst:0017}}
\and         F.~Jansen\inst{\ref{inst:0053}}
\and         D.~Katz\orcit{0000-0001-7986-3164}\inst{\ref{inst:0032}}
\and       M.G.~Lattanzi\orcit{0000-0003-0429-7748}\inst{\ref{inst:0017},\ref{inst:0056}}
\and         F.~van Leeuwen\inst{\ref{inst:0037}}
\and         J.~Bakker\inst{\ref{inst:0004}}
\and         C.~Cacciari\orcit{0000-0001-5174-3179}\inst{\ref{inst:0059}}
\and         F.~De Angeli\orcit{0000-0003-1879-0488}\inst{\ref{inst:0037}}
\and         M.~Fouesneau\orcit{0000-0001-9256-5516}\inst{\ref{inst:0051}}
\and         Y.~Fr\'{e}mat\orcit{0000-0002-4645-6017}\inst{\ref{inst:0062}}
\and         L.~Galluccio\orcit{0000-0002-8541-0476}\inst{\ref{inst:0003}}
\and         A.~Guerrier\inst{\ref{inst:0043}}
\and         U.~Heiter\orcit{0000-0001-6825-1066}\inst{\ref{inst:0065}}
\and         E.~Masana\orcit{0000-0002-4819-329X}\inst{\ref{inst:0018}}
\and         R.~Messineo\inst{\ref{inst:0067}}
\and         N.~Mowlavi\orcit{0000-0003-1578-6993}\inst{\ref{inst:0038}}
\and         C.~Nicolas\inst{\ref{inst:0043}}
\and         K.~Nienartowicz\orcit{0000-0001-5415-0547}\inst{\ref{inst:0070},\ref{inst:0071}}
\and         F.~Pailler\orcit{0000-0002-4834-481X}\inst{\ref{inst:0043}}
\and         P.~Panuzzo\orcit{0000-0002-0016-8271}\inst{\ref{inst:0032}}
\and         F.~Riclet\inst{\ref{inst:0043}}
\and         W.~Roux\orcit{0000-0002-7816-1950}\inst{\ref{inst:0043}}
\and       G.M.~Seabroke\orcit{0000-0003-4072-9536}\inst{\ref{inst:0076}}
\and         R.~Sordo\orcit{0000-0003-4979-0659}\inst{\ref{inst:0029}}
\and         F.~Th\'{e}venin\inst{\ref{inst:0003}}
\and         G.~Gracia-Abril\inst{\ref{inst:0079},\ref{inst:0006}}
\and         M.~Altmann\orcit{0000-0002-0530-0913}\inst{\ref{inst:0006},\ref{inst:0082}}
\and         R.~Andrae\orcit{0000-0001-8006-6365}\inst{\ref{inst:0051}}
\and         M.~Audard\orcit{0000-0003-4721-034X}\inst{\ref{inst:0038},\ref{inst:0071}}
\and         I.~Bellas-Velidis\inst{\ref{inst:0086}}
\and         K.~Benson\inst{\ref{inst:0076}}
\and         J.~Berthier\orcit{0000-0003-1846-6485}\inst{\ref{inst:0088}}
\and         R.~Blomme\orcit{0000-0002-2526-346X}\inst{\ref{inst:0062}}
\and       P.W.~Burgess\inst{\ref{inst:0037}}
\and         D.~Busonero\orcit{0000-0002-3903-7076}\inst{\ref{inst:0017}}
\and         G.~Busso\orcit{0000-0003-0937-9849}\inst{\ref{inst:0037}}
\and         H.~C\'{a}novas\orcit{0000-0001-7668-8022}\inst{\ref{inst:0005}}
\and         B.~Carry\orcit{0000-0001-5242-3089}\inst{\ref{inst:0003}}
\and         A.~Cellino\orcit{0000-0002-6645-334X}\inst{\ref{inst:0017}}
\and         N.~Cheek\inst{\ref{inst:0096}}
\and         G.~Clementini\orcit{0000-0001-9206-9723}\inst{\ref{inst:0059}}
\and         Y.~Damerdji\orcit{0000-0002-3107-4024}\inst{\ref{inst:0098},\ref{inst:0099}}
\and         P.~de Teodoro\inst{\ref{inst:0004}}
\and         M.~Nu\~{n}ez Campos\inst{\ref{inst:0040}}
\and         L.~Delchambre\orcit{0000-0003-2559-408X}\inst{\ref{inst:0098}}
\and         A.~Dell'Oro\orcit{0000-0003-1561-9685}\inst{\ref{inst:0046}}
\and         P.~Esquej\orcit{0000-0001-8195-628X}\inst{\ref{inst:0104}}
\and         J.~Fern\'{a}ndez-Hern\'{a}ndez\inst{\ref{inst:0105}}
\and         E.~Fraile\inst{\ref{inst:0104}}
\and         D.~Garabato\orcit{0000-0002-7133-6623}\inst{\ref{inst:0107}}
\and         P.~Garc\'{i}a-Lario\orcit{0000-0003-4039-8212}\inst{\ref{inst:0004}}
\and         E.~Gosset\inst{\ref{inst:0098},\ref{inst:0045}}
\and         R.~Haigron\inst{\ref{inst:0032}}
\and      J.-L.~Halbwachs\orcit{0000-0003-2968-6395}\inst{\ref{inst:0112}}
\and       N.C.~Hambly\orcit{0000-0002-9901-9064}\inst{\ref{inst:0021}}
\and       D.L.~Harrison\orcit{0000-0001-8687-6588}\inst{\ref{inst:0037},\ref{inst:0115}}
\and         D.~Hestroffer\orcit{0000-0003-0472-9459}\inst{\ref{inst:0088}}
\and       S.T.~Hodgkin\orcit{0000-0002-5470-3962}\inst{\ref{inst:0037}}
\and         B.~Holl\orcit{0000-0001-6220-3266}\inst{\ref{inst:0038},\ref{inst:0071}}
\and         K.~Jan{\ss}en\orcit{0000-0002-8163-2493}\inst{\ref{inst:0120}}
\and         G.~Jevardat de Fombelle\inst{\ref{inst:0038}}
\and         S.~Jordan\orcit{0000-0001-6316-6831}\inst{\ref{inst:0006}}
\and         A.~Krone-Martins\orcit{0000-0002-2308-6623}\inst{\ref{inst:0123},\ref{inst:0124}}
\and       A.C.~Lanzafame\orcit{0000-0002-2697-3607}\inst{\ref{inst:0125},\ref{inst:0126}}
\and         W.~L\"{ o}ffler\inst{\ref{inst:0006}}
\and         O.~Marchal\orcit{ 0000-0001-7461-892}\inst{\ref{inst:0112}}
\and       P.M.~Marrese\orcit{0000-0002-8162-3810}\inst{\ref{inst:0129},\ref{inst:0130}}
\and         A.~Moitinho\orcit{0000-0003-0822-5995}\inst{\ref{inst:0123}}
\and         K.~Muinonen\orcit{0000-0001-8058-2642}\inst{\ref{inst:0132},\ref{inst:0133}}
\and         P.~Osborne\inst{\ref{inst:0037}}
\and         E.~Pancino\orcit{0000-0003-0788-5879}\inst{\ref{inst:0046},\ref{inst:0130}}
\and         T.~Pauwels\inst{\ref{inst:0062}}
\and         A.~Recio-Blanco\orcit{0000-0002-6550-7377}\inst{\ref{inst:0003}}
\and         C.~Reyl\'{e}\orcit{0000-0003-2258-2403}\inst{\ref{inst:0139}}
\and         M.~Riello\orcit{0000-0002-3134-0935}\inst{\ref{inst:0037}}
\and         L.~Rimoldini\orcit{0000-0002-0306-585X}\inst{\ref{inst:0071}}
\and         T.~Roegiers\orcit{0000-0002-1231-4440}\inst{\ref{inst:0142}}
\and         J.~Rybizki\orcit{0000-0002-0993-6089}\inst{\ref{inst:0051}}
\and       L.M.~Sarro\orcit{0000-0002-5622-5191}\inst{\ref{inst:0144}}
\and         C.~Siopis\orcit{0000-0002-6267-2924}\inst{\ref{inst:0044}}
\and         M.~Smith\inst{\ref{inst:0076}}
\and         A.~Sozzetti\orcit{0000-0002-7504-365X}\inst{\ref{inst:0017}}
\and         E.~Utrilla\inst{\ref{inst:0040}}
\and         M.~van Leeuwen\orcit{0000-0001-9698-2392}\inst{\ref{inst:0037}}
\and         U.~Abbas\orcit{0000-0002-5076-766X}\inst{\ref{inst:0017}}
\and         P.~\'{A}brah\'{a}m\orcit{0000-0001-6015-646X}\inst{\ref{inst:0151},\ref{inst:0152}}
\and         A.~Abreu Aramburu\inst{\ref{inst:0105}}
\and         C.~Aerts\orcit{0000-0003-1822-7126}\inst{\ref{inst:0154},\ref{inst:0155},\ref{inst:0051}}
\and       J.J.~Aguado\inst{\ref{inst:0144}}
\and         M.~Ajaj\inst{\ref{inst:0032}}
\and         F.~Aldea-Montero\inst{\ref{inst:0004}}
\and         G.~Altavilla\orcit{0000-0002-9934-1352}\inst{\ref{inst:0129},\ref{inst:0130}}
\and       M.A.~\'{A}lvarez\orcit{0000-0002-6786-2620}\inst{\ref{inst:0107}}
\and         J.~Alves\orcit{0000-0002-4355-0921}\inst{\ref{inst:0163}}
\and       R.I.~Anderson\orcit{0000-0001-8089-4419}\inst{\ref{inst:0164}}
\and         E.~Anglada Varela\orcit{0000-0001-7563-0689}\inst{\ref{inst:0105}}
\and         T.~Antoja\orcit{0000-0003-2595-5148}\inst{\ref{inst:0018}}
\and         D.~Baines\orcit{0000-0002-6923-3756}\inst{\ref{inst:0005}}
\and       S.G.~Baker\orcit{0000-0002-6436-1257}\inst{\ref{inst:0076}}
\and         L.~Balaguer-N\'{u}\~{n}ez\orcit{0000-0001-9789-7069}\inst{\ref{inst:0018}}
\and         E.~Balbinot\orcit{0000-0002-1322-3153}\inst{\ref{inst:0170}}
\and         Z.~Balog\orcit{0000-0003-1748-2926}\inst{\ref{inst:0006},\ref{inst:0051}}
\and         C.~Barache\inst{\ref{inst:0082}}
\and         D.~Barbato\inst{\ref{inst:0038},\ref{inst:0017}}
\and         M.~Barros\orcit{0000-0002-9728-9618}\inst{\ref{inst:0123}}
\and       M.A.~Barstow\orcit{0000-0002-7116-3259}\inst{\ref{inst:0177}}
\and      J.-L.~Bassilana\inst{\ref{inst:0178}}
\and         N.~Bauchet\inst{\ref{inst:0032}}
\and         U.~Becciani\orcit{0000-0002-4389-8688}\inst{\ref{inst:0125}}
\and         M.~Bellazzini\orcit{0000-0001-8200-810X}\inst{\ref{inst:0059}}
\and         A.~Berihuete\orcit{0000-0002-8589-4423}\inst{\ref{inst:0182}}
\and         S.~Bertone\orcit{0000-0001-9885-8440}\inst{\ref{inst:0183},\ref{inst:0184},\ref{inst:0017}}
\and         L.~Bianchi\orcit{0000-0002-7999-4372}\inst{\ref{inst:0186}}
\and         A.~Binnenfeld\orcit{0000-0002-9319-3838}\inst{\ref{inst:0187}}
\and         S.~Blanco-Cuaresma\orcit{0000-0002-1584-0171}\inst{\ref{inst:0188}}
\and         T.~Boch\orcit{0000-0001-5818-2781}\inst{\ref{inst:0112}}
\and         D.~Bossini\orcit{0000-0002-9480-8400}\inst{\ref{inst:0190}}
\and         S.~Bouquillon\inst{\ref{inst:0082},\ref{inst:0192}}
\and         A.~Bragaglia\orcit{0000-0002-0338-7883}\inst{\ref{inst:0059}}
\and         L.~Bramante\inst{\ref{inst:0067}}
\and         E.~Breedt\orcit{0000-0001-6180-3438}\inst{\ref{inst:0037}}
\and         A.~Bressan\orcit{0000-0002-7922-8440}\inst{\ref{inst:0196}}
\and         N.~Brouillet\orcit{0000-0002-3274-7024}\inst{\ref{inst:0036}}
\and         E.~Brugaletta\orcit{0000-0003-2598-6737}\inst{\ref{inst:0125}}
\and         B.~Bucciarelli\orcit{0000-0002-5303-0268}\inst{\ref{inst:0017},\ref{inst:0056}}
\and         A.~Burlacu\inst{\ref{inst:0201}}
\and       A.G.~Butkevich\orcit{0000-0002-4098-3588}\inst{\ref{inst:0017}}
\and         R.~Buzzi\orcit{0000-0001-9389-5701}\inst{\ref{inst:0017}}
\and         E.~Caffau\orcit{0000-0001-6011-6134}\inst{\ref{inst:0032}}
\and         R.~Cancelliere\orcit{0000-0002-9120-3799}\inst{\ref{inst:0205}}
\and         T.~Cantat-Gaudin\orcit{0000-0001-8726-2588}\inst{\ref{inst:0018},\ref{inst:0051}}
\and         R.~Carballo\orcit{0000-0001-7412-2498}\inst{\ref{inst:0208}}
\and         T.~Carlucci\inst{\ref{inst:0082}}
\and       M.I.~Carnerero\orcit{0000-0001-5843-5515}\inst{\ref{inst:0017}}
\and       J.M.~Carrasco\orcit{0000-0002-3029-5853}\inst{\ref{inst:0018}}
\and         L.~Casamiquela\orcit{0000-0001-5238-8674}\inst{\ref{inst:0036},\ref{inst:0032}}
\and         M.~Castellani\orcit{0000-0002-7650-7428}\inst{\ref{inst:0129}}
\and         A.~Castro-Ginard\orcit{0000-0002-9419-3725}\inst{\ref{inst:0028}}
\and         L.~Chaoul\inst{\ref{inst:0043}}
\and         P.~Charlot\orcit{0000-0002-9142-716X}\inst{\ref{inst:0036}}
\and         L.~Chemin\orcit{0000-0002-3834-7937}\inst{\ref{inst:0218}}
\and         V.~Chiaramida\inst{\ref{inst:0067}}
\and         A.~Chiavassa\orcit{0000-0003-3891-7554}\inst{\ref{inst:0003}}
\and         N.~Chornay\orcit{0000-0002-8767-3907}\inst{\ref{inst:0037}}
\and         G.~Comoretto\inst{\ref{inst:0005},\ref{inst:0223}}
\and         G.~Contursi\orcit{0000-0001-5370-1511}\inst{\ref{inst:0003}}
\and       W.J.~Cooper\orcit{0000-0003-3501-8967}\inst{\ref{inst:0225},\ref{inst:0017}}
\and         T.~Cornez\inst{\ref{inst:0178}}
\and         S.~Cowell\inst{\ref{inst:0037}}
\and         F.~Crifo\inst{\ref{inst:0032}}
\and         M.~Cropper\orcit{0000-0003-4571-9468}\inst{\ref{inst:0076}}
\and         M.~Crosta\orcit{0000-0003-4369-3786}\inst{\ref{inst:0017},\ref{inst:0232}}
\and         C.~Crowley\inst{\ref{inst:0008}}
\and         C.~Dafonte\orcit{0000-0003-4693-7555}\inst{\ref{inst:0107}}
\and         A.~Dapergolas\inst{\ref{inst:0086}}
\and         P.~David\inst{\ref{inst:0088}}
\and         P.~de Laverny\orcit{0000-0002-2817-4104}\inst{\ref{inst:0003}}
\and         F.~De Luise\orcit{0000-0002-6570-8208}\inst{\ref{inst:0238}}
\and         R.~De March\orcit{0000-0003-0567-842X}\inst{\ref{inst:0067}}
\and         J.~De Ridder\orcit{0000-0001-6726-2863}\inst{\ref{inst:0154}}
\and         R.~de Souza\inst{\ref{inst:0241}}
\and       E.F.~del Peloso\inst{\ref{inst:0006}}
\and         E.~del Pozo\inst{\ref{inst:0040}}
\and         M.~Delbo\orcit{0000-0002-8963-2404}\inst{\ref{inst:0003}}
\and         A.~Delgado\inst{\ref{inst:0104}}
\and      J.-B.~Delisle\orcit{0000-0001-5844-9888}\inst{\ref{inst:0038}}
\and         C.~Demouchy\inst{\ref{inst:0247}}
\and       T.E.~Dharmawardena\orcit{0000-0002-9583-5216}\inst{\ref{inst:0051}}
\and         S.~Diakite\inst{\ref{inst:0249}}
\and         C.~Diener\inst{\ref{inst:0037}}
\and         E.~Distefano\orcit{0000-0002-2448-2513}\inst{\ref{inst:0125}}
\and         C.~Dolding\inst{\ref{inst:0076}}
\and         H.~Enke\orcit{0000-0002-2366-8316}\inst{\ref{inst:0120}}
\and         C.~Fabre\inst{\ref{inst:0254}}
\and         M.~Fabrizio\orcit{0000-0001-5829-111X}\inst{\ref{inst:0129},\ref{inst:0130}}
\and         S.~Faigler\orcit{0000-0002-8368-5724}\inst{\ref{inst:0257}}
\and         G.~Fedorets\orcit{0000-0002-8418-4809}\inst{\ref{inst:0132},\ref{inst:0259}}
\and         P.~Fernique\orcit{0000-0002-3304-2923}\inst{\ref{inst:0112},\ref{inst:0261}}
\and         A.~Fienga\orcit{0000-0002-4755-7637}\inst{\ref{inst:0262},\ref{inst:0088}}
\and         F.~Figueras\orcit{0000-0002-3393-0007}\inst{\ref{inst:0018}}
\and         Y.~Fournier\orcit{0000-0002-6633-9088}\inst{\ref{inst:0120}}
\and         C.~Fouron\inst{\ref{inst:0201}}
\and         F.~Fragkoudi\orcit{0000-0002-0897-3013}\inst{\ref{inst:0267},\ref{inst:0268},\ref{inst:0269}}
\and         M.~Gai\orcit{0000-0001-9008-134X}\inst{\ref{inst:0017}}
\and         A.~Garcia-Gutierrez\inst{\ref{inst:0018}}
\and         M.~Garcia-Reinaldos\inst{\ref{inst:0004}}
\and         M.~Garc\'{i}a-Torres\orcit{0000-0002-6867-7080}\inst{\ref{inst:0273}}
\and         A.~Garofalo\orcit{0000-0002-5907-0375}\inst{\ref{inst:0059}}
\and         A.~Gavel\orcit{0000-0002-2963-722X}\inst{\ref{inst:0065}}
\and         P.~Gavras\orcit{0000-0002-4383-4836}\inst{\ref{inst:0104}}
\and         P.~Giacobbe\orcit{0000-0001-7034-7024}\inst{\ref{inst:0017}}
\and         G.~Gilmore\orcit{0000-0003-4632-0213}\inst{\ref{inst:0037}}
\and         S.~Girona\orcit{0000-0002-1975-1918}\inst{\ref{inst:0279}}
\and         G.~Giuffrida\inst{\ref{inst:0129}}
\and         R.~Gomel\inst{\ref{inst:0257}}
\and         A.~Gomez\orcit{0000-0002-3796-3690}\inst{\ref{inst:0107}}
\and         J.~Gonz\'{a}lez-N\'{u}\~{n}ez\orcit{0000-0001-5311-5555}\inst{\ref{inst:0096},\ref{inst:0284}}
\and         I.~Gonz\'{a}lez-Santamar\'{i}a\orcit{0000-0002-8537-9384}\inst{\ref{inst:0107}}
\and         M.~Granvik\orcit{0000-0002-5624-1888}\inst{\ref{inst:0132},\ref{inst:0287}}
\and         P.~Guillout\inst{\ref{inst:0112}}
\and         J.~Guiraud\inst{\ref{inst:0043}}
\and         R.~Guti\'{e}rrez-S\'{a}nchez\inst{\ref{inst:0005}}
\and       L.P.~Guy\orcit{0000-0003-0800-8755}\inst{\ref{inst:0071},\ref{inst:0292}}
\and         D.~Hatzidimitriou\orcit{0000-0002-5415-0464}\inst{\ref{inst:0293},\ref{inst:0086}}
\and         M.~Hauser\inst{\ref{inst:0051},\ref{inst:0296}}
\and         M.~Haywood\orcit{0000-0003-0434-0400}\inst{\ref{inst:0032}}
\and         A.~Helmer\inst{\ref{inst:0178}}
\and         A.~Helmi\orcit{0000-0003-3937-7641}\inst{\ref{inst:0170}}
\and       M.H.~Sarmiento\orcit{0000-0003-4252-5115}\inst{\ref{inst:0040}}
\and       S.L.~Hidalgo\orcit{0000-0002-0002-9298}\inst{\ref{inst:0301},\ref{inst:0302}}
\and         N.~H\l{}adczuk\orcit{0000-0001-9163-4209}\inst{\ref{inst:0004},\ref{inst:0304}}
\and         G.~Holland\inst{\ref{inst:0037}}
\and       H.E.~Huckle\inst{\ref{inst:0076}}
\and         K.~Jardine\inst{\ref{inst:0307}}
\and         G.~Jasniewicz\inst{\ref{inst:0308}}
\and         A.~Jean-Antoine Piccolo\orcit{0000-0001-8622-212X}\inst{\ref{inst:0043}}
\and     \'{O}.~Jim\'{e}nez-Arranz\orcit{0000-0001-7434-5165}\inst{\ref{inst:0018}}
\and         J.~Juaristi Campillo\inst{\ref{inst:0006}}
\and         F.~Julbe\inst{\ref{inst:0018}}
\and         L.~Karbevska\inst{\ref{inst:0071},\ref{inst:0314}}
\and         P.~Kervella\orcit{0000-0003-0626-1749}\inst{\ref{inst:0315}}
\and         S.~Khanna\orcit{0000-0002-2604-4277}\inst{\ref{inst:0170},\ref{inst:0017}}
\and         G.~Kordopatis\orcit{0000-0002-9035-3920}\inst{\ref{inst:0003}}
\and       A.J.~Korn\orcit{0000-0002-3881-6756}\inst{\ref{inst:0065}}
\and      \'{A}~K\'{o}sp\'{a}l\orcit{\'{u}t 15-17, 1121 }\inst{\ref{inst:0151},\ref{inst:0051},\ref{inst:0152}}
\and         Z.~Kostrzewa-Rutkowska\inst{\ref{inst:0028},\ref{inst:0324}}
\and         K.~Kruszy\'{n}ska\orcit{0000-0002-2729-5369}\inst{\ref{inst:0325}}
\and         M.~Kun\orcit{0000-0002-7538-5166}\inst{\ref{inst:0151}}
\and         P.~Laizeau\inst{\ref{inst:0327}}
\and         S.~Lambert\orcit{0000-0001-6759-5502}\inst{\ref{inst:0082}}
\and       A.F.~Lanza\orcit{0000-0001-5928-7251}\inst{\ref{inst:0125}}
\and         Y.~Lasne\inst{\ref{inst:0178}}
\and      J.-F.~Le Campion\inst{\ref{inst:0036}}
\and         Y.~Lebreton\orcit{0000-0002-4834-2144}\inst{\ref{inst:0315},\ref{inst:0333}}
\and         T.~Lebzelter\orcit{0000-0002-0702-7551}\inst{\ref{inst:0163}}
\and         S.~Leccia\orcit{0000-0001-5685-6930}\inst{\ref{inst:0335}}
\and         N.~Leclerc\inst{\ref{inst:0032}}
\and         I.~Lecoeur-Taibi\orcit{0000-0003-0029-8575}\inst{\ref{inst:0071}}
\and         S.~Liao\orcit{0000-0002-9346-0211}\inst{\ref{inst:0338},\ref{inst:0017},\ref{inst:0340}}
\and       E.L.~Licata\orcit{0000-0002-5203-0135}\inst{\ref{inst:0017}}
\and     H.E.P.~Lindstr{\o}m\inst{\ref{inst:0017},\ref{inst:0343},\ref{inst:0344}}
\and       T.A.~Lister\orcit{0000-0002-3818-7769}\inst{\ref{inst:0345}}
\and         E.~Livanou\orcit{0000-0003-0628-2347}\inst{\ref{inst:0293}}
\and         A.~Lobel\orcit{0000-0001-5030-019X}\inst{\ref{inst:0062}}
\and         A.~Lorca\inst{\ref{inst:0040}}
\and         C.~Loup\inst{\ref{inst:0112}}
\and         P.~Madrero Pardo\inst{\ref{inst:0018}}
\and         A.~Magdaleno Romeo\inst{\ref{inst:0201}}
\and         S.~Managau\inst{\ref{inst:0178}}
\and       R.G.~Mann\orcit{0000-0002-0194-325X}\inst{\ref{inst:0021}}
\and         M.~Manteiga\orcit{0000-0002-7711-5581}\inst{\ref{inst:0354}}
\and       J.M.~Marchant\orcit{0000-0002-3678-3145}\inst{\ref{inst:0355}}
\and         M.~Marconi\orcit{0000-0002-1330-2927}\inst{\ref{inst:0335}}
\and         J.~Marcos\inst{\ref{inst:0005}}
\and     M.M.S.~Marcos Santos\inst{\ref{inst:0096}}
\and         D.~Mar\'{i}n Pina\orcit{0000-0001-6482-1842}\inst{\ref{inst:0018}}
\and         S.~Marinoni\orcit{0000-0001-7990-6849}\inst{\ref{inst:0129},\ref{inst:0130}}
\and         F.~Marocco\orcit{0000-0001-7519-1700}\inst{\ref{inst:0362}}
\and       D.J.~Marshall\orcit{0000-0003-3956-3524}\inst{\ref{inst:0363}}
\and         L.~Martin Polo\inst{\ref{inst:0096}}
\and       J.M.~Mart\'{i}n-Fleitas\orcit{0000-0002-8594-569X}\inst{\ref{inst:0040}}
\and         G.~Marton\orcit{0000-0002-1326-1686}\inst{\ref{inst:0151}}
\and         N.~Mary\inst{\ref{inst:0178}}
\and         A.~Masip\orcit{0000-0003-1419-0020}\inst{\ref{inst:0018}}
\and         D.~Massari\orcit{0000-0001-8892-4301}\inst{\ref{inst:0059}}
\and         A.~Mastrobuono-Battisti\orcit{0000-0002-2386-9142}\inst{\ref{inst:0032}}
\and         T.~Mazeh\orcit{0000-0002-3569-3391}\inst{\ref{inst:0257}}
\and         S.~Messina\orcit{0000-0002-2851-2468}\inst{\ref{inst:0125}}
\and         D.~Michalik\orcit{0000-0002-7618-6556}\inst{\ref{inst:0030}}
\and       N.R.~Millar\inst{\ref{inst:0037}}
\and         A.~Mints\orcit{0000-0002-8440-1455}\inst{\ref{inst:0120}}
\and         D.~Molina\orcit{0000-0003-4814-0275}\inst{\ref{inst:0018}}
\and         R.~Molinaro\orcit{0000-0003-3055-6002}\inst{\ref{inst:0335}}
\and         L.~Moln\'{a}r\orcit{0000-0002-8159-1599}\inst{\ref{inst:0151},\ref{inst:0379},\ref{inst:0152}}
\and         G.~Monari\orcit{0000-0002-6863-0661}\inst{\ref{inst:0112}}
\and         M.~Mongui\'{o}\orcit{0000-0002-4519-6700}\inst{\ref{inst:0018}}
\and         P.~Montegriffo\orcit{0000-0001-5013-5948}\inst{\ref{inst:0059}}
\and         A.~Montero\inst{\ref{inst:0040}}
\and         R.~Mor\orcit{0000-0002-8179-6527}\inst{\ref{inst:0018}}
\and         A.~Mora\inst{\ref{inst:0040}}
\and         R.~Morbidelli\orcit{0000-0001-7627-4946}\inst{\ref{inst:0017}}
\and         T.~Morel\orcit{0000-0002-8176-4816}\inst{\ref{inst:0098}}
\and         D.~Morris\inst{\ref{inst:0021}}
\and         T.~Muraveva\orcit{0000-0002-0969-1915}\inst{\ref{inst:0059}}
\and       C.P.~Murphy\inst{\ref{inst:0004}}
\and         I.~Musella\orcit{0000-0001-5909-6615}\inst{\ref{inst:0335}}
\and         Z.~Nagy\orcit{0000-0002-3632-1194}\inst{\ref{inst:0151}}
\and         L.~Noval\inst{\ref{inst:0178}}
\and         F.~Oca\~{n}a\inst{\ref{inst:0005},\ref{inst:0395}}
\and         A.~Ogden\inst{\ref{inst:0037}}
\and         C.~Ordenovic\inst{\ref{inst:0003}}
\and       J.O.~Osinde\inst{\ref{inst:0104}}
\and         C.~Pagani\orcit{0000-0001-5477-4720}\inst{\ref{inst:0177}}
\and         I.~Pagano\orcit{0000-0001-9573-4928}\inst{\ref{inst:0125}}
\and         L.~Palaversa\orcit{0000-0003-3710-0331}\inst{\ref{inst:0401},\ref{inst:0037}}
\and       P.A.~Palicio\orcit{0000-0002-7432-8709}\inst{\ref{inst:0003}}
\and         L.~Pallas-Quintela\orcit{0000-0001-9296-3100}\inst{\ref{inst:0107}}
\and         A.~Panahi\orcit{0000-0001-5850-4373}\inst{\ref{inst:0257}}
\and         S.~Payne-Wardenaar\inst{\ref{inst:0006}}
\and         X.~Pe\~{n}alosa Esteller\inst{\ref{inst:0018}}
\and         A.~Penttil\"{ a}\orcit{0000-0001-7403-1721}\inst{\ref{inst:0132}}
\and         B.~Pichon\orcit{0000 0000 0062 1449}\inst{\ref{inst:0003}}
\and       A.M.~Piersimoni\orcit{0000-0002-8019-3708}\inst{\ref{inst:0238}}
\and      F.-X.~Pineau\orcit{0000-0002-2335-4499}\inst{\ref{inst:0112}}
\and         E.~Plachy\orcit{0000-0002-5481-3352}\inst{\ref{inst:0151},\ref{inst:0379},\ref{inst:0152}}
\and         G.~Plum\inst{\ref{inst:0032}}
\and         E.~Poggio\orcit{0000-0003-3793-8505}\inst{\ref{inst:0003},\ref{inst:0017}}
\and         A.~Pr\v{s}a\orcit{0000-0002-1913-0281}\inst{\ref{inst:0417}}
\and         L.~Pulone\orcit{0000-0002-5285-998X}\inst{\ref{inst:0129}}
\and         E.~Racero\orcit{0000-0002-6101-9050}\inst{\ref{inst:0096},\ref{inst:0395}}
\and         S.~Ragaini\inst{\ref{inst:0059}}
\and         M.~Rainer\orcit{0000-0002-8786-2572}\inst{\ref{inst:0046},\ref{inst:0423}}
\and         N.~Rambaux\orcit{0000-0002-9380-271X}\inst{\ref{inst:0088}}
\and         P.~Ramos\orcit{0000-0002-5080-7027}\inst{\ref{inst:0018},\ref{inst:0112}}
\and         P.~Re Fiorentin\orcit{0000-0002-4995-0475}\inst{\ref{inst:0017}}
\and         S.~Regibo\inst{\ref{inst:0154}}
\and       P.J.~Richards\inst{\ref{inst:0429}}
\and         C.~Rios Diaz\inst{\ref{inst:0104}}
\and         V.~Ripepi\orcit{0000-0003-1801-426X}\inst{\ref{inst:0335}}
\and         A.~Riva\orcit{0000-0002-6928-8589}\inst{\ref{inst:0017}}
\and      H.-W.~Rix\orcit{0000-0003-4996-9069}\inst{\ref{inst:0051}}
\and         G.~Rixon\orcit{0000-0003-4399-6568}\inst{\ref{inst:0037}}
\and         N.~Robichon\orcit{0000-0003-4545-7517}\inst{\ref{inst:0032}}
\and       A.C.~Robin\orcit{0000-0001-8654-9499}\inst{\ref{inst:0139}}
\and         C.~Robin\inst{\ref{inst:0178}}
\and         M.~Roelens\orcit{0000-0003-0876-4673}\inst{\ref{inst:0038}}
\and     H.R.O.~Rogues\inst{\ref{inst:0247}}
\and         L.~Rohrbasser\inst{\ref{inst:0071}}
\and         M.~Romero-G\'{o}mez\orcit{0000-0003-3936-1025}\inst{\ref{inst:0018}}
\and         F.~Royer\orcit{0000-0002-9374-8645}\inst{\ref{inst:0032}}
\and         D.~Ruz Mieres\orcit{0000-0002-9455-157X}\inst{\ref{inst:0037}}
\and       K.A.~Rybicki\orcit{0000-0002-9326-9329}\inst{\ref{inst:0325}}
\and         G.~Sadowski\orcit{0000-0002-3411-1003}\inst{\ref{inst:0044}}
\and         A.~S\'{a}ez N\'{u}\~{n}ez\inst{\ref{inst:0018}}
\and         A.~Sagrist\`{a} Sell\'{e}s\orcit{0000-0001-6191-2028}\inst{\ref{inst:0006}}
\and         J.~Sahlmann\orcit{0000-0001-9525-3673}\inst{\ref{inst:0104}}
\and         E.~Salguero\inst{\ref{inst:0105}}
\and         N.~Samaras\orcit{0000-0001-8375-6652}\inst{\ref{inst:0062},\ref{inst:0451}}
\and         V.~Sanchez Gimenez\orcit{0000-0003-1797-3557}\inst{\ref{inst:0018}}
\and         N.~Sanna\orcit{0000-0001-9275-9492}\inst{\ref{inst:0046}}
\and         R.~Santove\~{n}a\orcit{0000-0002-9257-2131}\inst{\ref{inst:0107}}
\and         M.~Sarasso\orcit{0000-0001-5121-0727}\inst{\ref{inst:0017}}
\and        M.~Schultheis\orcit{0000-0002-6590-1657}\inst{\ref{inst:0003}}
\and         E.~Sciacca\orcit{0000-0002-5574-2787}\inst{\ref{inst:0125}}
\and         M.~Segol\inst{\ref{inst:0247}}
\and       J.C.~Segovia\inst{\ref{inst:0096}}
\and         D.~S\'{e}gransan\orcit{0000-0003-2355-8034}\inst{\ref{inst:0038}}
\and         D.~Semeux\inst{\ref{inst:0254}}
\and         S.~Shahaf\orcit{0000-0001-9298-8068}\inst{\ref{inst:0462}}
\and       H.I.~Siddiqui\orcit{0000-0003-1853-6033}\inst{\ref{inst:0463}}
\and         A.~Siebert\orcit{0000-0001-8059-2840}\inst{\ref{inst:0112},\ref{inst:0261}}
\and         L.~Siltala\orcit{0000-0002-6938-794X}\inst{\ref{inst:0132}}
\and         A.~Silvelo\orcit{0000-0002-5126-6365}\inst{\ref{inst:0107}}
\and         E.~Slezak\inst{\ref{inst:0003}}
\and         I.~Slezak\inst{\ref{inst:0003}}
\and       R.L.~Smart\orcit{0000-0002-4424-4766}\inst{\ref{inst:0017}}
\and       O.N.~Snaith\inst{\ref{inst:0032}}
\and         E.~Solano\inst{\ref{inst:0472}}
\and         F.~Solitro\inst{\ref{inst:0067}}
\and         D.~Souami\orcit{0000-0003-4058-0815}\inst{\ref{inst:0315},\ref{inst:0475}}
\and         J.~Souchay\inst{\ref{inst:0082}}
\and         A.~Spagna\orcit{0000-0003-1732-2412}\inst{\ref{inst:0017}}
\and         L.~Spina\orcit{0000-0002-9760-6249}\inst{\ref{inst:0029}}
\and         F.~Spoto\orcit{0000-0001-7319-5847}\inst{\ref{inst:0188}}
\and       I.A.~Steele\orcit{0000-0001-8397-5759}\inst{\ref{inst:0355}}
\and       C.A.~Stephenson\inst{\ref{inst:0005},\ref{inst:0482}}
\and         M.~S\"{ u}veges\orcit{0000-0003-3017-5322}\inst{\ref{inst:0483}}
\and         J.~Surdej\orcit{0000-0002-7005-1976}\inst{\ref{inst:0098},\ref{inst:0485}}
\and         L.~Szabados\orcit{0000-0002-2046-4131}\inst{\ref{inst:0151}}
\and         E.~Szegedi-Elek\orcit{0000-0001-7807-6644}\inst{\ref{inst:0151}}
\and         F.~Taris\inst{\ref{inst:0082}}
\and       M.B.~Taylor\orcit{0000-0002-4209-1479}\inst{\ref{inst:0489}}
\and         R.~Teixeira\orcit{0000-0002-6806-6626}\inst{\ref{inst:0241}}
\and         L.~Tolomei\orcit{0000-0002-3541-3230}\inst{\ref{inst:0067}}
\and         N.~Tonello\orcit{0000-0003-0550-1667}\inst{\ref{inst:0279}}
\and         G.~Torralba Elipe\orcit{0000-0001-8738-194X}\inst{\ref{inst:0107}}
\and         M.~Trabucchi\orcit{0000-0002-1429-2388}\inst{\ref{inst:0494},\ref{inst:0038}}
\and       A.T.~Tsounis\inst{\ref{inst:0496}}
\and         C.~Turon\orcit{0000-0003-1236-5157}\inst{\ref{inst:0032}}
\and         A.~Ulla\orcit{0000-0001-6424-5005}\inst{\ref{inst:0498}}
\and         N.~Unger\orcit{0000-0003-3993-7127}\inst{\ref{inst:0038}}
\and       M.V.~Vaillant\inst{\ref{inst:0178}}
\and         E.~van Dillen\inst{\ref{inst:0247}}
\and         W.~van Reeven\inst{\ref{inst:0502}}
\and         O.~Vanel\orcit{0000-0002-7898-0454}\inst{\ref{inst:0032}}
\and         A.~Vecchiato\orcit{0000-0003-1399-5556}\inst{\ref{inst:0017}}
\and         Y.~Viala\inst{\ref{inst:0032}}
\and         D.~Vicente\orcit{0000-0002-1584-1182}\inst{\ref{inst:0279}}
\and         S.~Voutsinas\inst{\ref{inst:0021}}
\and         M.~Weiler\inst{\ref{inst:0018}}
\and         T.~Wevers\orcit{0000-0002-4043-9400}\inst{\ref{inst:0037},\ref{inst:0510}}
\and      \L{}.~Wyrzykowski\orcit{0000-0002-9658-6151}\inst{\ref{inst:0325}}
\and         A.~Yoldas\inst{\ref{inst:0037}}
\and         P.~Yvard\inst{\ref{inst:0247}}
\and         H.~Zhao\orcit{0000-0003-2645-6869}\inst{\ref{inst:0003}}
\and         J.~Zorec\inst{\ref{inst:0515}}
\and         S.~Zucker\orcit{0000-0003-3173-3138}\inst{\ref{inst:0187}}
\and         T.~Zwitter\orcit{0000-0002-2325-8763}\inst{\ref{inst:0517}}
}
\institute{
     Lohrmann Observatory, Technische Universit\"{ a}t Dresden, Mommsenstra{\ss}e 13, 01062 Dresden, Germany\relax                                                                                                                                                                                                                                                 \label{inst:0001}
\and Lund Observatory, Department of Astronomy and Theoretical Physics, Lund University, Box 43, 22100 Lund, Sweden\relax                                                                                                                                                                                                                                          \label{inst:0002}\vfill
\and Universit\'{e} C\^{o}te d'Azur, Observatoire de la C\^{o}te d'Azur, CNRS, Laboratoire Lagrange, Bd de l'Observatoire, CS 34229, 06304 Nice Cedex 4, France\relax                                                                                                                                                                                              \label{inst:0003}\vfill
\and European Space Agency (ESA), European Space Astronomy Centre (ESAC), Camino bajo del Castillo, s/n, Urbanizacion Villafranca del Castillo, Villanueva de la Ca\~{n}ada, 28692 Madrid, Spain\relax                                                                                                                                                             \label{inst:0004}\vfill
\and Telespazio UK S.L. for European Space Agency (ESA), Camino bajo del Castillo, s/n, Urbanizacion Villafranca del Castillo, Villanueva de la Ca\~{n}ada, 28692 Madrid, Spain\relax                                                                                                                                                                              \label{inst:0005}\vfill
\and Astronomisches Rechen-Institut, Zentrum f\"{ u}r Astronomie der Universit\"{ a}t Heidelberg, M\"{ o}nchhofstr. 12-14, 69120 Heidelberg, Germany\relax                                                                                                                                                                                                         \label{inst:0006}\vfill
\and HE Space Operations BV for European Space Agency (ESA), Camino bajo del Castillo, s/n, Urbanizacion Villafranca del Castillo, Villanueva de la Ca\~{n}ada, 28692 Madrid, Spain\relax                                                                                                                                                                          \label{inst:0008}\vfill
\and INAF - Osservatorio Astrofisico di Torino, via Osservatorio 20, 10025 Pino Torinese (TO), Italy\relax                                                                                                                                                                                                                                                         \label{inst:0017}\vfill
\and Institut de Ci\`{e}ncies del Cosmos (ICCUB), Universitat  de  Barcelona  (IEEC-UB), Mart\'{i} i  Franqu\`{e}s  1, 08028 Barcelona, Spain\relax                                                                                                                                                                                                                \label{inst:0018}\vfill
\and DAPCOM for Institut de Ci\`{e}ncies del Cosmos (ICCUB), Universitat  de  Barcelona  (IEEC-UB), Mart\'{i} i  Franqu\`{e}s  1, 08028 Barcelona, Spain\relax                                                                                                                                                                                                     \label{inst:0020}\vfill
\and Institute for Astronomy, University of Edinburgh, Royal Observatory, Blackford Hill, Edinburgh EH9 3HJ, United Kingdom\relax                                                                                                                                                                                                                                  \label{inst:0021}\vfill
\and ATG Europe for European Space Agency (ESA), Camino bajo del Castillo, s/n, Urbanizacion Villafranca del Castillo, Villanueva de la Ca\~{n}ada, 28692 Madrid, Spain\relax                                                                                                                                                                                      \label{inst:0105}\vfill
\and Leiden Observatory, Leiden University, Niels Bohrweg 2, 2333 CA Leiden, The Netherlands\relax                                                                                                                                                                                                                                                                 \label{inst:0028}\vfill
\and INAF - Osservatorio astronomico di Padova, Vicolo Osservatorio 5, 35122 Padova, Italy\relax                                                                                                                                                                                                                                                                   \label{inst:0029}\vfill
\and European Space Agency (ESA), European Space Research and Technology Centre (ESTEC), Keplerlaan 1, 2201AZ, Noordwijk, The Netherlands\relax                                                                                                                                                                                                                    \label{inst:0030}\vfill
\and GEPI, Observatoire de Paris, Universit\'{e} PSL, CNRS, 5 Place Jules Janssen, 92190 Meudon, France\relax                                                                                                                                                                                                                                                      \label{inst:0032}\vfill
\and Univ. Grenoble Alpes, CNRS, IPAG, 38000 Grenoble, France\relax                                                                                                                                                                                                                                                                                                \label{inst:0033}\vfill
\and Laboratoire d'astrophysique de Bordeaux, Univ. Bordeaux, CNRS, B18N, all{\'e}e Geoffroy Saint-Hilaire, 33615 Pessac, France\relax                                                                                                                                                                                                                             \label{inst:0036}\vfill
\and Institute of Astronomy, University of Cambridge, Madingley Road, Cambridge CB3 0HA, United Kingdom\relax                                                                                                                                                                                                                                                      \label{inst:0037}\vfill
\and Department of Astronomy, University of Geneva, Chemin Pegasi 51, 1290 Versoix, Switzerland\relax                                                                                                                                                                                                                                                              \label{inst:0038}\vfill
\and Aurora Technology for European Space Agency (ESA), Camino bajo del Castillo, s/n, Urbanizacion Villafranca del Castillo, Villanueva de la Ca\~{n}ada, 28692 Madrid, Spain\relax                                                                                                                                                                               \label{inst:0040}\vfill
\and CNES Centre Spatial de Toulouse, 18 avenue Edouard Belin, 31401 Toulouse Cedex 9, France\relax                                                                                                                                                                                                                                                                \label{inst:0043}\vfill
\and Institut d'Astronomie et d'Astrophysique, Universit\'{e} Libre de Bruxelles CP 226, Boulevard du Triomphe, 1050 Brussels, Belgium\relax                                                                                                                                                                                                                       \label{inst:0044}\vfill
\and F.R.S.-FNRS, Rue d'Egmont 5, 1000 Brussels, Belgium\relax                                                                                                                                                                                                                                                                                                     \label{inst:0045}\vfill
\and INAF - Osservatorio Astrofisico di Arcetri, Largo Enrico Fermi 5, 50125 Firenze, Italy\relax                                                                                                                                                                                                                                                                  \label{inst:0046}\vfill
\and Max Planck Institute for Astronomy, K\"{ o}nigstuhl 17, 69117 Heidelberg, Germany\relax                                                                                                                                                                                                                                                                       \label{inst:0051}\vfill
\and European Space Agency (ESA, retired)\relax                                                                                                                                                                                                                                                                                                                    \label{inst:0053}\vfill
\and University of Turin, Department of Physics, Via Pietro Giuria 1, 10125 Torino, Italy\relax                                                                                                                                                                                                                                                                    \label{inst:0056}\vfill
\and INAF - Osservatorio di Astrofisica e Scienza dello Spazio di Bologna, via Piero Gobetti 93/3, 40129 Bologna, Italy\relax                                                                                                                                                                                                                                      \label{inst:0059}\vfill
\and Royal Observatory of Belgium, Ringlaan 3, 1180 Brussels, Belgium\relax                                                                                                                                                                                                                                                                                        \label{inst:0062}\vfill
\and Observational Astrophysics, Division of Astronomy and Space Physics, Department of Physics and Astronomy, Uppsala University, Box 516, 751 20 Uppsala, Sweden\relax                                                                                                                                                                                           \label{inst:0065}\vfill
\and ALTEC S.p.a, Corso Marche, 79,10146 Torino, Italy\relax                                                                                                                                                                                                                                                                                                       \label{inst:0067}\vfill
\and S\`{a}rl, Geneva, Switzerland\relax                                                                                                                                                                                                                                                                                                                           \label{inst:0070}\vfill
\and Department of Astronomy, University of Geneva, Chemin d'Ecogia 16, 1290 Versoix, Switzerland\relax                                                                                                                                                                                                                                                            \label{inst:0071}\vfill
\and Mullard Space Science Laboratory, University College London, Holmbury St Mary, Dorking, Surrey RH5 6NT, United Kingdom\relax                                                                                                                                                                                                                                  \label{inst:0076}\vfill
\and Gaia DPAC Project Office, ESAC, Camino bajo del Castillo, s/n, Urbanizacion Villafranca del Castillo, Villanueva de la Ca\~{n}ada, 28692 Madrid, Spain\relax                                                                                                                                                                                                  \label{inst:0079}\vfill
\and SYRTE, Observatoire de Paris, Universit\'{e} PSL, CNRS,  Sorbonne Universit\'{e}, LNE, 61 avenue de l'Observatoire 75014 Paris, France\relax                                                                                                                                                                                                                  \label{inst:0082}\vfill
\and National Observatory of Athens, I. Metaxa and Vas. Pavlou, Palaia Penteli, 15236 Athens, Greece\relax                                                                                                                                                                                                                                                         \label{inst:0086}\vfill
\and IMCCE, Observatoire de Paris, Universit\'{e} PSL, CNRS, Sorbonne Universit{\'e}, Univ. Lille, 77 av. Denfert-Rochereau, 75014 Paris, France\relax                                                                                                                                                                                                             \label{inst:0088}\vfill
\and Serco Gesti\'{o}n de Negocios for European Space Agency (ESA), Camino bajo del Castillo, s/n, Urbanizacion Villafranca del Castillo, Villanueva de la Ca\~{n}ada, 28692 Madrid, Spain\relax                                                                                                                                                                   \label{inst:0096}\vfill
\and Institut d'Astrophysique et de G\'{e}ophysique, Universit\'{e} de Li\`{e}ge, 19c, All\'{e}e du 6 Ao\^{u}t, B-4000 Li\`{e}ge, Belgium\relax                                                                                                                                                                                                                    \label{inst:0098}\vfill
\and CRAAG - Centre de Recherche en Astronomie, Astrophysique et G\'{e}ophysique, Route de l'Observatoire Bp 63 Bouzareah 16340 Algiers, Algeria\relax                                                                                                                                                                                                             \label{inst:0099}\vfill
\and RHEA for European Space Agency (ESA), Camino bajo del Castillo, s/n, Urbanizacion Villafranca del Castillo, Villanueva de la Ca\~{n}ada, 28692 Madrid, Spain\relax                                                                                                                                                                                            \label{inst:0104}\vfill
\and CIGUS CITIC - Department of Computer Science and Information Technologies, University of A Coru\~{n}a, Campus de Elvi\~{n}a s/n, A Coru\~{n}a, 15071, Spain\relax                                                                                                                                                                                             \label{inst:0107}\vfill
\and Universit\'{e} de Strasbourg, CNRS, Observatoire astronomique de Strasbourg, UMR 7550,  11 rue de l'Universit\'{e}, 67000 Strasbourg, France\relax                                                                                                                                                                                                            \label{inst:0112}\vfill
\and Kavli Institute for Cosmology Cambridge, Institute of Astronomy, Madingley Road, Cambridge, CB3 0HA\relax                                                                                                                                                                                                                                                     \label{inst:0115}\vfill
\and Leibniz Institute for Astrophysics Potsdam (AIP), An der Sternwarte 16, 14482 Potsdam, Germany\relax                                                                                                                                                                                                                                                          \label{inst:0120}\vfill
\and CENTRA, Faculdade de Ci\^{e}ncias, Universidade de Lisboa, Edif. C8, Campo Grande, 1749-016 Lisboa, Portugal\relax                                                                                                                                                                                                                                            \label{inst:0123}\vfill
\and Department of Informatics, Donald Bren School of Information and Computer Sciences, University of California, Irvine, 5226 Donald Bren Hall, 92697-3440 CA Irvine, United States\relax                                                                                                                                                                        \label{inst:0124}\vfill
\and INAF - Osservatorio Astrofisico di Catania, via S. Sofia 78, 95123 Catania, Italy\relax                                                                                                                                                                                                                                                                       \label{inst:0125}\vfill
\and Dipartimento di Fisica e Astronomia ""Ettore Majorana"", Universit\`{a} di Catania, Via S. Sofia 64, 95123 Catania, Italy\relax                                                                                                                                                                                                                               \label{inst:0126}\vfill
\and INAF - Osservatorio Astronomico di Roma, Via Frascati 33, 00078 Monte Porzio Catone (Roma), Italy\relax                                                                                                                                                                                                                                                       \label{inst:0129}\vfill
\and Space Science Data Center - ASI, Via del Politecnico SNC, 00133 Roma, Italy\relax                                                                                                                                                                                                                                                                             \label{inst:0130}\vfill
\and Department of Physics, University of Helsinki, P.O. Box 64, 00014 Helsinki, Finland\relax                                                                                                                                                                                                                                                                     \label{inst:0132}\vfill
\and Finnish Geospatial Research Institute FGI, Geodeetinrinne 2, 02430 Masala, Finland\relax                                                                                                                                                                                                                                                                      \label{inst:0133}\vfill
\and Institut UTINAM CNRS UMR6213, Universit\'{e} Bourgogne Franche-Comt\'{e}, OSU THETA Franche-Comt\'{e} Bourgogne, Observatoire de Besan\c{c}on, BP1615, 25010 Besan\c{c}on Cedex, France\relax                                                                                                                                                                 \label{inst:0139}\vfill
\and HE Space Operations BV for European Space Agency (ESA), Keplerlaan 1, 2201AZ, Noordwijk, The Netherlands\relax                                                                                                                                                                                                                                                \label{inst:0142}\vfill
\and Dpto. de Inteligencia Artificial, UNED, c/ Juan del Rosal 16, 28040 Madrid, Spain\relax                                                                                                                                                                                                                                                                       \label{inst:0144}\vfill
\and Konkoly Observatory, Research Centre for Astronomy and Earth Sciences, E\"{ o}tv\"{ o}s Lor{\'a}nd Research Network (ELKH), MTA Centre of Excellence, Konkoly Thege Mikl\'{o}s \'{u}t 15-17, 1121 Budapest, Hungary\relax                                                                                                                                     \label{inst:0151}\vfill
\and ELTE E\"{ o}tv\"{ o}s Lor\'{a}nd University, Institute of Physics, 1117, P\'{a}zm\'{a}ny P\'{e}ter s\'{e}t\'{a}ny 1A, Budapest, Hungary\relax                                                                                                                                                                                                                 \label{inst:0152}\vfill
\and Instituut voor Sterrenkunde, KU Leuven, Celestijnenlaan 200D, 3001 Leuven, Belgium\relax                                                                                                                                                                                                                                                                      \label{inst:0154}\vfill
\and Department of Astrophysics/IMAPP, Radboud University, P.O.Box 9010, 6500 GL Nijmegen, The Netherlands\relax                                                                                                                                                                                                                                                   \label{inst:0155}\vfill
\and University of Vienna, Department of Astrophysics, T\"{ u}rkenschanzstra{\ss}e 17, A1180 Vienna, Austria\relax                                                                                                                                                                                                                                                 \label{inst:0163}\vfill
\and Institute of Physics, Laboratory of Astrophysics, Ecole Polytechnique F\'ed\'erale de Lausanne (EPFL), Observatoire de Sauverny, 1290 Versoix, Switzerland\relax                                                                                                                                                                                              \label{inst:0164}\vfill
\and Kapteyn Astronomical Institute, University of Groningen, Landleven 12, 9747 AD Groningen, The Netherlands\relax                                                                                                                                                                                                                                               \label{inst:0170}\vfill
\and School of Physics and Astronomy / Space Park Leicester, University of Leicester, University Road, Leicester LE1 7RH, United Kingdom\relax                                                                                                                                                                                                                     \label{inst:0177}\vfill
\and Thales Services for CNES Centre Spatial de Toulouse, 18 avenue Edouard Belin, 31401 Toulouse Cedex 9, France\relax                                                                                                                                                                                                                                            \label{inst:0178}\vfill
\and Depto. Estad\'istica e Investigaci\'on Operativa. Universidad de C\'adiz, Avda. Rep\'ublica Saharaui s/n, 11510 Puerto Real, C\'adiz, Spain\relax                                                                                                                                                                                                             \label{inst:0182}\vfill
\and Center for Research and Exploration in Space Science and Technology, University of Maryland Baltimore County, 1000 Hilltop Circle, Baltimore MD, USA\relax                                                                                                                                                                                                    \label{inst:0183}\vfill
\and GSFC - Goddard Space Flight Center, Code 698, 8800 Greenbelt Rd, 20771 MD Greenbelt, United States\relax                                                                                                                                                                                                                                                      \label{inst:0184}\vfill
\and EURIX S.r.l., Corso Vittorio Emanuele II 61, 10128, Torino, Italy\relax                                                                                                                                                                                                                                                                                       \label{inst:0186}\vfill
\and Porter School of the Environment and Earth Sciences, Tel Aviv University, Tel Aviv 6997801, Israel\relax                                                                                                                                                                                                                                                      \label{inst:0187}\vfill
\and Harvard-Smithsonian Center for Astrophysics, 60 Garden St., MS 15, Cambridge, MA 02138, USA\relax                                                                                                                                                                                                                                                             \label{inst:0188}\vfill
\and Instituto de Astrof\'{i}sica e Ci\^{e}ncias do Espa\c{c}o, Universidade do Porto, CAUP, Rua das Estrelas, PT4150-762 Porto, Portugal\relax                                                                                                                                                                                                                    \label{inst:0190}\vfill
\and LFCA/DAS,Universidad de Chile,CNRS,Casilla 36-D, Santiago, Chile\relax                                                                                                                                                                                                                                                                                        \label{inst:0192}\vfill
\and SISSA - Scuola Internazionale Superiore di Studi Avanzati, via Bonomea 265, 34136 Trieste, Italy\relax                                                                                                                                                                                                                                                        \label{inst:0196}\vfill
\and Telespazio for CNES Centre Spatial de Toulouse, 18 avenue Edouard Belin, 31401 Toulouse Cedex 9, France\relax                                                                                                                                                                                                                                                 \label{inst:0201}\vfill
\and University of Turin, Department of Computer Sciences, Corso Svizzera 185, 10149 Torino, Italy\relax                                                                                                                                                                                                                                                           \label{inst:0205}\vfill
\and Dpto. de Matem\'{a}tica Aplicada y Ciencias de la Computaci\'{o}n, Univ. de Cantabria, ETS Ingenieros de Caminos, Canales y Puertos, Avda. de los Castros s/n, 39005 Santander, Spain\relax                                                                                                                                                                   \label{inst:0208}\vfill
\and Centro de Astronom\'{i}a - CITEVA, Universidad de Antofagasta, Avenida Angamos 601, Antofagasta 1270300, Chile\relax                                                                                                                                                                                                                                          \label{inst:0218}\vfill
\and DLR Gesellschaft f\"{ u}r Raumfahrtanwendungen (GfR) mbH M\"{ u}nchener Stra{\ss}e 20 , 82234 We{\ss}ling\relax                                                                                                                                                                                                                                               \label{inst:0223}\vfill
\and Centre for Astrophysics Research, University of Hertfordshire, College Lane, AL10 9AB, Hatfield, United Kingdom\relax                                                                                                                                                                                                                                         \label{inst:0225}\vfill
\and University of Turin, Mathematical Department ""G.Peano"", Via Carlo Alberto 10, 10123 Torino, Italy\relax                                                                                                                                                                                                                                                     \label{inst:0232}\vfill
\and INAF - Osservatorio Astronomico d'Abruzzo, Via Mentore Maggini, 64100 Teramo, Italy\relax                                                                                                                                                                                                                                                                     \label{inst:0238}\vfill
\and Instituto de Astronomia, Geof\`{i}sica e Ci\^{e}ncias Atmosf\'{e}ricas, Universidade de S\~{a}o Paulo, Rua do Mat\~{a}o, 1226, Cidade Universitaria, 05508-900 S\~{a}o Paulo, SP, Brazil\relax                                                                                                                                                                \label{inst:0241}\vfill
\and APAVE SUDEUROPE SAS for CNES Centre Spatial de Toulouse, 18 avenue Edouard Belin, 31401 Toulouse Cedex 9, France\relax                                                                                                                                                                                                                                        \label{inst:0247}\vfill
\and M\'{e}socentre de calcul de Franche-Comt\'{e}, Universit\'{e} de Franche-Comt\'{e}, 16 route de Gray, 25030 Besan\c{c}on Cedex, France\relax                                                                                                                                                                                                                  \label{inst:0249}\vfill
\and ATOS for CNES Centre Spatial de Toulouse, 18 avenue Edouard Belin, 31401 Toulouse Cedex 9, France\relax                                                                                                                                                                                                                                                       \label{inst:0254}\vfill
\and School of Physics and Astronomy, Tel Aviv University, Tel Aviv 6997801, Israel\relax                                                                                                                                                                                                                                                                          \label{inst:0257}\vfill
\and Astrophysics Research Centre, School of Mathematics and Physics, Queen's University Belfast, Belfast BT7 1NN, UK\relax                                                                                                                                                                                                                                        \label{inst:0259}\vfill
\and Centre de Donn\'{e}es Astronomique de Strasbourg, Strasbourg, France\relax                                                                                                                                                                                                                                                                                    \label{inst:0261}\vfill
\and Universit\'{e} C\^{o}te d'Azur, Observatoire de la C\^{o}te d'Azur, CNRS, Laboratoire G\'{e}oazur, Bd de l'Observatoire, CS 34229, 06304 Nice Cedex 4, France\relax                                                                                                                                                                                           \label{inst:0262}\vfill
\and Institute for Computational Cosmology, Department of Physics, Durham University, Durham DH1 3LE, UK\relax                                                                                                                                                                                                                                                     \label{inst:0267}\vfill
\and European Southern Observatory, Karl-Schwarzschild-Str. 2, 85748 Garching, Germany\relax                                                                                                                                                                                                                                                                       \label{inst:0268}\vfill
\and Max-Planck-Institut f\"{ u}r Astrophysik, Karl-Schwarzschild-Stra{\ss}e 1, 85748 Garching, Germany\relax                                                                                                                                                                                                                                                      \label{inst:0269}\vfill
\and Data Science and Big Data Lab, Pablo de Olavide University, 41013, Seville, Spain\relax                                                                                                                                                                                                                                                                       \label{inst:0273}\vfill
\and Barcelona Supercomputing Center (BSC), Pla\c{c}a Eusebi G\"{ u}ell 1-3, 08034-Barcelona, Spain\relax                                                                                                                                                                                                                                                          \label{inst:0279}\vfill
\and ETSE Telecomunicaci\'{o}n, Universidade de Vigo, Campus Lagoas-Marcosende, 36310 Vigo, Galicia, Spain\relax                                                                                                                                                                                                                                                   \label{inst:0284}\vfill
\and Asteroid Engineering Laboratory, Space Systems, Lule\aa{} University of Technology, Box 848, S-981 28 Kiruna, Sweden\relax                                                                                                                                                                                                                                    \label{inst:0287}\vfill
\and Vera C Rubin Observatory,  950 N. Cherry Avenue, Tucson, AZ 85719, USA\relax                                                                                                                                                                                                                                                                                  \label{inst:0292}\vfill
\and Department of Astrophysics, Astronomy and Mechanics, National and Kapodistrian University of Athens, Panepistimiopolis, Zografos, 15783 Athens, Greece\relax                                                                                                                                                                                                  \label{inst:0293}\vfill
\and TRUMPF Photonic Components GmbH, Lise-Meitner-Stra{\ss}e 13,  89081 Ulm, Germany\relax                                                                                                                                                                                                                                                                        \label{inst:0296}\vfill
\and IAC - Instituto de Astrofisica de Canarias, Via L\'{a}ctea s/n, 38200 La Laguna S.C., Tenerife, Spain\relax                                                                                                                                                                                                                                                   \label{inst:0301}\vfill
\and Department of Astrophysics, University of La Laguna, Via L\'{a}ctea s/n, 38200 La Laguna S.C., Tenerife, Spain\relax                                                                                                                                                                                                                                          \label{inst:0302}\vfill
\and Faculty of Aerospace Engineering, Delft University of Technology, Kluyverweg 1, 2629 HS Delft, The Netherlands\relax                                                                                                                                                                                                                                          \label{inst:0304}\vfill
\and Radagast Solutions, Leiden, The Netherlands\relax                                                                                                                                                                                                                                                                                                             \label{inst:0307}\vfill
\and Laboratoire Univers et Particules de Montpellier, CNRS Universit\'{e} Montpellier, Place Eug\`{e}ne Bataillon, CC72, 34095 Montpellier Cedex 05, France\relax                                                                                                                                                                                                 \label{inst:0308}\vfill
\and Universit\'{e} de Caen Normandie, C\^{o}te de Nacre Boulevard Mar\'{e}chal Juin, 14032 Caen, France\relax                                                                                                                                                                                                                                                     \label{inst:0314}\vfill
\and LESIA, Observatoire de Paris, Universit\'{e} PSL, CNRS, Sorbonne Universit\'{e}, Universit\'{e} de Paris, 5 Place Jules Janssen, 92190 Meudon, France\relax                                                                                                                                                                                                   \label{inst:0315}\vfill
\and SRON Netherlands Institute for Space Research, Niels Bohrweg 4, 2333 CA Leiden, The Netherlands\relax                                                                                                                                                                                                                                                         \label{inst:0324}\vfill
\and Astronomical Observatory, University of Warsaw,  Al. Ujazdowskie 4, 00-478 Warszawa, Poland\relax                                                                                                                                                                                                                                                             \label{inst:0325}\vfill
\and Scalian for CNES Centre Spatial de Toulouse, 18 avenue Edouard Belin, 31401 Toulouse Cedex 9, France\relax                                                                                                                                                                                                                                                    \label{inst:0327}\vfill
\and Universit\'{e} Rennes, CNRS, IPR (Institut de Physique de Rennes) - UMR 6251, 35000 Rennes, France\relax                                                                                                                                                                                                                                                      \label{inst:0333}\vfill
\and INAF - Osservatorio Astronomico di Capodimonte, Via Moiariello 16, 80131, Napoli, Italy\relax                                                                                                                                                                                                                                                                 \label{inst:0335}\vfill
\and Shanghai Astronomical Observatory, Chinese Academy of Sciences, 80 Nandan Road, Shanghai 200030, People's Republic of China\relax                                                                                                                                                                                                                             \label{inst:0338}\vfill
\and University of Chinese Academy of Sciences, No.19(A) Yuquan Road, Shijingshan District, Beijing 100049, People's Republic of China\relax                                                                                                                                                                                                                       \label{inst:0340}\vfill
\and Niels Bohr Institute, University of Copenhagen, Juliane Maries Vej 30, 2100 Copenhagen {\O}, Denmark\relax                                                                                                                                                                                                                                                    \label{inst:0343}\vfill
\and DXC Technology, Retortvej 8, 2500 Valby, Denmark\relax                                                                                                                                                                                                                                                                                                        \label{inst:0344}\vfill
\and Las Cumbres Observatory, 6740 Cortona Drive Suite 102, Goleta, CA 93117, USA\relax                                                                                                                                                                                                                                                                            \label{inst:0345}\vfill
\and CIGUS CITIC, Department of Nautical Sciences and Marine Engineering, University of A Coru\~{n}a, Paseo de Ronda 51, 15071, A Coru\~{n}a, Spain\relax                                                                                                                                                                                                          \label{inst:0354}\vfill
\and Astrophysics Research Institute, Liverpool John Moores University, 146 Brownlow Hill, Liverpool L3 5RF, United Kingdom\relax                                                                                                                                                                                                                                  \label{inst:0355}\vfill
\and IPAC, Mail Code 100-22, California Institute of Technology, 1200 E. California Blvd., Pasadena, CA 91125, USA\relax                                                                                                                                                                                                                                           \label{inst:0362}\vfill
\and IRAP, Universit\'{e} de Toulouse, CNRS, UPS, CNES, 9 Av. colonel Roche, BP 44346, 31028 Toulouse Cedex 4, France\relax                                                                                                                                                                                                                                        \label{inst:0363}\vfill
\and MTA CSFK Lend\"{ u}let Near-Field Cosmology Research Group, Konkoly Observatory, MTA Research Centre for Astronomy and Earth Sciences, Konkoly Thege Mikl\'{o}s \'{u}t 15-17, 1121 Budapest, Hungary\relax                                                                                                                                                    \label{inst:0379}\vfill
\and Departmento de F\'{i}sica de la Tierra y Astrof\'{i}sica, Universidad Complutense de Madrid, 28040 Madrid, Spain\relax                                                                                                                                                                                                                                        \label{inst:0395}\vfill
\and Ru{\dj}er Bo\v{s}kovi\'{c} Institute, Bijeni\v{c}ka cesta 54, 10000 Zagreb, Croatia\relax                                                                                                                                                                                                                                                                     \label{inst:0401}\vfill
\and Villanova University, Department of Astrophysics and Planetary Science, 800 E Lancaster Avenue, Villanova PA 19085, USA\relax                                                                                                                                                                                                                                 \label{inst:0417}\vfill
\and INAF - Osservatorio Astronomico di Brera, via E. Bianchi, 46, 23807 Merate (LC), Italy\relax                                                                                                                                                                                                                                                                  \label{inst:0423}\vfill
\and STFC, Rutherford Appleton Laboratory, Harwell, Didcot, OX11 0QX, United Kingdom\relax                                                                                                                                                                                                                                                                         \label{inst:0429}\vfill
\and Charles University, Faculty of Mathematics and Physics, Astronomical Institute of Charles University, V Holesovickach 2, 18000 Prague, Czech Republic\relax                                                                                                                                                                                                   \label{inst:0451}\vfill
\and Department of Particle Physics and Astrophysics, Weizmann Institute of Science, Rehovot 7610001, Israel\relax                                                                                                                                                                                                                                                 \label{inst:0462}\vfill
\and Department of Astrophysical Sciences, 4 Ivy Lane, Princeton University, Princeton NJ 08544, USA\relax                                                                                                                                                                                                                                                         \label{inst:0463}\vfill
\and Departamento de Astrof\'{i}sica, Centro de Astrobiolog\'{i}a (CSIC-INTA), ESA-ESAC. Camino Bajo del Castillo s/n. 28692 Villanueva de la Ca\~{n}ada, Madrid, Spain\relax                                                                                                                                                                                      \label{inst:0472}\vfill
\and naXys, University of Namur, Rempart de la Vierge, 5000 Namur, Belgium\relax                                                                                                                                                                                                                                                                                   \label{inst:0475}\vfill
\and CGI Deutschland B.V. \& Co. KG, Mornewegstr. 30, 64293 Darmstadt, Germany\relax                                                                                                                                                                                                                                                                               \label{inst:0482}\vfill
\and Institute of Global Health, University of Geneva\relax                                                                                                                                                                                                                                                                                                        \label{inst:0483}\vfill
\and Astronomical Observatory Institute, Faculty of Physics, Adam Mickiewicz University, Pozna\'{n}, Poland\relax                                                                                                                                                                                                                                                  \label{inst:0485}\vfill
\and H H Wills Physics Laboratory, University of Bristol, Tyndall Avenue, Bristol BS8 1TL, United Kingdom\relax                                                                                                                                                                                                                                                    \label{inst:0489}\vfill
\and Department of Physics and Astronomy G. Galilei, University of Padova, Vicolo dell'Osservatorio 3, 35122, Padova, Italy\relax                                                                                                                                                                                                                                  \label{inst:0494}\vfill
\and CERN, Geneva, Switzerland\relax                                                                                                                                                                                                                                                                                                                               \label{inst:0496}\vfill
\and Applied Physics Department, Universidade de Vigo, 36310 Vigo, Spain\relax                                                                                                                                                                                                                                                                                     \label{inst:0498}\vfill
\and Association of Universities for Research in Astronomy, 1331 Pennsylvania Ave. NW, Washington, DC 20004, USA\relax                                                                                                                                                                                                                                             \label{inst:0502}\vfill
\and European Southern Observatory, Alonso de C\'ordova 3107, Casilla 19, Santiago, Chile\relax                                                                                                                                                                                                                                                                    \label{inst:0510}\vfill
\and Sorbonne Universit\'{e}, CNRS, UMR7095, Institut d'Astrophysique de Paris, 98bis bd. Arago, 75014 Paris, France\relax                                                                                                                                                                                                                                         \label{inst:0515}\vfill
\and Faculty of Mathematics and Physics, University of Ljubljana, Jadranska ulica 19, 1000 Ljubljana, Slovenia\relax                                                                                                                                                                                                                                               \label{inst:0517}\vfill
}

   \date{ }

 
\abstract
    {
    \gcrf{3} is the celestial reference frame for positions and proper motions in the third
    release of data from the \gaia\ mission,  \gdr{3} (and for the early third release, \gedr{3}, which 
    contains identical astrometric results). The reference frame is defined by the positions and proper
    motions at epoch 2016.0 for a specific set of extragalactic sources in the (E)DR3 catalogue.
    }
    {
     We describe the construction of \gcrf{3} and its properties in terms of the distributions in
     magnitude, colour, and astrometric quality. 
    }
    {
     Compact extragalactic sources in \gdr{3} were identified by positional cross-matching with 
     17 external catalogues of quasi-stellar objects (QSO) and active galactic nuclei (AGN), 
     followed by astrometric filtering designed to remove stellar 
     contaminants. Selecting a clean sample was favoured over including a higher 
     number of extragalactic sources. 
     For the final sample, the random and systematic errors in the proper motions are analysed, 
     as well as the radio--optical offsets in position for sources in the third realisation of the
     International Celestial Reference Frame (ICRF3).
    }
    {
    \gcrf{3} comprises about 1.6~million QSO-like sources, of which 
    1.2~million have five-parameter astrometric solutions in \gdr{3} and 0.4~million have 
    six-parameter solutions. The sources span the magnitude range $G=13$ to 21 with a peak density
    at 20.6~mag, at which the typical positional uncertainty is about 1~mas. The proper motions show
    systematic errors on the level of 12~\muasyr\ on angular scales greater than 15~deg. For the
    3142 optical counterparts of ICRF3 sources in the S/X frequency bands, 
    the median offset from the radio positions is about 
    0.5~mas, but it exceeds 4~mas in either coordinate for 127 sources. We outline the future of 
    \gcrf{} in the next \gaia\ data releases. Appendices give further details on the external
    catalogues used, how to extract information about the \gcrf{3} sources, potential (Galactic) 
    confusion sources, and the estimation of the spin and orientation of an astrometric solution. 
    }
  {}


   \titlerunning{{\textit{Gaia}} Early Data Release 3 -- \textit{Gaia}-CRF3} 
   \authorrunning{Gaia Collaboration et al.}

   \maketitle

%

\section{Introduction}
\label{sec:intro}

\gaia\ is the second highly successful astrometric space mission of
the European Space Agency (ESA)%
\footnote{Some further acronyms used in this paper: 
EDR3 -- \textit{Gaia} Early third Data Release; 
DR3 -- \textit{Gaia} third Data Release; 
QSO -- quasi stellar object (here understood as all compact extragalactic sources); 
AGN -- active galactic nuclei; 
ICRS -- International Celestial Reference System; 
ICRF3 -- third realisation of the International Celestial Reference Frame;
\textit{Gaia}-CRF$n$ -- $n$th realisation of the \textit{Gaia} Celestial Reference Frame; 
VLBI -- very long baseline interferometry;
S/X, K, X/Ka (after ICRF3) denote the radio frequency bands used in VLBI.}
\citepads{2016A&A...595A...1G}, vastly extending and improving
the achievements of its predecessor \textsc{Hipparcos}
(\citeads{1997ESASP1200.....E}; \citeads{1997A&A...323L..49P}; \citeads{2007A&A...474..653V}). 
While the
latter delivered astrometry at the level of a milliarcsecond (mas) for nearly  120\,000 preselected
stars at the mean epoch 1991.25, the latest data release \gedr{3} contains
astrometry for about 1.8~billion sources with accuracies from a few
tens of microarcseconds ($\mu$as) to $\sim$1~mas, mainly depending 
on the brightness \citepads{2021A&A...649A...1G,2021A&A...649A...2L}.

It is well known that the observational principles of \gaia\ (and
\textsc{Hipparcos}) do not allow one to fully fix the reference frame, that is
the coordinate system of positions and proper motions on the sky. 
\gaia\ effectively measures angular
separations of sources on the sky, and these separations remain
unchanged if one rotates the whole solution by an arbitrary angle 
at each moment of time. Although the degeneracy with rotation is not mathematically
exact, owing to the effects of aberration and gravitational light bending
which are coupled to the fixed solar system and \gaia\ ephemerides, those 
effects are too small to be of any practical help in fixing the reference frame.
In the standard astrometric solution, the
apparent motion of a source on the sky is modelled by five freely adjusted 
parameters: two angular components of its position at some reference epoch, 
the parallax, and two components of the mean proper motion over the 
observed time interval. This model gives a six-fold near-degeneracy in the
astrometric solution: three constant angles of rotation for the positions at the
reference epoch (the `orientation' of the solution), and three components of the constant 
angular velocity of rotation 
that affect the proper motions (the `spin' of the solution). 
In short, although there are many more observations than unknowns to fit, 
the resulting system always has a rank deficiency of six.
To fix the reference
frame of a \gaia\ solution, external information should be used to
constrain precisely these six degrees of freedom, and nothing more.

\gaia\ routinely delivers astrometry for sources as faint as $G=21$~mag
and, unlike \textsc{Hipparcos}, it can observe a large number of
extragalactic sources. This capability plays an important role for the
whole project.  By directly linking the astrometric solution to a set
of remote extragalactic sources, \gaia\ implements the principles set
forth in the International Celestial Reference System (ICRS;
\citeads{1995A&A...303..604A}), a concept originally devised 
for the use in radio astrometry observing radio-loud quasars at
cosmological distances. In practice, the term `extragalactic' here
refers to objects more distant than about 50~Mpc, that is well beyond
the Local Group.  The reasoning for choosing remote extragalactic
sources is simple: because they are not affected by the complex
motions in our Galaxy and have negligible apparent motions on the sky
(thanks to their large distances), they provide reliable and stable
beacons for astrometry. The suggestion to rely on very distant and nearly motionless
sources can be traced back to W.~Herschel and P.S.~Laplace two
centuries ago, well before quasars were spotted (see Sect.~6 of
\citeads{1985CeMec..36..207F} for a brief historical account). 

The ICRS embraces two ideas 
of a very different nature. The first one concerns the
orientation of a celestial reference frame at a given epoch while
the second one deals with the rotational state (spin) of the reference frame.
The orientation of a celestial reference frame at a given epoch is a pure convention, with
no physical significance to the particular choice of axes. In this respect the
choice of origin for the right ascension in ICRS is as arbitrary as, for example, the
choice of Greenwich as the Prime Meridian for the geographic longitude in 1884.  
\gaia\ uses its
own astrometric observations of the optical counterparts to sources
from the well-established radio astrometric catalogue ICRF3
\citepads{2020A&A...644A.159C} to fix the orientation of the
\gaia\ catalogue. 
This choice is of a practical nature, namely to ensure consistency between the optical
and radio frames and continuity of the orientation used in astronomy
over past decades.

The second aspect, much more significant, is the use of remote extragalactic sources to fix
the rotational state of the reference frame.  Individual sources may
show some (even non-linear) positional variations in time, owing to
their peculiar motions or intrinsic variations of their structure 
(e.g.\ \citeads{2018A&A...611A..52T}), but it is assumed that objects at
cosmological distances, on the average, show `no rotation'. 
%

In the framework of Newtonian physics, the assumption of no
rotation is interpreted as no rotation with respect to Newtonian
absolute space.  In the modern framework of the 
General Theory of Relativity, the procedure results in a reference
frame in which the Hubble flow has no rotational component.
One can argue that the no rotation assumption is related to Mach's principle
(e.g.\ \citeads{1964RvMP...36..510S}). However, we stress that
\gaia's observations of distant extragalactic objects can neither 
prove nor disprove that principle. To test Mach's principle is a separate 
task beyond the scope of the current discussion.

Besides the task of fixing the \gaia\ reference frame, having a set of astrometrically
well-behaved extragalactic sources is also important for assessing 
systematic errors in the \gaia\ astrometric solution.
Again, this is possible because we can assume that the true values of the proper 
motions and parallaxes of these sources are all very close to zero.
Such applications of the \gcrf{3} sample are found, for example, in
\citetads{2021A&A...649A...4L} and \citetads{2021A&A...649A...2L}.

Based on the way \gaia\ automatically detects and observes all point-like
objects of sufficient brightness \citepads{2015A&A...576A..74D}, \gedr{3} is
expected to contain a few million extragalactic objects suitable for defining 
a reference frame. However, it is not easy to reliably identify 
those sources in the \gaia\ catalogue.  As was the case for \gcrf{1}
\citepads{2016A&A...595A...5M} and \gcrf{2}
\citepads{2018A&A...616A..14G}, the identification of well-behaved 
extragalactic sources for \gedr{3} relies on a dedicated cross-matching 
with external catalogues of quasi-stellar objects (QSOs;
also often dubbed `quasars' irrespective of their radio loudness)
and active galactic nuclei (AGNs).  For the purpose of this work the
actual nature of the compact extragalactic sources plays no role, and
we refer to them as `QSO-like' sources, whether they are genuine 
QSOs of various kinds, galaxies with astrometrically compact nuclei, 
or other types of extragalactic objects. The resulting
set of QSO-like objects selected on this basis constitutes \gcrf{3} -- the third version of
the \gcrf{}. This paper describes the procedure used to select the sources of \gcrf{3}
and presents its most important properties and its relation to the radio ICRF.

According to the principles of the ICRS, \gcrf{3} is entirely defined by the set of ${\sim\,}1.6$~million 
QSO-like sources described in this paper. It is implicitly assumed that the positions and
proper motions of all other (mostly stellar) sources in \gdr{3} are expressed in the same
reference frame, and thus constitute a secondary, much denser reference frame that extends 
to the brightest magnitudes. For stars of similar magnitude and colour as the QSO-like sources, 
the consistency of the reference frames should be guaranteed by the methods of observation
and data analysis in \gaia, which are exactly the same for both kinds of objects. For much
brighter sources, and perhaps also for objects of extreme colours, this may not be the case
owing to various instrumental effects and deficiencies in their modelling. Indeed,
\citetads{2021A&A...649A.124C} have shown that the proper motions of bright 
($G\lesssim 13$~mag) sources in \gedr{3} have a residual spin with respect to the fainter
stars by up to $80\,\muasyr$, and provide a recipe for their correction. Similarly, the 
alignment errors for the bright sources in \gedr{3} were investigated by 
\citetads{2021EGUGA..23.8604L}, who found differences of about 0.5~mas. 
The properties of the secondary (stellar) reference frame in \gedr{3} are not further 
discussed in this paper.

The structure of the paper is as follows. Section~\ref{sec:selection}
describes the multi-step procedure used to select the \gcrf{3}\ sources.
The properties of those sources are discussed in Sect.~\ref{sec:properties}.
The \gcrf{3}\ sources matched to the radio sources of ICRF3 give an important link to
radio astrometry and are discussed in Sect.~\ref{sec:gcrf3}.
Section~\ref{sec:framerotator} reviews the role of the \gcrf{3}\ sources in the 
`frame rotator' of the primary astrometric solution for \gedr{3}, which is where 
the reference frame was actually fixed. 
Some conclusions and prospects for future versions of the \gcrf{}\ are given
in Sect.~\ref{sec:summary}.

The appendices provide more detailed information on a few more
technical issues. Appendix~\ref{sec:external} provides some
additional details of the external catalogues used for \gcrf{3}.
Appendix~\ref{sec:SQL-requests} demonstrates how to select the
\gcrf{3} sources in the \gaia\ Archive and trace which external
catalogues have matches of those sources.  In
Appendix~\ref{sec:confusion} we discuss the `confusion sources' in
\gedr{3}, that is sources, mainly in the Galaxy or Local Group, whose
astrometric parameters are roughly consistent with them being at
cosmological distances, and that could therefore potentially be
confused with QSO-like sources. Finally, Appendix~\ref{sec:estimating}
describes the algorithm used by the frame rotator to estimate the spin
and orientation corrections for astrometric solutions.


\section{Selection of \gaia\ sources}
\label{sec:selection}

The goal of the source selection process for the \gcrf{}\ is to find
a set of QSO-like sources that combines three conflicting requirements:
it should be as large as possible, as clean as possible (that is, free of stellar contaminants), 
and homogeneously distributed on the sky. 
We expect that this selection can eventually be made exclusively from \gaia's own 
observations, using the astrophysical characterisation of sources based on a combination 
of astrometric and photometric information 
(e.g.\ \citeads{2013A&A...559A..74B}; \citeads{2018MNRAS.473.1785D}; 
\citeads{2018A&A...615L...8H}; \citeads{2019MNRAS.490.5615B}).
This was not possible for \gedr{3}, and similarly to what was done for the first two
\gaia\ data releases, \gedr{3}\ uses external catalogues of QSOs and AGNs to identify 
QSO-like sources among the nearly two billion \gaia\ sources. As detailed below, the
\gcrf{3} content is the result of cross-matches with 17 external catalogues, followed 
by a two-stage astrometric filtering. These external catalogues were all publicly available 
at the time of the final construction of \gcrf{3} in mid-June 2020. 

\subsection{Cross-matching}
\label{sect:cross-matching}

For the construction of \gcrf{3} we cross-matched the \gedr{3}
astrometric catalogue with the external catalogues listed in
Table~\ref{tab:gaiacrf3-matches} (see also
Appendix~\ref{sec:external}). Only astrometric information was used
for the cross-matches.
The first four entries in the table --
three ICRF3 catalogues and OCARS -- are based on VLBI astrometry in
the radio domain and have an astrometric quality comparable to that of
\gedr{3}. For those catalogues the maximal cross-match radius was set
to $\Delta_{\max}=100$~mas. The other catalogues have various levels
of astrometric quality and for them the cross-match radius was set to
$\Delta_{\max}=2$~arcsec. For virtually all catalogues the
cross-matching results in a (relatively small) number of ambiguous
matches, where more than one \gaia\ source was matched to a source in
the external catalogue, or more than one source in the external
catalogue was matched to the same \gaia\ source. In all such cases
only the closest positional match was retained.
For each external
catalogue this gives a list of `unique' matches in the form of pairs
`\gaia\ source identifier' -- `external source identifier', in which
all identifiers appear only once. Table~\ref{tab:gaiacrf3-matches}
gives the total number of sources in the catalogue and, in column~3,
the number of unique matches in \gedr{3}, to which the
two-stage filtering described below was applied.

This cross-matching procedure adopted for \gcrf{3} is not perfect and 
could be refined in future releases. For example, the maximum cross-matching
radius could be better tuned to the precisions of the individual external catalogues
and the expected densities of stars and quasars in the different situations.
A particular case concerns multiple lensed quasar images, where ideally all the 
images should be retained instead of just the closest match, as is presently done. 
In some ambiguous cases the cross-matching could benefit from using also the 
photometric information available in the external catalogues. This is however 
far from straightforward owing to the expected variability of the QSOs, differences 
in wavelength bands and resolution between the instruments, etc., which may
even require a manual check of individual cases. For the current cross-matching we 
have chosen not to use such information.

\subsection{Astrometric filtering: Individual sources}
\label{sect:filtering-individual}

The first stage of the astrometric filtering ensures that the retained \gaia\ sources
have five- and six-parameter astrometric solutions in \gedr{3} compatible with the assumed
extragalactic nature of the source. For all the catalogues in Table~\ref{tab:gaiacrf3-matches}
we applied the three criteria in Eqs.~(\ref{eq:filterA})--(\ref{eq:X2}) below. This was
found to be sufficient for the VLBI catalogues (the first four entries in the table), but for the
remaining catalogues the two criteria in Eqs.~(\ref{eq:filterB}) and (\ref{eq:filterRho})
were also applied. Some of the criteria are alternatively formulated as parts of a query
in the \gaia\ Archive%
\footnote{\href{https://gea.esac.esa.int/archive/}{https://gea.esac.esa.int/archive/}}
using the Astronomical Data Query Language (ADQL).

The first criterion is that the source has a valid parallax and proper motion. This is
ensured by selecting sources with five- or six-parameter astrometric solutions:
\begin{equation}\label{eq:filterA}
\begin{minipage}{0.8\hsize}
{\small
\begin{verbatim}  
astrometric_params_solved = 31 OR 
astrometric_params_solved = 95 
\end{verbatim}
}
\end{minipage}
\end{equation}
We note that {\small\texttt{astrometric\_params\_solved}} is a binary flag indicating which astrometric
parameters are provided for the source, that is $11_2$, $11111_2$, and $1011111_2$ 
for two-, five-, and six-parameter solutions, with the seventh bit representing pseudo-colour.

Secondly, the measured parallax $\varpi$ of the source, corrected for the median parallax bias,
should be zero within five times the formal uncertainty $\sigma_\varpi$:
\begin{equation}
    \label{eq:Xvarpi}
    \left|\frac{\varpi+0.017\,{\rm mas}}{\sigma_{\varpi}}\right|<5\,,
\end{equation}
which translates to the ADQL query
{\small
\begin{verbatim}  
abs((parallax + 0.017)/parallax_error) < 5 
\end{verbatim}
}\noindent
This takes into account the median parallax of $-0.017$~mas obtained from a provisional 
selection of QSO-like sources in \gedr{3}, but ignores the complicated dependence of the 
parallax bias on the magnitude and other parameters discussed by \citetads{2021A&A...649A...4L}.
As shown by their Fig.~5, the variation of the parallax bias for the QSO-like sources 
is typically less than $\pm 0.010$~mas. Because this is many times smaller than the individual
uncertainties (for example, 99\% of the \gcrf{3} sources have $\sigma_\varpi>0.1$~mas), 
these variations were deemed uncritical for the astrometric filtering. Assuming Gaussian errors, 
$\left(\varpi+0.017\,{\rm mas}\right)/\sigma_{\varpi}$ is then expected to follow the 
standard normal distribution, and the probability that Eq.~(\ref{eq:Xvarpi}) is violated 
by random errors is, nominally, $\erfc\left(5/\!\sqrt{2}\,\right)\simeq 5.7\times 10^{-7}$.

The third criterion is that the proper motion components $\mu_{\alpha*}$ and
$\mu_{\delta}$ should be small compared to their uncertainties $\sigma_{\mu\alpha*}$ and
$\sigma_{\mu\delta}$. Here we take into account the correlation
$\rho(\mu_{\alpha*},\mu_\delta)$ between the two parameters. We require
\begin{equation}\label{eq:X2}
X_\mu^2\equiv\begin{bmatrix} ~\mu_{\alpha*} && \mu_{\delta}~ \end{bmatrix}\,
\text{Cov}(\vec{\mu})^{-1}
  \begin{bmatrix} ~\mu_{\alpha*} \\[6pt] 
  \mu_{\delta} \end{bmatrix}\,<25\,,
\end{equation}
where $\text{Cov}(\vec{\mu})$ is the covariance matrix given by Eq.~(\ref{eq:covPm}).
This corresponds to the query
{\small
\begin{verbatim}  
( power(pmra/pmra_error,2)
  +power(pmdec/pmdec_error,2)
  -2*pmra_pmdec_corr
    *pmra/pmra_error
    *pmdec/pmdec_error )
/(1-power(pmra_pmdec_corr,2)) < 25
\end{verbatim}
}\noindent
For Gaussian errors of the proper motion components, $X_\mu^2$ is expected to follow the
chi-squared distribution with two degrees of freedom (see Appendix~B
of \citeads{2016A&A...595A...5M}). In this case the nominal
probability that a source violates Eq.~(\ref{eq:X2}) is 
$\exp\left(-25/2\right)\simeq 3.7\times10^{-6}$.
We neglect here the systematic proper motions of extragalactic objects caused by the
acceleration of the solar system barycentre (e.g.\ \citeads{2021A&A...649A...9G}).
The amplitude of this effect, about 0.005~mas~yr$^{-1}$, is much smaller than the 
typical proper-motion uncertainty of QSO-like sources in \gedr{3} (see below) and 
is therefore ignored in the filtering. We note that $X_\mu$ is independent
of the reference frame, so that, in particular, it plays no role if the equatorial
or, say, galactic components of the proper motions are used in Eq.~(\ref{eq:X2}).

For the VLBI catalogues based on high-accuracy radio astrometry (the
first four entries in Table~\ref{tab:gaiacrf3-matches}), the application of
Eqs.~(\ref{eq:filterA})--(\ref{eq:X2}) results in filtered samples that show no 
traces of stellar contamination. This is to be expected thanks to the high 
positional accuracy of the VLBI data and the small cross-match radius used 
(100~mas). Consequently, no additional filtering was considered for these 
catalogues. 

The situation is very different for the remaining (non-VLBI) catalogues in 
Table~\ref{tab:gaiacrf3-matches}. An analysis of their matches,
filtered by Eqs.~(\ref{eq:filterA})--(\ref{eq:X2}), revealed that a significant 
number of chance matches to (stellar) sources in \gedr{3} happen through
a combination of two factors: (i) the high density of potentially matching stellar 
sources in certain areas of the sky; and (ii) the lower positional accuracy of these
catalogues which motivated the use of a much larger cross-matching radius 
(2~arcsec) than for the VLBI catalogues.

To investigate the likely effects of chance matches we make use of the  
`confusion sources' discussed in Appendix~\ref{sec:confusion}. These are 
all the sources in \gedr{3} that satisfy all three conditions in 
Eqs.~(\ref{eq:filterA})--(\ref{eq:X2}), irrespective of their true nature. 
The total number of confusion sources in \gedr{3} is approximately 213.5~million,
of which 30.7~million have five-parameter solutions and 182.8~million have
six-parameter solutions. 

The galactic plane is especially problematic for the cross-matching: 23\%
of the confusion sources with five-parameter astrometric solutions and
50\% of the confusion sources with six-parameter solutions are
located in the 10\% of the celestial sphere with the galactic
latitude $b$ such that $\left|\,\sin b\,\right|<0.1$. The analysis of the
matches from non-VLBI catalogues showed that the resulting samples of
QSO candidates in that area are too much contaminated by stars and
cannot reasonably be further cleaned by means of astrometric criteria alone. 
We therefore decided to completely avoid this area of the sky for \gcrf{3}
for the matches to the non-VLBI catalogues.

Considering the distribution of the \gcrf{3} sources and the confusion sources
in galactic latitude shown on Figs.~\ref{fig:gcrf3-galactic-coordinates-hist}
and \ref{fig:histograms-confusion}, respectively,
our results for \gedr{3} generally agree with the discussion of the stellar contamination
at various galactic latitudes provided by \citetads{2020A&A...644A..17H} (see their Fig.~9).


%
%
%
%
%
%
%
%
%
%
%

The second factor important for the chance matches -- the lower quality
of the astrometric data in the non-VLBI catalogues -- is also more
problematic in the areas of higher density of confusion sources.
Fig.~\ref{fig:histograms-confusion} shows
that the density of confusion sources is a strong function of the Galactic latitude $b$.
To avoid too many chance matches at low $|\,b\,|$ we require a
smaller matching distance $\Delta$ towards the Galactic plane.
For the matches to the non-VLBI catalogues we therefore add the two criteria
\begin{equation}\label{eq:filterB}
\left|\,\sin b\,\right|>0.1\,
\end{equation}
and
\begin{equation}\label{eq:filterRho}
\Delta<(2~\text{arcsec})\times\,\left|\,\sin b\,\right|\,.
\end{equation}
We note that, after the initial cross-matching, this last criterion is the only one 
that uses astrometric data (positions) from the external catalogue. 
The number of unique matches that survived the filtering by 
Eqs.~(\ref{eq:filterA})--(\ref{eq:filterRho}), or Eqs.~(\ref{eq:filterA})--(\ref{eq:X2}) for
the VLBI catalogues, is given in the fourth column of Table~\ref{tab:gaiacrf3-matches}.

\begin{table*}
  \caption{Statistics of the cross-matches of \gedr{3} to external QSO catalogues used in the
  construction of \gcrf{3}.
\label{tab:gaiacrf3-matches}}
\footnotesize\setlength{\tabcolsep}{6pt}
\begin{center}
\begin{tabular}{lrrrrll}
\hline\hline
\noalign{\smallskip}
catalogue & \multicolumn{1}{r}{sources} & \multicolumn{1}{r}{unique matches} & \multicolumn{1}{r}{filtered sources}
& \multicolumn{1}{r}{retained sources}
&\multicolumn{1}{l}{code} & \multicolumn{1}{l}{reference} \\
\noalign{\smallskip}\hline\noalign{\smallskip}
ICRF3 S/X & 4536 & 3477 & 3142 & 3142 & icrf3sx & \hbox{\citetads{2020A&A...644A.159C}} \\
ICRF3 K & 824 & 715 & 660 & 660 & icrf3k & \hbox{\citetads{2020A&A...644A.159C}} \\
ICRF3 X/Ka & 678 & 611 & 576 & 576 & icrf3xka & \hbox{\citetads{2020A&A...644A.159C}} \\
OCARS & 7607 & 5337 & 4723 & 4723 & ocars & \hbox{\citetads{2018ApJS..239...20M}} \\
AllWISE & 1\,354\,775 & 733\,462 & 580\,403 & 580\,403 & aw15 & \hbox{\citetads{2015ApJS..221...12S}} \\
Milliquas v6.5 & 1\,980\,903 & 1\,347\,414 & 1\,065\,936\rlap{*} & 1\,039\,610 & m65 & \hbox{\citetads{2019arXiv191205614F}} \\
R90 & 4\,543\,530 & 1\,331\,547 & 1\,022\,081 & 1\,022\,081 & r90 & \hbox{\citetads{2018ApJS..234...23A}} \\
C75 & 20\,907\,127 & 2\,068\,813 & 1\,265\,419\rlap{*} & 1\,169\,431 & c75 & \hbox{\citetads{2018ApJS..234...23A}} \\
SDSS DR14Q & 526\,356 & 368\,013 & 308\,608 & 308\,608 & dr14q & \hbox{\citetads{2018A&A...613A..51P}} \\
LQAC-5 & 592\,809 & 421\,289 & 348\,085 & 348\,085 & lqac5 & \hbox{\citetads{2019A&A...624A.145S}} \\
LAMOST phase 1, DR1-5 & 42\,578 & 42\,255 & 39\,886 & 39\,886 & lamost5 & \hbox{\citetads{2019ApJS..240....6Y}} \\
LQRF & 100\,165 & 98\,902 & 94\,839 & 94\,839 & lqrf & \hbox{\citetads{2009A&A...505..385A}} \\
2QZ & 28\,495 & 24\,192 & 21\,569 & 21\,569 & cat2qz & \hbox{\citetads{2004MNRAS.349.1397C}} \\
Roma-BZCAT, release 5 & 3561 & 3343 & 2986 & 2986 & bzcat5 & \hbox{\citetads{2015Ap&SS.357...75M}} \\
2WHSP & 1691 & 982 & 413 & 413 & cat2whsp & \hbox{\citetads{2017A&A...598A..17C}} \\
ALMA calibrators & 3361 & 2746 & 2209 & 2209 & alma19 & \hbox{\citetads{2019MNRAS.485.1188B}} \\
Gaia-unWISE C75 & 2\,734\,464 & 2\,726\,201 & 2\,061\,253\rlap{*} & 1\,569\,359 & guw & \hbox{\citetads{2019MNRAS.489.4741S}} \\
\noalign{\smallskip}
\hline
\noalign{\smallskip}
\multicolumn{4}{l}{total retained matches} & 1\,614\,173 & & \\ 
\noalign{\smallskip}
\hline
\end{tabular}
\tablefoot{The external QSO catalogues used to construct \gcrf{3}\ are identified in columns 1 and 7.
Column~2 is the total number of sources in the catalogue. 
Column~3 gives the number of unique matches in \gedr{3}\ to the catalogue, using a matching 
radius of $\Delta_{\rm max}=100$~mas for the four VLBI-based catalogues (ICRF3 and OCARS) 
and $\Delta_{\rm max}=2$~arcsec for the other catalogues. 
Column~4 contains the number of matched sources satisfying Eqs.~(\ref{eq:filterA})--(\ref{eq:filterRho}) 
(where the last two criteria were not used for the VLBI-based catalogues). An asterisk means that
the filtered sources show evidence of stellar contamination.
Column~5 gives the number of sources retained in \gcrf{3} after the second stage of the filtering
described in Sect.~\ref{sect:filtering-sample}.
Column~6 gives the code for the catalogue in the cross-match table \texttt{gaiadr3.gaia\_crf3\_xm}
(see Appendix~\ref{sec:SQL-requests}).
}
\end{center}
\end{table*}

\subsection{Astrometric filtering: Sample statistics}
\label{sect:filtering-sample}

The second stage of the filtering is based on an assessment of the 
level of stellar contamination among the filtered unique matches in the
different external catalogues. To this end we analysed histograms of 
the normalised parallaxes $\varpi/\sigma_{\varpi}$ and normalised 
proper motion components $\mu_{\alpha*}/\sigma_{\mu_{\alpha*}}$ 
and $\mu_{\delta}/\sigma_{\mu_{\delta}}$. Ignoring the small systematic
and random offsets discussed in Sect.~\ref{sect:filtering-individual},
we expect that, for a clean sample of QSOs, the normalised quantities 
should have Gaussian distributions with mean values close to zero
and standard deviations only slightly greater than one (reflecting the slight 
underestimation of the formal uncertainties; see Sect.~\ref{sec:properties}). 
Conversely, if their distributions deviate significantly from the expected 
Gaussian, it indicates that the sample is contaminated by stellar objects 
with non-zero parallaxes and/or proper motions.

Although the criteria in Eqs.~(\ref{eq:filterA})--(\ref{eq:X2}) are based
on the Gaussian hypothesis, they absolutely do not guarantee that the 
normalised parallaxes and proper motions are Gaussian after the filtering. 
Figures~\ref{fig:histograms-normalized-Galaxy} and 
\ref{fig:histograms-normalized-LMCSMC} show examples of the
distributions of the normalised parameters for general \gedr{3} sources
satisfying the criteria. While a fraction of the sources in these samples 
are QSOs (including \gcrf{3} sources), most of them are stars in our Galaxy 
or in nearby satellite galaxies such as the Large Magellanic Cloud (LMC) and 
Small Magellanic Cloud (SMC), and the histograms 
of the normalised quantities are very far from Gaussian. 

A  high level of stellar contamination in a sample can be directly seen 
in the histograms of the normalised parameters as a distortion of the
expected Gaussian distribution. 
Any sub-sample large enough to create representative histograms
can be checked
separately. This gives a flexible tool for screening samples of QSO
candidates against stellar contamination.

\begin{figure}[t]
  \begin{center}
  \includegraphics[width=1\hsize]{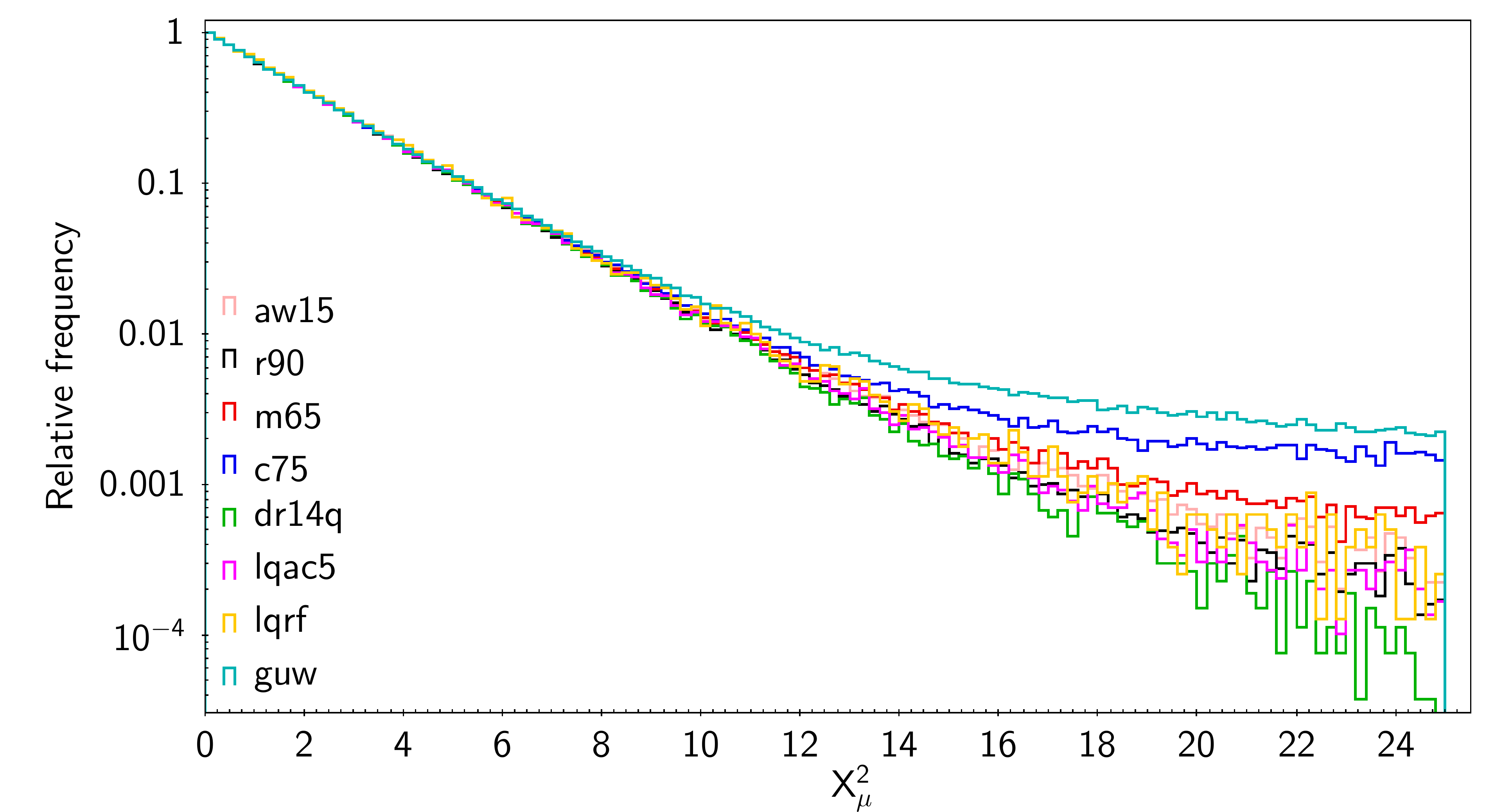}
    
  \includegraphics[width=1\hsize]{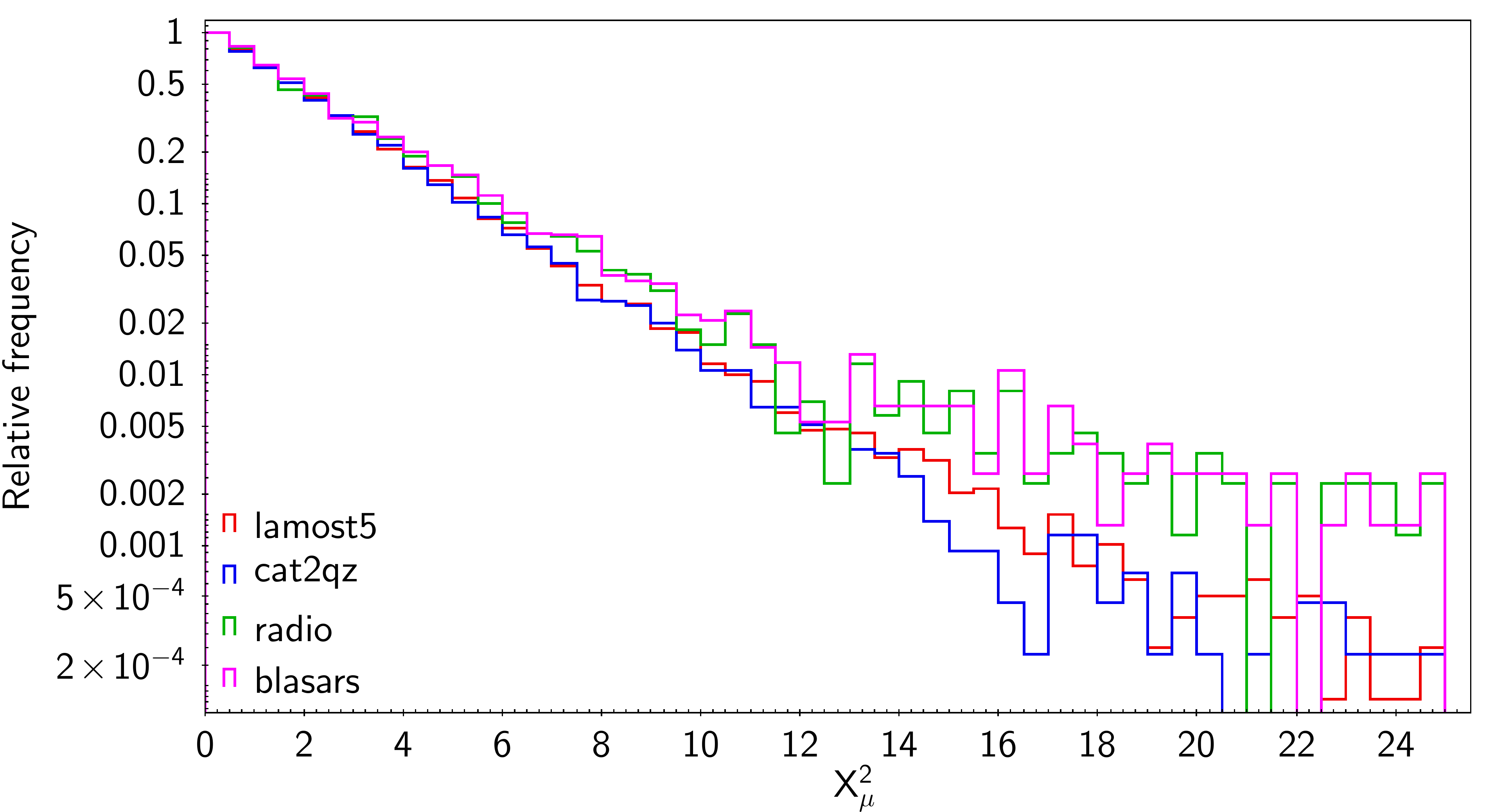}
  \caption{Relative frequencies of $X_\mu^2$ for the
    filtered matches to the catalogues in
    Table~\ref{fig:X2-histograms}. The catalogues are identified by
    the codes given in column~6 of the table, with the eight larger catalogues
    in the upper panel and the smaller catalogues in the lower panel. 
    The four radio catalogues (ICRF3 S/X, ICRF3 K, ICRF X/Ka, and OCARS) 
    are combined under `radio', and the three blasar catalogues (Roma-BZCAT,
    2WHSP, and ALMA calibrators) are combined under `blasars'.}
  \label{fig:X2-histograms}
\end{center}
\end{figure}

For the proper motions the same effect is illustrated in Fig.~\ref{fig:X2-histograms}.
This shows the relative frequencies of $X_\mu^2$ from Eq.~(\ref{eq:X2}) 
for the filtered matches to the catalogues in 
Table~\ref{fig:X2-histograms}. If $X_\mu^2$ comes purely from the Gaussian errors of
the \gaia\ proper motions, it should follow the $\chi^2$ distribution with two degrees 
of freedom, with normalised frequency $\exp(-\frac{1}{2}x/s^2)$. 
Here $s$ is a scaling factor ($\simeq 1$) that takes into account that the uncertainties
of the astrometric parameters in \gedr{3} are slightly underestimated.
In Fig.~\ref{fig:X2-histograms} this expected frequency is a straight
line through unity at $X_\mu^2=0$. 

Applying this tool to the full sets of filtered matches to each external 
catalogue in Table~\ref{tab:gaiacrf3-matches} reveals
that only the sources matched to three of the catalogues 
(C75, 
Milliquas~v6.5, 
and Gaia-unWISE) 
show clear evidence of stellar
contamination. These catalogues are marked with an asterisk in the table.
The filtered matches of the remaining 14 catalogues show no obvious traces 
of stellar contamination and therefore qualify as being part of \gcrf{3}. 

The same \gaia\ source usually appears in the filtered cross-matches to
several external catalogues. One reason for this, as discussed in 
Appendix~\ref{sec:external}, is that many of the catalogues are at least 
partially based on the same observational material (the IR photometry 
of WISE, certain data releases of SDSS, etc.), although their criteria 
for the selection of QSOs may be different.  But even when the
observational material is different, it is no wonder that different
approaches and algorithms result in strongly overlapping samples,
namely if they tend to find the same objects.
It is therefore reasonable to expect that the circumstance that a \gaia\ source
has matches in several external catalogues increases the reliability
of its classification as a QSO. If we split the set of
\gaia\ sources matching a given external catalogue (after the
astrometric filtering) into subsets that are present also in other
catalogues and a subset that appears only in the cross-match with this
particular catalogue, we can therefore use an analysis of these subsets to
improve the purity of the final QSO selection.

\begin{figure}[t]
  \begin{center}
  \includegraphics[width=1\hsize]{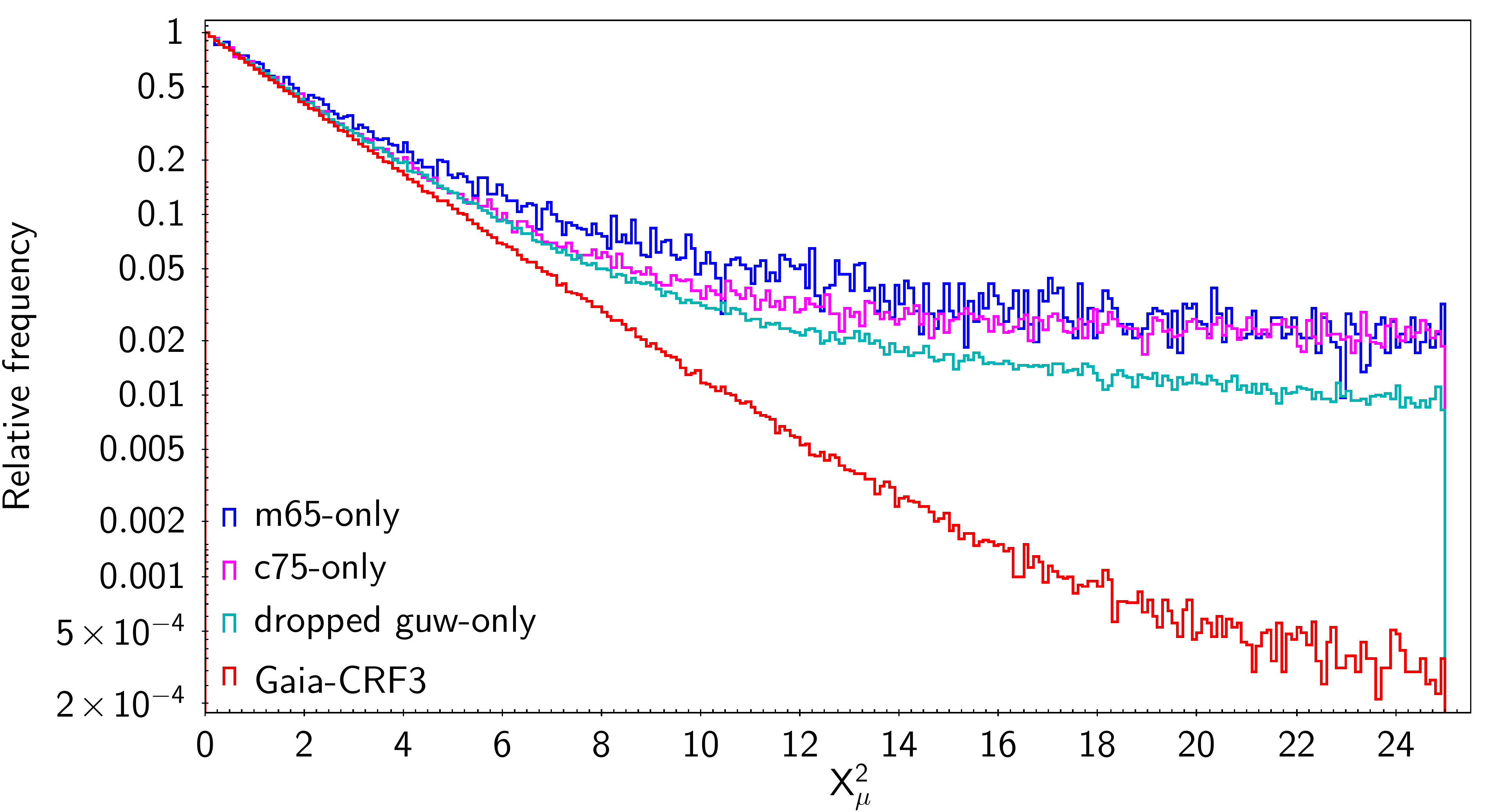}  
  \caption{Relative frequencies of $X_\mu^2$ for the final selection of \gcrf{3} sources,
  and for the filtered matches to C75, Milliquas~v6.5, and Gaia-unWISE that were 
  dropped in the final selection.
  }
\label{fig:X2-GaiaCRF3-dropped-histograms}
\end{center}
\end{figure}

Using this idea, we selected three sets of \gaia\ sources that appear
only among the filtered matches of C75, Milliquas~v6.5, and
Gaia-unWISE, but do not appear in any of the other catalogues: we call
these the c75-only, m65-only, and guw-only sets.  All three sets show
significant stellar contamination. We found no obvious way to clean
the c75-only sources and this whole set of 95\,988 sources was
therefore dropped from further consideration. The Milliquas~v6.5 and
Gaia-unWISE catalogues include estimates of the probability that a
source is a QSO.  Selecting from the m65-only set only the sources
with the highest probability of being a QSO did not remove the
contamination and that set of 26\,326 sources was therefore also
dropped. For the guw-only set we found, on the contrary, that the
subset of sources with the highest QSO probability
($\texttt{PROB\_RF}=1$) appears to be free of stellar
contaminants. This gave an additional set of 229\,914 sources for
\gcrf{3}. The other 491\,894 guw-only sources were dropped (any relaxation
of $\texttt{PROB\_RF}=1$ immediately gave samples with clear signs of stellar contamination).
The
number of sources from the cross-matches of each catalogue that passed
also this second stage of astrometric filtering is given in the fifth
column of Table~\ref{tab:gaiacrf3-matches}, summing up to a total of 1\,614\,173 sources.
The relative frequencies of $X_\mu^2$ in \gcrf{3} and in the three sets of
sources dropped from C75, Milliquas~v6.5, and Gaia-unWISE are shown in
Fig.~\ref{fig:X2-GaiaCRF3-dropped-histograms}. Deviations from the
exponential distribution expected for Gaussian errors are very obvious
for the dropped sets; for the final selection the distribution is briefly discussed 
below (Sect.~\ref{sect:final_selection}).  

\begin{figure*}[t]
  \begin{center}
  \includegraphics[width=\scalethree\hsize]{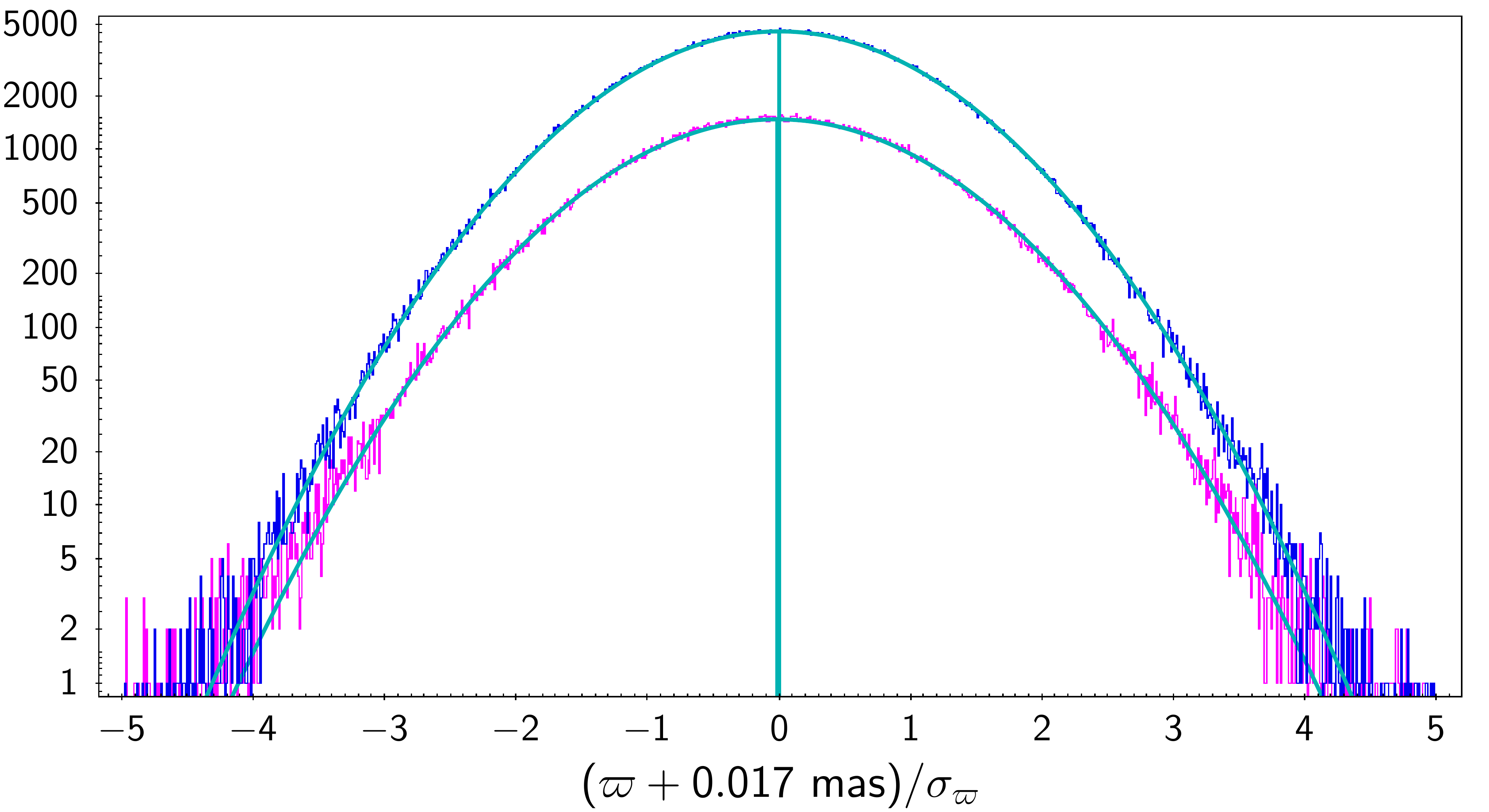}  
  \includegraphics[width=\scalethree\hsize]{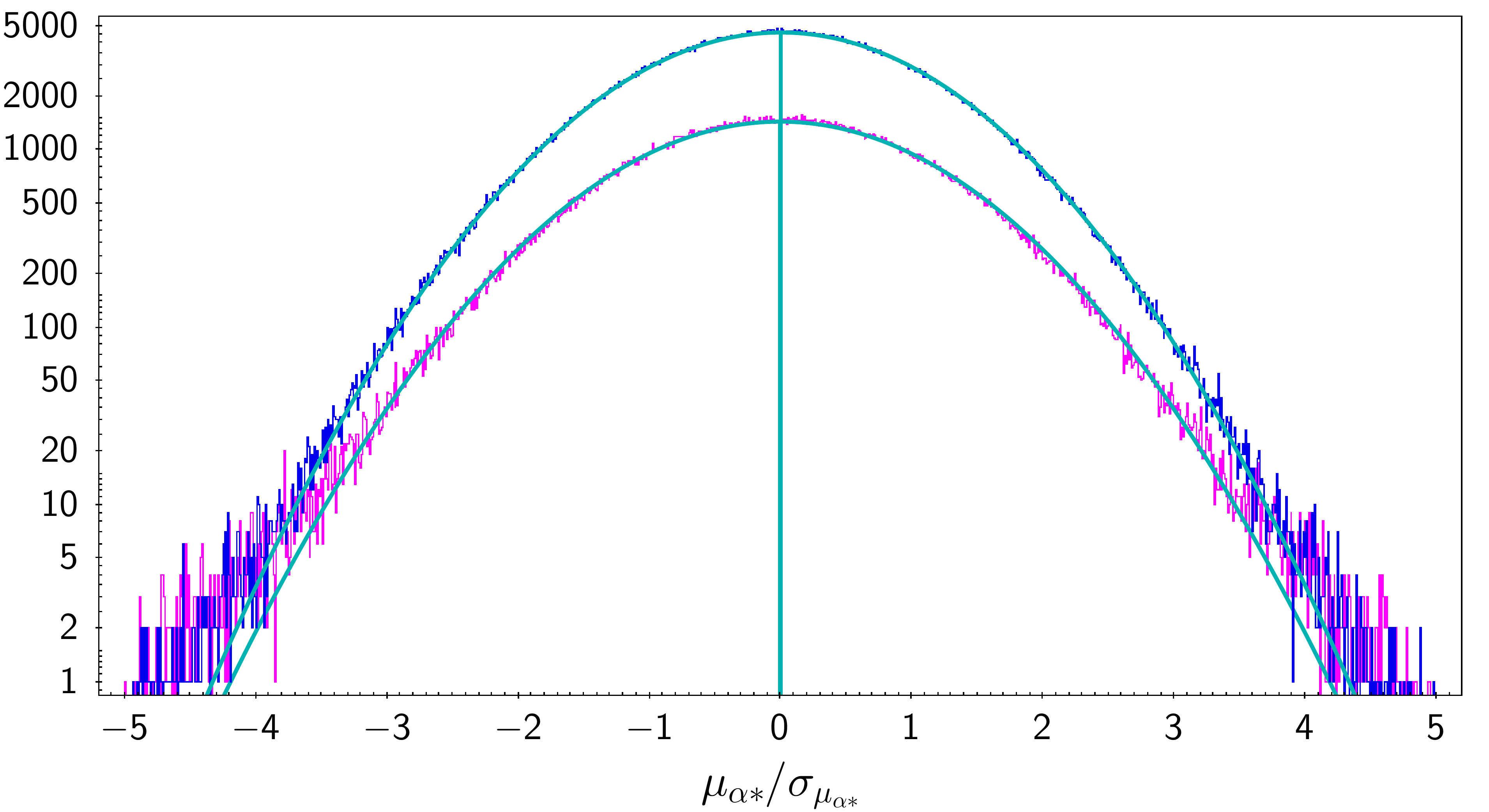}
  \includegraphics[width=\scalethree\hsize]{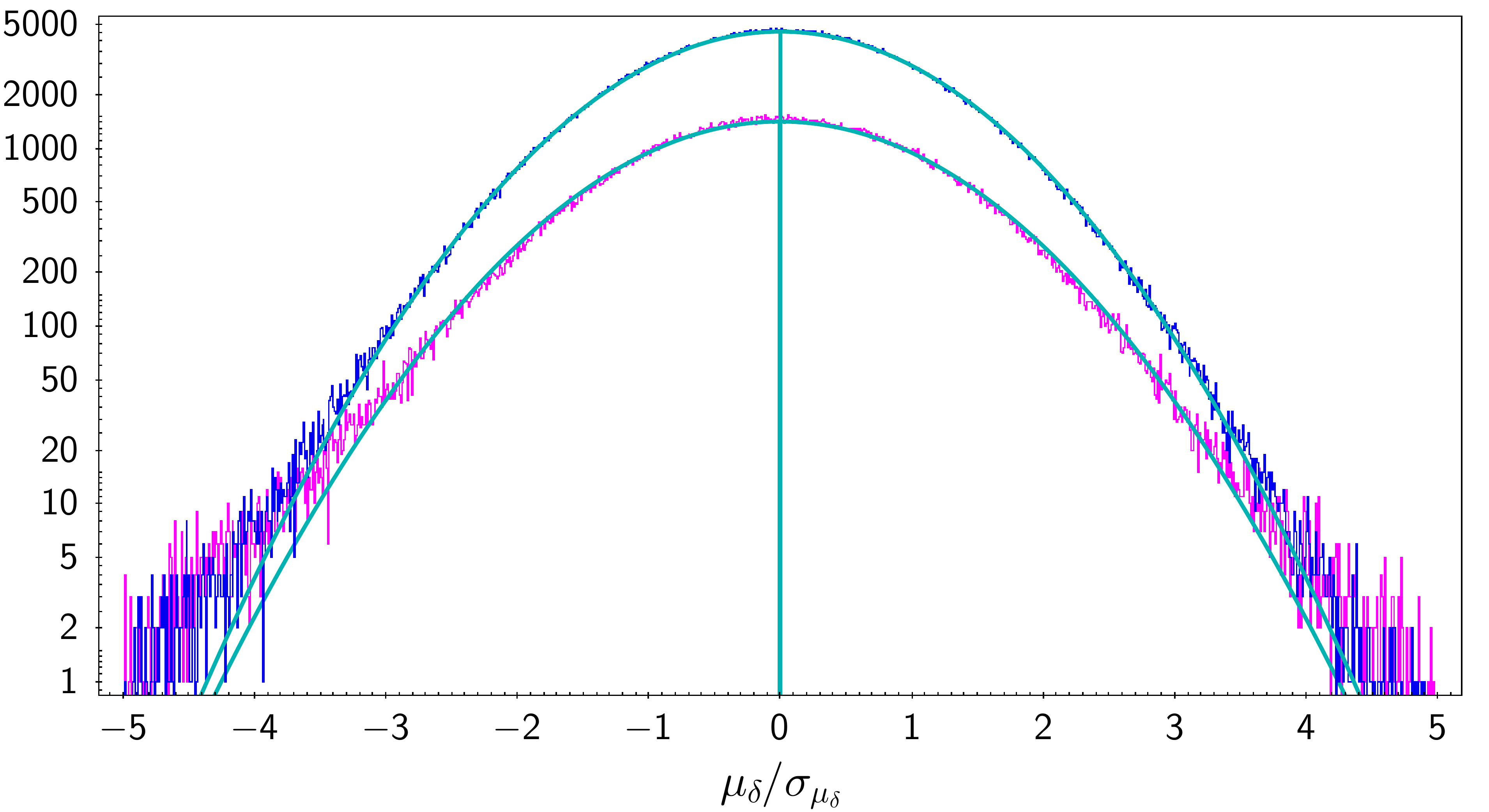}
  \caption{Distributions of the normalised parallaxes and proper motion components 
    for the \gcrf{3} sources with five-parameter (blue) and six-parameter (magenta) solutions. 
    The turquoise curves show the corresponding best-fit Gaussian distributions. 
  }
\label{fig:pm-histograms}
\end{center}
\end{figure*}

A separate investigation was done for the QSO candidates in the areas
of the LMC and SMC, which have a very high density of confusion sources
(Fig.~\ref{fig:sky-distribution-confusionSources}) and therefore a higher
risk of chance matches. The statistics of confusion sources in these areas 
are discussed in Appendix~\ref{sec:confusion}. The \gcrf{3} contains 
5411 and 2779 sources, respectively, in the LMC and SMC areas (as defined
in the appendix). The distributions of the normalised quantities indicate a
low level of stellar contamination in these sets.

Further checks of the selection were done by splitting the \gcrf{3}\ sources
into subsets according to magnitude, colour, and Galactic or ecliptic latitude,
and examining the distributions of normalised parallaxes and proper motions.
No clear contamination was detected in any of the subsets.

\gcrf{3} contains only 195 sources with $\left|\,\sin b\,\right|<0.1$.
They all come from the VLBI-based catalogues, for which Eqs.~(\ref{eq:filterB}) 
and (\ref{eq:filterRho}) were not applied. We made an attempt to find a clean set 
of QSO-like sources also in the Galactic plane area, that is without applying the criterion 
in Eq.~(\ref{eq:filterB}). Using the cross-matches and the remaining four criteria, 
we found nearly 100\,000 additional QSO candidates with $\left|\,\sin b\,\right|<0.1$. 
However, the distributions of their normalised parameters indicate a high level of stellar 
contamination. This demonstrates that the cross-matches and/or external astrophysical 
classifications are unreliable in this crowded part of the sky, and none of these 
sources were further considered for \gcrf{3}. However, we note that this set
should contain a significant percentage of genuine QSOs, part of which may be 
recovered in future versions of the \gcrf{}\ (see Sect.~\ref{sect:selection_discussion}).

\subsection{Final selection}
\label{sect:final_selection}

The union of all \gaia\ sources that passed both stages of the filtering contains
1\,614\,173 sources and constitutes the final source list of
\gcrf{3}. Of these, 1\,215\,942 (75\%) have five-parameter solutions in \gedr{3} and
398\,231 (25\%) have six-parameter solution. 
Appendix~\ref{sec:SQL-requests} explains how to find the 
astrometric parameters of these sources and how to trace the
matches of these sources in the external catalogues.

Figure~\ref{fig:pm-histograms} shows the distributions of the normalised parallaxes and
proper motion components for both kinds of solutions. The standard deviations of the three
normalised quantities are 1.052, 1.055, and 1.063 for the five-parameter solutions and
1.073, 1.099, and 1.116 for the six-parameter solutions. All distributions are very symmetric
and close to the Gaussian shape (i.e.\ parabolic in the logarithmic plots), suggesting a low level
of stellar contamination. The small excess of sources with normalised parameters beyond $\pm 4$
is not necessarily a sign of contamination, but could be produced by source structure or 
instrumental effects as well as by statistical fluctuations in the sample.

The number of (non-QSO) chance matches in \gcrf{3} can be estimated by
counting the confusion sources that satisfy Eqs.~(\ref{eq:filterB})
and (\ref{eq:filterRho}) around positions that are slightly offset
from the \gcrf{3} positions. Using offsets of $\pm 0.1$~deg in
$\Delta\alpha\cos\delta$ or $\Delta\delta$ we conclude that \gcrf{3} is 
expected to contain up to 420 chance matches, of which 150 have
five-parameter solutions and 270 have six-parameter solutions. This
corresponds to a probable contamination of 0.012\% and 0.068\%,
respectively, of the five- and six-parameter sources in \gcrf{3}.  

The chance matches are obviously not the only cause of stellar
contamination in \gcrf{3}.  Further contaminants may come from erroneous
classification of some sources in the external catalogues, even among
the sources that passed the astrometric filtering in
Eqs.~(\ref{eq:filterA})--(\ref{eq:filterRho}). We have found no means to
quantify this effect, but expect it to be quite small.

A certain hint on the possible (total) level of contamination could be given by 
the relative frequency of $X_\mu^2$ for \gcrf{3}, as shown by the red curve in
Fig.~\ref{fig:X2-GaiaCRF3-dropped-histograms}. This distribution can be 
rather well described as the mixture of two Gaussian distributions: one with 
standard deviation of 1.051 containing 98.0\% of the sources, the other with 
standard deviation of 2.038 containing 2.0\% of the sources. We stress that this 
cannot directly be interpreted in terms of the purity of the \gcrf{3}: on one hand,
genuine QSOs may show deviations from Gaussianity for other reasons (as mentioned
above); on the other hand, some contaminants will have measured proper motion
that are perfectly consistent with the extragalactic hypothesis. Nevertheless,
we consider it unlikely that the level of contamination in \gcrf{3} is higher than 2\%.  
Further investigations are needed to show if this claim is justified.

\subsection{Discussion of the selection}
\label{sect:selection_discussion}


Assuming that QSO-like sources have zero true values for the parallaxes and 
proper motions, and that the astrometric errors are Gaussian with uncertainties
as given in \gedr{3}, we expect that very few actual QSOs in a sample of a few 
million candidates will violate Eqs.~(\ref{eq:Xvarpi}) or (\ref{eq:X2}). The numerical 
boundaries in Eqs.~(\ref{eq:Xvarpi}) and (\ref{eq:X2}) were
selected empirically, based on the data themselves, and represent a
compromise taking into account possible non-Gaussian systematic errors
and effects of complex and sometimes time-dependent source structure.
The latter are to be expected for QSOs and could produce apparent
proper motions over several years of observations, while parallaxes are 
much less affected by such effects. This justifies setting separate criteria for 
parallaxes and proper motions as described above. In future \gaia\ data releases, 
when these effects may be better known and statistically quantified, it will be 
possible to adopt a mathematically more consistent filtering approach based on the 
$3\times3$ covariance matrix of the parallax and proper motion components.

As already mentioned, we expect that future versions of \gcrf{} will to a higher degree
depend on \gaia's own (astrometric, photometric, and spectroscopic) observations,
and that external catalogues of AGNs will consequently become less important for the 
selection of QSO-like sources. These catalogues will continue to be important for 
validation purposes, but their uses in the selection process may become more complex 
if the astrometry in external catalogues is already \gaia-based. For example, a criterion
such as Eq.~(\ref{eq:filterRho}) may not be meaningful if the position in the external
catalogue comes from a previous \gaia\ data release.

The search for the QSOs in the Galactic plane is an important scientific problem that 
requires further investigation. We note that more advanced methods to combine different
datasets in the search for QSOs in the Galactic plane are under development
(e.g., \citeads{2021ApJS..254....6F}). Combined with the expected use of \gaia's own 
object classification, one can hope that these methods will result in a much improved 
completeness of the \gcrf{}\ in the Galactic plane in the future.

The selection of QSO-like objects described here is mainly
driven by the reliability of the resulting set of sources. The
resulting set of QSO-like sources can be considered `reliable', but
by no means `complete'. It is clear that the selection algorithm
rejects some genuine quasars, for example if they show observable mean proper
motions for the time span of the \gedr{3}\ data. A prominent example
here is 3C273 -- the brightest quasar on the sky, which has
significant proper motion in declination in \gedr{3} and  $X_\mu^2=29.7$ (Eq.~\ref{eq:X2}).

For the purposes of aligning the reference frame of \gedr{3} with the ICRF, and
especially for ensuring that it has negligible spin with respect to distant galaxies, 
one can use a much smaller set of QSO-like sources than the full \gcrf{3} without 
significantly increasing the statistical uncertainties of the global system. Indeed, this
system was defined in the final stages of the primary astrometric solution, as
described in Sect.~\ref{sec:framerotator}, using only about 0.4~million `frame rotation' 
sources. However, the purpose of \gcrf{3} is not just to fix the system for the \gaia\ 
astrometric solution, but to provide the community with a direct access to this system
with sources as faint as permitted by the sensitivity of \gaia, while still having a very 
low level of stellar contamination.

\section{Statistical properties of the \gcrf{3} sources}
\label{sec:properties}


In this section we present the overall properties of \gcrf{3} primarily in terms of its 
distributions in position, magnitude, colour, and astrometric quality.

\subsection{Distribution of the sources on the sky}
\label{sect:distribution}

\begin{figure}[t]
\begin{center}
  \includegraphics[width=1\hsize]{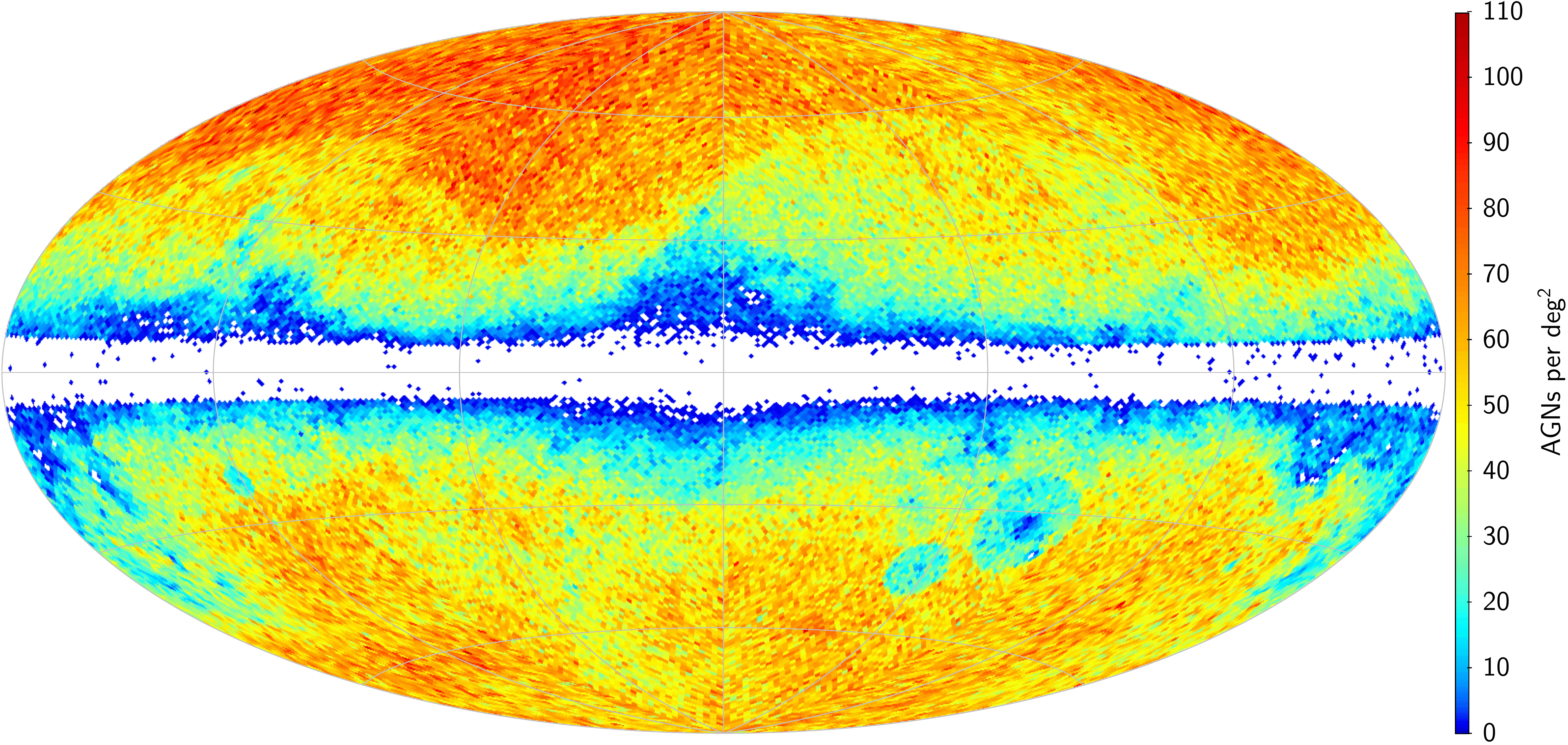}
  \caption{Sky distribution of the \gcrf{3} sources. The plot shows the density of sources per
    square degree computed from the source counts per pixel using
    HEALPix of level 6 (pixel size $\simeq 0.84$\,deg$^2$). The magnitude distribution and further properties
    of the sources are shown on Fig.~\ref{fig:histograms}. This
    full-sky map uses a Hammer–Aitoff projection in galactic
    coordinates, with $l = b = 0$ at the centre, north up, and $l$
    increasing from right to left.}
\label{fig:sky-distribution-gcrf3}
\end{center}
\end{figure}

\begin{figure}[t]
\begin{center}
  \includegraphics[width=1\hsize]{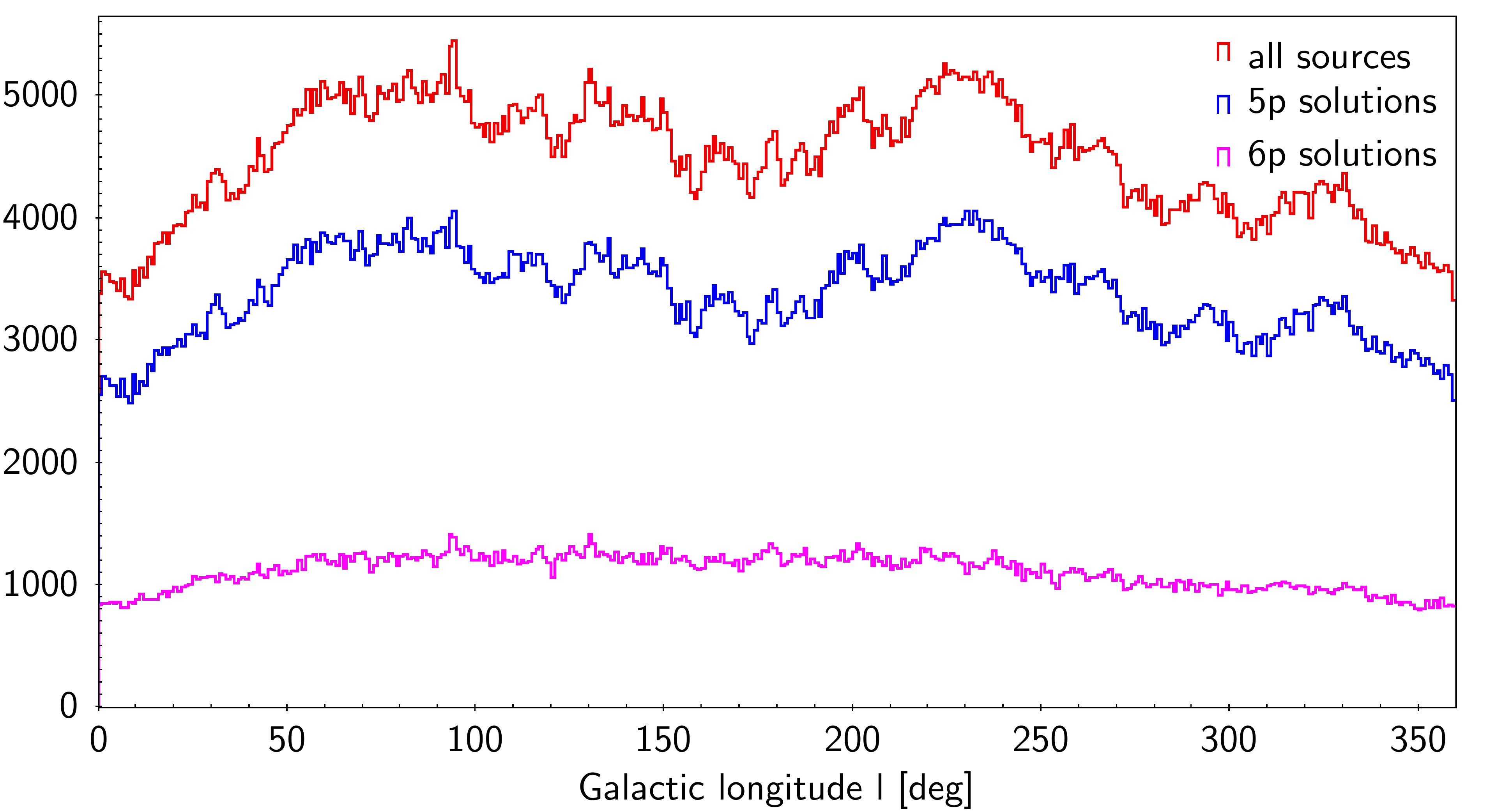}
  \includegraphics[width=1\hsize]{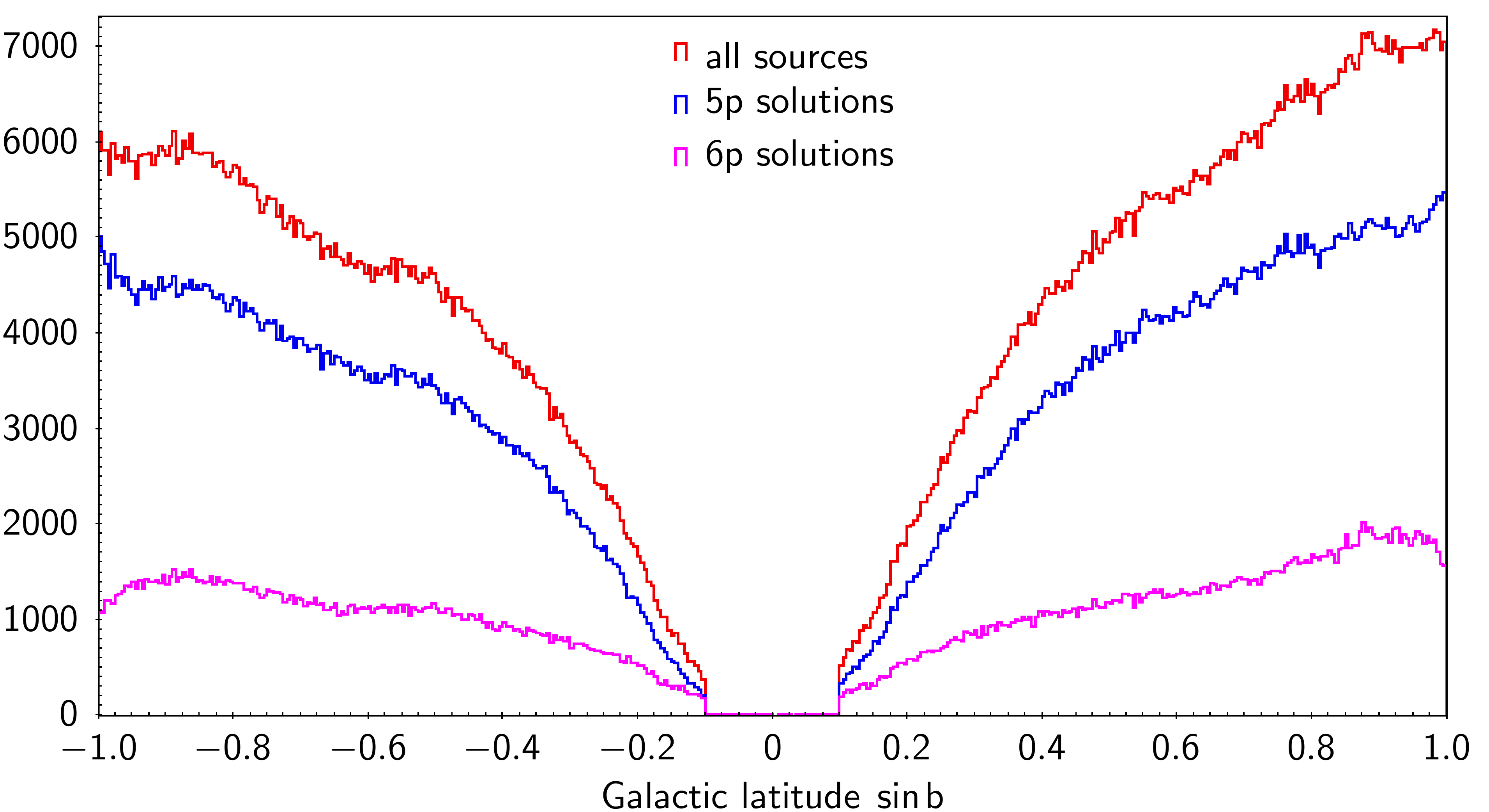}
  \caption{Distribution of the \gcrf{3} sources in galactic coordinates. There are only
    195 sources in the avoidance zone $\left|\,\sin b\,\right| < 0.1$. The distribution is shown
  for the whole \gcrf{3} and separately for the sources with five- and six-parameter astrometric solutions.}
\label{fig:gcrf3-galactic-coordinates-hist}
\end{center}
\end{figure}

The spatial distribution of the ${\sim}1.6$~million \gcrf{3} sources is shown in 
Fig.~\ref{fig:sky-distribution-gcrf3}. Their distributions in galactic longitude and latitude
are shown in Fig.~\ref{fig:gcrf3-galactic-coordinates-hist}.
In the avoidance zone  $\left|\,\sin b\,\right| < 0.1$
\gcrf{3} has only 195 sources matched to the radio catalogues as described in Sect.~\ref{sec:selection}. 
Outside the avoidance zone the 
average density is about 42~deg$^{-2}$, but with significant variations with galactic latitude;
the density is typically below average for $|\,b\,|\lesssim 30\degr$ and above at higher 
latitudes. This comes from the source catalogues used in our compilation combined 
with the selection function in Eq.~(\ref{eq:filterB}), designed to avoid stellar contaminants 
near the Galactic plane. 
From Fig.~\ref{fig:skymaps-external} one can see that also most of the external
catalogues avoid crowded regions on the sky.
We note that the density of \gcrf{3} reaches about 110~deg$^{-2}$ in some areas on the sky. 
Given the expected 
isotropic distribution of QSOs and the effects of Galactic extinction, a complete version of a \gcrf{} 
could therefore potentially comprise about 4~million sources. This estimate is also suggested by 
recent studies
(e.g.\ \citeads{2019MNRAS.489.4741S}; \citeads{2019MNRAS.490.5615B}; \citeads{2021ApJS..254....6F})
of the QSOs as well as by the extragalactic results of \gdr{3}.



\begin{table*}
  \caption{Characteristics of the \gcrf{3} sources.   
\label{tab:gaiacrf3-characteristics}}
\footnotesize\setlength{\tabcolsep}{6pt}
\begin{center}
\begin{tabular}{crccccccc}
\hline\hline
\noalign{\smallskip}
\multicolumn{1}{c}{type} 
& \multicolumn{1}{c}{number} 
& \multicolumn{1}{c}{$G$} 
&\multicolumn{1}{c}{$G_\text{BP}-G_\text{RP}$} 
& \multicolumn{1}{c}{$\nu_{\rm eff}$} 
&\multicolumn{1}{c}{RUWE} 
& \multicolumn{1}{c}{$\sigma_\text{pos,max}$}
& \multicolumn{1}{c}{$\sigma_{\mu\alpha*}$} 
& \multicolumn{1}{c}{$\sigma_{\mu\delta}$}\\
\multicolumn{1}{c}{of solution} 
& \multicolumn{1}{c}{of sources} 
& \multicolumn{1}{c}{[mag]} 
&\multicolumn{1}{c}{[mag]} 
& \multicolumn{1}{c}{[$\mu\text{m}^{-1}$]} 
&\multicolumn{1}{c}{} 
& \multicolumn{1}{c}{[$\muas$]} 
& \multicolumn{1}{c}{[$\muasyr$]} 
& \multicolumn{1}{c}{[$\muasyr$] }\\
\noalign{\smallskip}\hline\noalign{\smallskip}
five-parameter & 1\,215\,942 & 19.92 & 0.64 & 1.589 & 1.013 & 385 & 457 & 423 \\ 
six-parameter &  398\,231 & 20.46 & 0.92 & 1.541 & 1.044 & 749 & 892 & 832 \\
all         & 1\,614\,173 & 20.06 & 0.68 & 1.585 & 1.019 & 447 & 531 & 493 \\  
\noalign{\smallskip}
\hline
\end{tabular}
\tablefoot{Columns~3--9 give median values
    of the $G$ magnitude, the $G_\text{BP}-G_\text{RP}$ colour index, the effective
    wavenumber $\nu_{\rm eff}$ (or the pseudo-colour for six-parameter solutions), 
    the astrometric quality indicator RUWE (see text), the positional uncertainty at
    epoch J2016.0 (semi-major axis of the error ellipse),
    and the uncertainties of the proper motion components in $\alpha$ and $\delta$. 
   The last line (`all') is for the whole set of \gcrf{3} sources. 
}
\end{center}
\end{table*}

\begin{figure*}[t]
\begin{center}
  \includegraphics[width=\scalethree\hsize]{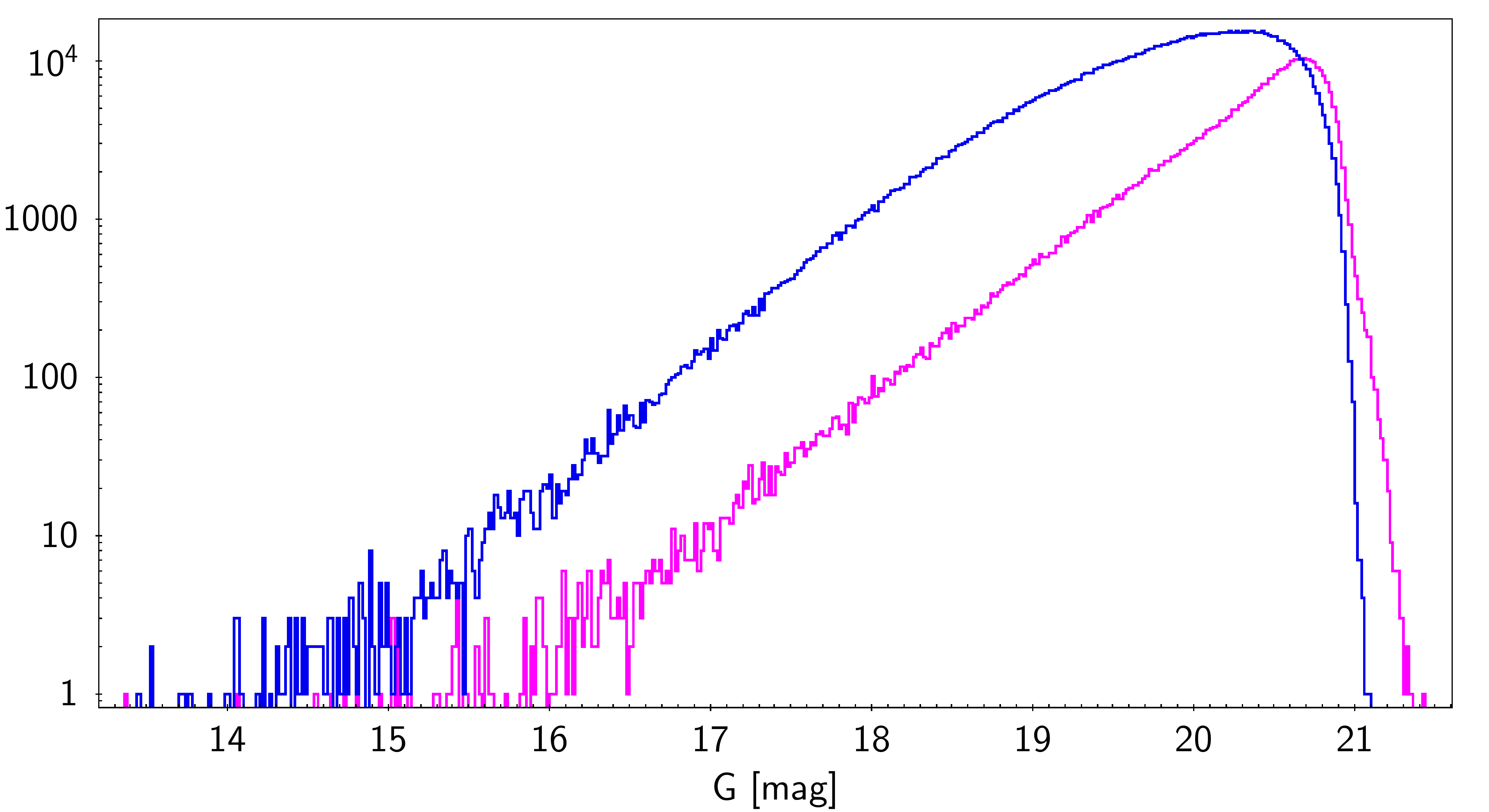}
  \includegraphics[width=\scalethree\hsize]{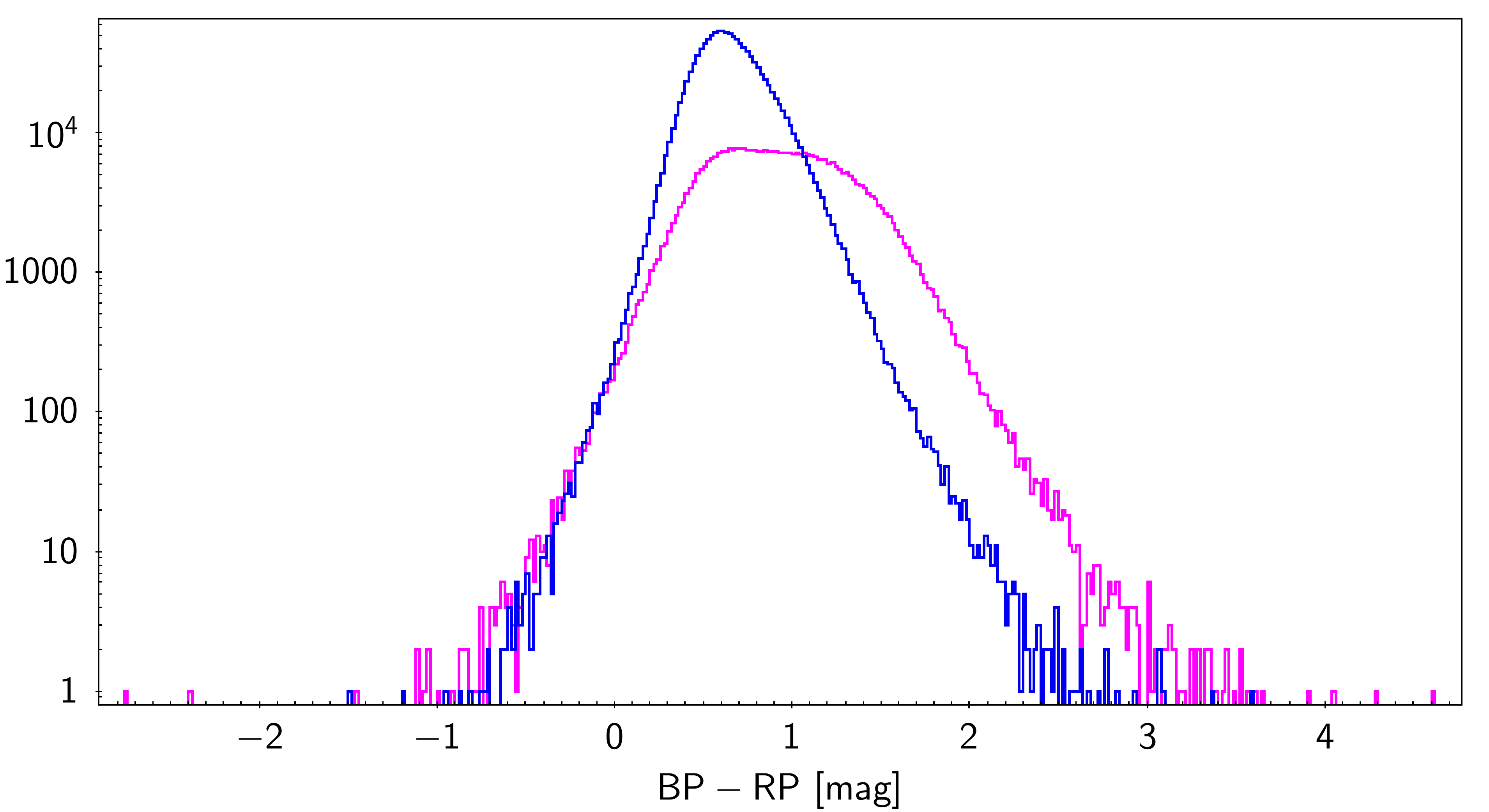}
  \includegraphics[width=\scalethree\hsize]{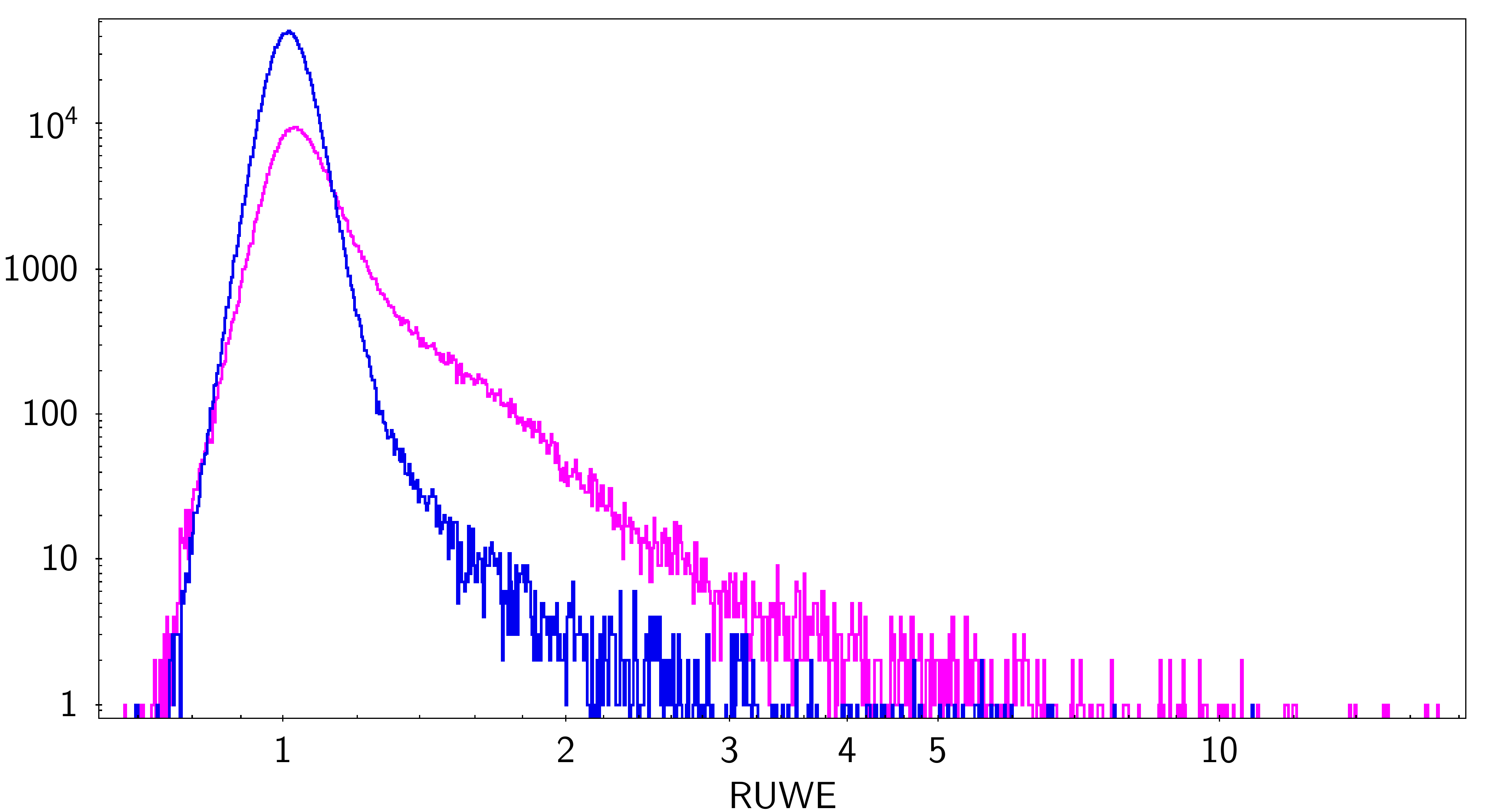}
  \caption{Histograms of some important characteristics of the \gcrf{3} sources 
  with five-parameter solutions (blue) and six-parameter solutions (magenta). \textit{Left to right:} 
  $G$ magnitudes, colour index $G_\text{BP}-G_\text{RP}$,
    and the astrometric quality indicator RUWE.
  }
\label{fig:histograms}
\end{center}
\end{figure*}

\begin{figure}[t]
\begin{center}
  \includegraphics[width=1\hsize]{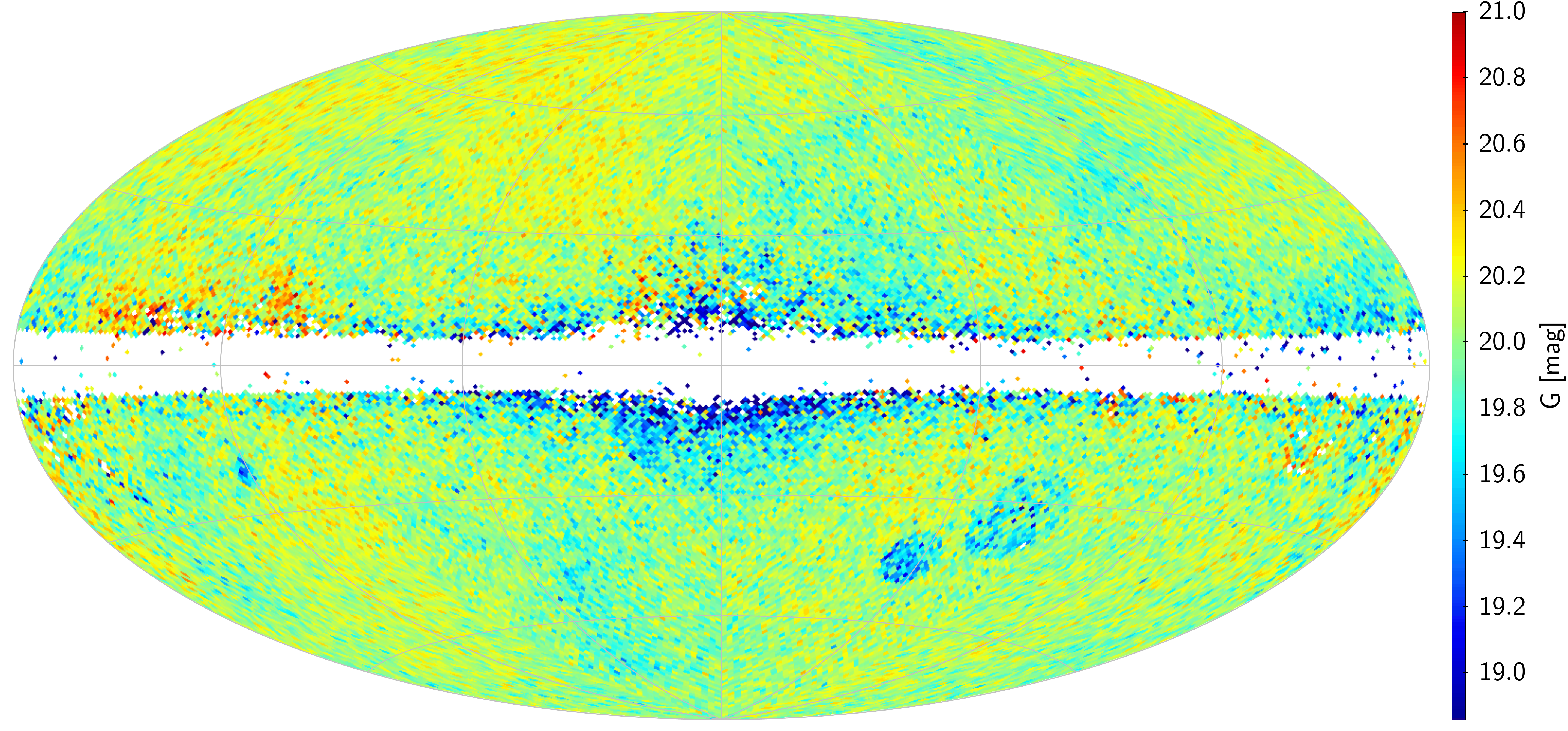}
  \includegraphics[width=1\hsize]{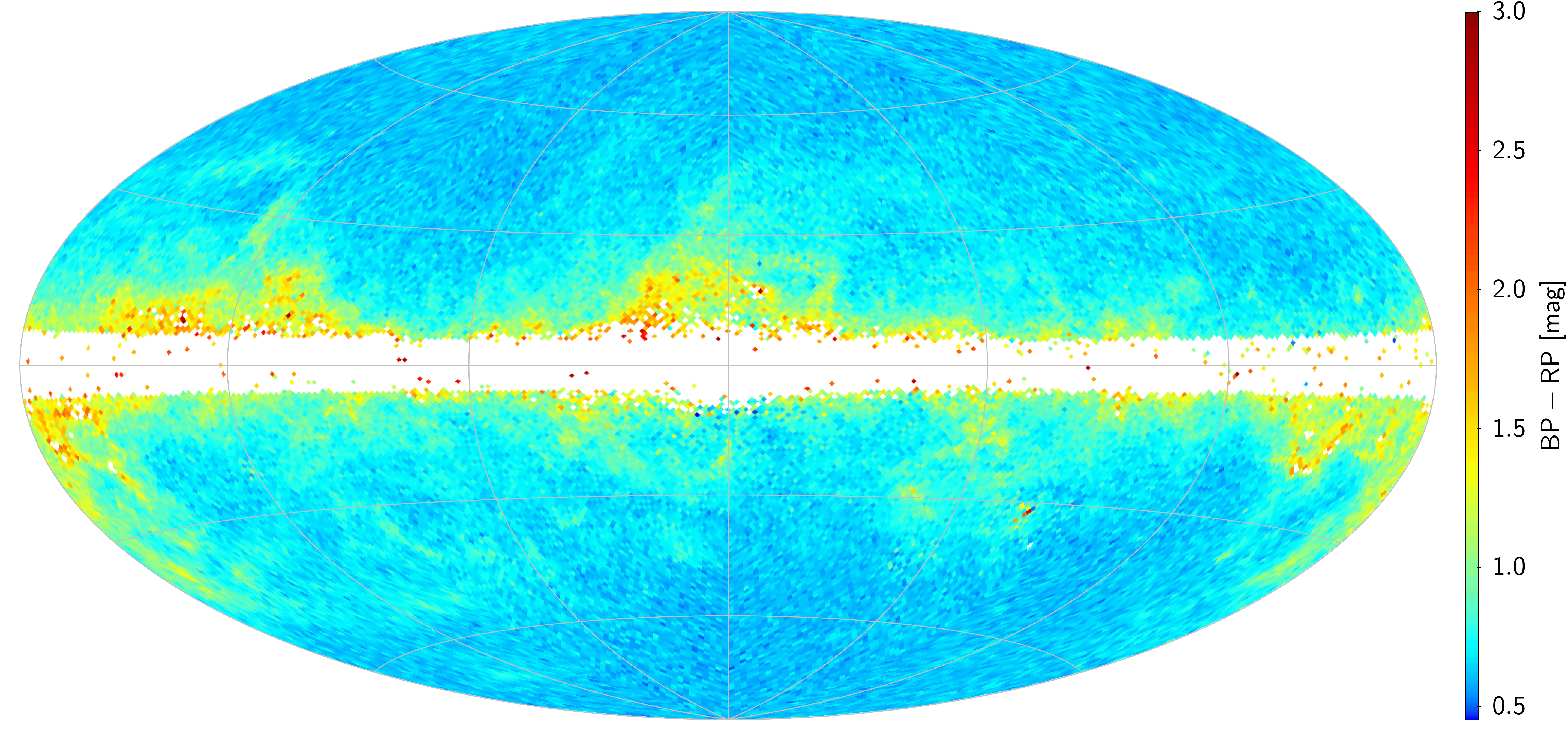}
  \caption{Median $G$ magnitude (\textit{top}) and median
    colour index \hbox{$G_\text{BP}-G_\text{RP}$} (\textit{bottom}) of the \gcrf{3} sources.
    These maps use a Hammer–Aitoff projection in galactic
    coordinates, with $l = b = 0$ at the centre, north up, and $l$
    increasing from right to left.
  }
\label{fig:g-bpMinusRp-skymaps}
\end{center}
\end{figure}

\subsection{Main characteristics of the sources}

Table~\ref{tab:gaiacrf3-characteristics} and Fig.~\ref{fig:histograms} show the median values
and distributions of several important properties of the \gcrf{3} sources, split according to the 
nature of the solution (five- or six-parameter solution): magnitude ($G$), colour index 
($G_\text{BP}-G_\text{RP}$), and the renormalised unit weight error (RUWE; \citeads{2021A&A...649A...2L}).
The RUWE is a measure of the goodness-of-fit of the astrometric model 
to the observations of the source. The expected value for a good fit is 1.0. A higher value could 
indicate that the source is not point-like at the optical resolution of {\gaia} (${\simeq}\,0.1\arcsec$), 
or that it has a time-variable structure. As seen in the left panel of Fig.~\ref{fig:histograms}, the 
proportion of six-parameter solutions increases for $G\gtrsim 19$, a feature that also implies
that these sources are on average of lower astrometric accuracy, not only because they are fainter but
also because the model fit is often worse, as shown in the right panel. The sources with 
six-parameter solutions are also on average redder (middle panel).

The sky distributions of the median magnitude $G$ and the colour
index $G_\text{BP}-G_\text{RP}$ (Fig.~\ref{fig:g-bpMinusRp-skymaps}) clearly show the effects
of the Galactic extinction and reddening. We also see that the selection 
favours brighter sources in the crowded areas such as the Galactic plane and the LMC and SMC areas.

\subsection{Astrometric parameters of the sources}

\begin{figure*}[t]
\begin{center}
  \includegraphics[width=\scalethree\hsize]{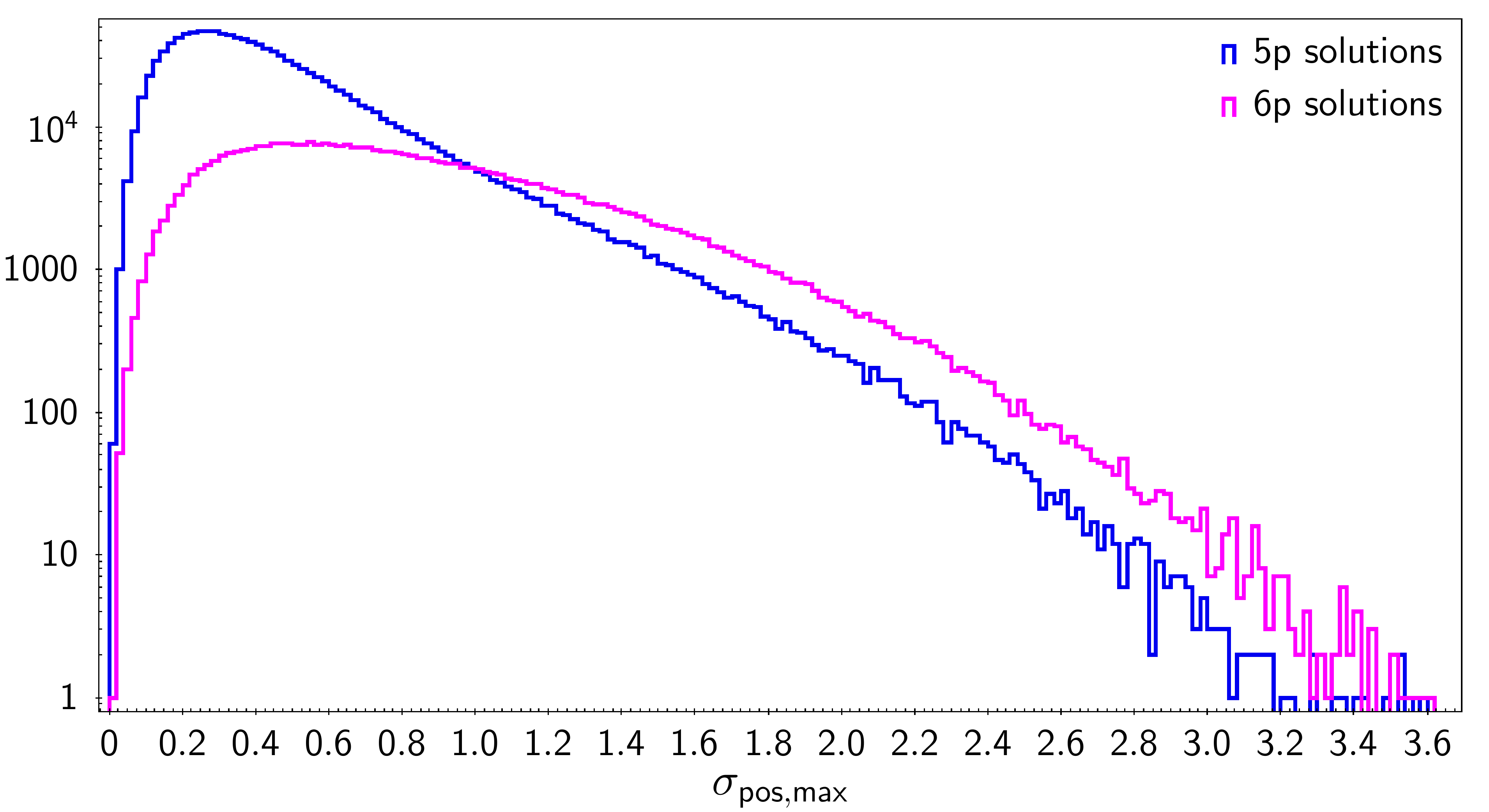}
  \includegraphics[width=\scalethree\hsize]{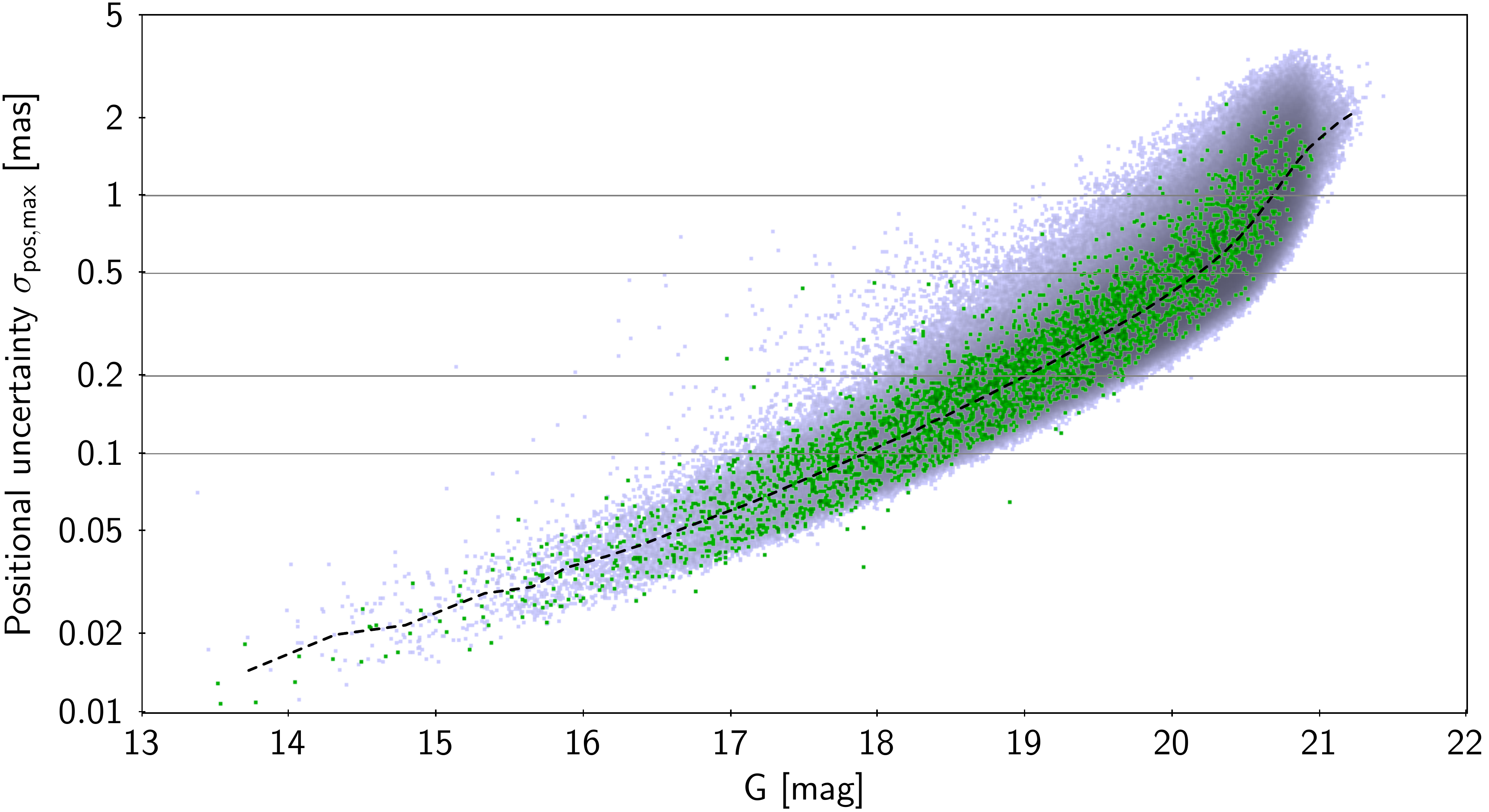}
  \includegraphics[width=\scalethree\hsize]{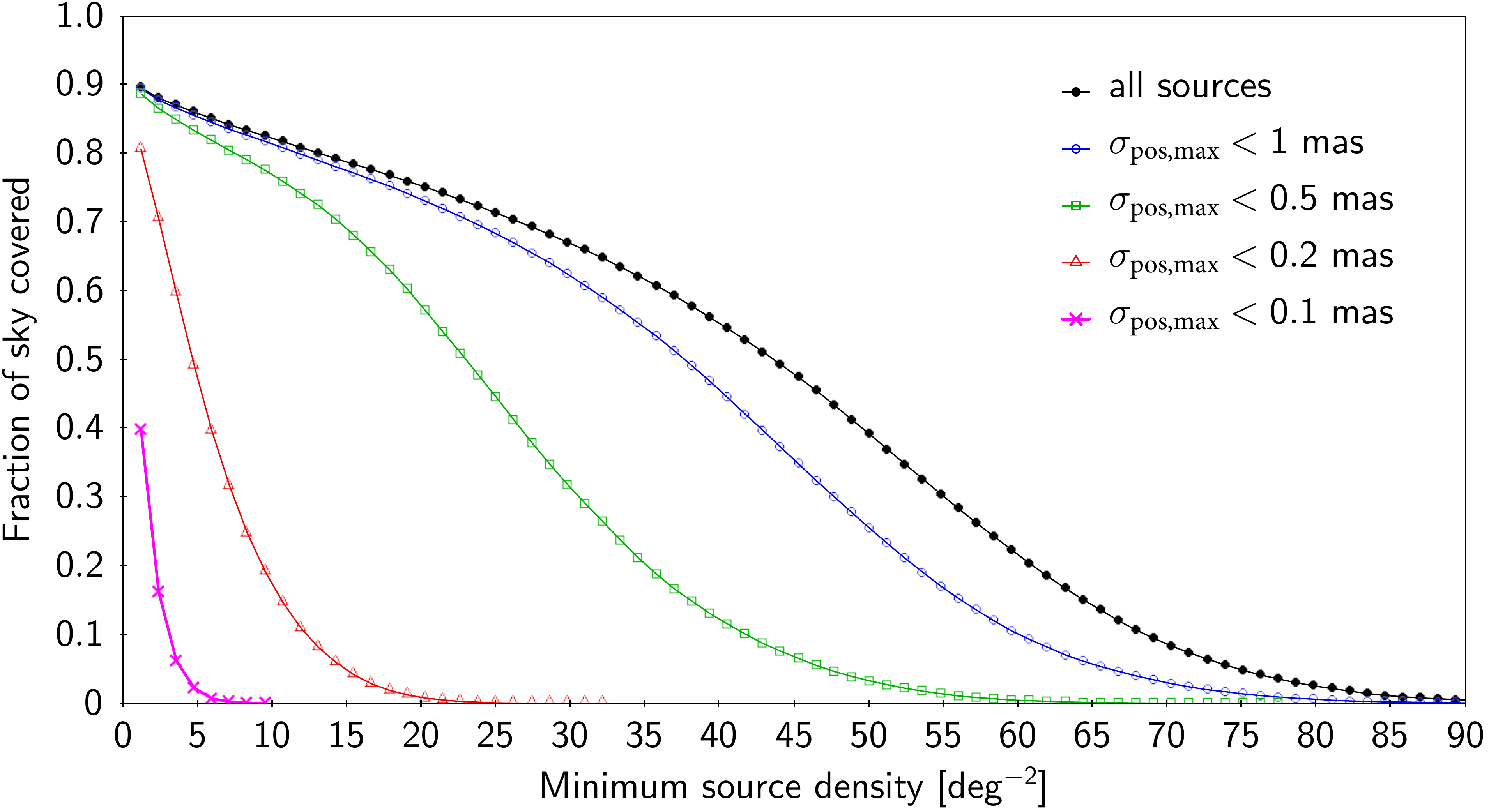}
  \caption{Positional uncertainty and source density distribution in \gcrf{3}. 
  \textit{Left:} histogram of the positional uncertainty $\sigma_\text{pos,max}$ for
  \gcrf{3} sources with five-parameter solutions (blue) and six-parameter solutions (magenta).
  \textit{Middle:} scatter plot of $\sigma_\text{pos,max}$ versus the $G$ magnitude. 
  The dashed curve is the smoothed median uncertainty versus magnitude. 
  The green points are the sources matched to ICRF3 S/X.
  \textit{Right:} fraction of sky covered by a given minimum density of \gcrf{3} sources 
  at different levels of positional uncertainty (see text for further explanation).}
\label{fig:GCRF-accuracy}
\end{center}
\end{figure*}

The left panel of Fig.~\ref{fig:GCRF-accuracy} shows the distribution of the formal positional 
uncertainty in \gcrf{3} separately for the five- and six-parameter solution. The positional
uncertainty, $\sigma_\text{pos,max}$, is defined as the semi-major axis of the uncertainty ellipse 
(see Eq.~(1) in \citeads{2018A&A...616A..14G}) at the reference epoch J2016.0.
The middle panel is a scatter plot of $\sigma_\text{pos,max}$ versus $G$, with the smoothed median 
indicated by the dashed curve. In the right panel, the black curve (labeled `all sources') gives 
the fraction of the full sky that has a density of \gcrf{3} sources exceeding the value on the
horizontal axis. The other curves give the fractions when only sources with $\sigma_\text{pos,max}$ 
below a certain level are counted. In the 10\% of the sky nearest to the Galactic plane 
($\left|\,\sin b\,\right|<0.1$) the source density is negligible (cf.\ Fig.~\ref{fig:sky-distribution-gcrf3}), 
which explains why none of the curves reaches above 90\% sky coverage. To compute the fractions,
the celestial sphere was divided into 49\,152 pixels of equal area (${\simeq\,}0.8393~\text{deg}^2$, 
i.e.\ HEALPix level 6) and the density computed in each pixel. It should be noted that the curves
therefore refer to the source densities at a spatial resolution of about 1~deg and would not be
the same at a different resolution (pixel size). 
The plot shows, for example, that for a minimum density of 35~\gcrf{3} sources per square
degree, about 62\% of the sky could be covered; this fraction is reduced to 55\% or 20\% if we require, 
in addition, that $\sigma_\text{pos,max}$ is below 1.0 or 0.5~mas.
Conversely, at 62\% coverage there could be at least 30, 18, and 3~sources per square degree
with $\sigma_\text{pos,max}<1$, 0.5, and 0.2~mas, if in each case the most populated pixels 
are chosen. The corresponding magnitude distributions can be gleaned from the scatter plot
in the middle panel, where the different levels of positional uncertainty have been marked.

\begin{figure}[t]
\begin{center}
  \includegraphics[width=1\hsize]{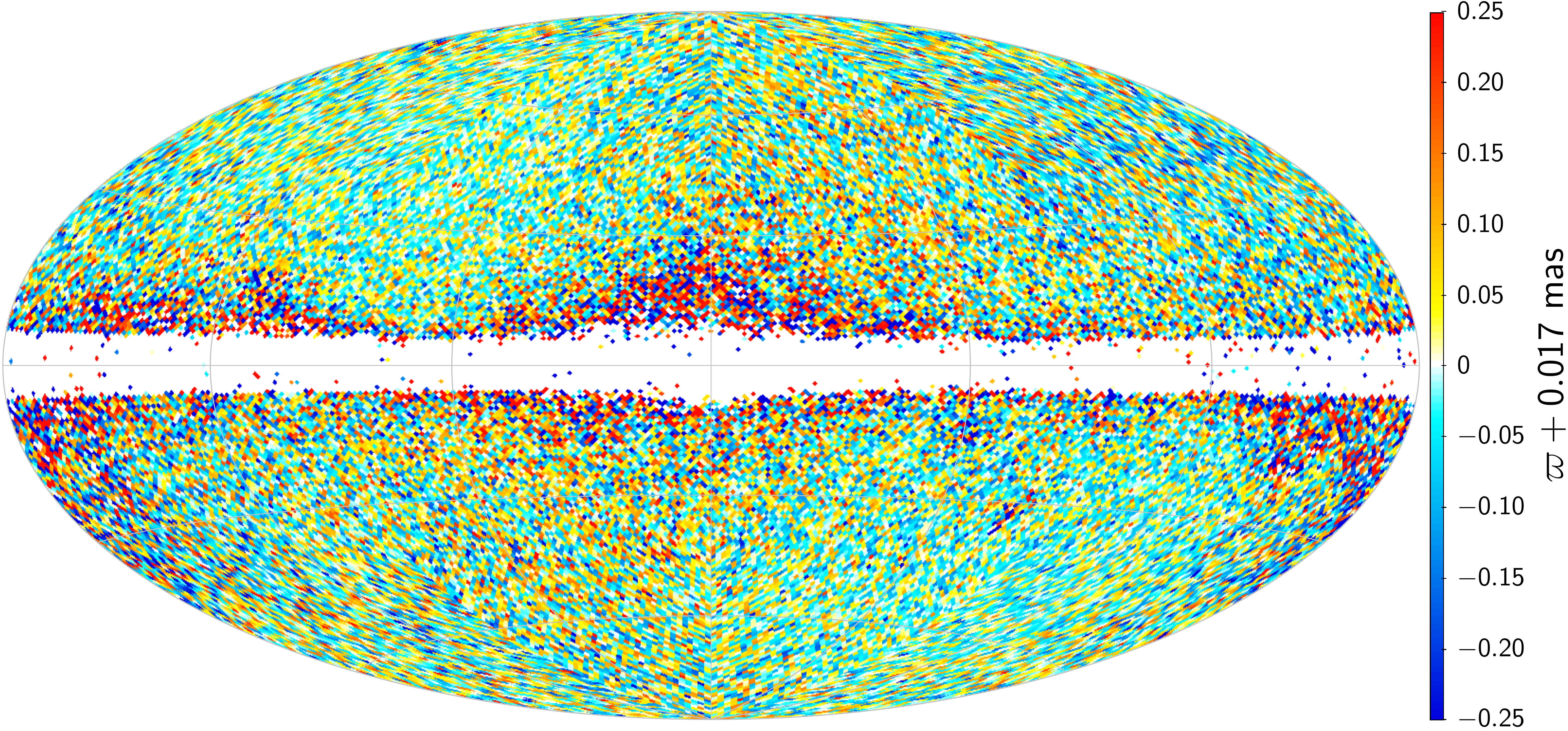}
  \includegraphics[width=1\hsize]{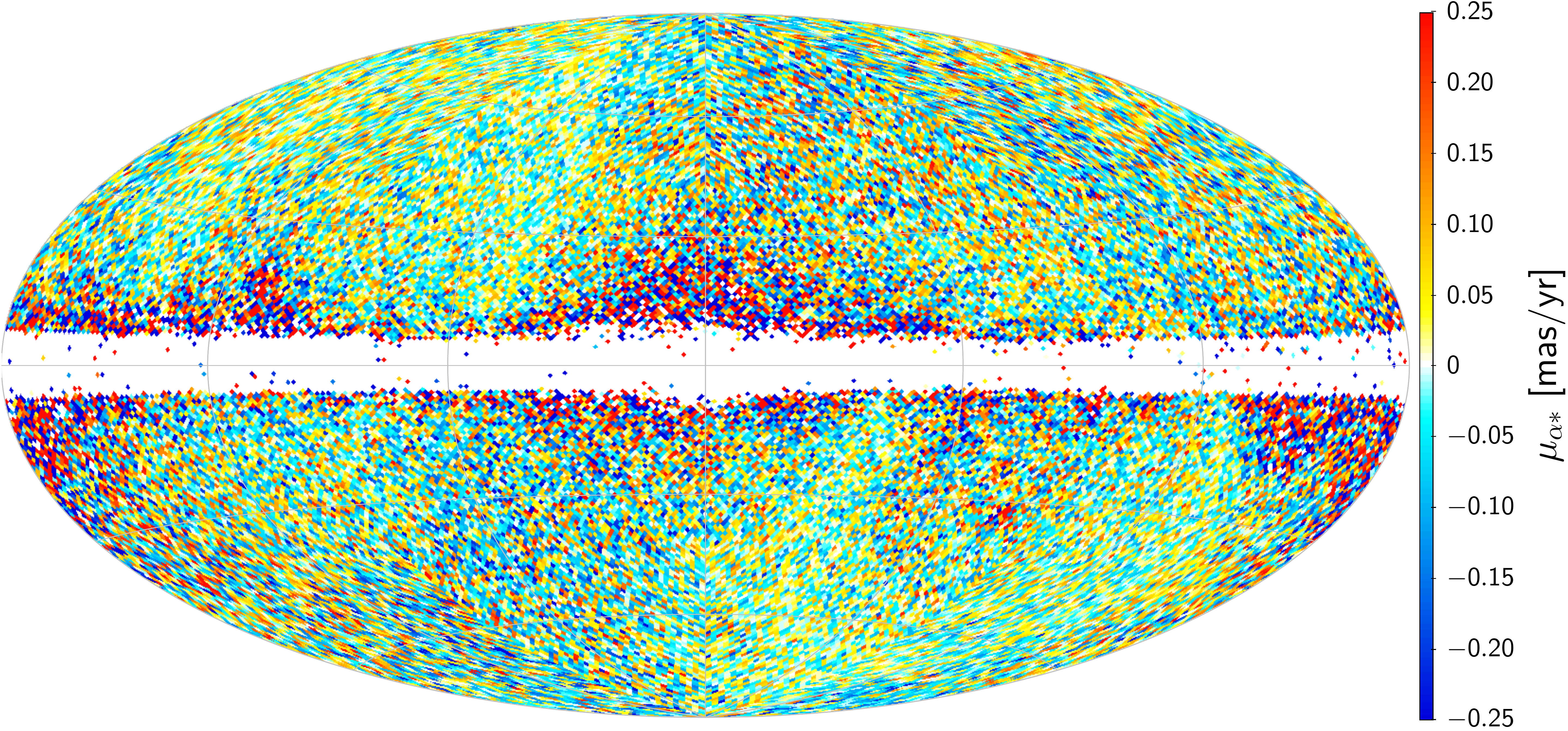}
  \includegraphics[width=1\hsize]{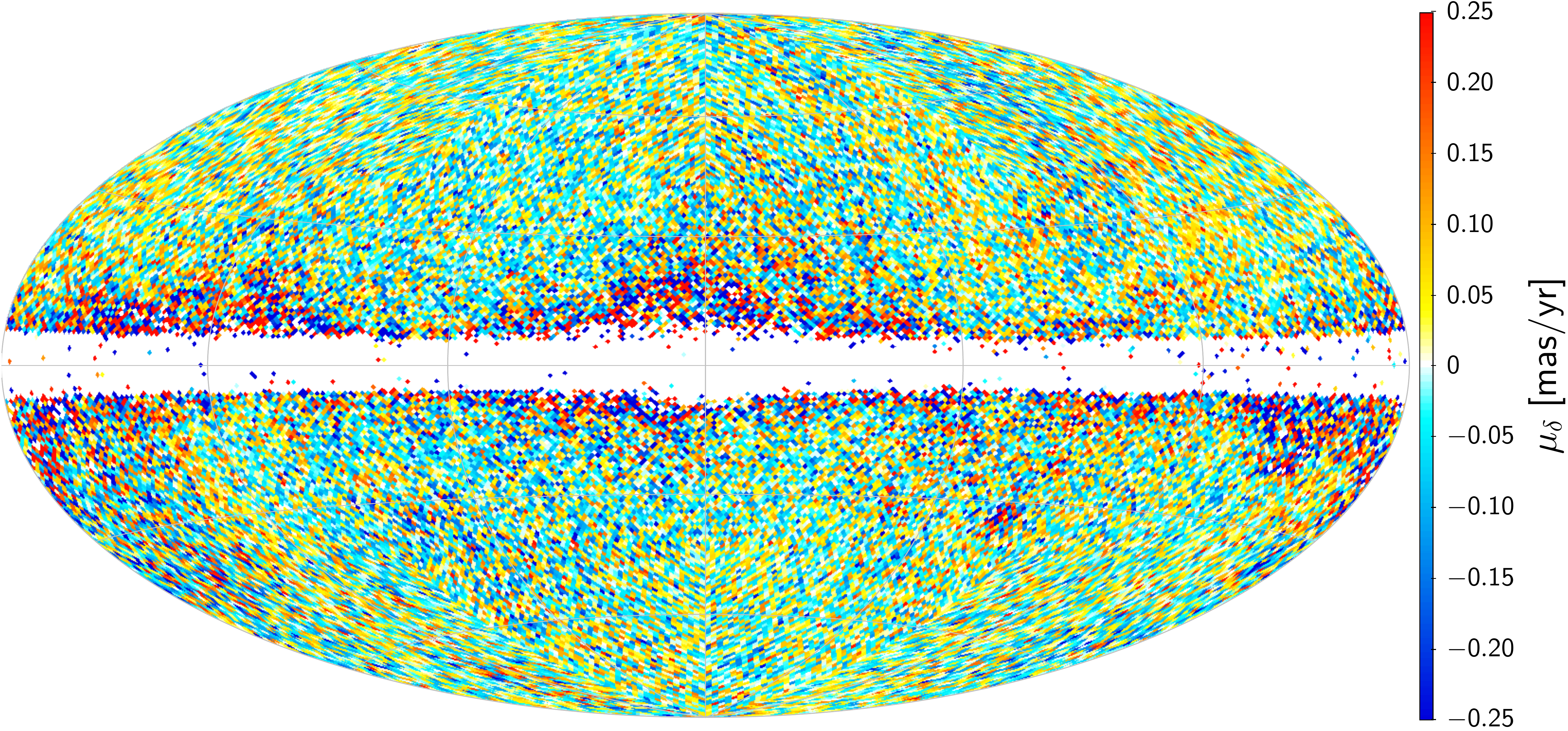}  
  \caption{Sky distributions of the parallax $\varpi$ and the components of proper 
  motion $\mu_{\alpha*}$ and $\mu_\delta$. Median values over the pixels of HEALPix 
  level~6 (pixel size $\simeq 0.84$\,deg$^2$) are shown.  
  These maps use a Hammer–Aitoff projection in galactic
    coordinates, with $l = b = 0$ at the centre, north up, and $l$
    increasing from right to left.
  }
\label{fig:astrometry-distributions}
\end{center}
\end{figure}

Figure~\ref{fig:astrometry-distributions} shows the distributions of the
median parallax and proper motion components on the sky. 
Both large-scale systematics and small-scale inhomogeneities are visible
in these maps. 
The large-scale systematics are discussed in Sect.~\ref{sect:systematics}.
Small-scale inhomogeneities, seen as fluctuating median values between 
neighbouring HEALPix cells in the maps, are mainly caused by the statistical
scatter of the astrometric parameters in each cell. They are generally 
stronger in areas of low source density (Fig.~\ref{fig:sky-distribution-gcrf3}) or 
fainter median magnitude (Fig.~\ref{fig:g-bpMinusRp-skymaps}, top), and also in
the ecliptic belt (ecliptic latitude $\left|\,\beta\,\right|\lesssim 45^\circ$), 
where the scanning law usually results in less accurate astrometric solutions 
than at high ecliptic latitudes (see, for example,
Sect.~5.4 and Figs.~8 and 9 in \citeads{2021A&A...649A...2L}). Numerous streaks,
edges, and other small-scale features can also be traced to the scanning law.
The overall scatter of the median values in the three maps is about
0.11~mas ($\varpi$), 0.12~{\masyr} ($\mu_{\alpha*}$), and 0.11~{\masyr} ($\mu_\delta$), 
if only the cells with at least five QSOs are counted (covering 85\% of the sky). 
The positional errors are expected to be of a roughly similar size as the parallax scatter.
Thus, we conclude that \gcrf{3}, on an angular scale of about $1^\circ$ and over most 
of the sky, defines a local reference frame accurate to better than 0.2~mas in position 
and 0.2~{\masyr} in proper motion.

\subsection{Systematic errors}
\label{sect:systematics}

Systematic errors in \gcrf{3} were discussed already in 
\citetads[][Section 5.6]{2021A&A...649A...2L}. In
particular, the left column of their Fig.~13 contains smoothed maps
of the parallaxes and proper motions of the \gcrf{3}\ sources that have
five-parameter astrometric solutions. These maps were obtained by 
a special smoothing procedure applied to the parameters of the 
individual sources. Including also the less numerous and fainter sources 
with six-parameter solutions does not noticeably change these maps.

Here we present the results of a different way to analyse systematics 
in the \gcrf{3} proper motions, namely by fitting an expansion of 
vector spherical harmonics (VSH; \citeads{2012A&A...547A..59M})
up to a certain degree $\ell$. 
Although the
VSH provide a smooth model of the systematic errors also in
the regions where only a few sources are present, such as in the Galactic plane,
the higher-degree harmonics ($\ell\gtrsim 15$) are only weakly constrained 
in these areas, resulting in unreliable estimates of the VSH coefficients
also in other parts of the sky because of the global scope of the fitted
VSH functions.

Analogous to the discussion of parallax systematics in Sect.~5.7 of 
\citetads{2021A&A...649A...2L}, using (scalar) spherical harmonics, 
the VSH allow us to estimate the angular power spectrum of 
the systematics in the \gcrf{3} proper motions. To this end 
we used the set of 1\,215\,942 \gcrf{3} sources with
five-parameter solutions. Given the vector field of proper motions of
those sources we computed the VSH expansion as given by Eq.~(30) of
\citetads{2012A&A...547A..59M}, including all the terms of degree 
$1\le\ell\le L$. It was found that the VSH fit is reasonably
stable for $L\le 12$. When higher degrees are included, the fits 
become questionable, mainly because of the inhomogeneous distribution 
of the sources on the sky (see Sect.~\ref{sect:distribution} above and 
Figs.~3 and 4 of \citeads{2021A&A...649A...9G}). All subsequent
computations used $L=12$.
The fitted VSH coefficients confirmed the maps of systematic errors in proper motion
shown on Fig.~13 of \citetads{2021A&A...649A...2L}. 
From the same coefficients we obtained the powers 
$P^t_\ell$, $P^s_\ell$ of the toroidal and spheroidal signals of degree 
$\ell$, as defined by Eq.~(76) of \citetads{2012A&A...547A..59M}.  
The total power of degree $\ell$ is $P_\ell=P^t_\ell+P^s_\ell$. 
The mean powers of the toroidal and spheroidal
harmonics of degree $\ell$ are then
$C^t_\ell=P^t_\ell/(2\ell+1)$ and $C^s_\ell=P^s_\ell/(2\ell+1)$,
where $2\ell+1$ is the number of harmonics at the degree $\ell$.
This is fully analogous to the definitions for the scalar spherical harmonics 
given by Eq.~(31) of \citetads{2021A&A...649A...2L}.

We note that the VSH components of degree $\ell=1$ in the proper motions 
should be analysed separately from other systematic effects. The
toroidal harmonic of degree $\ell=1$ represents a residual spin of the 
reference frame, and should in principle be negligible thanks to the way 
the reference frame is constructed (see Sect.~\ref{sec:framerotator} below).
The spheroidal harmonic of degree $\ell=1$ represents a `glide' component
of the proper motions that is expected as a consequence of the acceleration 
of the Solar system in the rest frame of remote sources 
\citepads{2021A&A...649A...9G}. Although these terms need to be included in the 
VSH fit, the subsequent discussion of systematics in the \gcrf{3} proper motions
only considers the VSH components of degree $\ell=2,\dots,12$.

Formally, the powers and the mean powers are weighted sums of the
squared VSH coefficients. Appendix~A of \citetads{2021A&A...649A...9G} demonstrates
that such quantities are biased (overestimated) when the VSH coefficient are estimated
in a least squares fit. Equation~(A.1) of that Appendix gives the exact formula to correct the bias.
In particular, the corrected (unbiased) values of the mean powers are
\begin{eqnarray}
  \hat{C}_\ell^t&=&C_\ell^t-N_\ell^t\,,\\
  \hat{C}_\ell^s&=&C_\ell^s-N_\ell^s\,,
\end{eqnarray}
\noindent
where the noise corrections $N_\ell^t$ and $N_\ell^s$
are obtained by the same formulas as $C_\ell^t$ and $C_\ell^s$ but
using the uncertainties of the VSH coefficients instead of the
coefficients themselves. Explicit computations for the present data show that 
$N_\ell^t \simeq N_\ell^s \simeq 
3.44-0.07\,\ell-0.01\,\ell^2~\mu\text{as}^2\,{\text{yr}}^{-2}$
($2\le\ell\le 12$).

\begin{figure}[t]
\begin{center}
  \includegraphics[width=1\hsize]{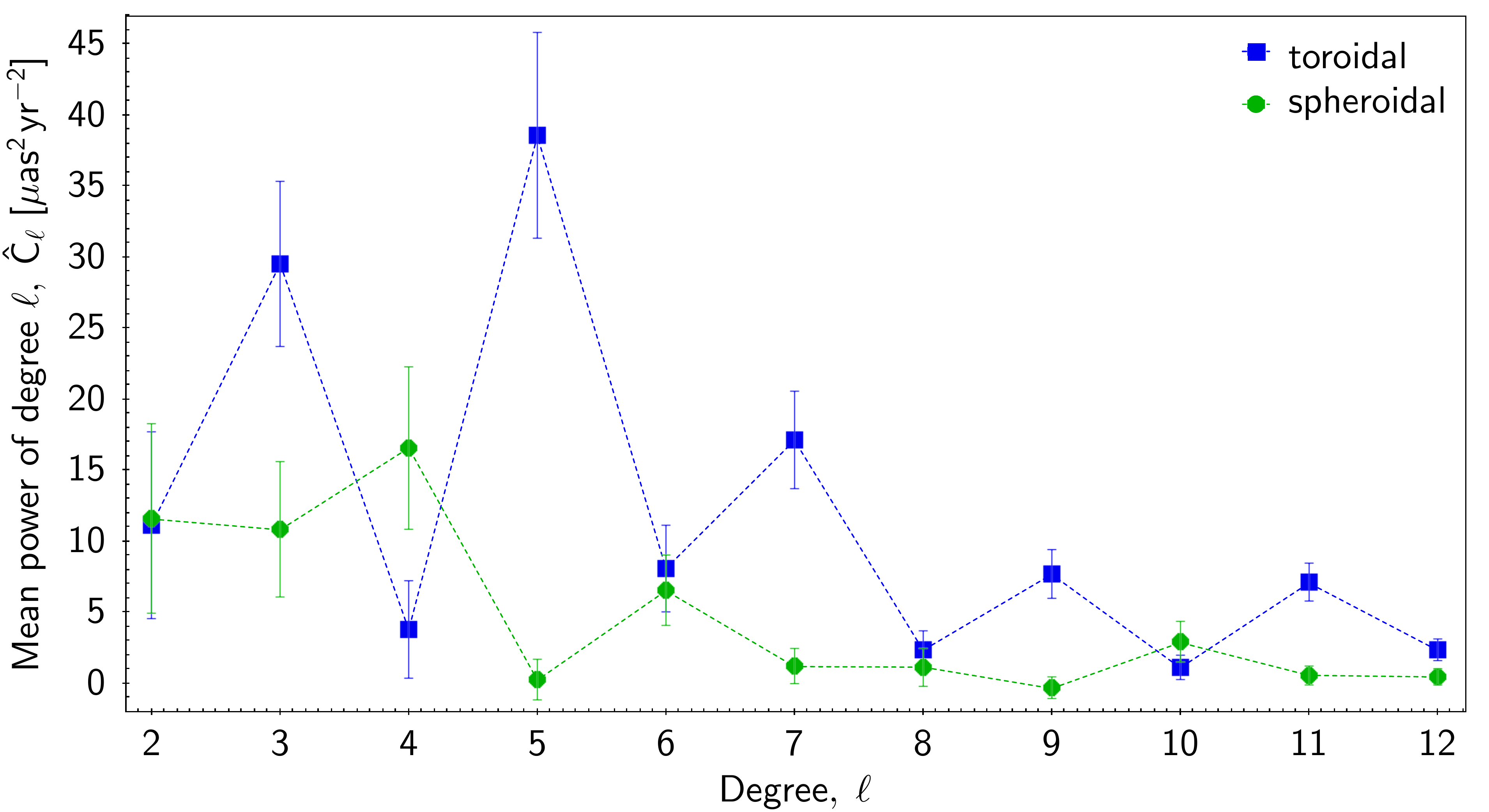}
  \caption{Corrected mean toroidal and spheroidal powers $\hat{C}^t_\ell$ and $\hat{C}^s_\ell$ of 
  the systematics in proper motions of the \gcrf{3} sources with five-parameter solutions.
  The error bars show the $\pm 1$~standard deviation of each point, calculated on the assumption 
  that the $2\ell+1$ contributing VSH coefficients have Gaussian errors.
  }
\label{fig:Clt-Cls}
\end{center}
\end{figure}

Figure~\ref{fig:Clt-Cls} shows the corrected estimates $\hat{C}^t_\ell$
and $\hat{C}^s_\ell$ for $\ell=2,\dots,12$. The ragged shape of (especially)
the toroidal spectrum is very real, as shown by the error bars. 
One sees that the toroidal powers 
are substantially larger than the spheroidal ones for the odd degrees, while they
are more similar for even degrees. We have no explanation for this but presume 
it is related to the specific parameters of the scanning law and basic angle. 

\begin{figure}[t]
\begin{center}
  \includegraphics[width=1\hsize]{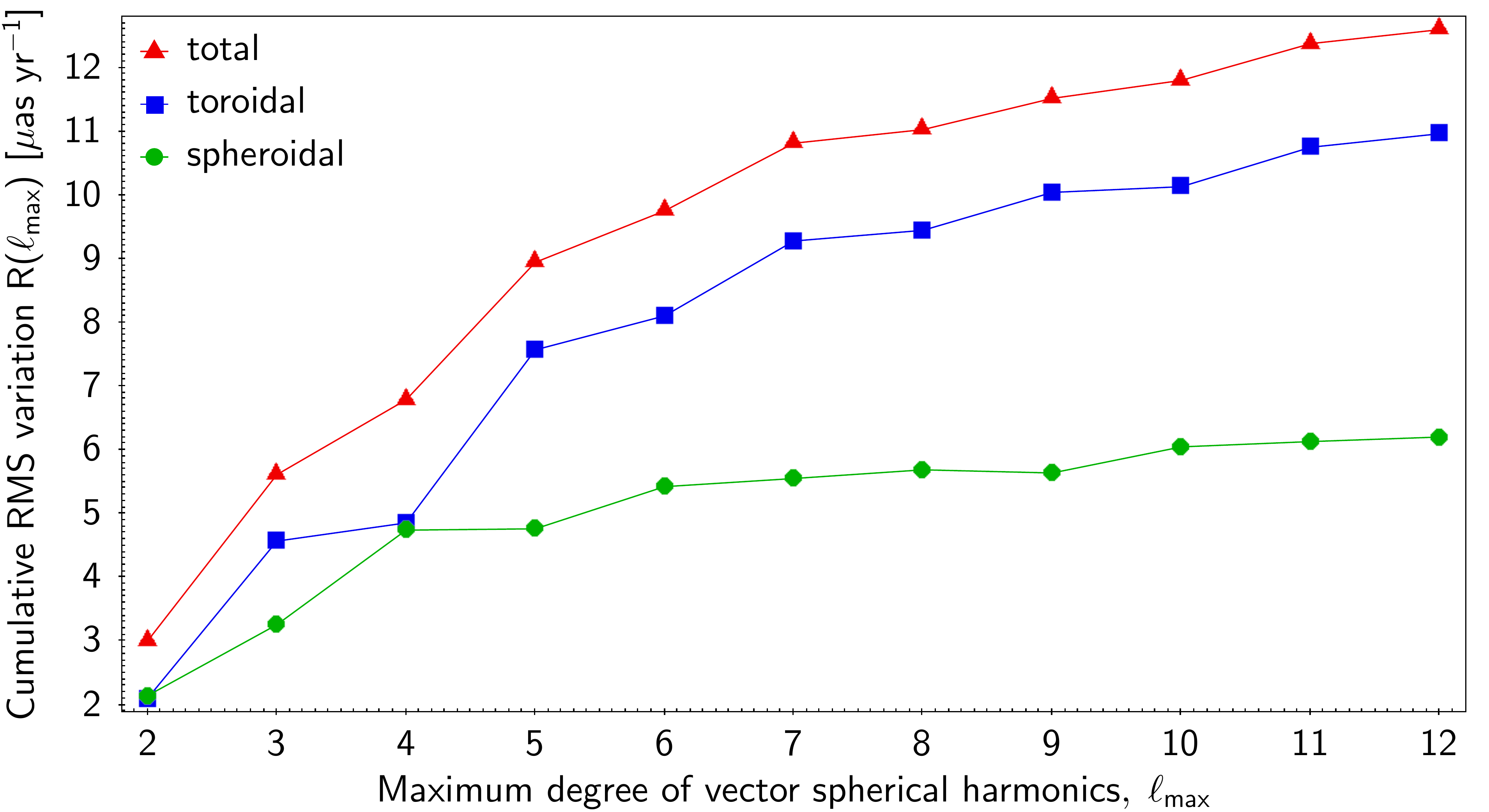}
  \caption{Corrected cumulative angular power spectrum of the systematic in proper motions 
  of the \gcrf{3} sources with five-parameter solutions.
  }
\label{fig:Rlt-Rls-Rl} 
\end{center}
\end{figure}

In analogy with Eq.~(32) of \citetads{2021A&A...649A...2L}, it is convenient to 
consider the cumulative characteristics
\begin{eqnarray}
R^t(\ell_\text{max})&=&\sqrt{{1\over 4\pi}\sum_{\ell=2}^{\ell_\text{max}}(2\ell+1)\,\hat{C}_\ell^t}\,,\\
R^s(\ell_\text{max})&=&\sqrt{{1\over 4\pi}\sum_{\ell=2}^{\ell_\text{max}}(2\ell+1)\,\hat{C}_\ell^s}\,,\\
R(\ell_\text{max})&=&\sqrt{{1\over 4\pi}\sum_{\ell=2}^{\ell_\text{max}}(2\ell+1)\,\left(\hat{C}_\ell^t+\hat{C}_\ell^s\right)}\,
\end{eqnarray}
quantifying the RMS variations of the toroidal, spheroidal and total proper motion systematics 
on angular scales $\gtrsim 180^\circ/\ell_\text{max}$ (Fig.~\ref{fig:Rlt-Rls-Rl}).  
We note that the toroidal components completely dominate the systematics for $\ell\ge 5$.

Summarising all this information we can claim that the
frame-independent RMS of the vector field representing systematic
errors in proper motions is $\lesssim 12$~\muasyr down to angular
scales of ${\sim\,}15^\circ$.  This level of systematic errors is low compared
to the random errors of individual sources given in
Table~\ref{tab:gaiacrf3-characteristics} and made it possible to
reliably measure the subtle physical effect of the acceleration of the
Solar system in \gedr{3} \citepads{2021A&A...649A...9G}.

\section{\gcrf{3} counterparts of ICRF3 sources}
\label{sec:gcrf3}

Radio-loud sources in \gcrf{3} deserve special attention as they provide a direct link
between the ICRF implementations in the radio and optical regimes. 
In this section we focus on the properties of the \gcrf{3} sources matched to 
ICRF3 \citepads{2020A&A...644A.159C} and a statistical comparison of the optical
and radio positions.

\subsection{Statistics of ICRF3 sources in \gcrf{3}}
\label{sec:gcrf3stats}

\gcrf{3} contains a total of 4723 radio-loud
sources, of which 3181 
are optical counterparts of radio sources in
ICRF3 \citepads{2020A&A...644A.159C}, while an additional 1542 sources
are found only in OCARS. 
Because the ICRF3 sources have the best quality of the astrometry in the radio domain
and represent the official realisation of the ICRF in the radio, we restrict the discussion 
in this section to the 3181 \gcrf3 sources in common with ICRF3.
This set is the union of 3142, 660, and 576 sources matching the 
ICRF3 sources in the S/X, K, and X/Ka bands, respectively (69.3\%, 80.1\%
and 85.0\% of the corresponding catalogues). 
Most of the analysis below is further limited to the 3142 sources in ICRF3~S/X. 
Very few additional \gedr{3} sources are found only in the K or X/Ka bands of
the ICRF3 (24 and 31 sources, respectively), and their overall properties are not too 
different from those of the S/X set.

\begin{figure}
\begin{center}
  \includegraphics[width=1.00\hsize]{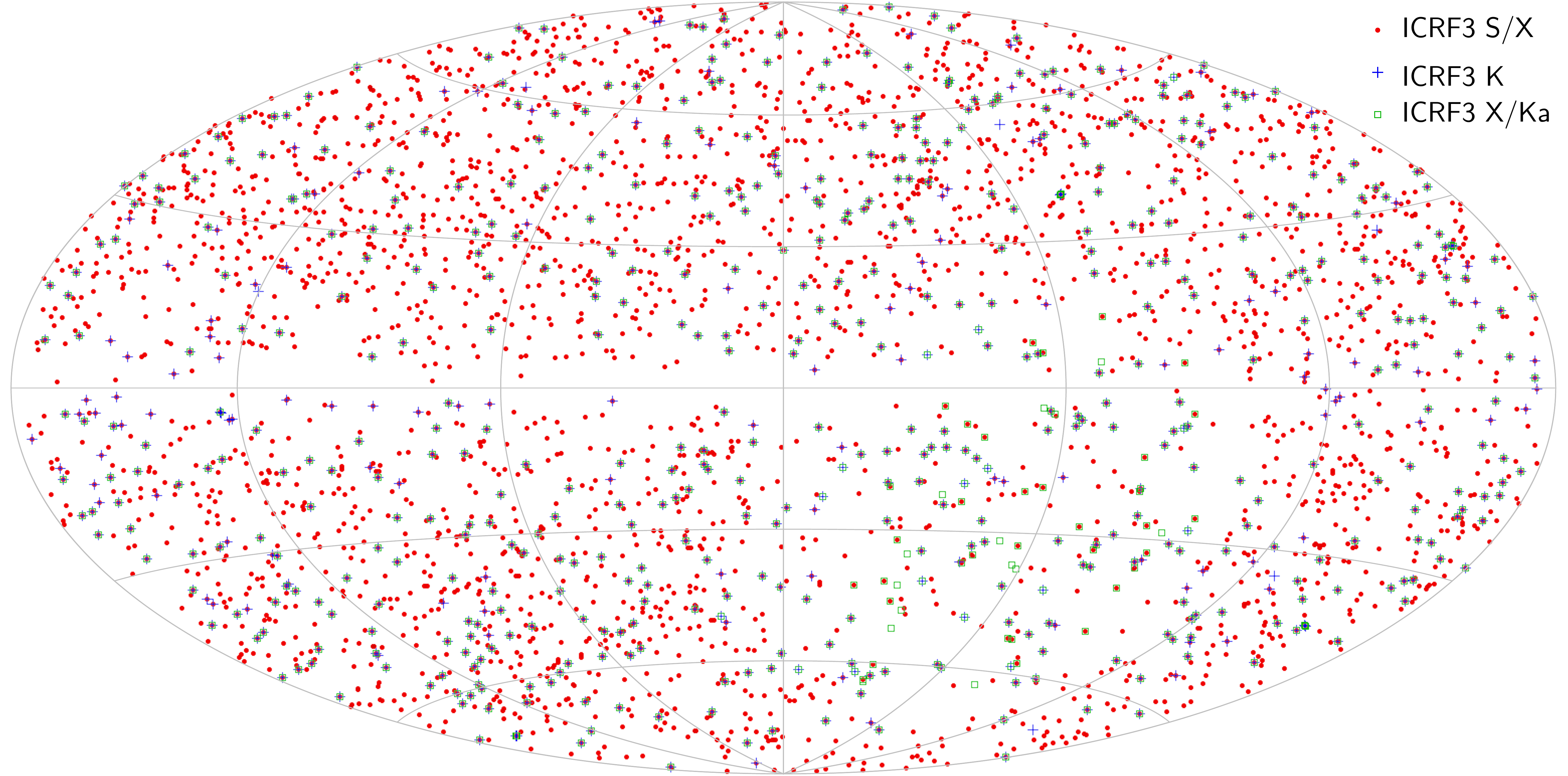}
  \caption{Distribution of the ICRF3 sources in \gcrf{3} in Galactic coordinates.
    Sources from ICRF3~S/X are shown by red dots, from ICRF3~K by blue crosses, and 
    from ICRF3~X/Ka by turquoise squares.
    This map uses a Hammer–Aitoff projection in galactic
    coordinates, with $l = b = 0$ at the centre, north up, and $l$
    increasing from right to left.
  }
\label{fig:gcrf3-icrf3}
\end{center}
\end{figure}

From the sky distribution of the 3181 ICRF3 
sources in \gcrf{3} (Fig.~\ref{fig:gcrf3-icrf3}) we see that a band along the 
Galactic plane and the region with $\delta<-30^\circ$ (part of the 
fourth quadrant of the map in the Galactic coordinates) are underpopulated. This reflects the 
strong optical extinction in the Galactic plane and the fact that ICRF3 itself has a lower
density of sources for $\delta<-30^\circ$.

\begin{figure*}
\begin{center}
  \includegraphics[width=\scalethree\hsize]{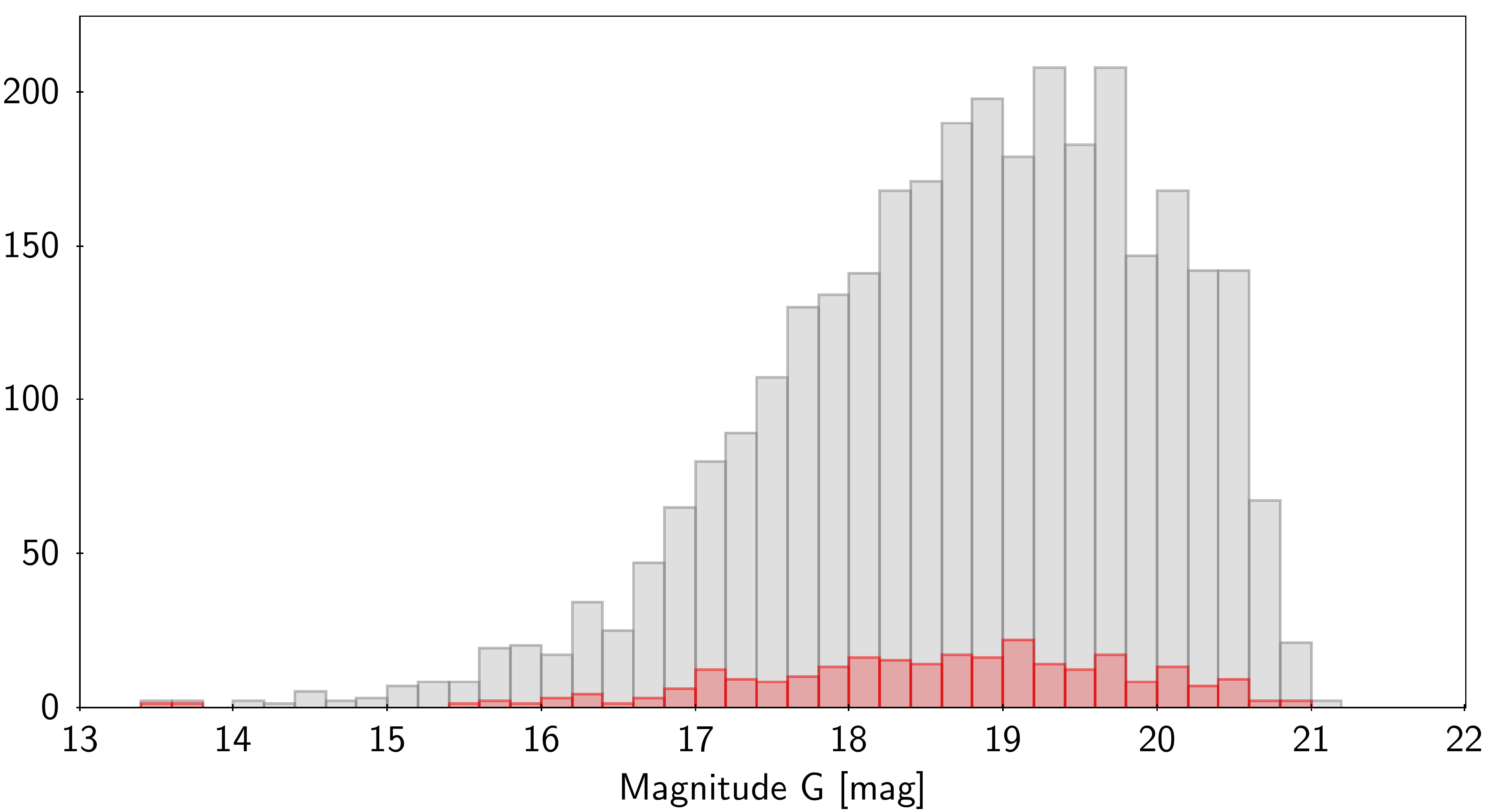}
  \includegraphics[width=\scalethree\hsize]{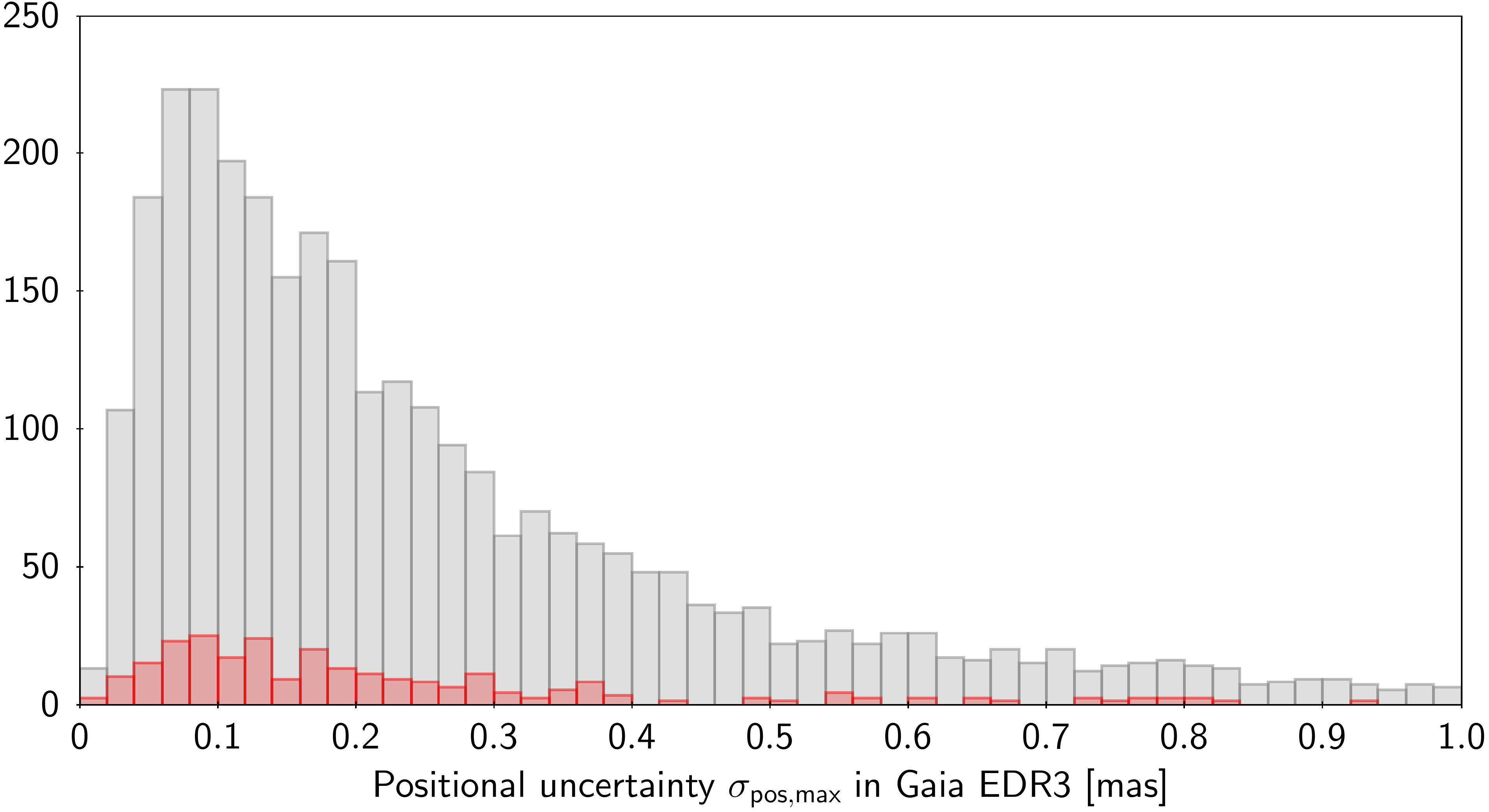}
  \includegraphics[width=\scalethree\hsize]{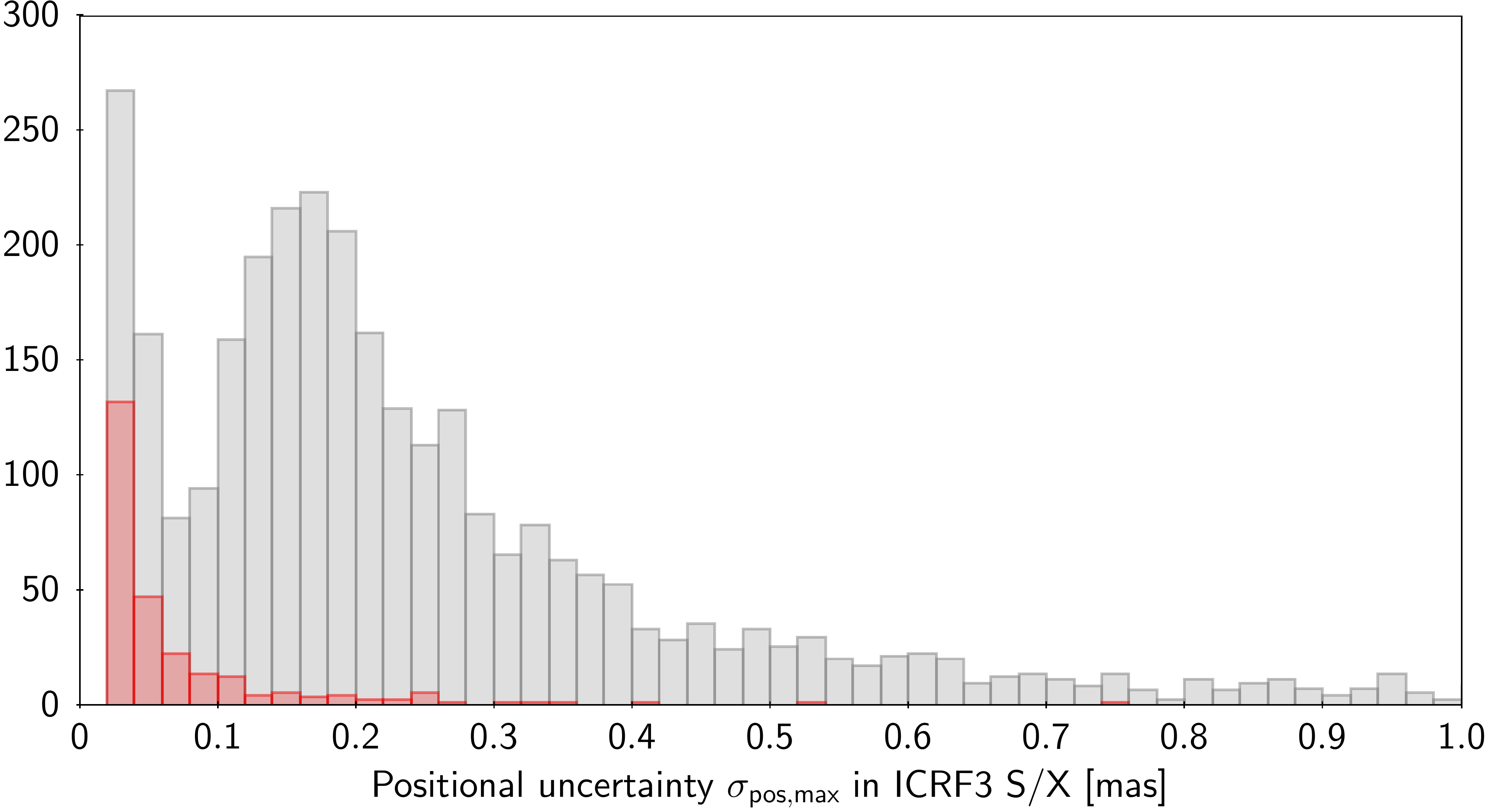}
  \caption{Distribution of magnitudes and formal positional uncertainties for the sources common
  to \gcrf{3} and ICRF3 S/X. The full set of 3142 sources is shown by the grey histograms, and
  the 259 defining sources by the red histograms. \textit{Left:} distribution of magnitudes $G$.
  \textit{Middle:} distribution of positional uncertainties in \gedr{3}. There are 126 sources with
  $\sigma_\text{pos,max}^\text{EDR3}>1$~mas, of which eight are defining sources in ICRF3.
  \textit{Right:} distribution of positional uncertainties in ICRF3 S/X. There are 155 sources with
  $\sigma_\text{pos,max}^\text{ICRF3}>1$~mas, of which one is a defining source.
  }
\label{fig:histo-sigpos-common}
\end{center}
\end{figure*}

The left panel in Fig.~\ref{fig:histo-sigpos-common} shows the magnitude distribution of
the 3142 \gcrf3 sources in common with ICRF3~S/X and for the subset of 259 sources 
that are defining sources in ICRF3. 
As expected, there is no marked difference between the two sets, since the optical brightness 
plays no role in the ICRF3 selection of defining sources. The median $G=18.9$ is about one 
magnitude brighter than for the whole \gcrf{3} (Table~\ref{tab:gaiacrf3-characteristics}).


The central panel in Fig.~\ref{fig:histo-sigpos-common} shows the distribution of the positional 
uncertainty $\sigma_\text{pos,max}$ in \gcrf{3} for the same selections. The median uncertainty 
is 194~$\muas$ for the whole set of ICRF3~S/X sources, and 162~$\muas$ for the 
defining sources. These values are less than one half of the median positional uncertainty in 
the full \gcrf{3} sample, which can be attributed to their different distributions in $G$. Indeed,
as shown by the middle panel of Fig.~\ref{fig:GCRF-accuracy}, the ICRF3 sources follow the 
same general variation of $\sigma_\text{pos,max}$ with $G$ as other \gcrf{3} sources. In other words,
nothing in the \gaia\ astrometry suggests that the ICRF3 are better or worse than 
other \gcrf{3} sources of comparable magnitude.

The right panel in Fig.~\ref{fig:histo-sigpos-common} shows the distribution of the positional 
uncertainty in the radio domain for the same sources. By coincidence, the median 
positional uncertainty of the 3142 sources in ICRF3, 197~$\muas$, is practically the same 
as in \gcrf{3}. For the subset of defining sources, however, the positional uncertainties in ICRF3 
are much smaller (the median is 39~$\muas$),  
as these sources were selected precisely for their high radio-astrometric quality. In \gcrf{3} 
the defining sources are only marginally more accurate than the non-defining sources, 
consistent with the former being, on average, about 0.2~mag brighter 
(cf.\ Fig.~\ref{fig:histo-sigpos-common}, left panel).



\subsection{Positional differences}
\label{sec:gcrf3posdiff}

We now compare the individual positions of ICRF3~S/X sources in ICRF3 and 
\gcrf{3}. Thanks to the high positional accuracy in both catalogues and the small 
cross-matching radius $\Delta_\text{max}=100$~mas used, cross-matching errors 
between the two catalogues are highly unlikely and we can assume that 
the optical and radio emissions being compared originate in broadly the same extragalactic object. 
Positional differences $\Delta\alpha{*}=(\alpha_\text{EDR3}-\alpha_\text{ICRF3})\cos\delta$ and
$\Delta\delta=\delta_\text{EDR3}-\delta_\text{ICRF3}$ are computed after propagating
the ICRF3 positions from the ICRF3 epoch (J2015.0) to the \gcrf3 epoch (J2016.0), using the
precepts described in \citetads[][Section~5.3]{2020A&A...644A.159C}; this takes into
account the small shifts ($\lesssim 5\,\muas$) expected from the Galactic acceleration.
 
\begin{figure}
\begin{center}
  \includegraphics[width=1.00\hsize]{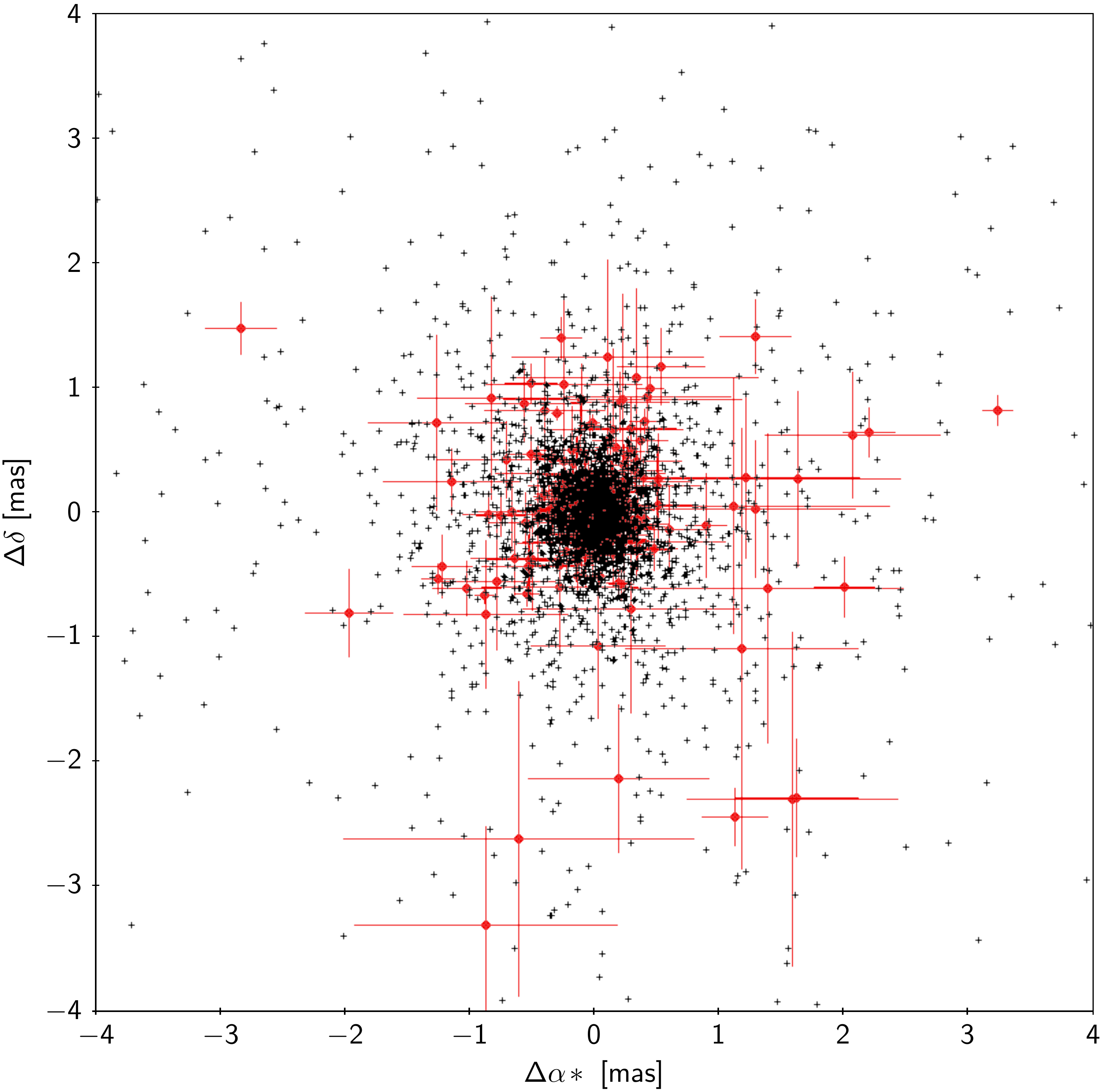}
  \includegraphics[width=1.00\hsize]{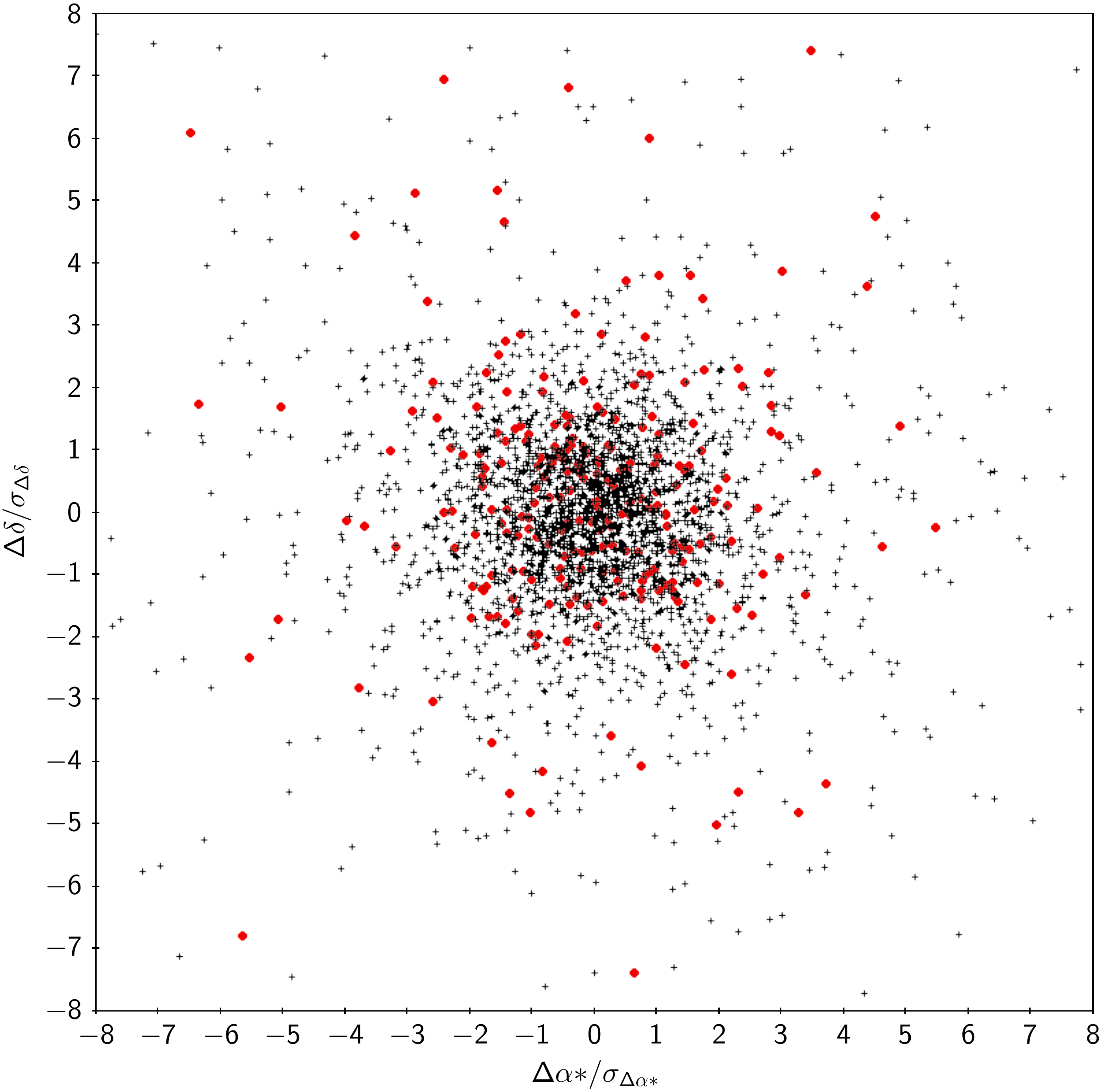}
  \caption{Positional differences in $\alpha$ and $\delta$ between \gcrf{3} and 
  ICRF3 S/X for the 3142 common sources. The 259 defining sources in ICRF3 
  are shown as filled red circles, other sources as black crosses.
  \textit{Top:} differences in mas. Errors bars representing the quadratically combined 
  uncertainties in the two catalogues are shown only for the defining sources.  
  \textit{Bottom:} the same differences divided by the quadratically combined uncertainties 
  in the two catalogues. No error bars are shown in this plot, as they would all be $\pm 1$ unit.
  }
\label{fig:scatter-draddec}
\end{center}
\end{figure}

\begin{figure}
\begin{center}
  \includegraphics[width=1.00\hsize]{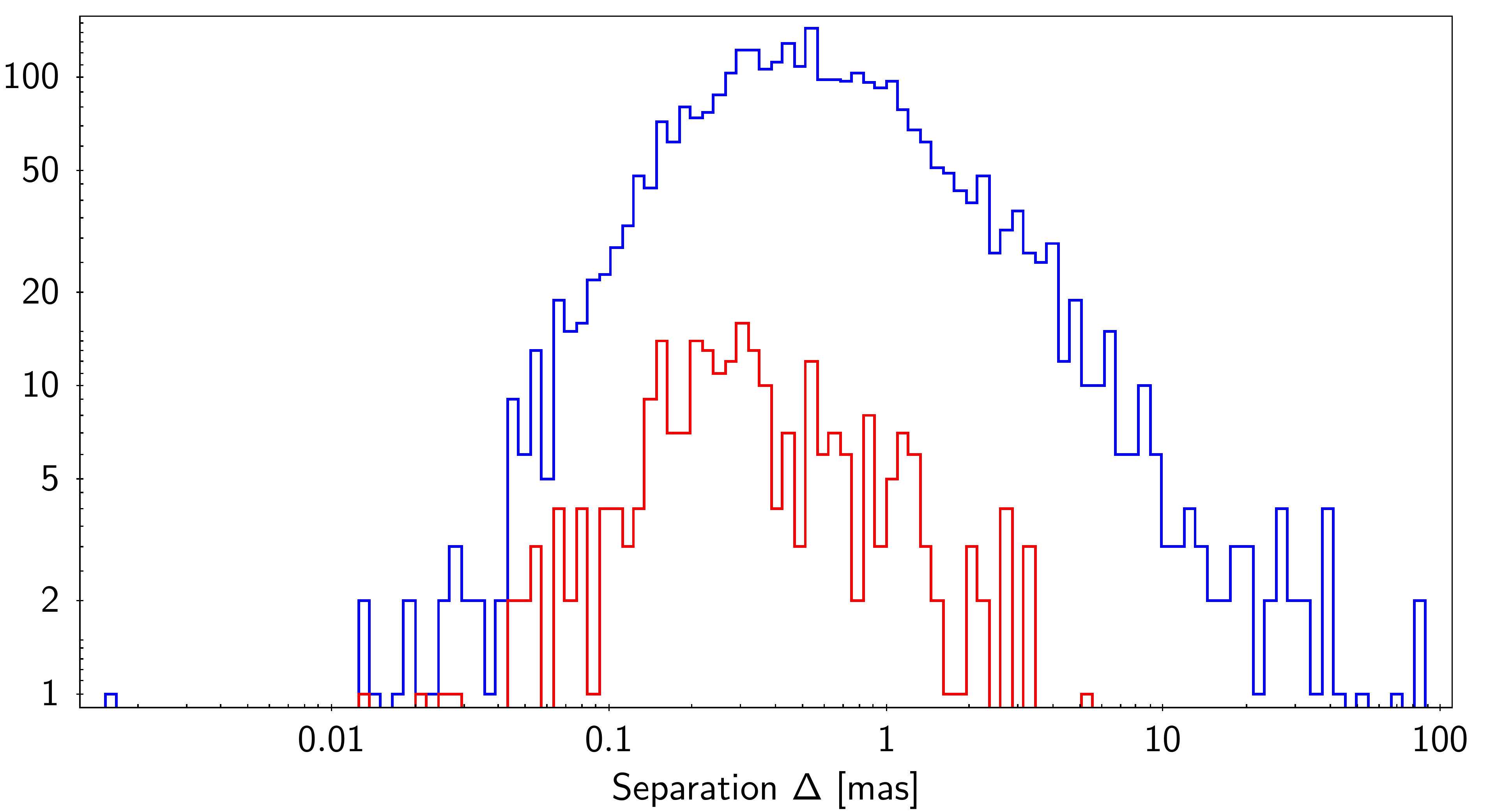}
  \caption{Histograms of the separations $\Delta$
  between \gcrf{3} and ICRF3 S/X for the full set of 3142 common sources (blue; the median is 0.516 mas) and the 
  259 defining sources (red; the median is 0.292 mas).
  }
\label{fig:histo-rho}
\end{center}
\end{figure}

The top panel of Fig.~\ref{fig:scatter-draddec} is a scatter diagram of the positional differences.
Of the 3142 sources, 127 (4\%) have a positional difference $>4$~mas in either coordinate and 
fall outside the boundaries of the plot. The median separation 
$\Delta=(\Delta\alpha{*}^2+\Delta\delta^2)^{1/2}$ is 0.516\,mas for all 3142 sources, 
and 0.292\,~mas for the 259 defining sources (Fig.~\ref{fig:histo-rho}).

\begin{figure*}
\begin{center}
  \includegraphics[width=0.49\hsize]{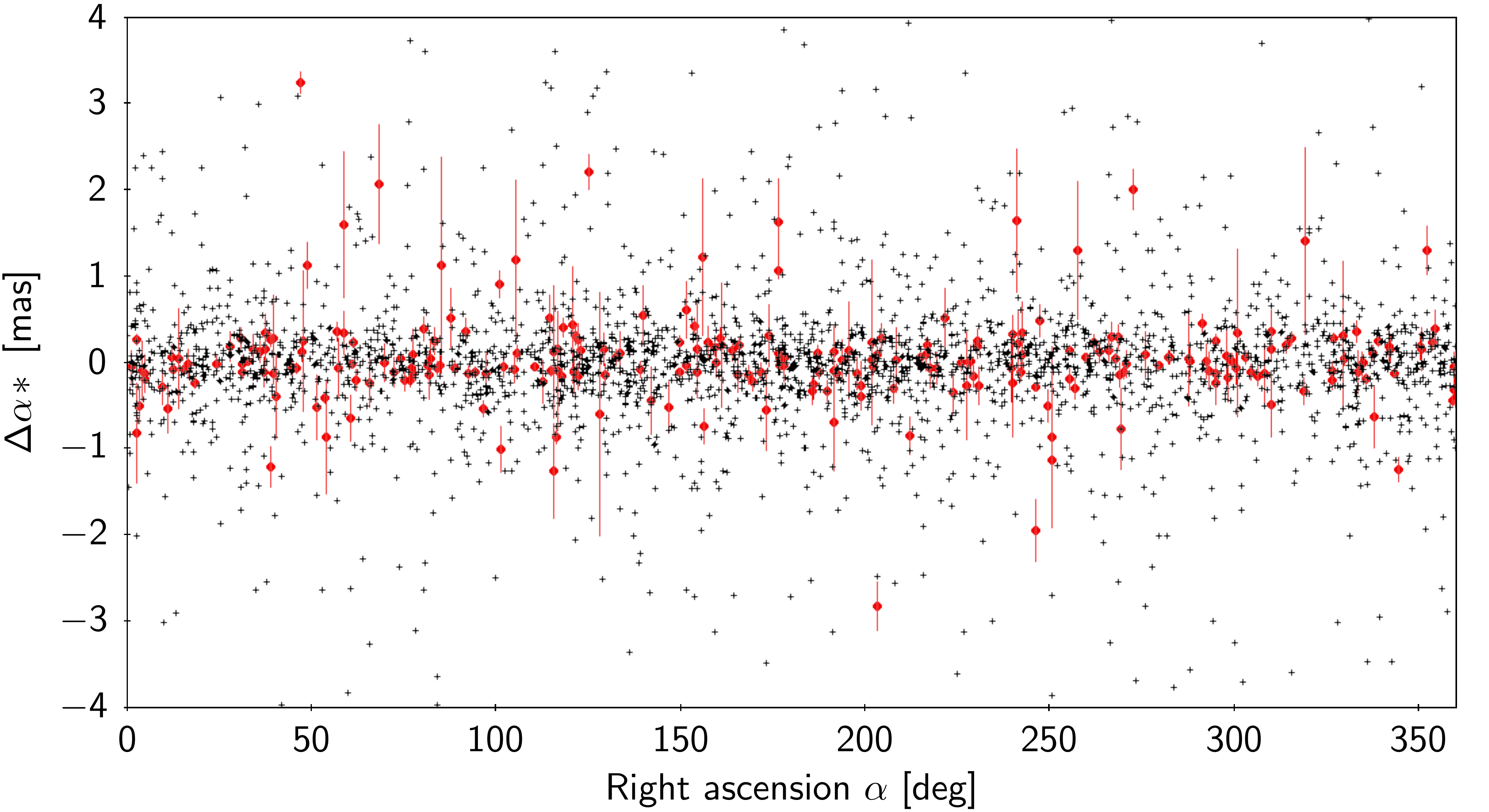}
  \includegraphics[width=0.49\hsize]{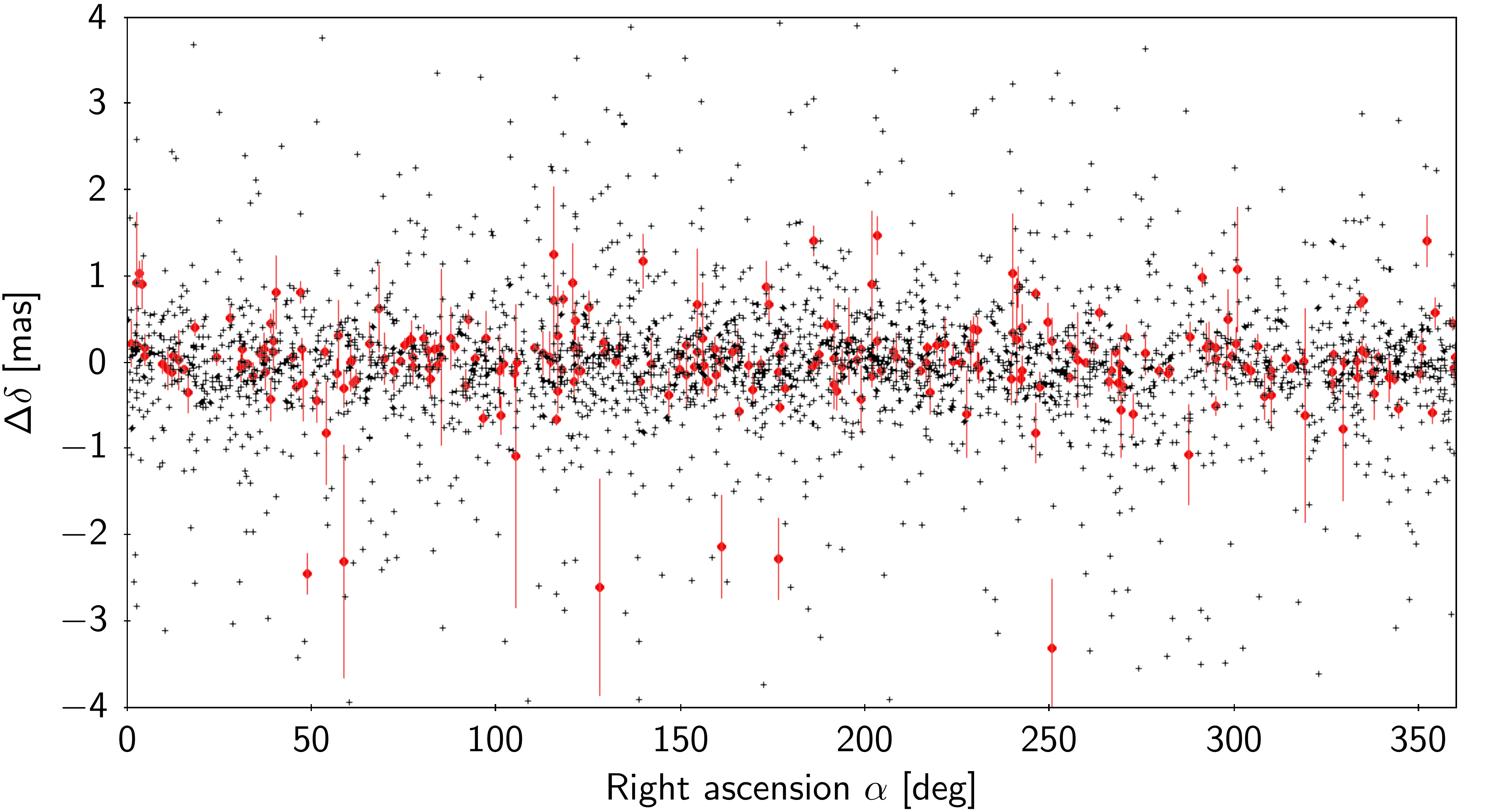}
  \includegraphics[width=0.49\hsize]{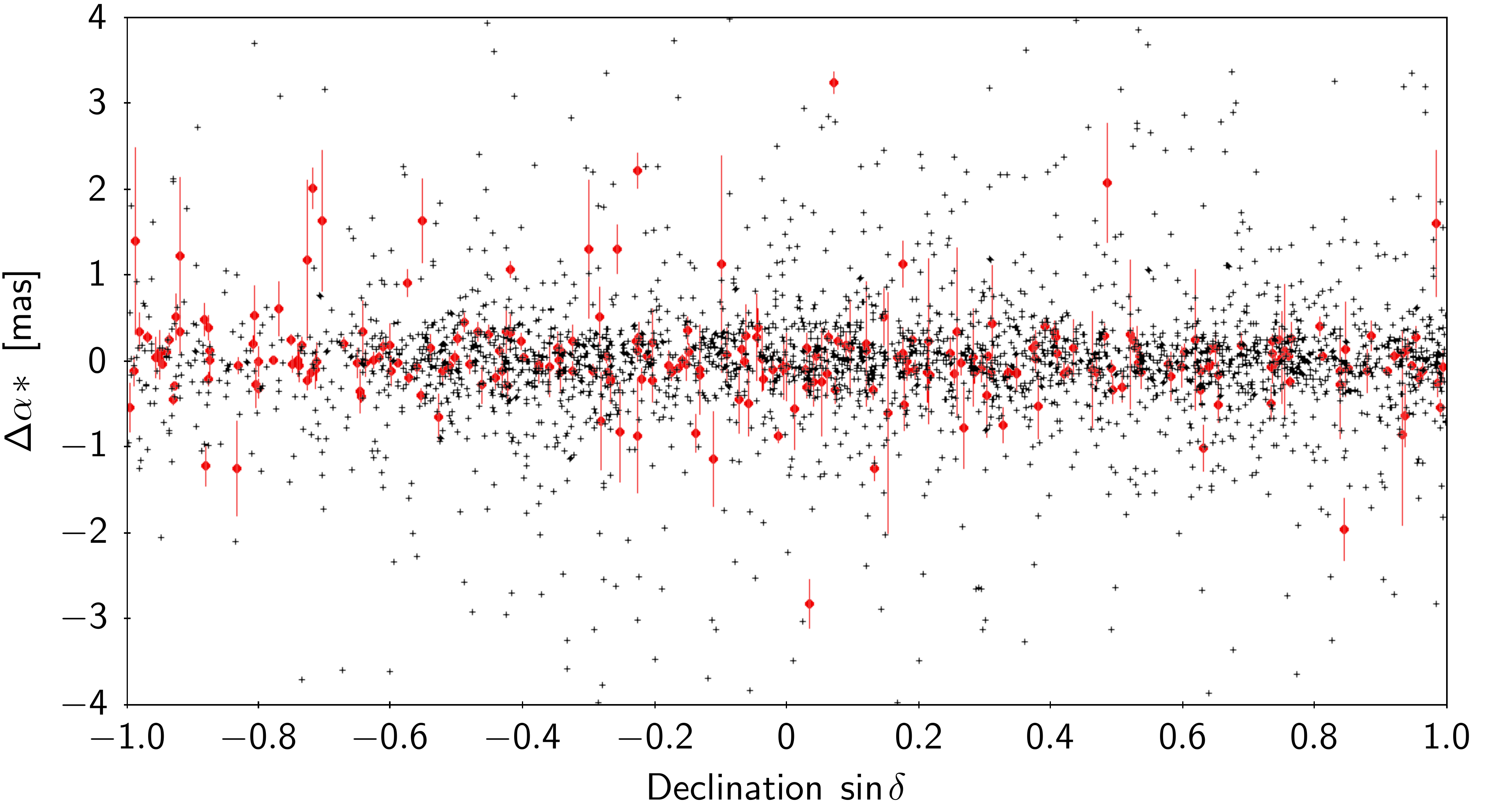}
  \includegraphics[width=0.49\hsize]{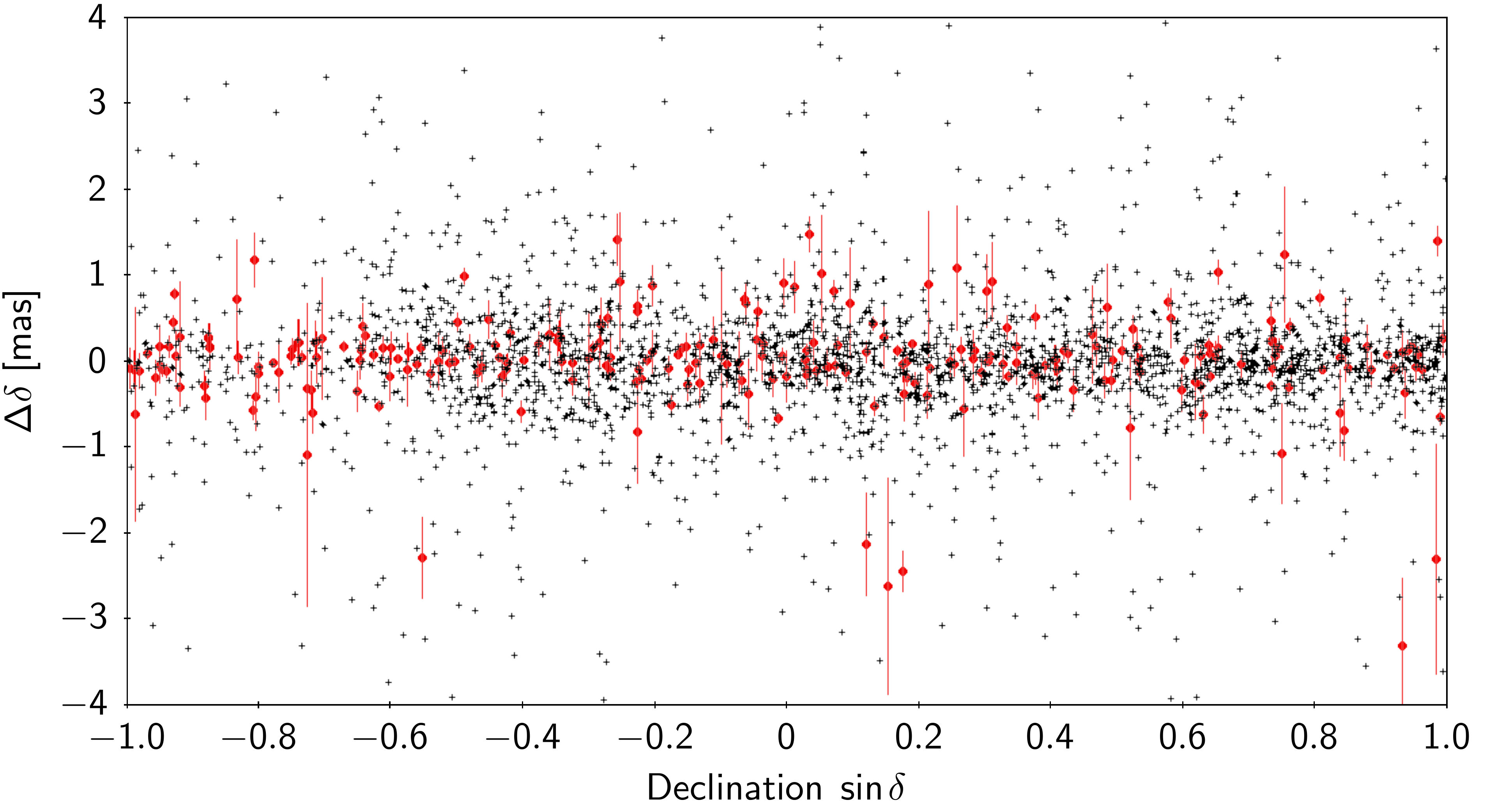}
  \caption{Positional differences between \gcrf{3} and ICRF3 S/X versus $\alpha$ (top) 
  and $\sin\delta$ (bottom). Defining sources in ICRF3 are shown as filled red circles, with
  errors bars representing the combined positional uncertainties in the two catalogues; other
  sources are shown as black crosses. 
  }
\label{fig:draddec_vsradec}
\end{center}
\end{figure*}

The distributions of $\Delta\alpha{*}$ and $\Delta\delta$ are roughly symmetric around
zero with a median offset of $+4\,\muas$ in right ascension and $-6\,\muas$ in declination.
Similarly, the scatter diagrams of the positional differences versus right ascension and declination 
(Fig.~\ref{fig:draddec_vsradec}) show no significant systematics or trends in either coordinate. 
The distributions are very regular with a pronounced  central concentration of width below 1~mas,
plus a scatter of several hundred outliers at several mas.
Furthermore, an analysis of the positional differences using VSH 
does not reveal any significant systematic effects. The statistical significance of the fitted coefficients
is low, while the largest effects represented by the VSH expansion are located in the areas with 
low source density (around the Galactic plane and for $\delta<-30^\circ$).

The lower panel of Fig.~\ref{fig:scatter-draddec} shows the same position differences as in the
upper panel, but normalised by the quadratic combination of the uncertainties in the ICRF3 
and \gcrf{3} catalogues. 
If the optical and radio centres of emission coincided, we would expect the normalised position 
differences to follow a normal distribution with standard deviation only slightly above 1.0. 
In fact, the distributions are significantly wider (the RSE%
\footnote{The robust scatter estimate (RSE) is defined as 
$\bigl[2\sqrt{2}\,\text{erf}^{-1}(4/5)\bigr]^{-1}\approx0.390152$ times the difference between 
the 90th and 10th percentiles of the distribution of the variable. For a Gaussian distribution it 
equals the standard deviation. The RSE is widely used in \gaia\ as a standardised, robust measure of dispersion.\label{fn:RSE}}
is 1.92 in right ascension and 1.80 in declination), with fatter tails than expected for a Gaussian:
225 sources fall outside the plotted frame, 
and more than 100 are outside $\pm5\times\text{RSE}$ in each coordinate.

\begin{figure}
\begin{center}
  \includegraphics[width=1.00\hsize]{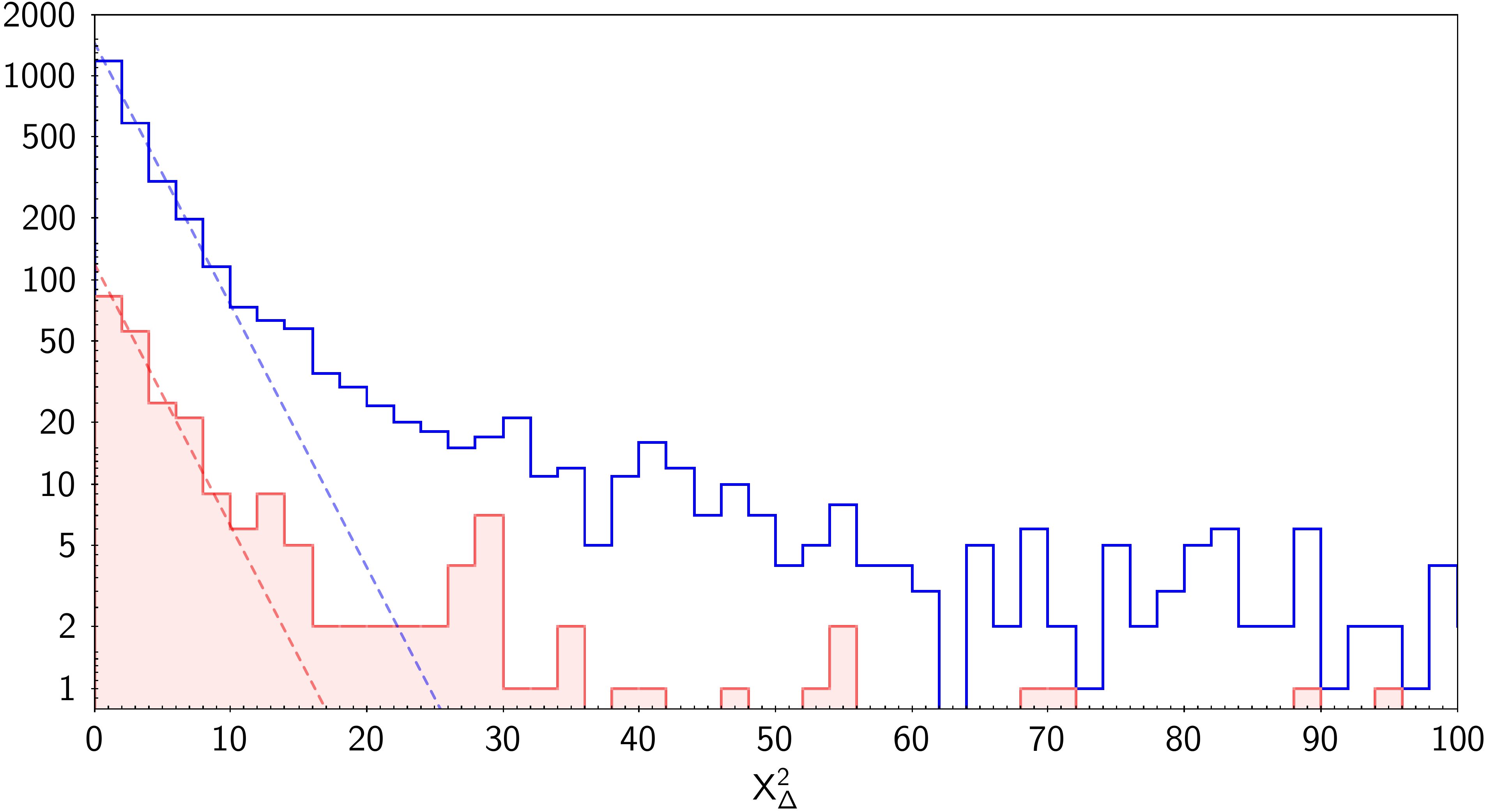}
  \caption{Histograms of the normalised separations $X_\Delta^2$ between \gcrf{3} and ICRF3 S/X
  for the full set of 3142 common sources (blue) and the 
  259 defining sources (red). The dashed curves are the exponential functions described in
  the text. A total of 195 sources have $X_\Delta>10$ and fall outside this plot (10 of them are defining sources
  in ICRF3).
  }
\label{fig:histo-X2sep}
\end{center}
\end{figure}

Both in ICRF3 and in \gcrf3 there are significant correlations between the errors in right ascension 
and declination, and a useful statistic that takes this into account is
\begin{equation}\label{eq:X2sep}
X_\Delta^2\equiv\begin{bmatrix} ~\Delta\alpha{*} && \Delta\delta~ \end{bmatrix}\,
\text{Cov}(\vec{\Delta})^{-1}
  \begin{bmatrix} ~\Delta\alpha{*} \\[6pt] 
  \Delta\delta \end{bmatrix}\,,
\end{equation}
where $\text{Cov}(\vec{\Delta})$ is the covariance matrix given by Eq.~(\ref{eq:covPos}).
This is completely analogous to the statistic $X_\mu^2$ for the proper motions, defined in 
Eq.~(\ref{eq:X2}), and like $\Delta$ it is independent of the reference frame used.%
\footnote{The quantity $X_\Delta^2$ is important also for another reason: as explained
in Appendix~\ref{sec:estimating} we consider the optical and radio frames to be aligned if
the sum of $X_\Delta^2$ over the `good' sources is minimal; cf.\ Eq.~(\ref{eq:obseqGen}). 
The equivalent statistic was introduced by \citetads{2016A&A...595A...5M}
already for the analysis of the reference frame in \gdr{1}; see their Eqs.~(1) and (3)--(5).}
The distributions of $X_\Delta^2$ are shown in Fig.~\ref{fig:histo-X2sep}. For Gaussian errors
we expect exponential distributions. The dashed lines are the functions 
$1450\exp[-\frac{1}{2}(X_\Delta/1.3)^2]$ and 
$120\exp[-\frac{1}{2}(X_\Delta/1.3)^2]$ fitted by eye to the 
inner parts of the histograms ($X_\Delta^2\lesssim 10$). The integrals of these exponential
are 2450.5 and 202.8, that is 78\% of respective sample. One possible interpretation
of this is that the (combined) uncertainties are generally underestimated by a factor 1.3, 
and that about 22\% of the objects in the samples have radio--optical positional offsets that are
statistically significant at the current level of accuracy in the two catalogues; this fraction 
is, moreover, the same for the defining sources as for the full sample. 
Using a criterion such as $X_\Delta^2>25$ to pick out the individual sources that have 
a significant offset yields 433 sources, 36 of which are defining ones, or 14\% of the respective samples.
The angular offsets of these sources range from 0.17~mas to tens of mas, with a median
value of 2.3~mas for the 433 sources from the full sample, and 0.84~mas for the 36 
defining sources.

We note that the median $X_\mu^2$ is slightly higher (2.36) for the 433 sources with 
$X_\Delta^2>25$ than for the full sample of 3142 sources (1.82). This could indicate that
the positional offsets are not constant over the time scale of the \gaia\ observations.


These results confirm and amplify the conclusion reached already with \gdr{1} and 
\gdr{2}, namely that radio--optical offsets are significant for a large fraction of the ICRF3
sources.  The correlation with jets has been well established, for example by 
\citetads{2019MNRAS.482.3023P} and \citetads{2019ApJ...871..143P}, and the physical 
signature of offsets is also supported in 
the studies by \citetads{2021A&A...651A..64L} and \citetads{2021A&A...652A..87L}. 


The question now is therefore not so much the existence of such offsets but rather
what consequences they have for the relation between the radio and optical
frames. A plain match between the two catalogues leaves sources with 
true differences in position, much larger than the expected random scatter.
These differences will not decrease with future versions of the radio and 
optical frames. Although this does not impact the quality of either realisation, 
it could limit the attainable accuracy of their mutual alignment. To circumvent
this problem, specific procedures may be required, such as the selection of a
common set of reference sources for the alignment. More generally, the 
importance of physical deviations between the reference frames at different 
wavelength bands needs to be better understood.
This problem will be even more important for the future high-accuracy astrometric observations
in the near infrared \citepads{2021ExA....51..783H}.

A closely related question concerns the interpretation and use of the
reference frame at different epochs.
ICRF3 is the first realisation of the VLBI frame that has
an epoch (2015.0) and an official procedure to transform from this
epoch to another \citepads{2020A&A...644A.159C}. The reason for the
epoch transformation is the effect of the acceleration of the Solar system
relative to the remote celestial sources, often referred to as 
`Galactic acceleration', although it may contain other components as well.
If ignored, the Galactic acceleration will slowly distort the frame at a rate 
of about 6~\muasyr. The effect should therefore be taken into account when 
comparing ICRF3 with the \gcrf{}, as was indeed done for this paper.
The transformation depends on three parameters that can be determined by 
observation; for ICRF3, however, a set of fixed values are specified.
In case of \gaia, full astrometric solutions including
parallaxes and proper motions are computed for all \gcrf{} sources.
Each version of \gcrf{} has a particular epoch common to that of the astrometric 
catalogue of the corresponding \gdr{}; for \gcrf{3} this is 2016.0. 
No specific correction for the Galactic acceleration is applied or recommended 
for \gcrf{}. However, the effect is implicitly contained in the measured proper 
motions and a normal epoch transformation would take also this effect into
account, albeit with a positional accuracy that degrades with the epoch difference. 
In \citetads{2021A&A...649A...9G} the components of the acceleration 
of the Solar system were determined from the proper motions of \gcrf{3}.
Although that determination differs from the ICRF3 prescription by less than 
1~\muasyr, there thus is a conceptual difference between how the effect is
treated in the two reference frames.


\section{Use of \gcrf{3} sources in the frame rotator of AGIS~3.2}
\label{sec:framerotator}

In the astrometric core solution for \textit{Gaia} (the Astrometric Global Iterative 
Solution, AGIS; \citeads{2012A&A...538A..78L}), the frame rotator ensures that 
the reference frame of the astrometric solution complies, in a global sense, with 
the ICRS requirements. The frame rotator uses a pre-selected set of QSO-like 
sources, including optical counterparts of ICRF3 sources, to estimate the six rotation 
parameters (spin and orientation) for the solutions obtained in successive iterations, 
and corrects them in such a way that the QSO-like sources have no net rotation 
and are globally aligned with the ICRF3 sources. 

The frame rotator sources are a subset of the much larger set of primary sources 
(about 14.3~million for EDR3) that contribute to the estimation of attitude, calibration, 
and global parameters in AGIS. Special selection rules apply to the primary sources; in 
particular, they must have colour information from \textit{Gaia}'s BP and RP photometers 
of adequate quality for the calibration of colour-dependent instrumental effects. 
The selection of primary sources, including the frame rotator sources, is revised with 
each new release of astrometric data, but for logistic reasons it is mainly based on the
astrometric and photometric data of the previous release. Thus, for \gedr{3} the 
selection of frame rotator sources was primarily based on the 556\,869 QSO-like 
sources, including 2820 ICRF3 counterparts, that constitute the \gcrf{2} 
\citepads{2018A&A...616A..14G}. However, not all of them had adequate colour 
information or passed the more stringent astrometric criteria for primary sources 
in \gedr{3}, and the number of frame rotator sources used for EDR3 is therefore 
substantially smaller (429\,249 sources). All of them have five-parameter solutions 
in \gedr{3} and were filtered by a procedure similar to the one used for \gcrf{3} 
but using a preliminary version of the \gedr{3} catalogue known as AGIS~3.1
\citepads{2021A&A...649A...2L}. 

Details of the frame rotator as used for \gedr{3} are given in Appendix~\ref{sec:estimating}.
As described in the appendix, the orientation ($\vec{\varepsilon}$) and spin ($\vec{\omega}$)
are estimated by separate least-squares solutions to the position differences of the ICRF3
sources (for the orientation) and to the proper motions of the many more QSO-like sources
(for the spin). The algorithm includes an iterative procedure to identify and eliminate outliers, 
that is sources that deviate very significantly from the overall fit in one or both coordinates. 
Thus, we need to distinguish between the sources that were `considered' by the algorithm, and
the subset of non-outliers that was actually `used' to determine the orientation or spin.
The auxiliary table \texttt{gaiaedr3.frame\_rotator\_source} in the \textit{Gaia} Archive lists 
the 429\,249 frame rotator sources for \gedr{3} and contains flags to indicate which of them
were considered and used for the orientation and spin solutions. 
As specified in the table, 428\,034 of the sources considered for the spin were actually used 
to determine the spin in the final iteration of the primary solution, while 1215 (0.3\%) were 
rejected as outliers. Of the 2269 sources with counterparts in ICRF3 S/X considered for the 
orientation, 2007 sources were actually used to determine the orientation, while 262 (11.5\%) 
were rejected as outliers. Eight of the sources were rejected in both the spin and orientation 
solutions.


Because the preliminary AGIS~3.1 astrometry had to be used to select the frame rotator 
sources, it turns out that 46 of the latter do not pass the stricter criteria for the \gcrf{3}
selection described in Sect.~\ref{sec:selection}. These sources are thus listed in 
\texttt{gaiaedr3.frame\_rotator\_source} although they are not part of the \gcrf{3}. 
This slight inconsistency has minimal impact on the resulting reference frame, as only two 
of the 46 sources were accepted by the frame rotator algorithm for the determination of 
the spin, and none of them for the determination of the orientation.

If the frame rotator algorithm is applied to this subset of sources, using the astrometric
data as given in \gedr{3}, the resulting estimates of the orientation and spin vectors are not
strictly zero, as might be expected. For the spin, we obtain 
$\left|\,\vec{\omega}\,\right|\simeq 0.8\,\muasyr$ with an uncertainty of about 
$0.4~\muasyr$ in each axis; for the orientation, the result is  
$\left|\,\vec{\varepsilon}(2016.0)\,\right|\simeq 6.2~\muas$ with an uncertainty of about 
$7~\muas$ per axis. These deviations from zero can be traced to small differences in the positions 
and proper motions between the last iteration of the primary solution and the secondary source 
updates, from which the final data are taken. The reason for this (undesirable) effect is currently 
under investigation.

If the same algorithm is instead applied to the full set of \gcrf{3} sources, the 
result for the spin is
\begin{equation}\label{eq:omegaFR}
\vec{{\omega}} =
\begin{bmatrix} -3.44 \pm 0.30 \\ +1.57 \pm 0.28 \\ -1.24 \pm 0.32 \end{bmatrix}
~\text{$\mu$as~yr$^{-1}$,}
\end{equation}
in which 1\,607\,289 (99.6\%) of the sources are accepted; for the
orientation correction the result is
\begin{equation}\label{eq:epsilonFR}
\vec{{\varepsilon}}(2016.0) =
\begin{bmatrix} +3.41 \pm 6.71 \\ +8.99 \pm 6.50 \\ -1.47 \pm 6.04 \end{bmatrix}
~\text{$\mu$as,}
\end{equation}
in which 2738 (87.1\%) of the 3142 ICRF3 S/X sources are accepted. Generally
speaking, the result is sensitive to the precise selection of sources at the level of 
a few $\muasyr$ in spin and several $\muas$ in the orientation.

\section{Conclusions and prospects}
\label{sec:summary}

In this paper we have reported on the construction and properties of
the {\gcrf{3}}, a celestial reference frame materialised in the
optical domain resulting from the \gaia\ all-sky survey. Although
built on only 33 months of data (from August 2014 to May 2017), the
astrometric accuracy is already very similar to what was expected for
the 5-year nominal mission, and with a larger set of QSOs than
originally anticipated.

The \gcrf{3} consists of an astrometric catalogue of more than 1.6~million QSOs
selected using a number of existing catalogues, with further filtering based on the
\gaia\ data to ensure that the sample is as much as possible free of
stellar contaminants. The $G$ magnitude extends from ${\sim\,}13$ to 
${\sim\,}21$~mag, with a peak density at $G\simeq 20.5$~mag. 
The formal uncertainty is primarily ruled by the $G$ magnitude,
with a typical precision of 1~mas at $G=20.6$, 0.4~mas at $G=20$ and
0.1~mas at $G=17.9$. There are 32\,000 sources with formal position
uncertainty ${<\,}0.1$~mas and 210\,000 with ${<\,}0.2$~mas. \gcrf{3} is aligned 
to the radio ICRF3 by minimising the position differences for about 2000 common 
sources with high-quality solutions.

The comparison with ICRF3 shows that the floor formal accuracies are 
comparable in the optical and radio domains and better than 0.05~mas for 
the best measured sources. However the normalised positional differences are
not compatible with a standard normal distribution and an estimated
${\sim\,}20\%$ of the common sources have statistically significant 
positional radio--optical offsets. For many sources, if not all, these 
offsets are probably real, although this is hard to prove without a detailed
investigation at the source level
\citepads{2019MNRAS.482.3023P,2019ApJ...871..143P}. The radio and
optical CRFs are both internally very consistent and materialise the
ICRS in their respective wavelength range, but in view of the detected 
offsets it remains unclear how much their mutual consistency can be
improved without a better knowledge of the physical processes responsible 
for the light and radio emissions. It is hoped that high-resolution observations 
and improved radio and optical positions will further characterise 
these offsets and lead to a better understanding of the underlying 
mechanisms \citepads{2019BAAS...51c.273J}.

It should be understood that \gcrf{3} is entirely defined by the \gedr{3} data for the
$\sim\,$1.6~million QSO-like sources discussed in this paper. These data include
not only the barycentric positions of the sources at the reference epoch 2016.0,
but also their proper motions (and parallaxes). Although individually insignificant
in the vast majority of cases, these data provide invaluable information on the 
quality of the data and on possible systematic effects in the reference frame, both 
of instrumental and astrophysical origin.  
The reference frame defined by the proper motions of the \gcrf{3} sources is 
globally non-rotating at a level of a few $\muasyr$. It is not possible to quantify
more precisely than this how well the \gcrf{3} complies with the `no rotation' condition, 
because different subsets of the \gcrf{3} lead to determinations of the spin that differ 
on this level, which is also consistent with the estimated RMS level of systematics on 
large angular scales (low values of $\ell_\text{max}$) depicted in Fig.~\ref{fig:Rlt-Rls-Rl}.
As mentioned in Sect.~\ref{sec:intro}, the secondary frame defined by the positions
and proper motions of stars in \gedr{3}, in particular at bright magnitudes, may have
significantly higher systematics.

\gcrf{3} is by far the best realisation to date of the celestial reference 
frame in the visible. It meets the ICRS requirements for a realisation 
based on extragalactic sources. The IAU, during its
XXXIst General Assembly in August 2021, resolved (Resolution B3)
that ICRF3 at radio wavelengths and \gcrf{3} in the optical are
realisations (ICRF) of the International Celestial Reference System
(ICRS) to be used as a fundamental standard in their respective domain.






A new version of the {\gcrf{}} comes out with every release of a new astrometric 
solution based on a larger volume of \gaia\ data. The current version, \gcrf{3},
is common to \gedr3 and \gdr{3}, for which the basic astrometric data are identical.
Two more releases are expected, \gdr{4} based on 66 months of 
observations, and \gdr{5} using up to 120 months of data. 
The positional accuracy improves roughly as $T^{-1/2}$,
where $T$ is the time baseline, leading to an improvement by almost 
a factor two for the final version compared with \gcrf{3}. 

It is expected that the systematics will also be improved in the future releases, 
although this cannot easily be quantified. A dramatic illustration of the improvement
from \gcrf{2} to \gcrf{3} is the successful use of the latter to determine the acceleration 
of the Solar System \citepads{2021A&A...649A...9G}. Of paramount importance for
the quality of the reference frame is the accuracy of proper motions, which impacts
both the astrometric filtering of the sources (thus reducing stellar contamination) and
the accuracy of the positions at earlier and later epochs. Proper motions generally 
improve as $T^{-3/2}$, which could give a factor seven more precise proper motions 
in the final release. Corresponding improvements can be expected for the determination 
of the reference frame (especially, its spin) as well as for the various physical effects that
may be seen in the proper motions (e.g.\ the solar system acceleration; see
Sect.~8 in \citeads{2021A&A...649A...9G} for a brief discussion of other effects).

As mentioned in Sect.~\ref{sec:properties}, up to 4~million QSOs could
be recognised by \gaia. In future releases the identification of
the QSOs will be refined using a combination of all kinds of
\gaia\ data: an improved astrometry, spectrophotometry, and
variability analysis. Special efforts will be made to better identify
QSOs in the crowded areas and to improve the homogeneity of the sky coverage
(a reduced source density at the Galactic plane will however persist due to the effects of Galactic extinction).
The prospect for the
future versions of the \gcrf{} therefore lies not only in a further
improvement of positional accuracy, but even more in the number of
sources, a better homogeneity of the source distribution of the sky,
the purity of the selection and in the overall consistency of the
solution.
 




\begin{acknowledgements}
The acknowledgements are available in Appendix \ref{appendix-acklow}.
\end{acknowledgements}  


\bibliographystyle{aa} 
\bibliography{galacc} 

\begin{appendix}

%

\section{Acknowledgments}
\label{appendix-acklow}


This work presents results from the European Space Agency (ESA) space mission \gaia. \gaia\ data are being processed by the \gaia\ Data Processing and Analysis Consortium (DPAC). Funding for the DPAC is provided by national institutions, in particular the institutions participating in the \gaia\ MultiLateral Agreement (MLA). The \gaia\ mission website is \url{https://www.cosmos.esa.int/gaia}. The \gaia\ archive website is \url{https://archives.esac.esa.int/gaia}.

The \gaia\ mission and data processing have financially been supported by, in alphabetical order by country:

 the Algerian Centre de Recherche en Astronomie, Astrophysique et G\'{e}ophysique of Bouzareah Observatory;
 the Austrian Fonds zur F\"{o}rderung der wissenschaftlichen Forschung (FWF) Hertha Firnberg Programme through grants T359, P20046, and P23737;
 the BELgian federal Science Policy Office (BELSPO) through various PROgramme de D\'{e}veloppement d'Exp\'{e}riences scientifiques (PRODEX) grants and the Polish Academy of Sciences - Fonds Wetenschappelijk Onderzoek through grant VS.091.16N, and the Fonds de la Recherche Scientifique (FNRS), and the Research Council of Katholieke Universiteit (KU) Leuven through grant C16/18/005 (Pushing AsteRoseismology to the next level with TESS, GaiA, and the Sloan DIgital Sky SurvEy -- PARADISE);  
 the Brazil-France exchange programmes Funda\c{c}\~{a}o de Amparo \`{a} Pesquisa do Estado de S\~{a}o Paulo (FAPESP) and Coordena\c{c}\~{a}o de Aperfeicoamento de Pessoal de N\'{\i}vel Superior (CAPES) - Comit\'{e} Fran\c{c}ais d'Evaluation de la Coop\'{e}ration Universitaire et Scientifique avec le Br\'{e}sil (COFECUB);
 the Chilean Agencia Nacional de Investigaci\'{o}n y Desarrollo (ANID) through Fondo Nacional de Desarrollo Cient\'{\i}fico y Tecnol\'{o}gico (FONDECYT) Regular Project 1210992 (L.~Chemin);
 the National Natural Science Foundation of China (NSFC) through grants 11573054, 11703065, and 12173069, the China Scholarship Council through grant 201806040200, and the Natural Science Foundation of Shanghai through grant 21ZR1474100;  
 the Tenure Track Pilot Programme of the Croatian Science Foundation and the \'{E}cole Polytechnique F\'{e}d\'{e}rale de Lausanne and the project TTP-2018-07-1171 `Mining the Variable Sky', with the funds of the Croatian-Swiss Research Programme;
 the Czech-Republic Ministry of Education, Youth, and Sports through grant LG 15010 and INTER-EXCELLENCE grant LTAUSA18093, and the Czech Space Office through ESA PECS contract 98058;
 the Danish Ministry of Science;
 the Estonian Ministry of Education and Research through grant IUT40-1;
 the European Commission’s Sixth Framework Programme through the European Leadership in Space Astrometry (\href{https://www.cosmos.esa.int/web/gaia/elsa-rtn-programme}{ELSA}) Marie Curie Research Training Network (MRTN-CT-2006-033481), through Marie Curie project PIOF-GA-2009-255267 (Space AsteroSeismology \& RR Lyrae stars, SAS-RRL), and through a Marie Curie Transfer-of-Knowledge (ToK) fellowship (MTKD-CT-2004-014188); the European Commission's Seventh Framework Programme through grant FP7-606740 (FP7-SPACE-2013-1) for the \gaia\ European Network for Improved data User Services (\href{https://gaia.ub.edu/twiki/do/view/GENIUS/}{GENIUS}) and through grant 264895 for the \gaia\ Research for European Astronomy Training (\href{https://www.cosmos.esa.int/web/gaia/great-programme}{GREAT-ITN}) network;
 the European Cooperation in Science and Technology (COST) through COST Action CA18104 `Revealing the Milky Way with \gaia\ (MW-Gaia)';
 the European Research Council (ERC) through grants 320360, 647208, and 834148 and through the European Union’s Horizon 2020 research and innovation and excellent science programmes through Marie Sk{\l}odowska-Curie grant 745617 (Our Galaxy at full HD -- Gal-HD) and 895174 (The build-up and fate of self-gravitating systems in the Universe) as well as grants 687378 (Small Bodies: Near and Far), 682115 (Using the Magellanic Clouds to Understand the Interaction of Galaxies), 695099 (A sub-percent distance scale from binaries and Cepheids -- CepBin), 716155 (Structured ACCREtion Disks -- SACCRED), 951549 (Sub-percent calibration of the extragalactic distance scale in the era of big surveys -- UniverScale), and 101004214 (Innovative Scientific Data Exploration and Exploitation Applications for Space Sciences -- EXPLORE);
 the European Science Foundation (ESF), in the framework of the \gaia\ Research for European Astronomy Training Research Network Programme (\href{https://www.cosmos.esa.int/web/gaia/great-programme}{GREAT-ESF});
 the European Space Agency (ESA) in the framework of the \gaia\ project, through the Plan for European Cooperating States (PECS) programme through contracts C98090 and 4000106398/12/NL/KML for Hungary, through contract 4000115263/15/NL/IB for Germany, and through PROgramme de D\'{e}veloppement d'Exp\'{e}riences scientifiques (PRODEX) grant 4000127986 for Slovenia;  
 the Academy of Finland through grants 299543, 307157, 325805, 328654, 336546, and 345115 and the Magnus Ehrnrooth Foundation;
 the French Centre National d’\'{E}tudes Spatiales (CNES), the Agence Nationale de la Recherche (ANR) through grant ANR-10-IDEX-0001-02 for the `Investissements d'avenir' programme, through grant ANR-15-CE31-0007 for project `Modelling the Milky Way in the \gaia\ era’ (MOD4Gaia), through grant ANR-14-CE33-0014-01 for project `The Milky Way disc formation in the \gaia\ era’ (ARCHEOGAL), through grant ANR-15-CE31-0012-01 for project `Unlocking the potential of Cepheids as primary distance calibrators’ (UnlockCepheids), through grant ANR-19-CE31-0017 for project `Secular evolution of galxies' (SEGAL), and through grant ANR-18-CE31-0006 for project `Galactic Dark Matter' (GaDaMa), the Centre National de la Recherche Scientifique (CNRS) and its SNO \gaia\ of the Institut des Sciences de l’Univers (INSU), its Programmes Nationaux: Cosmologie et Galaxies (PNCG), Gravitation R\'{e}f\'{e}rences Astronomie M\'{e}trologie (PNGRAM), Plan\'{e}tologie (PNP), Physique et Chimie du Milieu Interstellaire (PCMI), and Physique Stellaire (PNPS), the `Action F\'{e}d\'{e}ratrice \gaia' of the Observatoire de Paris, the R\'{e}gion de Franche-Comt\'{e}, the Institut National Polytechnique (INP) and the Institut National de Physique nucl\'{e}aire et de Physique des Particules (IN2P3) co-funded by CNES;
 the German Aerospace Agency (Deutsches Zentrum f\"{u}r Luft- und Raumfahrt e.V., DLR) through grants 50QG0501, 50QG0601, 50QG0602, 50QG0701, 50QG0901, 50QG1001, 50QG1101, 50\-QG1401, 50QG1402, 50QG1403, 50QG1404, 50QG1904, 50QG2101, 50QG2102, and 50QG2202, and the Centre for Information Services and High Performance Computing (ZIH) at the Technische Universit\"{a}t Dresden for generous allocations of computer time;
 the Hungarian Academy of Sciences through the Lend\"{u}let Programme grants LP2014-17 and LP2018-7 and the Hungarian National Research, Development, and Innovation Office (NKFIH) through grant KKP-137523 (`SeismoLab');
 the Science Foundation Ireland (SFI) through a Royal Society - SFI University Research Fellowship (M.~Fraser);
 the Israel Ministry of Science and Technology through grant 3-18143 and the Tel Aviv University Center for Artificial Intelligence and Data Science (TAD) through a grant;
 the Agenzia Spaziale Italiana (ASI) through contracts I/037/08/0, I/058/10/0, 2014-025-R.0, 2014-025-R.1.2015, and 2018-24-HH.0 to the Italian Istituto Nazionale di Astrofisica (INAF), contract 2014-049-R.0/1/2 to INAF for the Space Science Data Centre (SSDC, formerly known as the ASI Science Data Center, ASDC), contracts I/008/10/0, 2013/030/I.0, 2013-030-I.0.1-2015, and 2016-17-I.0 to the Aerospace Logistics Technology Engineering Company (ALTEC S.p.A.), INAF, and the Italian Ministry of Education, University, and Research (Ministero dell'Istruzione, dell'Universit\`{a} e della Ricerca) through the Premiale project `MIning The Cosmos Big Data and Innovative Italian Technology for Frontier Astrophysics and Cosmology' (MITiC);
 the Netherlands Organisation for Scientific Research (NWO) through grant NWO-M-614.061.414, through a VICI grant (A.~Helmi), and through a Spinoza prize (A.~Helmi), and the Netherlands Research School for Astronomy (NOVA);
 the Polish National Science Centre through HARMONIA grant 2018/30/M/ST9/00311 and DAINA grant 2017/27/L/ST9/03221 and the Ministry of Science and Higher Education (MNiSW) through grant DIR/WK/2018/12;
 the Portuguese Funda\c{c}\~{a}o para a Ci\^{e}ncia e a Tecnologia (FCT) through national funds, grants SFRH/\-BD/128840/2017 and PTDC/FIS-AST/30389/2017, and work contract DL 57/2016/CP1364/CT0006, the Fundo Europeu de Desenvolvimento Regional (FEDER) through grant POCI-01-0145-FEDER-030389 and its Programa Operacional Competitividade e Internacionaliza\c{c}\~{a}o (COMPETE2020) through grants UIDB/04434/2020 and UIDP/04434/2020, and the Strategic Programme UIDB/\-00099/2020 for the Centro de Astrof\'{\i}sica e Gravita\c{c}\~{a}o (CENTRA);  
 the Slovenian Research Agency through grant P1-0188;
 the Spanish Ministry of Economy (MINECO/FEDER, UE), the Spanish Ministry of Science and Innovation (MICIN), the Spanish Ministry of Education, Culture, and Sports, and the Spanish Government through grants BES-2016-078499, BES-2017-083126, BES-C-2017-0085, ESP2016-80079-C2-1-R, ESP2016-80079-C2-2-R, FPU16/03827, PDC2021-121059-C22, RTI2018-095076-B-C22, and TIN2015-65316-P (`Computaci\'{o}n de Altas Prestaciones VII'), the Juan de la Cierva Incorporaci\'{o}n Programme (FJCI-2015-2671 and IJC2019-04862-I for F.~Anders), the Severo Ochoa Centre of Excellence Programme (SEV2015-0493), and MICIN/AEI/10.13039/501100011033 (and the European Union through European Regional Development Fund `A way of making Europe') through grant RTI2018-095076-B-C21, the Institute of Cosmos Sciences University of Barcelona (ICCUB, Unidad de Excelencia `Mar\'{\i}a de Maeztu’) through grant CEX2019-000918-M, the University of Barcelona's official doctoral programme for the development of an R+D+i project through an Ajuts de Personal Investigador en Formaci\'{o} (APIF) grant, the Spanish Virtual Observatory through project AyA2017-84089, the Galician Regional Government, Xunta de Galicia, through grants ED431B-2021/36, ED481A-2019/155, and ED481A-2021/296, the Centro de Investigaci\'{o}n en Tecnolog\'{\i}as de la Informaci\'{o}n y las Comunicaciones (CITIC), funded by the Xunta de Galicia and the European Union (European Regional Development Fund -- Galicia 2014-2020 Programme), through grant ED431G-2019/01, the Red Espa\~{n}ola de Supercomputaci\'{o}n (RES) computer resources at MareNostrum, the Barcelona Supercomputing Centre - Centro Nacional de Supercomputaci\'{o}n (BSC-CNS) through activities AECT-2017-2-0002, AECT-2017-3-0006, AECT-2018-1-0017, AECT-2018-2-0013, AECT-2018-3-0011, AECT-2019-1-0010, AECT-2019-2-0014, AECT-2019-3-0003, AECT-2020-1-0004, and DATA-2020-1-0010, the Departament d'Innovaci\'{o}, Universitats i Empresa de la Generalitat de Catalunya through grant 2014-SGR-1051 for project `Models de Programaci\'{o} i Entorns d'Execuci\'{o} Parallels' (MPEXPAR), and Ramon y Cajal Fellowship RYC2018-025968-I funded by MICIN/AEI/10.13039/501100011033 and the European Science Foundation (`Investing in your future');
 the Swedish National Space Agency (SNSA/Rymdstyrelsen);
 the Swiss State Secretariat for Education, Research, and Innovation through the Swiss Activit\'{e}s Nationales Compl\'{e}mentaires and the Swiss National Science Foundation through an Eccellenza Professorial Fellowship (award PCEFP2\_194638 for R.~Anderson);
 the United Kingdom Particle Physics and Astronomy Research Council (PPARC), the United Kingdom Science and Technology Facilities Council (STFC), and the United Kingdom Space Agency (UKSA) through the following grants to the University of Bristol, the University of Cambridge, the University of Edinburgh, the University of Leicester, the Mullard Space Sciences Laboratory of University College London, and the United Kingdom Rutherford Appleton Laboratory (RAL): PP/D006511/1, PP/D006546/1, PP/D006570/1, ST/I000852/1, ST/J005045/1, ST/K00056X/1, ST/\-K000209/1, ST/K000756/1, ST/L006561/1, ST/N000595/1, ST/N000641/1, ST/N000978/1, ST/\-N001117/1, ST/S000089/1, ST/S000976/1, ST/S000984/1, ST/S001123/1, ST/S001948/1, ST/\-S001980/1, ST/S002103/1, ST/V000969/1, ST/W002469/1, ST/W002493/1, ST/W002671/1, ST/W002809/1, and EP/V520342/1.

Further acknowledgments can be found at \url{https://gea.esac.esa.int/archive/documentation/GEDR3/Miscellaneous/sec_acknowl/}.


\clearpage

\section{Source distributions in external catalogues}
\label{sec:external}

\def\scaleext{0.28}

\begin{sidewaysfigure*}
\begin{center}
%
\includegraphics[width=\scaleext\hsize]{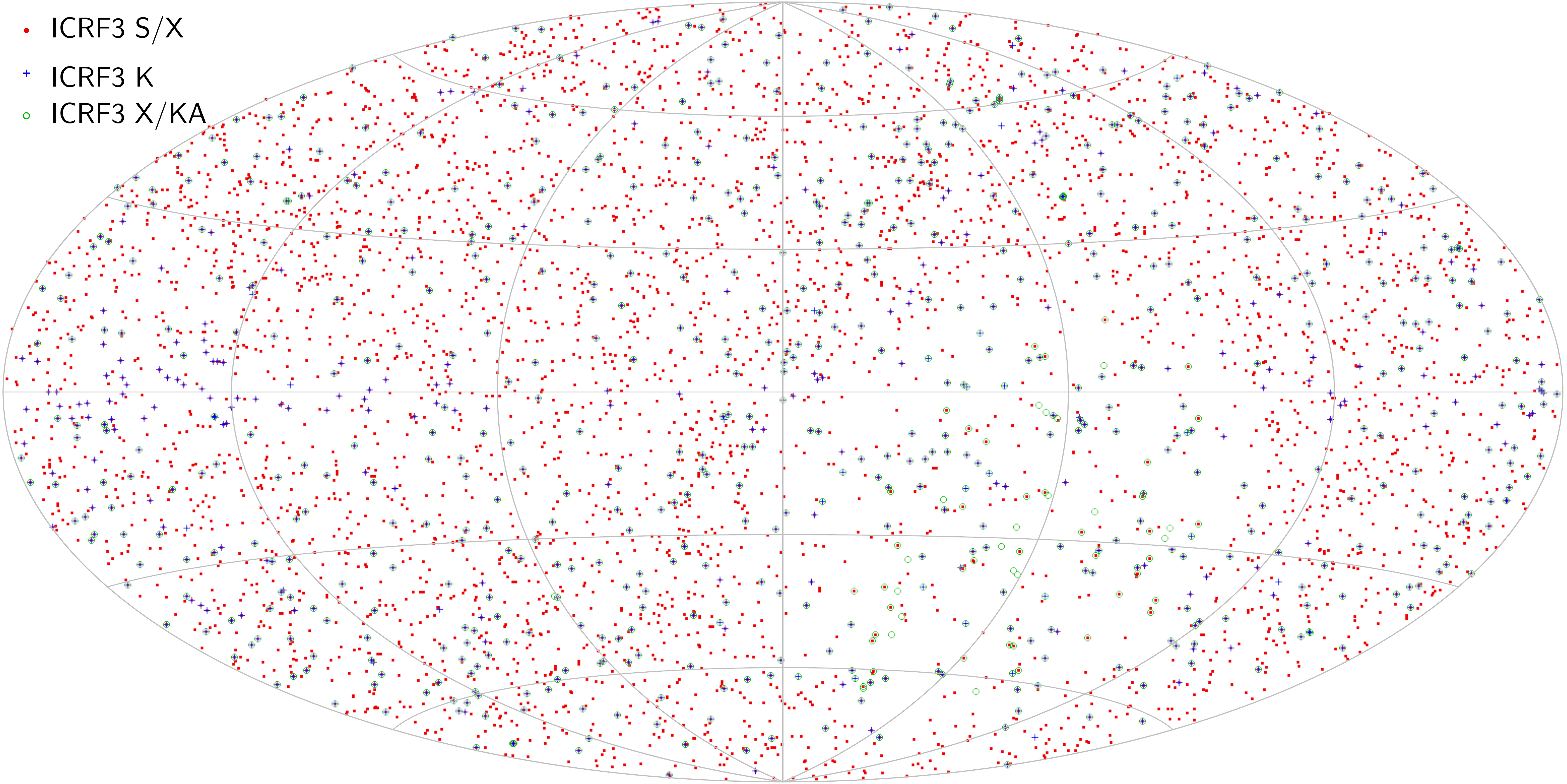}
\includegraphics[width=\scaleext\hsize]{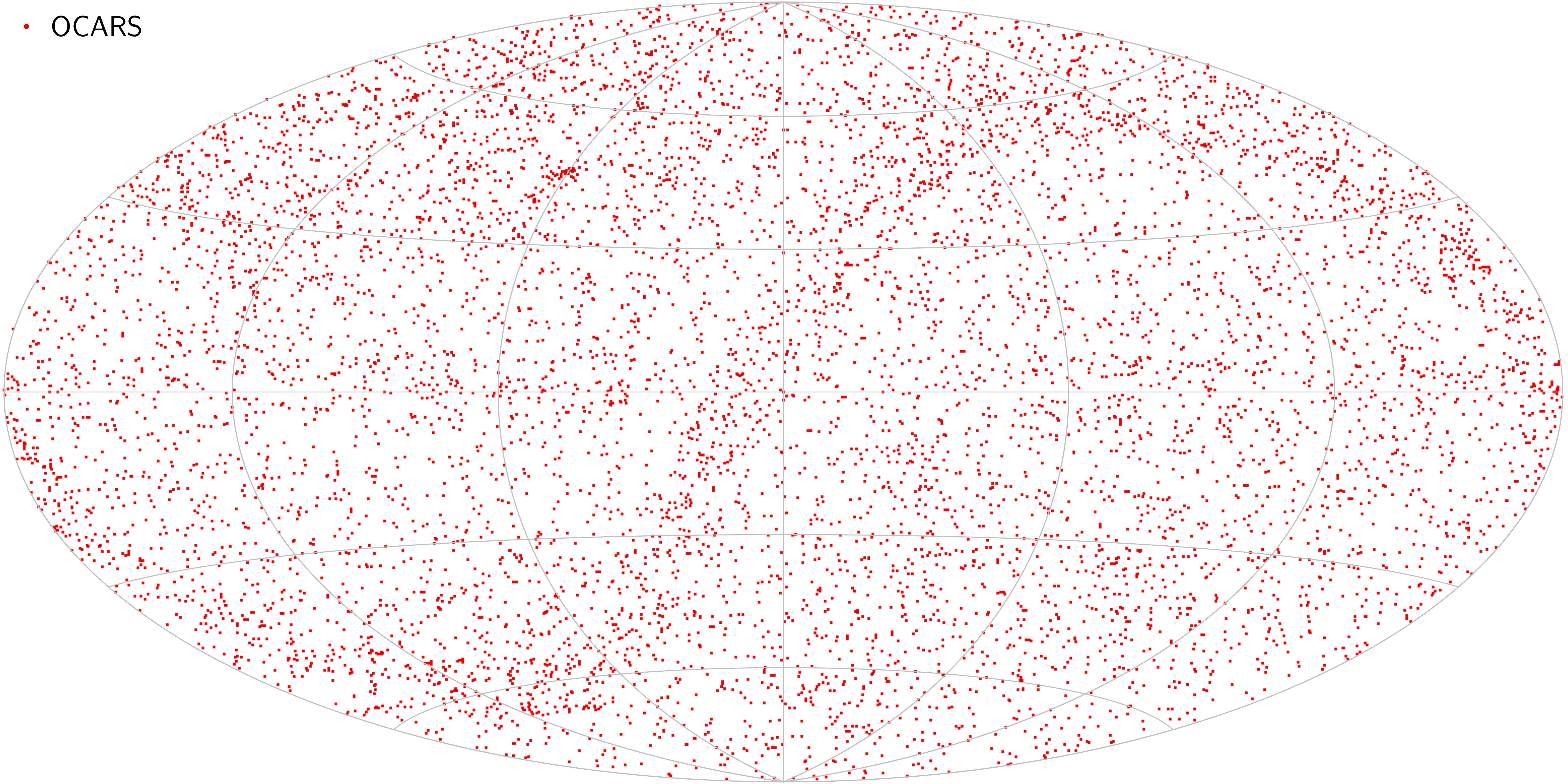}
\includegraphics[width=\scaleext\hsize]{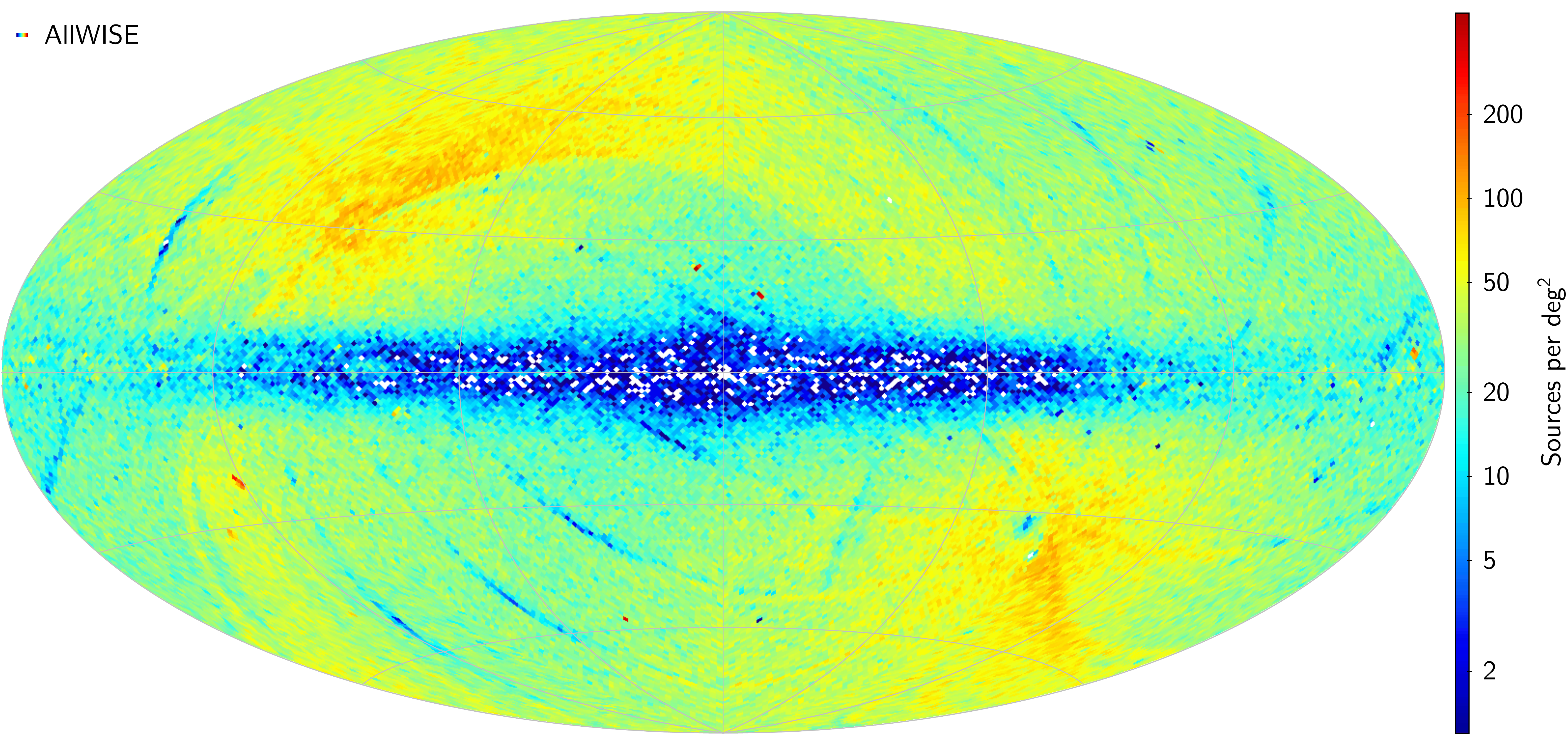}
\includegraphics[width=\scaleext\hsize]{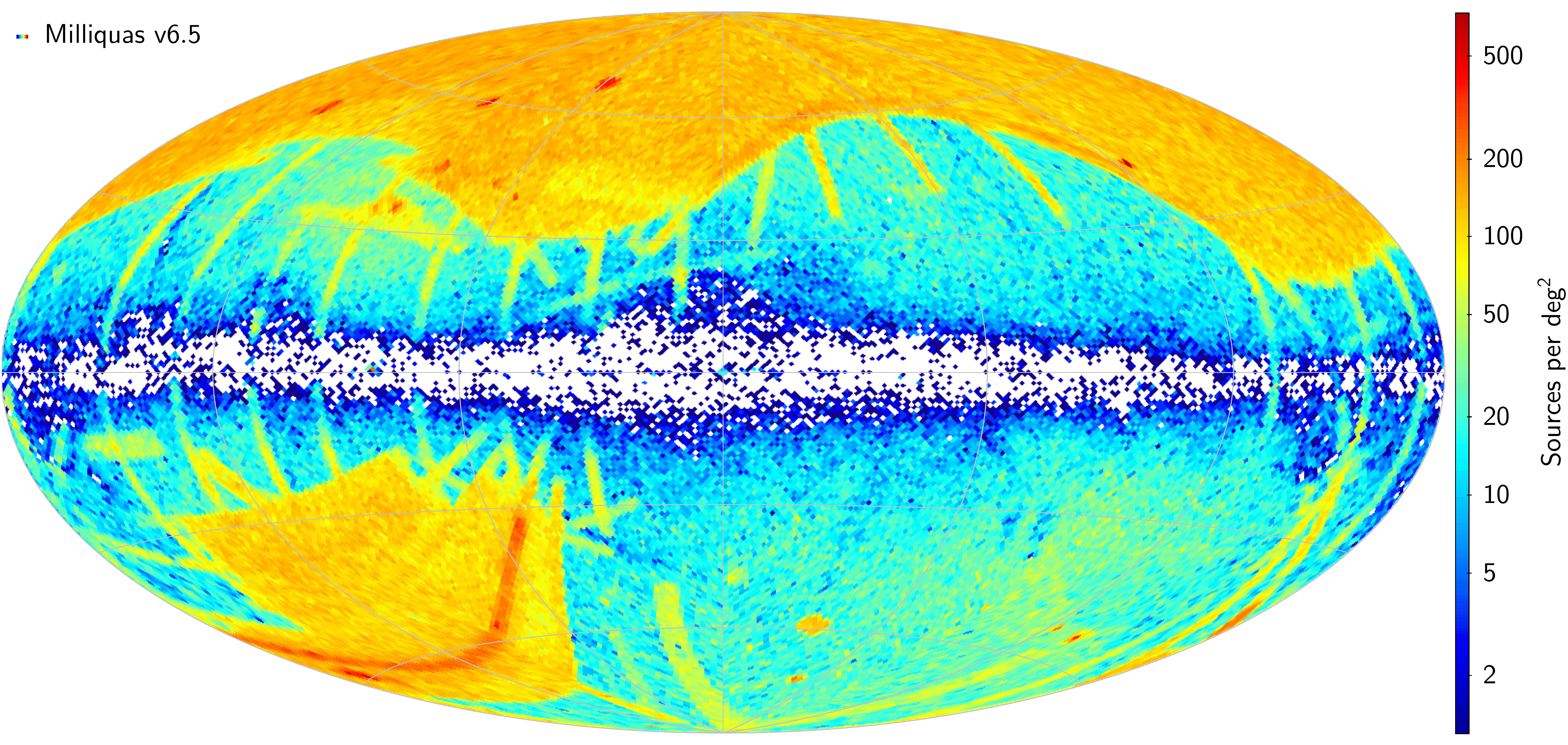}
\includegraphics[width=\scaleext\hsize]{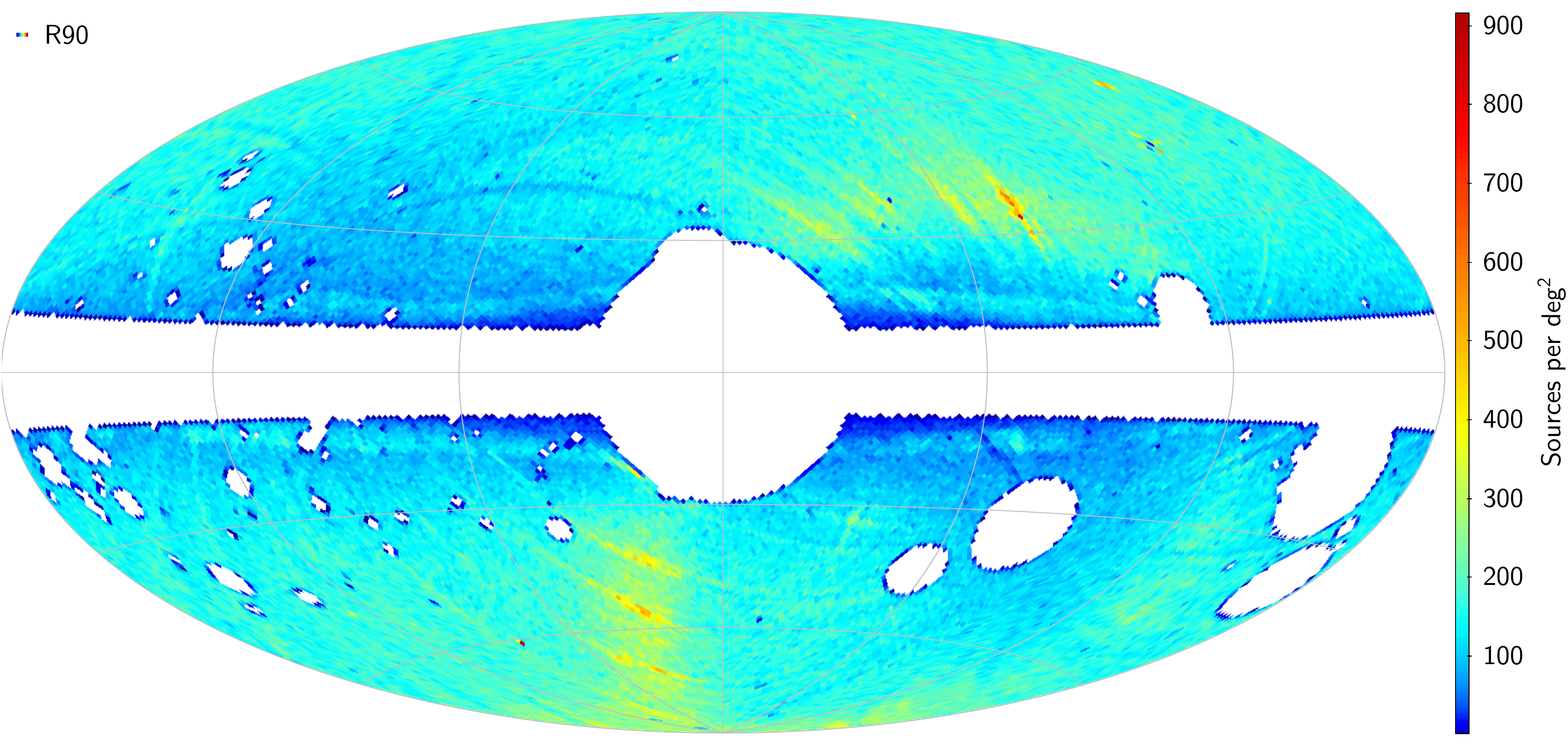}
\includegraphics[width=\scaleext\hsize]{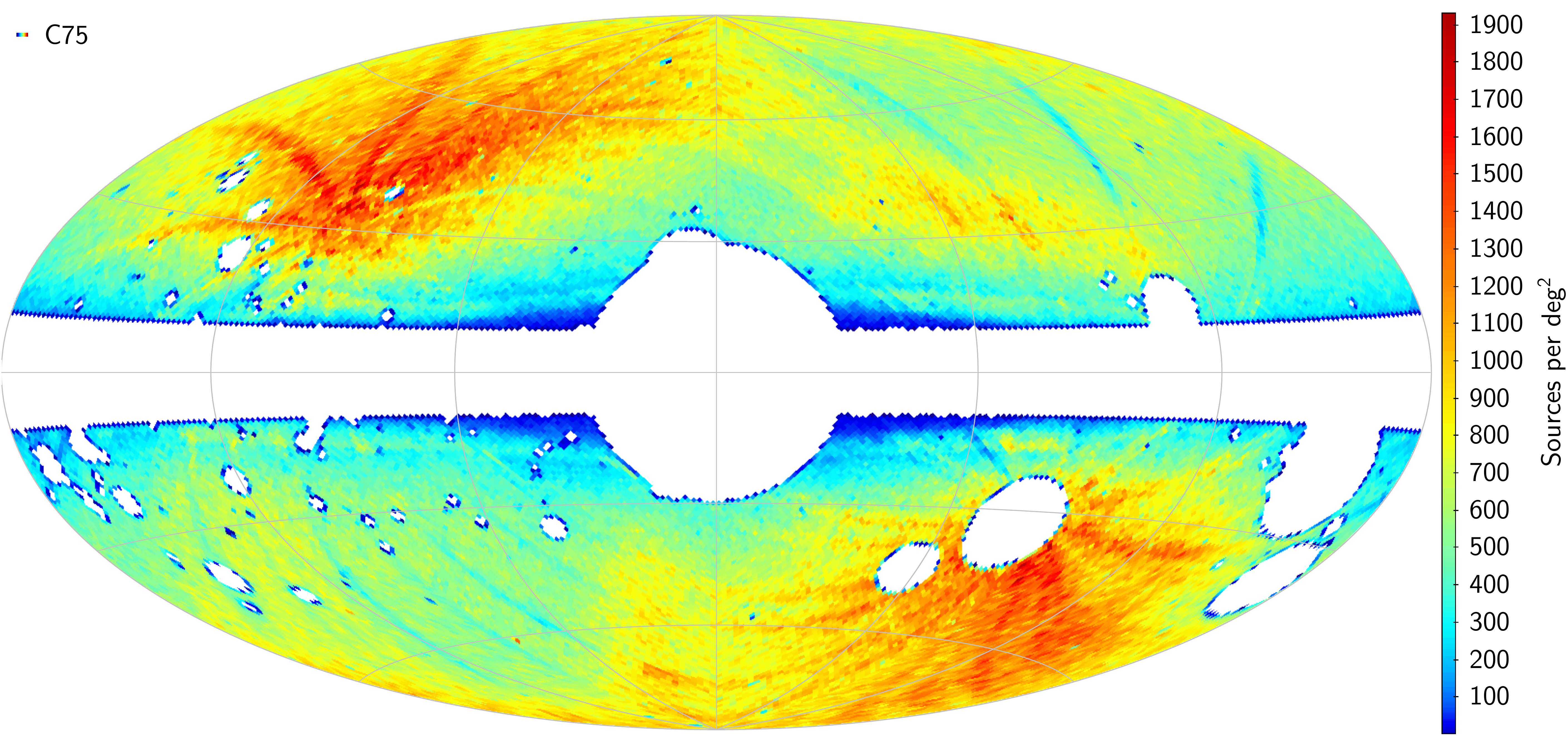}
\includegraphics[width=\scaleext\hsize]{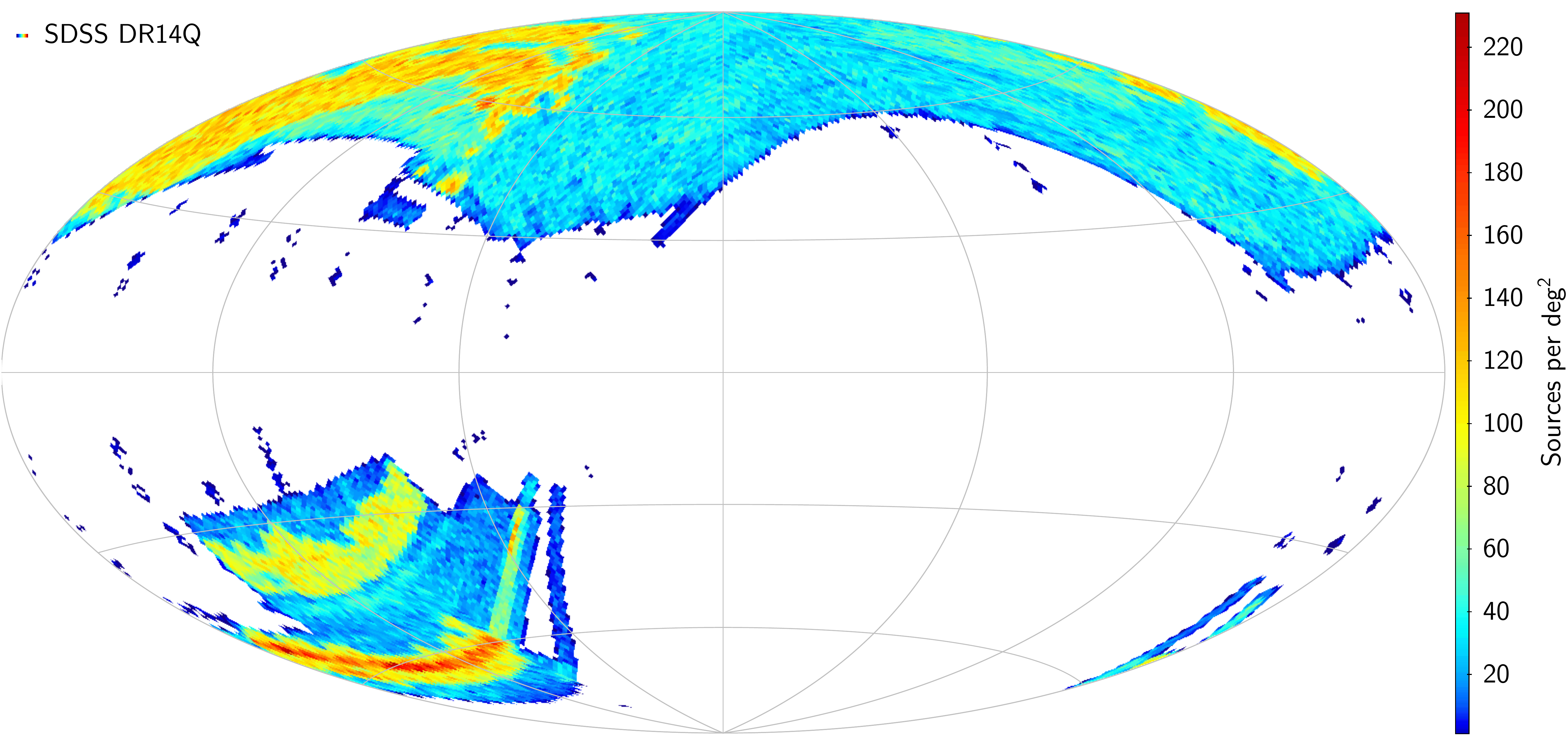}
\includegraphics[width=\scaleext\hsize]{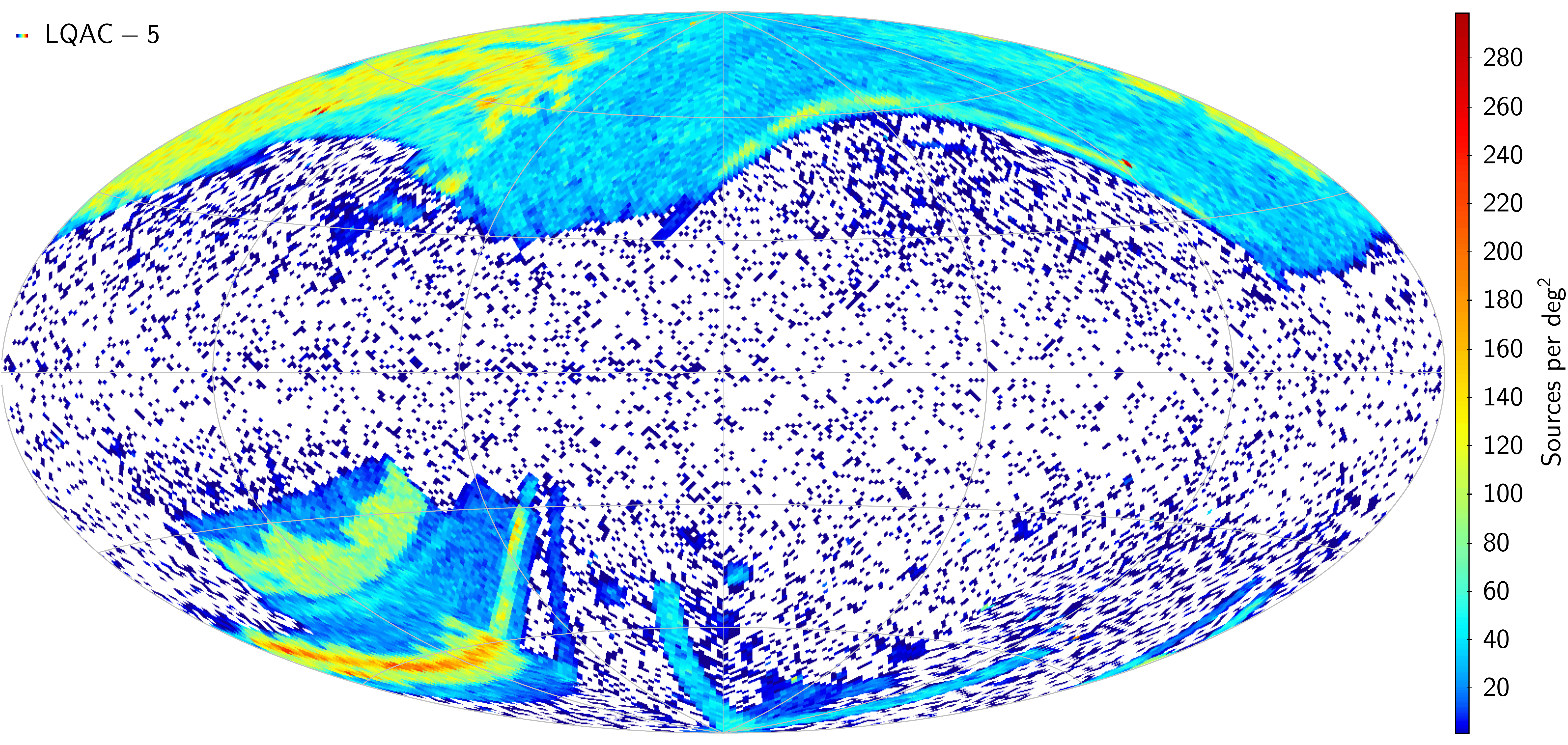}
\includegraphics[width=\scaleext\hsize]{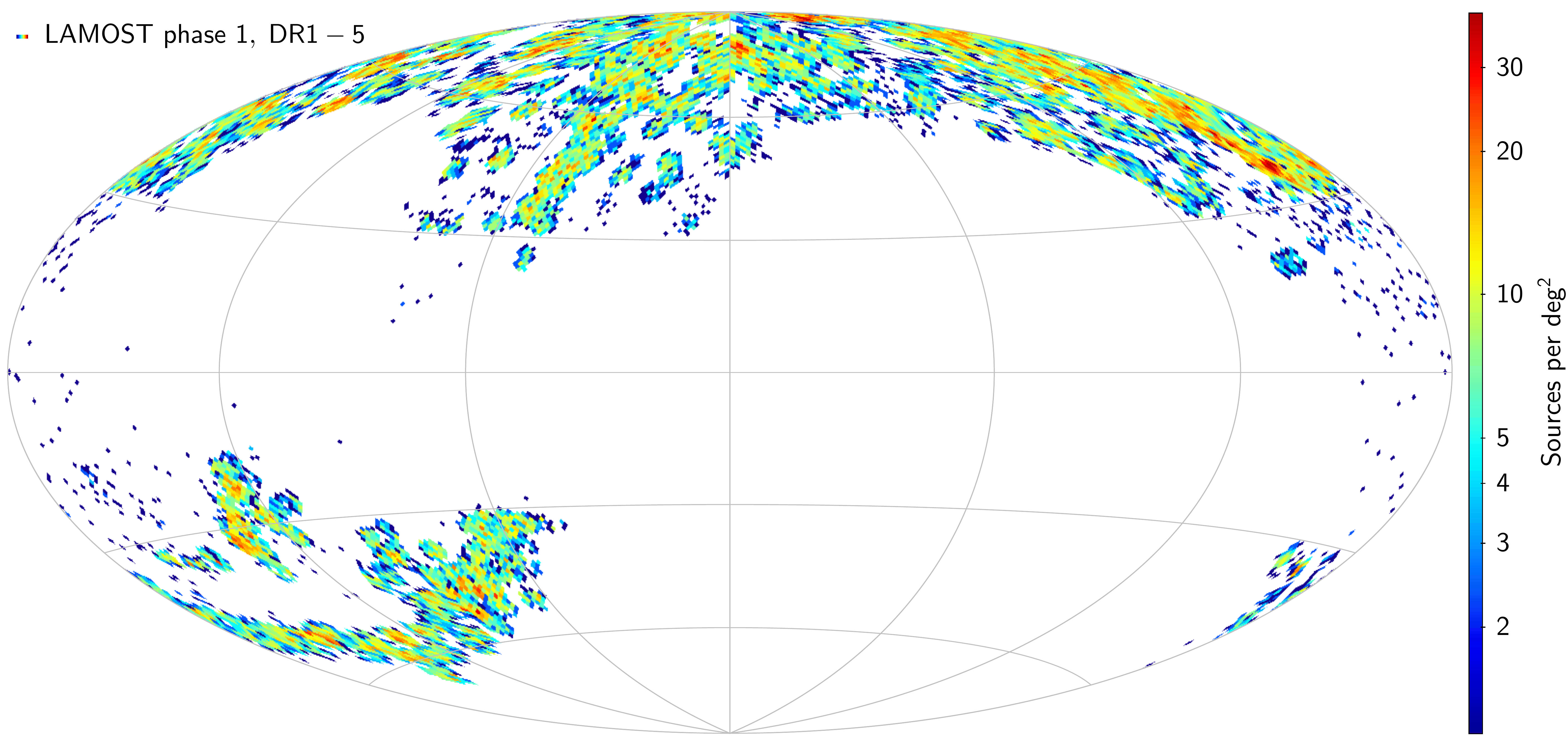}
\includegraphics[width=\scaleext\hsize]{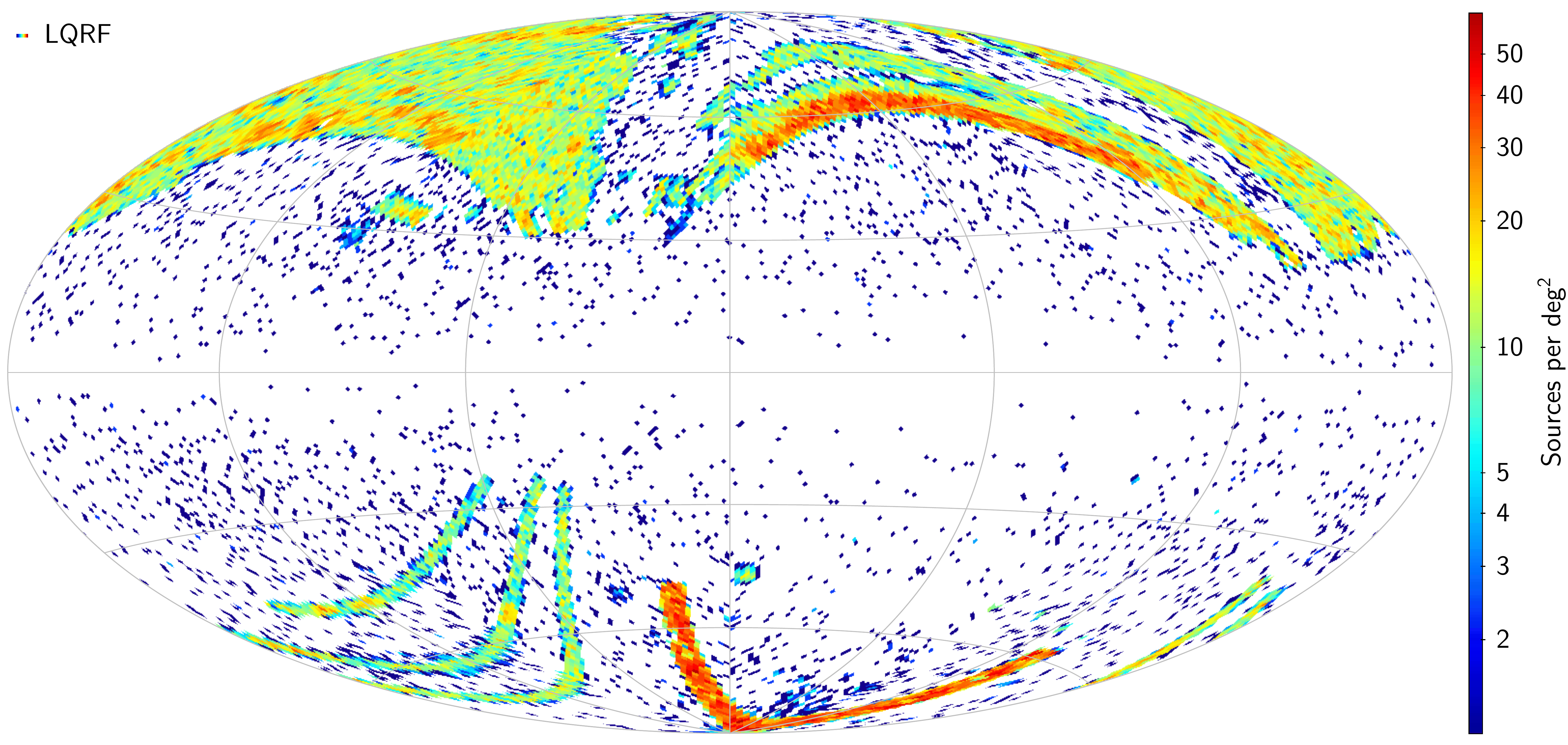}
\includegraphics[width=\scaleext\hsize]{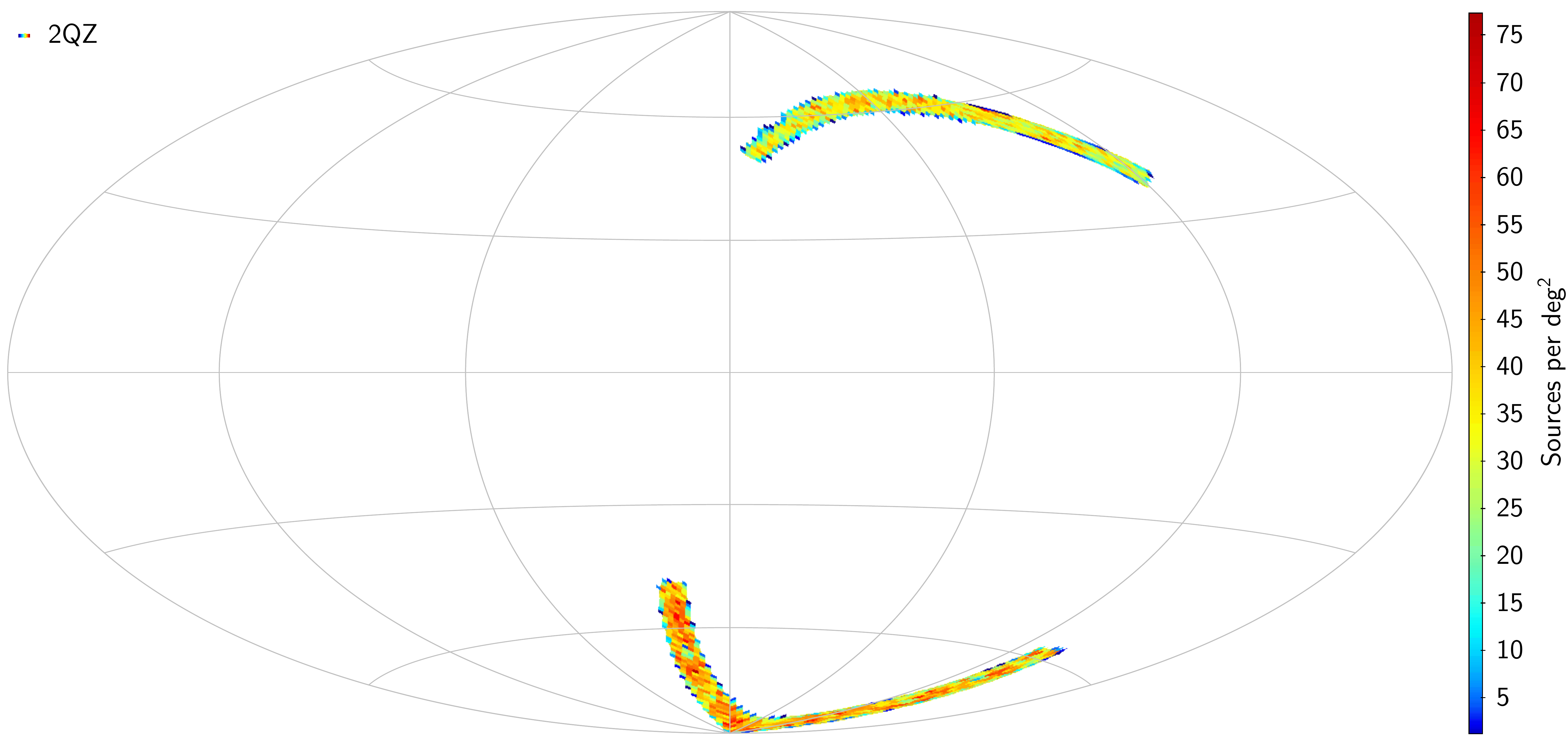}
\includegraphics[width=\scaleext\hsize]{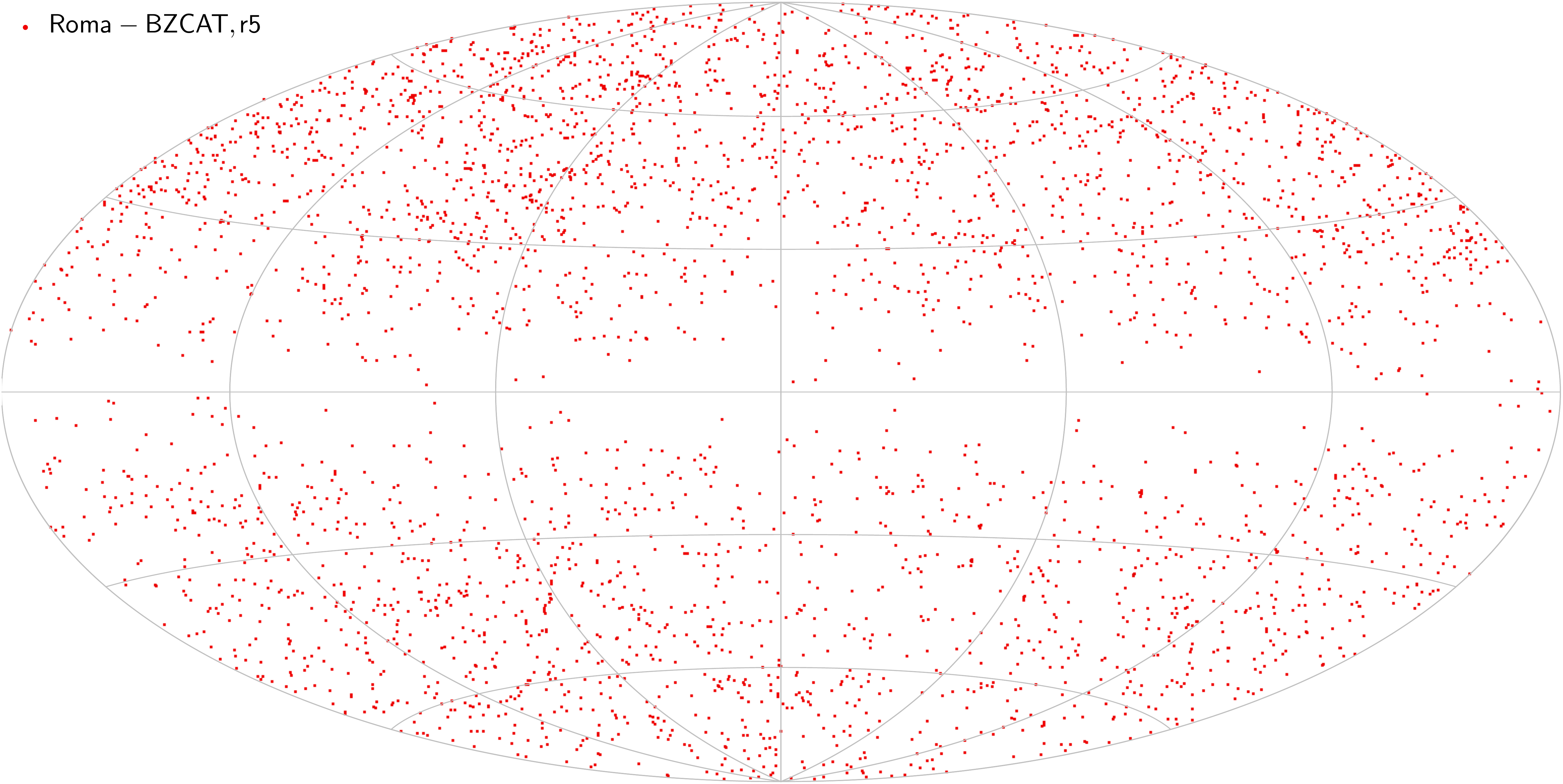}
\includegraphics[width=\scaleext\hsize]{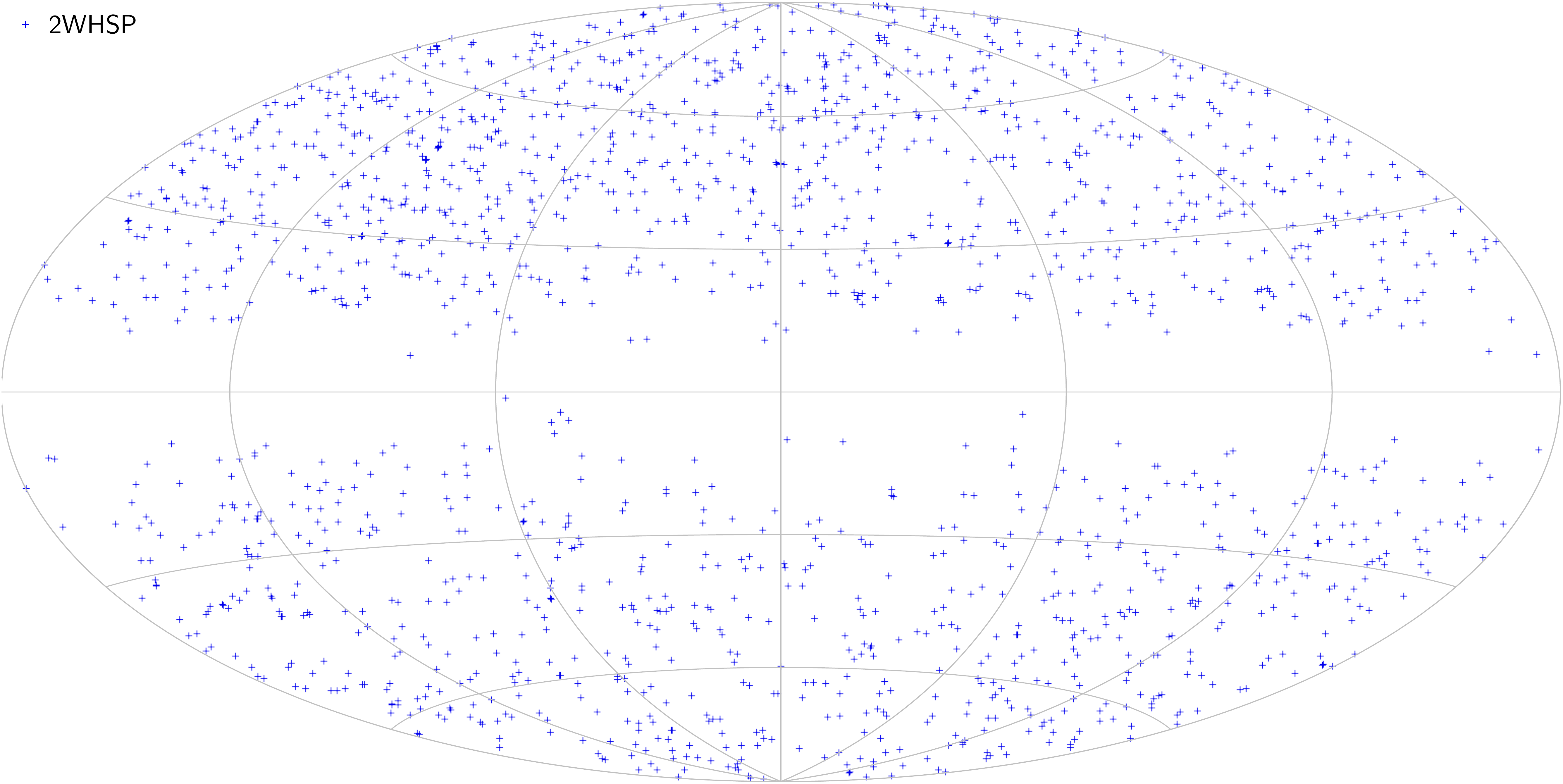} 
\includegraphics[width=\scaleext\hsize]{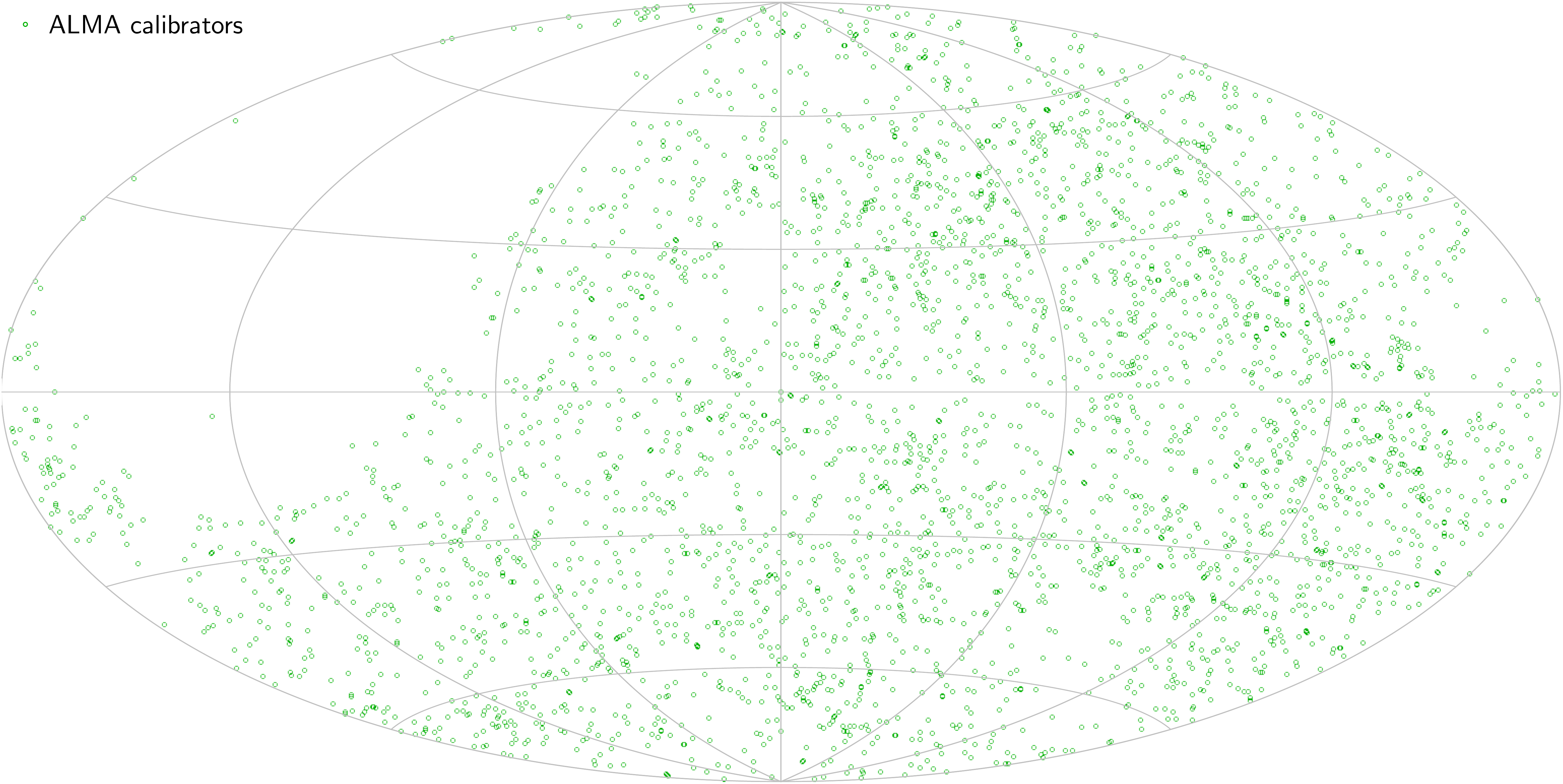}
\includegraphics[width=\scaleext\hsize]{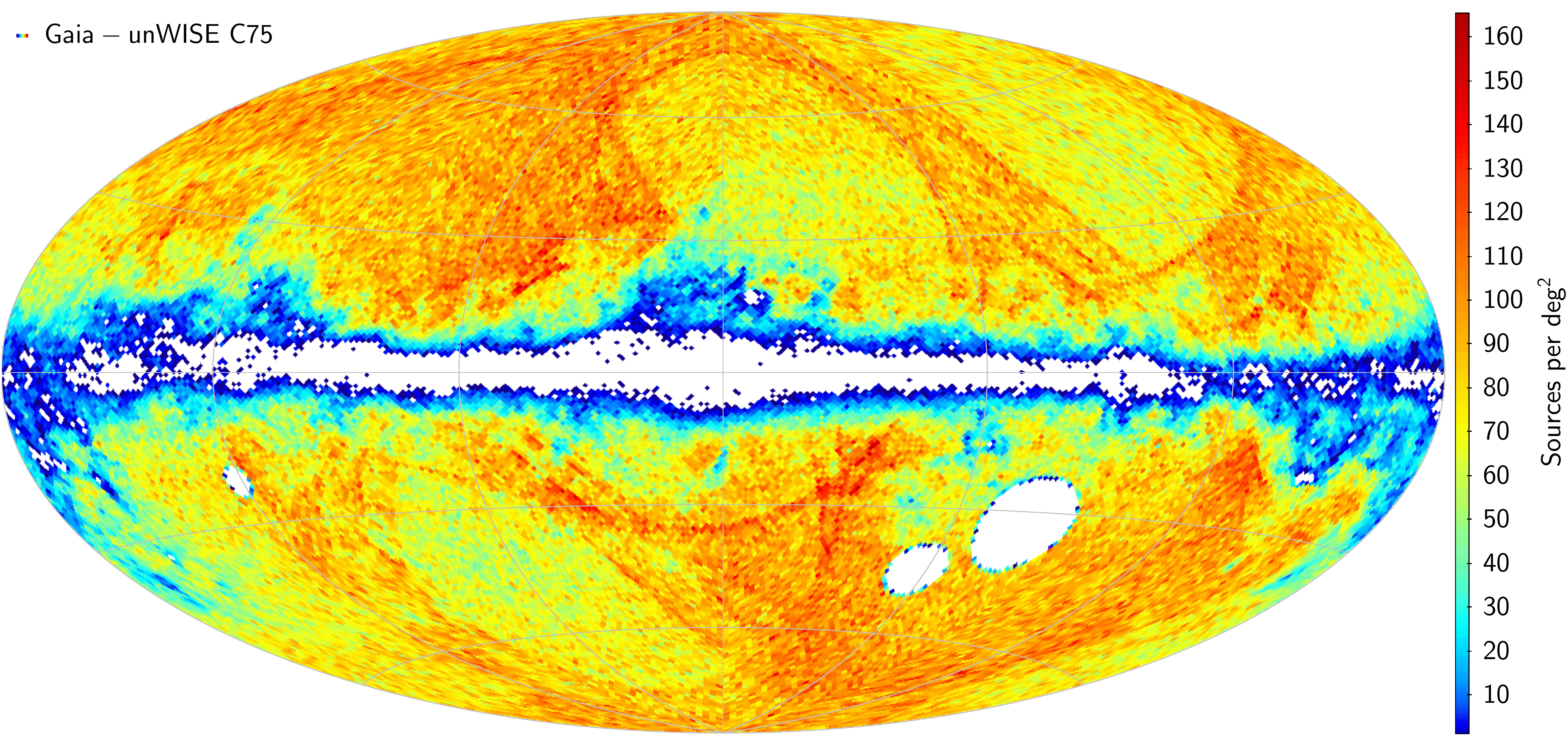}
\end{center}
  \caption{Distribution of the sources in the external catalogues listed in
   Table~\ref{tab:gaiacrf3-matches}. These maps use a Hammer–Aitoff projection in galactic
    coordinates, with $l = b = 0$ at the centre, north up, and $l$
    increasing from right to left.}
\label{fig:skymaps-external}
\end{sidewaysfigure*}

Figure~\ref{fig:skymaps-external} shows the distribution of the sources
in the 17 external catalogues used for \gcrf{3} as discussed in
Sect.~\ref{sec:selection} and listed in
Table~\ref{tab:gaiacrf3-matches}. The maps show all the sources in the
respective catalogue, irrespective of whether they are also in \gcrf{3}.
For the VLBI and blazar catalogues
(ICRF3 S/X, ICRF3 K, ICRF3 X/Ka, Roma-BZCAT, 2WHSP, and the ALMA
calibrators) that have a small source density on the sky,
individual sources are plotted. All three parts of ICRF3 are shown on
one plot using different colours. For the other catalogues, the maps
show the density of sources per square degree, computed from the source
counts per pixel using HEALPix of level 6 (pixel size $\simeq
0.84$\,deg$^2$). One sees the strong inhomogeneity of the input
catalogues caused by the special observational programmes not covering
the whole sky (e.g.\ SDSS and LAMOST), by the observational schedule of the
WISE satellite (AllWISE, R90, and C75), and by the artificial de-selection of certain 
crowded areas applied by the authors of some catalogues (e.g.\ the 
Galactic plane and the LMC and SMC areas in R90, C75, and Gaia-unWISE).  
It is also obvious from the maps that not all catalogues
are independent.  For example, one directly sees that LQRF and \hbox{LQAC-5}
considered the results of 2QZ and that SDSS DR14Q was used in \hbox{LQAC-5}.
Detailed information on all these aspects can be found in the
corresponding original publications cited in
Table~\ref{tab:gaiacrf3-matches}.  All these inhomogeneities
contribute to the resulting inhomogeneity of \gcrf3, the density of
which is shown in Fig.~\ref{fig:sky-distribution-gcrf3}.

\clearpage

\section{Access to the \gcrfthree\ data} 
\label{sec:SQL-requests}

Information on the \gcrfthree\ sources is distributed over several tables in the \gaia\ Archive.%
\footnote{\url{https://gea.esac.esa.int/archive/}}
Table \texttt{gaiaedr3.agn\_cross\_id}
lists all sources (\texttt{source\_id}) of \gcrfthree\ together with the name of one external catalogue where
the source can also be found as well as the name of the source in that catalogue. Table
\texttt{gaiaedr3.frame\_rotator\_source} contains information on the sources used
by the frame rotator (see Sect.~\ref{sec:framerotator}) to define the orientation and spin of the 
\gedr3\ catalogue. Together with the \gedr3\ source identifier
\texttt{source\_id}, it contains the boolean flags
\texttt{considered\_for\_reference\_frame\_orientation} and
\texttt{considered\_for\_reference\_frame\_spin} to indicate if the source was considered
by the frame rotator algorithm to fix the orientation and spin of the \gedr3\ catalogue, respectively.
The frame rotator algorithm has a mechanism to detect outliers (see Appendix~\ref{sec:estimating}),
and another pair of boolean flags, \texttt{used\_for\_reference\_frame\_orientation} and
\texttt{used\_for\_reference\_frame\_spin}, to specify which sources were used to fix, respectively,
the orientation and rotation of the \gedr3\ catalogue. The sources that were considered but not used
are the ones identified as outliers by the frame rotator algorithm.

Using these two tables and the main \gedr3\ astrometric table
\texttt{gaiaedr3.gaia\_source}, one can retrieve astrometric data of the \gcrfthree\ sources and
their subsets. We give here a few useful examples of ADQL queries that can be directly used 
in the \gaia\ Archive. The query
\vspace{0.25cm}
\noindent {\small
\begin{verbatim}
SELECT agn.source_name_in_catalogue,
       agn.catalogue_name, edr3.*
FROM gaiaedr3.agn_cross_id AS agn
INNER JOIN gaiaedr3.gaia_source AS edr3
USING (source_id)
\end{verbatim}
}
\noindent
gives a table containing astrometric data for all 1\,614\,173 \gcrfthree\ sources.
Adding a conditional clause at the end of the same query,
\vspace{0.25cm}
\noindent {\small
\begin{verbatim}  
SELECT agn.source_name_in_catalogue, 
       agn.catalogue_name, edr3.*
FROM gaiaedr3.agn_cross_id AS agn
INNER JOIN gaiaedr3.gaia_source AS edr3
USING (source_id)
WHERE edr3.astrometric_params_solved = 31
\end{verbatim}
}
\noindent
allows one to retrieve the astrometric data for only the 1\,215\,942 \gcrfthree\ sources
that have five-parameter astrometric solutions. This dataset was used
for example in \citetads{2021A&A...649A...9G} to measure the acceleration of the Solar System.
Changing `31' to `95' in the last line,
one gets instead the 398\,231 \gcrfthree\ sources with six-parameter astrometric solutions.

For the sources used to define the orientation and spin of the \gedr3\ catalogue
the following two queries are useful. The query
\vspace{0.25cm}
\noindent {\small
\begin{verbatim}  
SELECT sub.source_name_in_catalogue,
       sub.catalogue_name, edr3.*
FROM (
SELECT frame.source_id, agn.source_name_in_catalogue,
       agn.catalogue_name
FROM gaiaedr3.frame_rotator_source AS frame
LEFT OUTER JOIN gaiaedr3.agn_cross_id AS agn 
USING (source_id)
WHERE frame.used_for_reference_frame_orientation='true'
) AS sub
INNER JOIN gaiaedr3.gaia_source AS edr3
USING (source_id)
\end{verbatim}  
}
\vspace{0.25cm}
\noindent
gives data for the 2007 sources that were used to define the orientation of the 
\gedr3\ catalogue, while
\vspace{0.25cm}
\noindent {\small
\begin{verbatim}
SELECT sub.source_name_in_catalogue,
       sub.catalogue_name, edr3.*
FROM (
SELECT frame.source_id, agn.source_name_in_catalogue,
       agn.catalogue_name
FROM gaiaedr3.frame_rotator_source AS frame
LEFT OUTER JOIN gaiaedr3.agn_cross_id AS agn 
USING (source_id)
WHERE frame.used_for_reference_frame_spin = 'true'
) AS sub
INNER JOIN gaiaedr3.gaia_source AS edr3
USING (source_id)
\end{verbatim}  
}
\vspace{0.25cm}
\noindent
gives data for the 428\,034 sources used to define the rotation (spin) of 
\gedr3. The field names {\tt used\_for\_\dots} can be replaced by {\tt considered\_for\_\dots} 
to get the data of all sources considered by the reference frame algorithm.
As discussed in Sect.~\ref{sec:framerotator}, all sources that were
considered by the frame rotator algorithm for fixing the orientation of
the \gedr3\ catalogue also belong to \gcrf{3}. However, among the sources that were
considered for fixing the spin, there are 46 sources that are not part of \gcrf{3}. 
Two of them were used to fix the spin. For those 46 or 2 sources, the
respective queries return empty names of the source and the external catalogue.

As described in Sect.~\ref{sec:selection}, the \gcrfthree\ sources resulted from a filtered 
cross-match between the \gedr3\ catalogue and a number of external AGN/QSO catalogues. 
The information on which sources were found in the different external catalogues is given 
in table \verb!gaiadr3.gaia_crf3_xm!. We note that \verb!gaiaedr3.agn_cross_id! cannot be 
used to retrieve all \gcrfthree\ sources from a given external catalogue, because
a \gcrfthree\ source may be found in several external catalogues but \verb!gaiaedr3.agn_cross_id! 
only gives its identifier in one of them.
To obtain a list of the \gcrfthree\ sources that were selected from a specific external catalogue 
one can use the query:\footnote{Table \texttt{gaiadr3.gaia\_crf3\_xm} will be published in 
the \gaia\ Archive together with \gdr{3}. Alternatively, the table can be downloaded from 
\url{https://www.cosmos.esa.int/web/gaia/gaia-crf3-cross-match-table} and uploaded to the \gaia\ 
Archive as a user table. In that case the table name in the ADQL query should be modified accordingly.}
\vspace{0.25cm}
\noindent {\small
\begin{verbatim}
SELECT xm.$$$_name, edr3.*
FROM gaiaedr3.gaia_source AS edr3
INNER JOIN gaiadr3.gaia_crf3_xm AS xm 
USING (source_id)
WHERE xm.$$$ = 'true'
\end{verbatim}  
}
\vspace{0.25cm}
\noindent
where `\verb!$$$!' should be replaced by the catalogue code from column~6 of 
Table~\ref{tab:gaiacrf3-matches}. For the sources in the three ICRF3 catalogues (the
catalogue codes {\tt icrf3sx}, {\tt icrf3k}, {\tt icrf3xka}) \verb!gaiadr3.gaia_crf3_xm! provides
two source designations as also given in the original ICRF3 catalogues: the ICRF and IERS
designations as {\tt icrf\_name} and {\tt iers\_name}, respectively.
Additionally, the above ADQL query with the catalogue code '{\tt b19}' can be used to see the
920\,599 \gcrfthree\ sources that were also found in the catalogue of
\citetads{2019MNRAS.490.5615B}.

We note that when considering an external catalogue `\verb!$$$!' some
sources were rejected only because a lower quality of astrometry used
in that catalogue resulted in the cross-match distance $\Delta$ being too
large to satisfy Eq.~(\ref{eq:filterRho}) in
Sect.~\ref{sec:selection}.  For those sources the field
\verb!gaiadr3.gaia_crf3_xm.$$$! is set to false.  In some cases,
however, the same \gaia\ source was found also in other external
catalogues and eventually selected for \gcrfthree. In those cases, the
name \verb!gaiadr3.gaia_crf3_xm.$$$_name! of that \gaia\ source in
catalogue `\verb!$$$!' is also given. To obtain all
\gcrfthree\ sources that appear in an external catalogue one should
replace the last line in the previous ADQL query by
`{\small\verb!WHERE xm.$$$_name IS NOT NULL!}'.  This helps to
mitigate the effect of a lower quality astrometry in some of the
external catalogues and improve the overall cross-match quality for
those catalogues.

\clearpage

\section{Confusion sources in \gedr{3}}
\label{sec:confusion}

In this appendix we give the number and distributions of sources in \gedr{3} that satisfy 
Eqs.~(\ref{eq:filterA})--(\ref{eq:X2}), ensuring that the  
\gaia\ astrometry is statistically compatible with the hypothesis that a source is extragalactic 
with zero parallax and proper motion. (As discussed in Sect.~\ref{sec:selection}, we ignore 
the small effects of the Galactic acceleration and possible apparent proper motions induced by variable 
source structure, but take into account an overall parallax offset of $-0.017$~mas.)
In this paper we call these the `confusion sources', because most of them are in fact distant
stars in our Galaxy or its satellite galaxies, and stellar contaminants resulting from erroneous
cross-matching with external QSO catalogue (or from stellar contamination in these catalogues)
can be expected to have similar characteristics as the confusion sources.

\begin{figure*}[th!]
\begin{center}
  \includegraphics[width=\hsize]{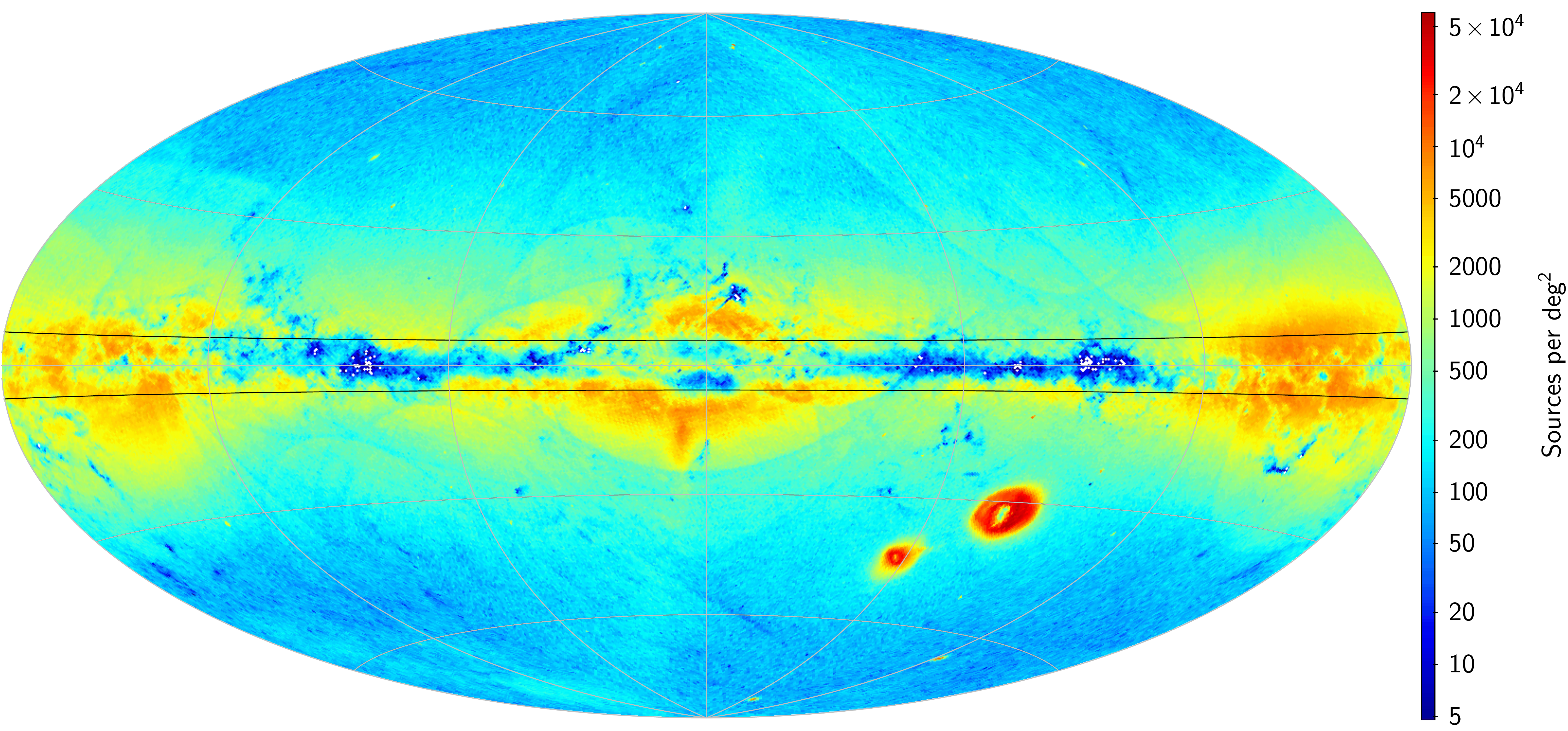}
  \includegraphics[width=\hsize]{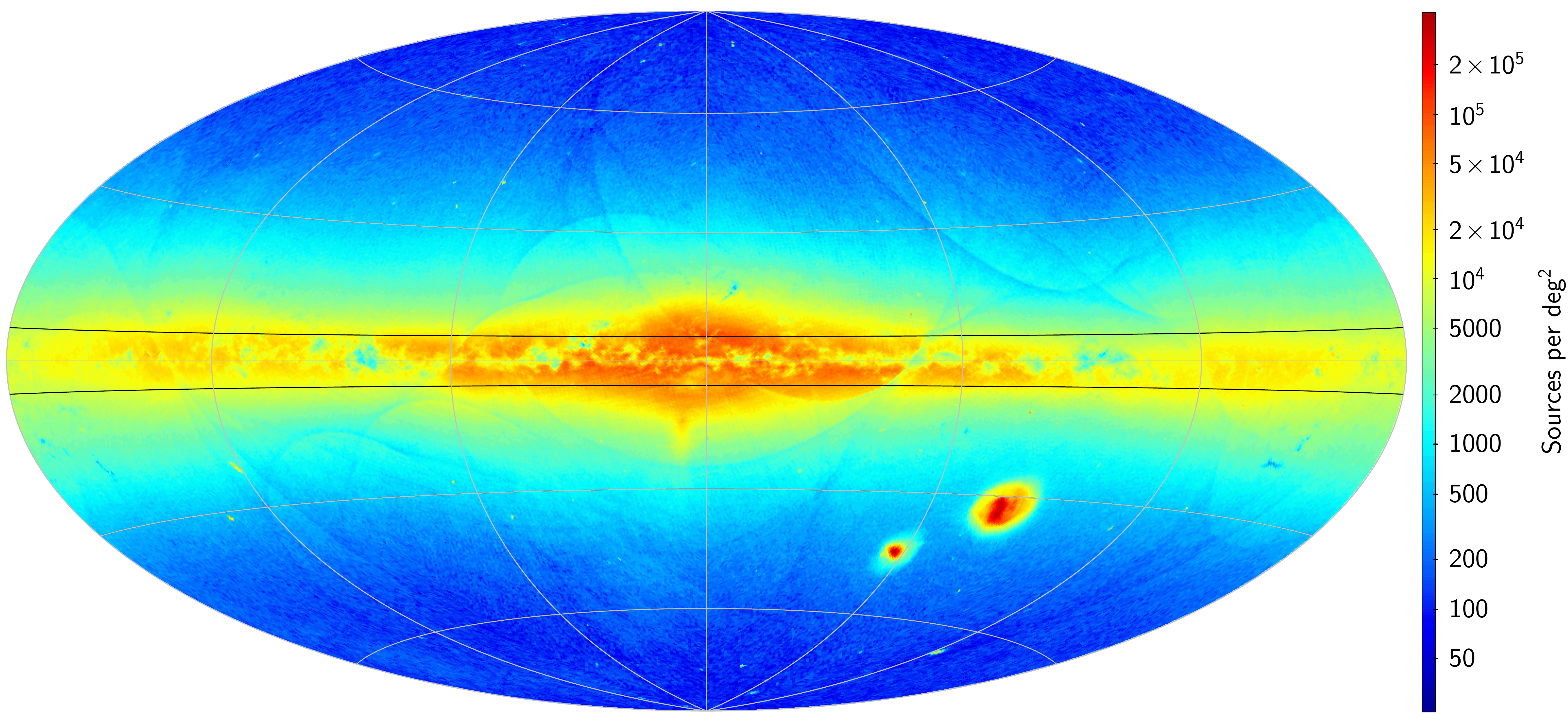}
  \caption{Distribution of the confusion sources in \gedr{3}.
    These maps use a Hammer–Aitoff projection in galactic
    coordinates, with $l = b = 0$ at the centre, north up, and $l$
    increasing from right to left.
    The black lines 
    bound the Galactic zone $\left|\,\sin b\,\right|<0.1$ where sources are normally excluded by
    Eq.~(\ref{eq:filterB}). \textit{Top:} sources with five-parameter solutions.
    \textit{Bottom:} sources with six-parameter solutions.
}    
\label{fig:sky-distribution-confusionSources}
\end{center}
\end{figure*}

We give separate statistics for confusion sources with five- and six-parameter solutions
in \gedr{3}. For the five-parameter solutions, an explicit query of the \gedr{3} catalogue\footnote{The queries
discussed in this Appendix are costly and result in very big tables. To cope with the technical limitations, 
one can either split the results in chunks by using some conditions on
'\texttt{random\_index}' (the random index assigned for each source in the \gaia\ Archive) or
execute the equivalent queries offline using a local copy of the whole \gedr{3} catalogue.}
{\small\begin{verbatim}  
SELECT * FROM gaiaedr3.gaia_source AS g 
WHERE g.astrometric_params_solved = 31 
  AND abs((g.parallax + 0.017)/g.parallax_error) < 5 
  AND ( power(g.pmra/g.pmra_error,2)
       +power(g.pmdec/g.pmdec_error,2)
       -2*g.pmra_pmdec_corr*
          g.pmra/g.pmra_error*
          g.pmdec/g.pmdec_error)/
      (1-power(g.pmra_pmdec_corr,2)) < 25
\end{verbatim}}\noindent
results in 30\,723\,995 sources satisfying Eqs.~(\ref{eq:filterA})--(\ref{eq:X2}).
Their sky density is shown in the top panel of Fig.~\ref{fig:sky-distribution-confusionSources}.  
Of those, 7\,090\,844 sources are in the Galactic zone with 
$\left|\,\sin b\,\right|<0.1$, thus violating the fourth criterion in Eq.~(\ref{eq:filterB}). 
The remaining 23\,633\,151 confusion sources with five-parameter astrometric solutions in 
\gedr{3} satisfy all four criteria and could, based on their astrometry alone, be potential
candidates for the \gcrf{}.

A corresponding query of \gedr{3} for confusion sources with six-parameter
solutions (obtained by changing the second line of the query to
`\texttt{WHERE g.astrometric\_params\_solved = 95}') returns
182\,814\,959 sources. Their sky density is shown in the lower panel of 
Fig.~\ref{fig:sky-distribution-confusionSources}).
90\,979\,403 of the sources have $\left|\,\sin b\,\right|>0.1$ and thus satisfy 
all four criteria in Eqs.~(\ref{eq:filterA})--(\ref{eq:filterB}).

%
     
Among the many features seen in Fig.~\ref{fig:sky-distribution-confusionSources} 
we recognise the expected density distribution due to the overall distribution of
Galactic stars and a strong over-density in the areas of the LMC and SMC.
There are several more compact high-density 
areas outside the Galactic plane. A detailed (but not exhaustive) inspection 
of the density map revealed about 20 stellar (mostly globular) clusters and 
10 (dwarf) galaxies on the maps. Not counting the LMC and SMC areas, the total 
number of sources in these clusters and galaxies is however only about 5000, which 
is ignored in the statistical considerations below. We note that the Sagittarius stream 
is clearly visible (mainly in the top panel), as well as numerous streaks of under-density 
related to the scanning law.

A prominent feature in Fig.~\ref{fig:sky-distribution-confusionSources} is the 
relatively low density of sources with five-parameter solutions along the Galactic 
equator and in the centres of the Magellanic Clouds. This is caused by the difficulty to 
obtain reliable colour information from the BP and RP spectra on faint sources in 
these very crowded areas \citepads{2021A&A...649A...3R}. As a result, the astrometric 
solution did not use a spectrophotometrically determined colour (effective
wavenumber) for many of these sources, but instead determined a pseudo-colour 
as the sixth astrometric parameter.  

Because the confusion sources in the LMC and SMC tend to have distinctly different
kinematics from the Galactic stars, it is advantageous to separate out the sources
in the LMC and SMC areas in the further analysis. Guided by the density maps,
we take a circle of radius $9^\circ$ centred on $(\alpha,\delta)=(81.3^\circ,-68.7^\circ)$
to represent the LMC, and a circle of radius $6^\circ$ centred on 
$(\alpha,\delta)=(16.0^\circ,-72.8^\circ)$ to represent the SMC.
These areas contain, respectively, 2\,657\,059 and 496\,520 confusion sources
with five-parameter solutions (6\,817\,915 and 1\,307\,591 with six-parameter solutions).
This simplistic selection does not pretend to any deeper physical significance,
and only partially coincides with the carefully selected samples of \gedr{3}\ sources
in the LMC and SMC discussed in \citetads{2021A&A...649A...7G}. 

\begin{figure*}
\begin{center}
  \includegraphics[width=\scalethree\hsize]{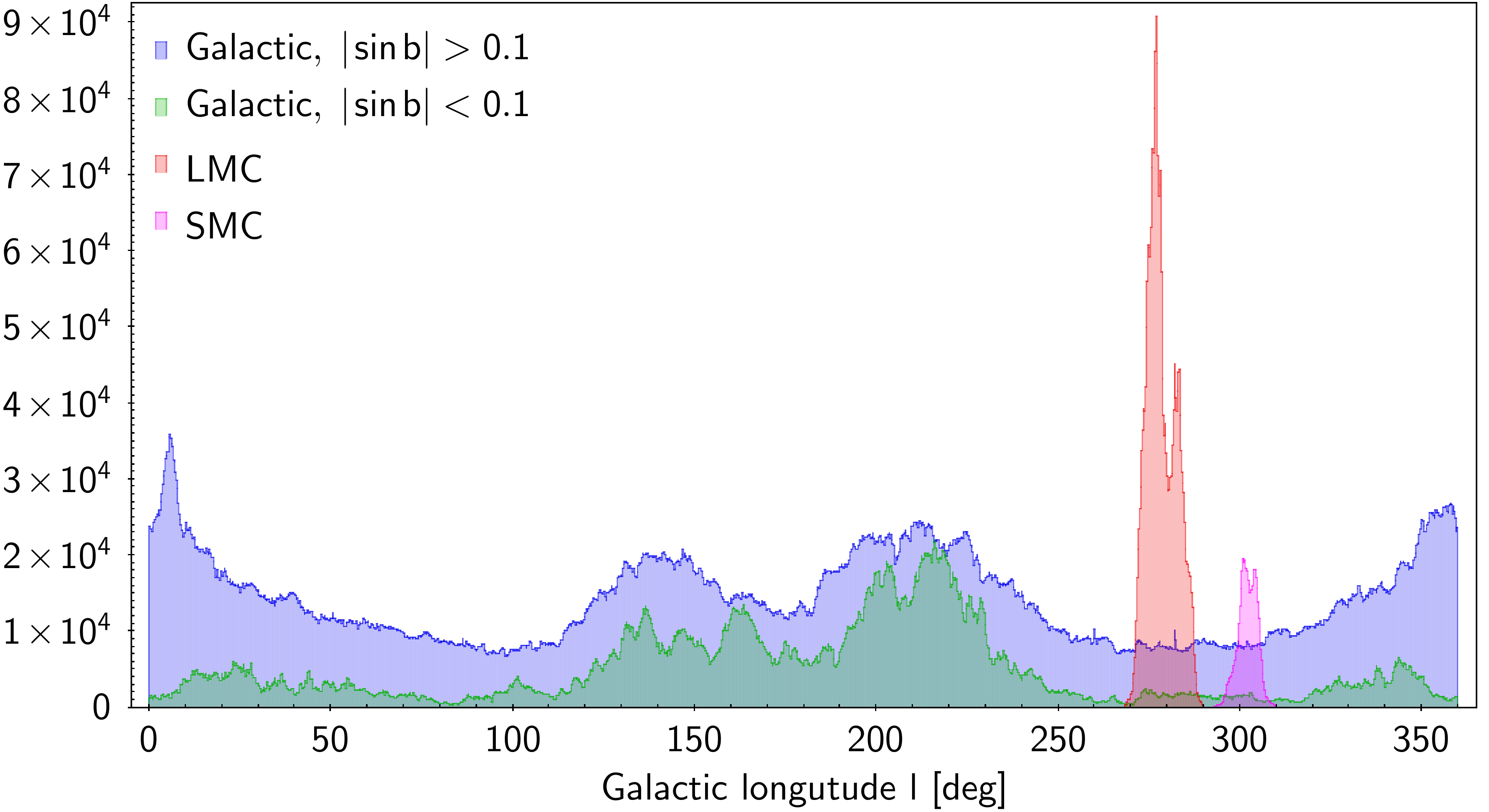}
  \includegraphics[width=\scalethree\hsize]{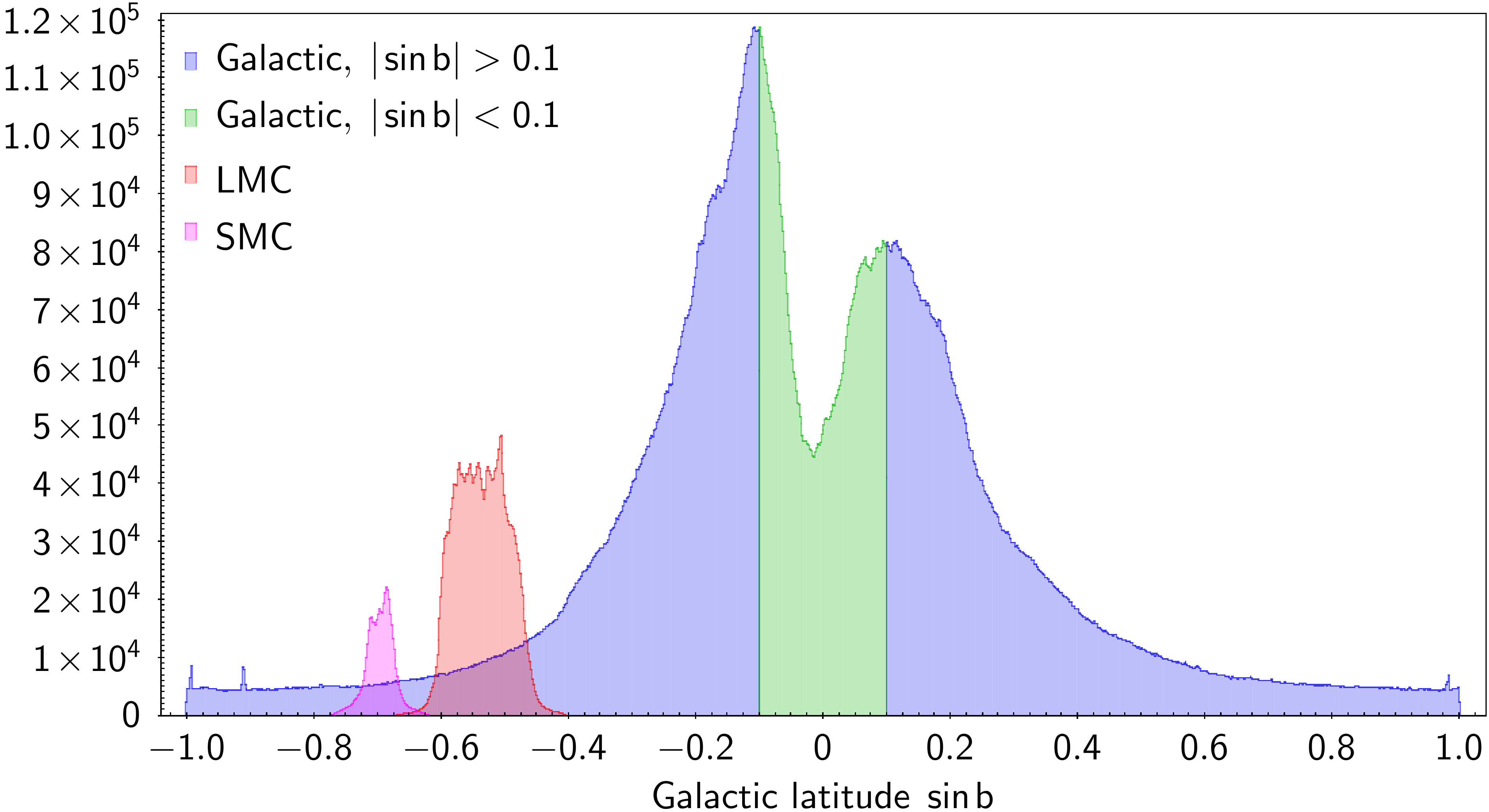}
  \includegraphics[width=\scalethree\hsize]{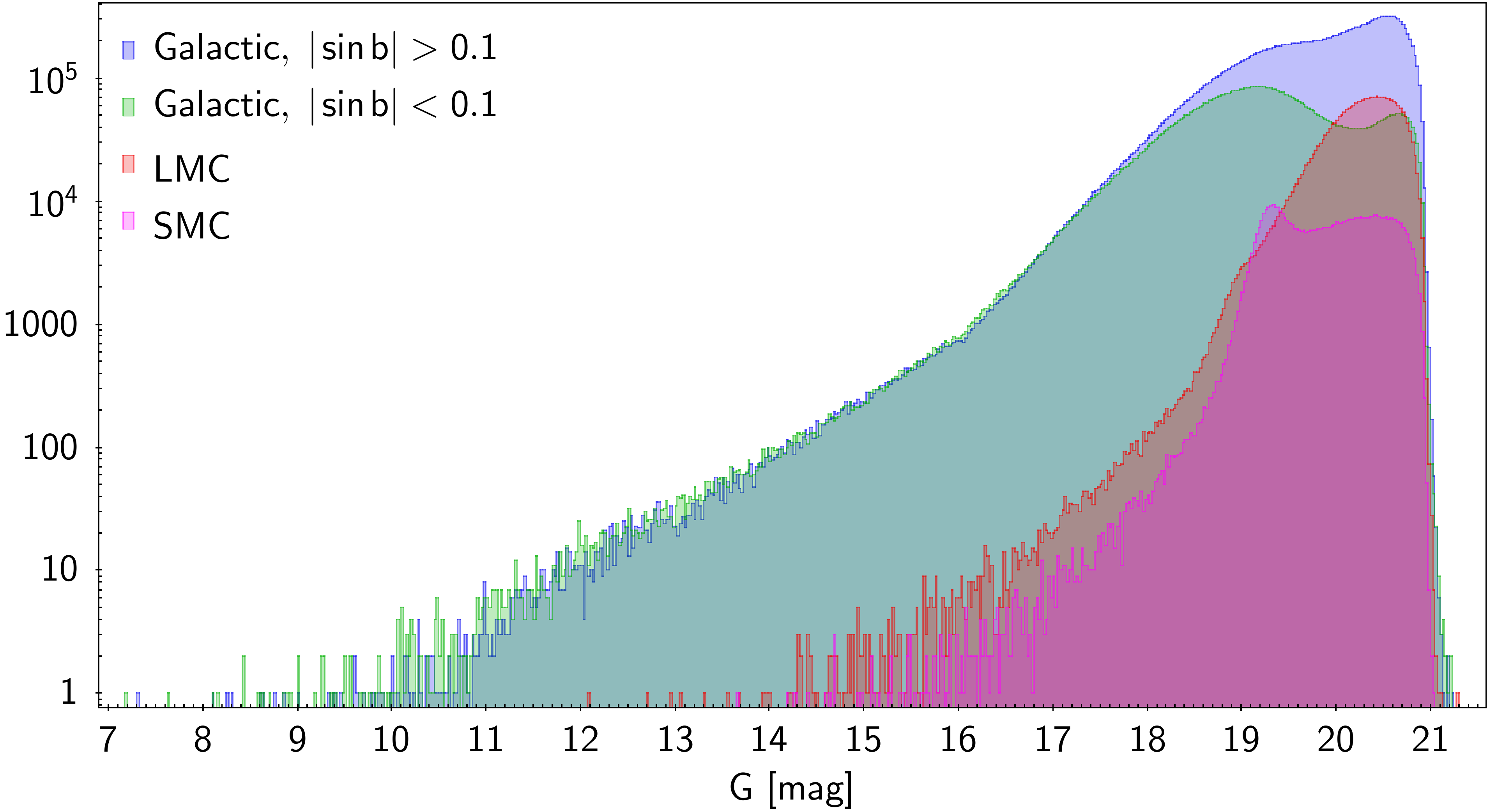}
  \includegraphics[width=\scalethree\hsize]{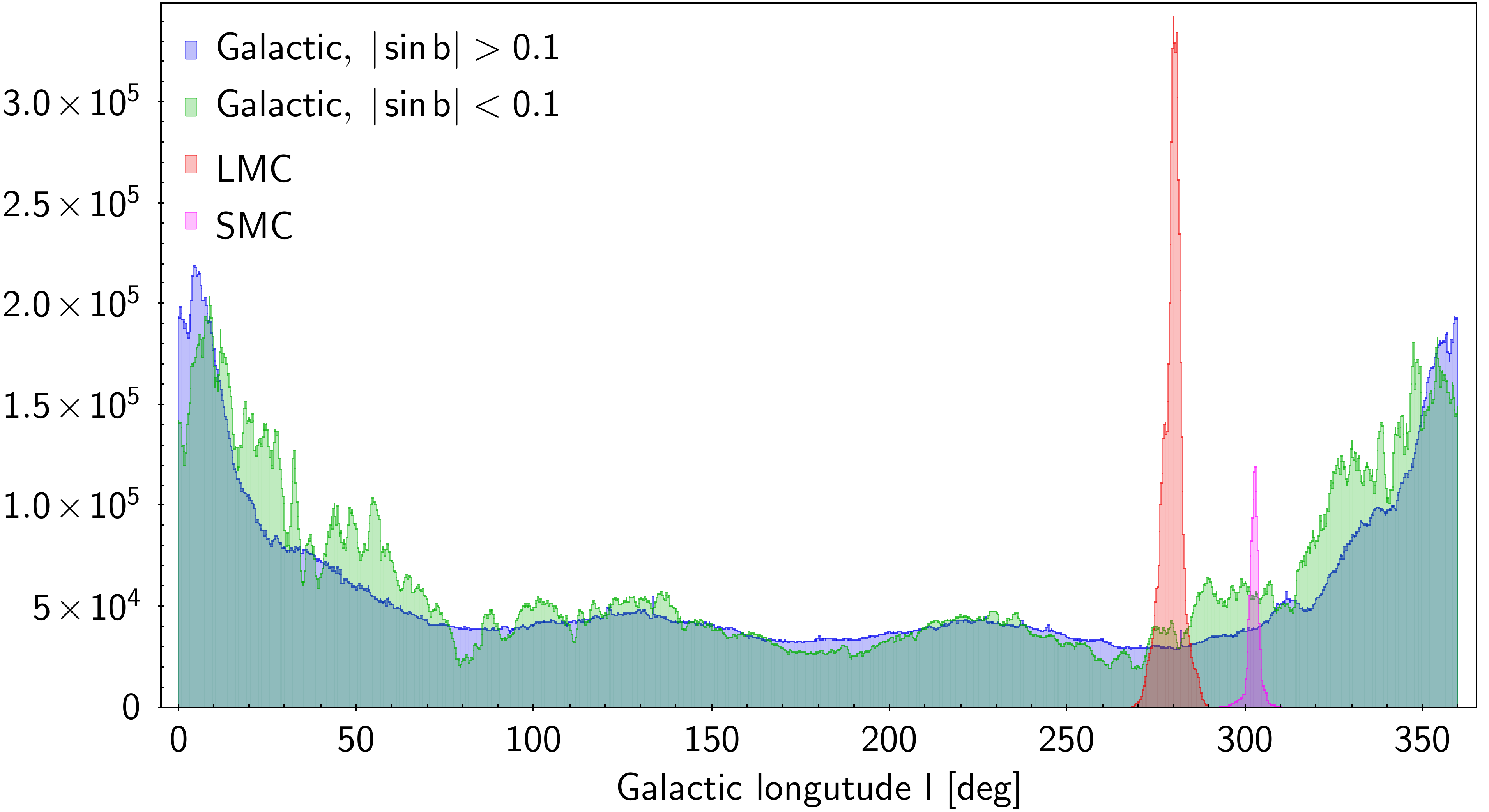}
  \includegraphics[width=\scalethree\hsize]{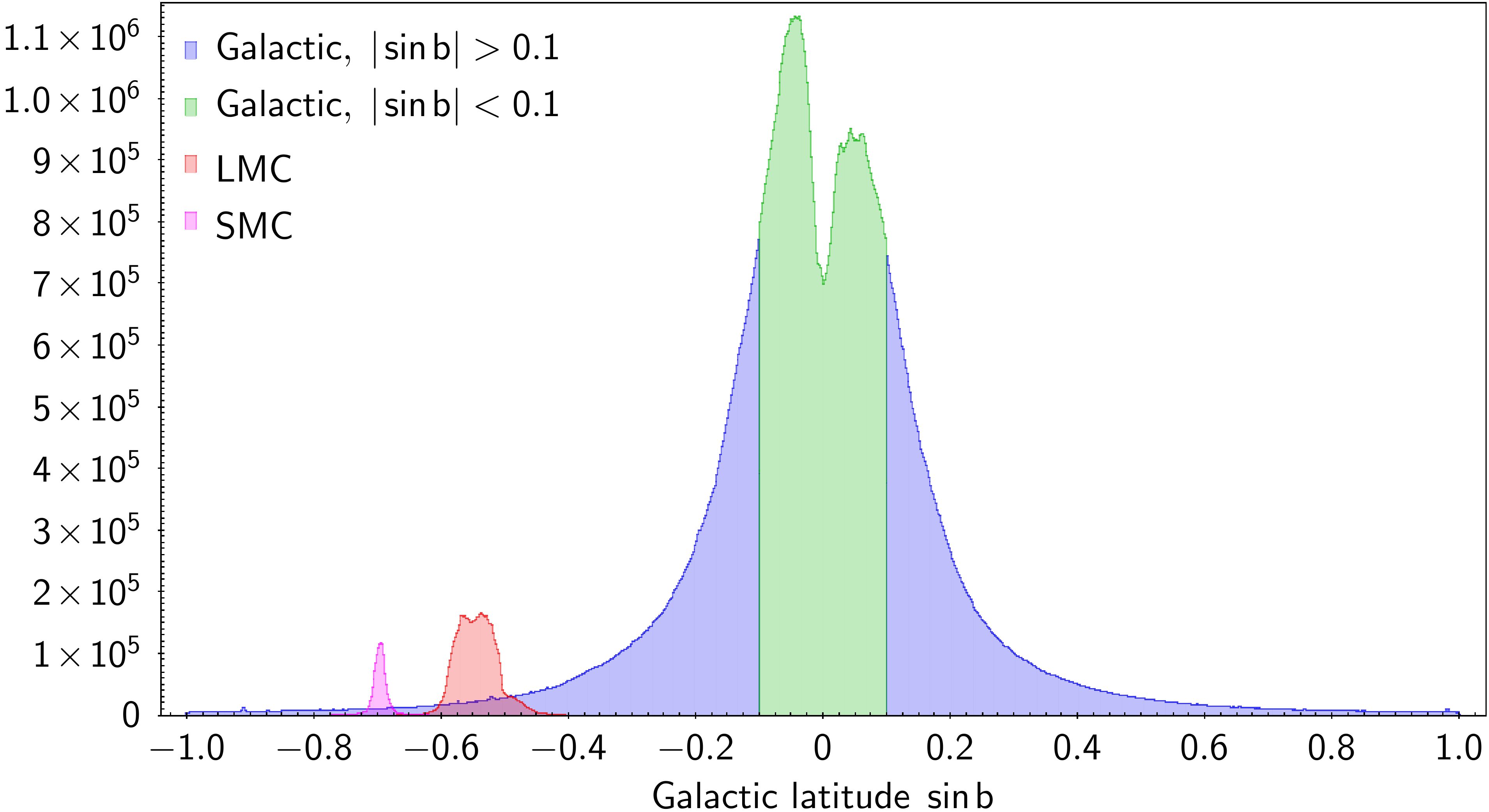}
  \includegraphics[width=\scalethree\hsize]{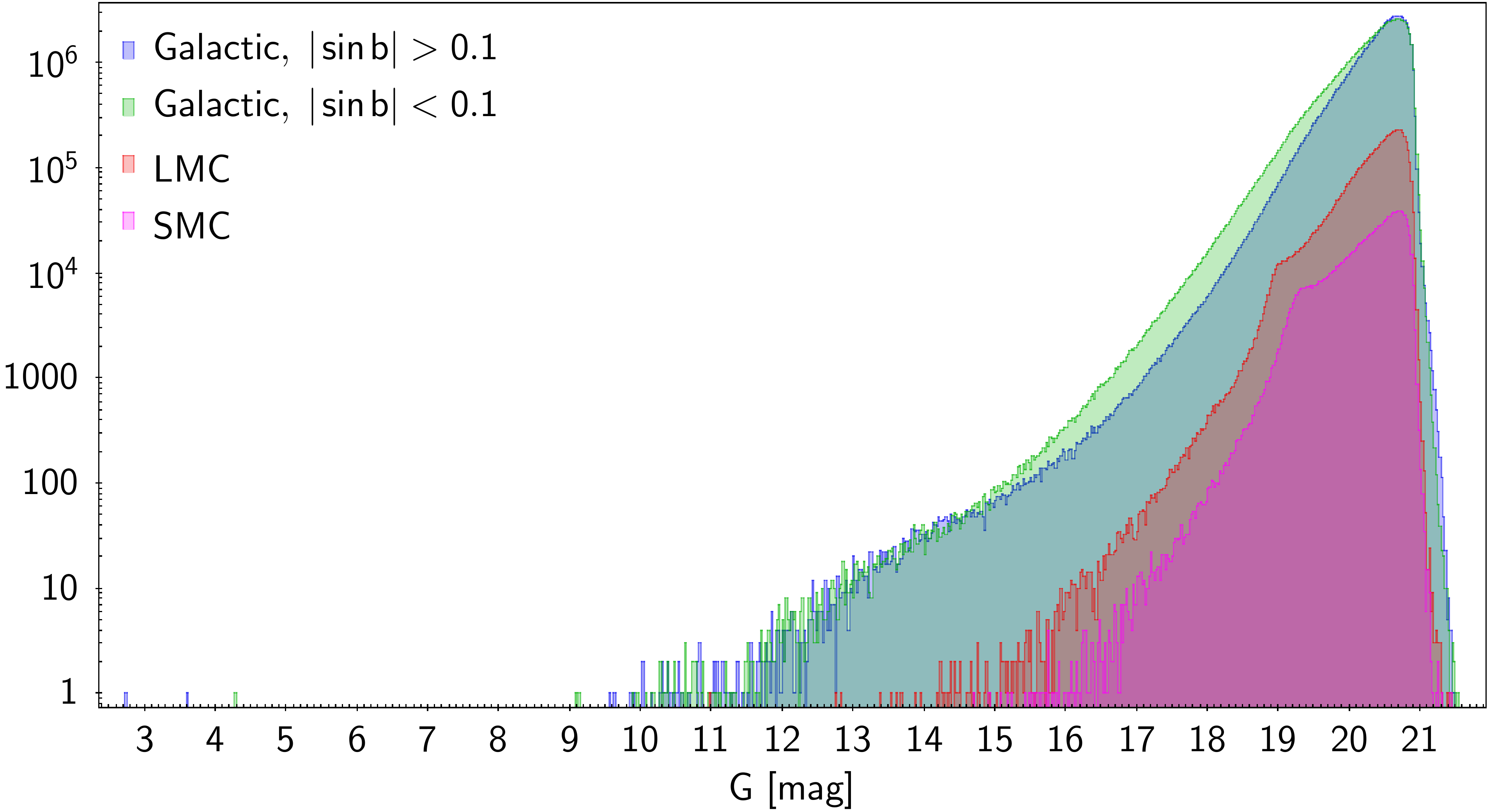}
  \caption{Distribution of confusion sources in (from \textit{left to right}) galactic 
  longitude, sine of galactic latitude, and magnitude. 
  \textit{Top:} sources with five-parameter solutions. \textit{Bottom:} sources with six-parameter solutions.
  The selections indicated in the legend are further explained in the text.}
\label{fig:histograms-confusion}
\end{center}
\end{figure*}

Figure~\ref{fig:histograms-confusion} shows the distribution of the
confusion sources in galactic coordinates and magnitude, split by the kind
of solution (five or six parameters) and further divided into four disjoint groups: 
sources in the LMC and SMC areas defined above, and for the remaining sources 
according to $\left|\,\sin b\,\right|\gtrless 0.1$. The confusion sources outside
the LMC and SMC areas are called `Galactic' below, although it is clear that this
set contains also many extragalactic objects.


\begin{figure*}
\begin{center}
  \includegraphics[width=\scalethree\hsize]{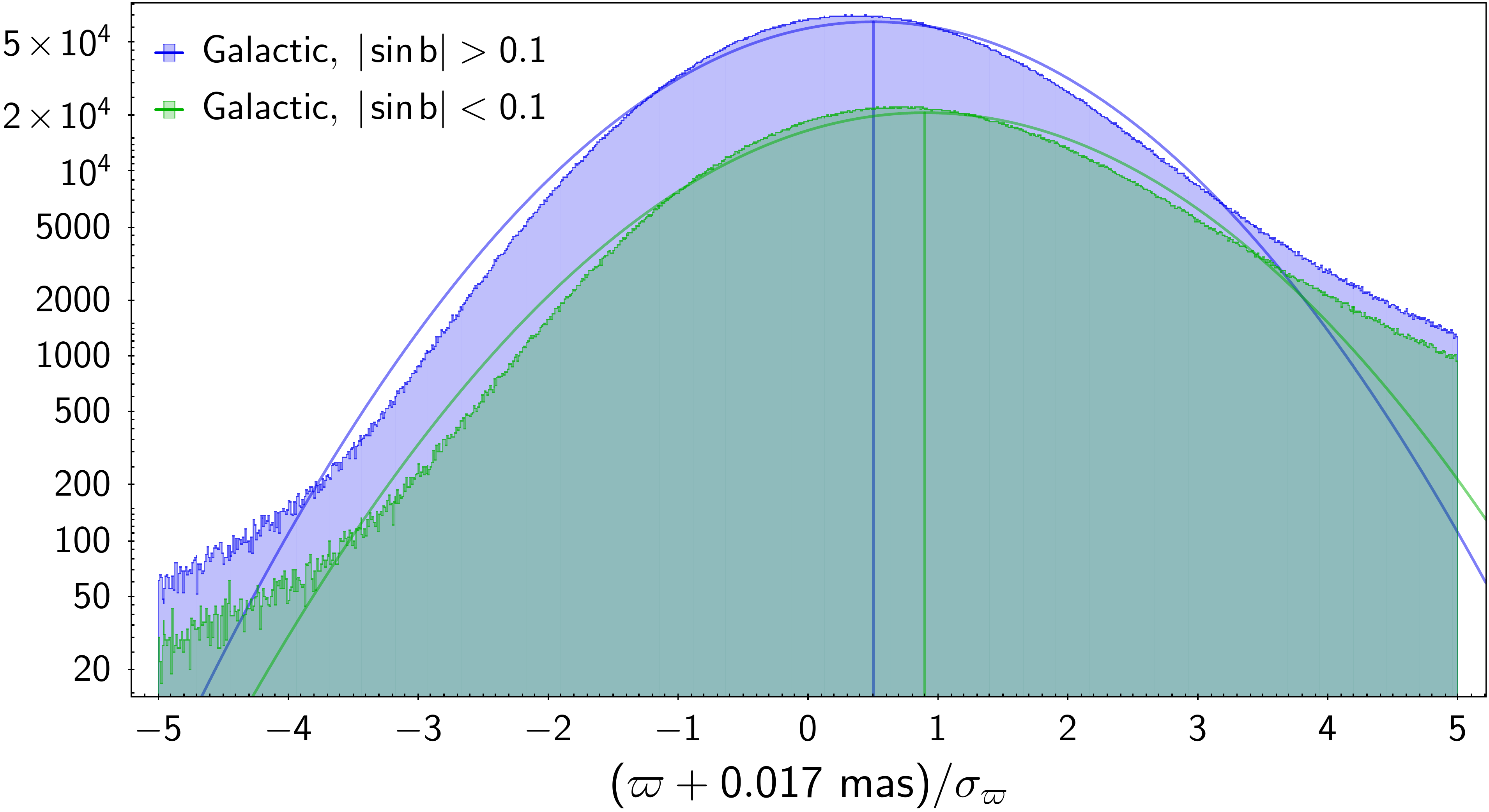}
  \includegraphics[width=\scalethree\hsize]{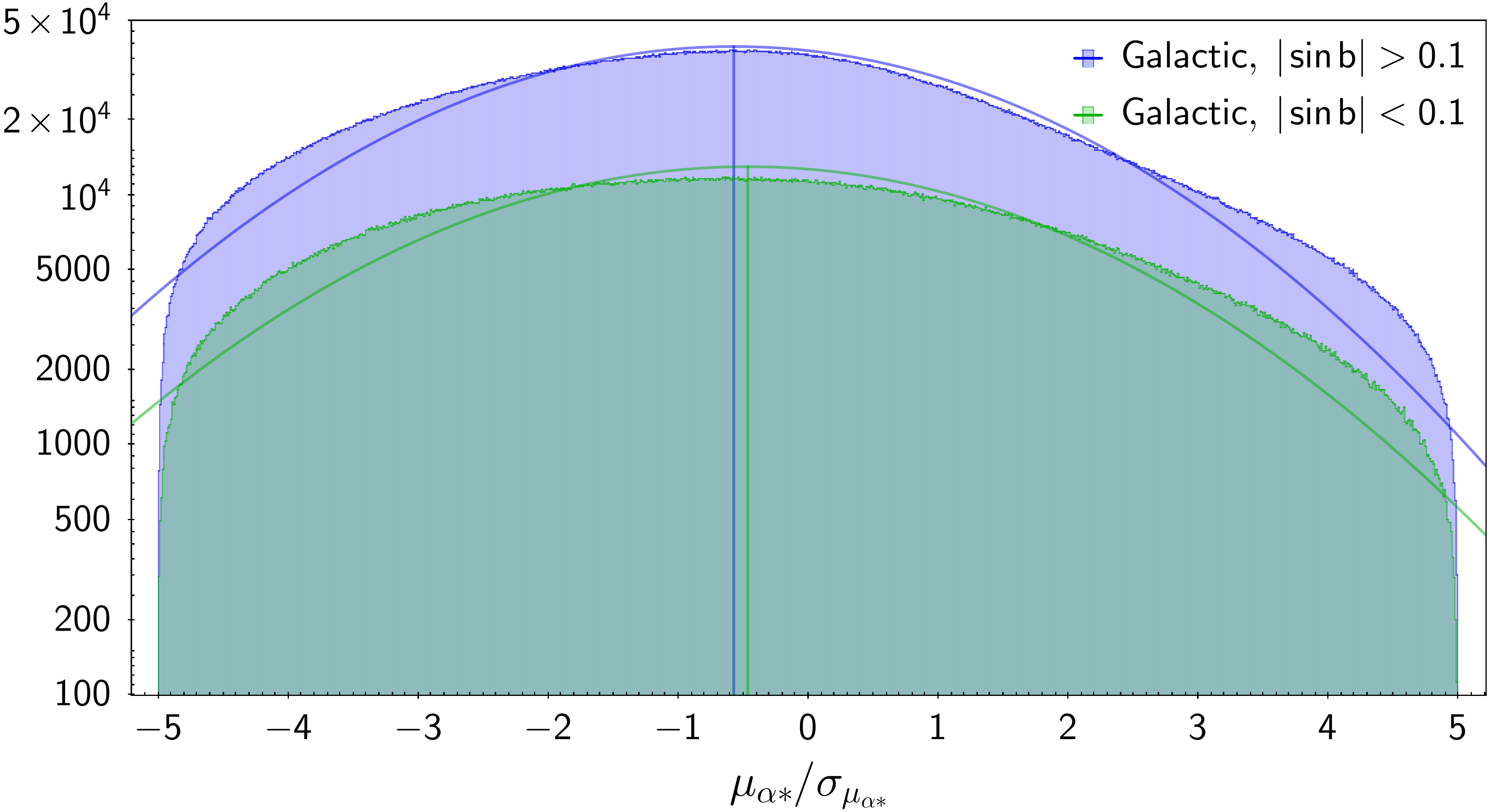}
  \includegraphics[width=\scalethree\hsize]{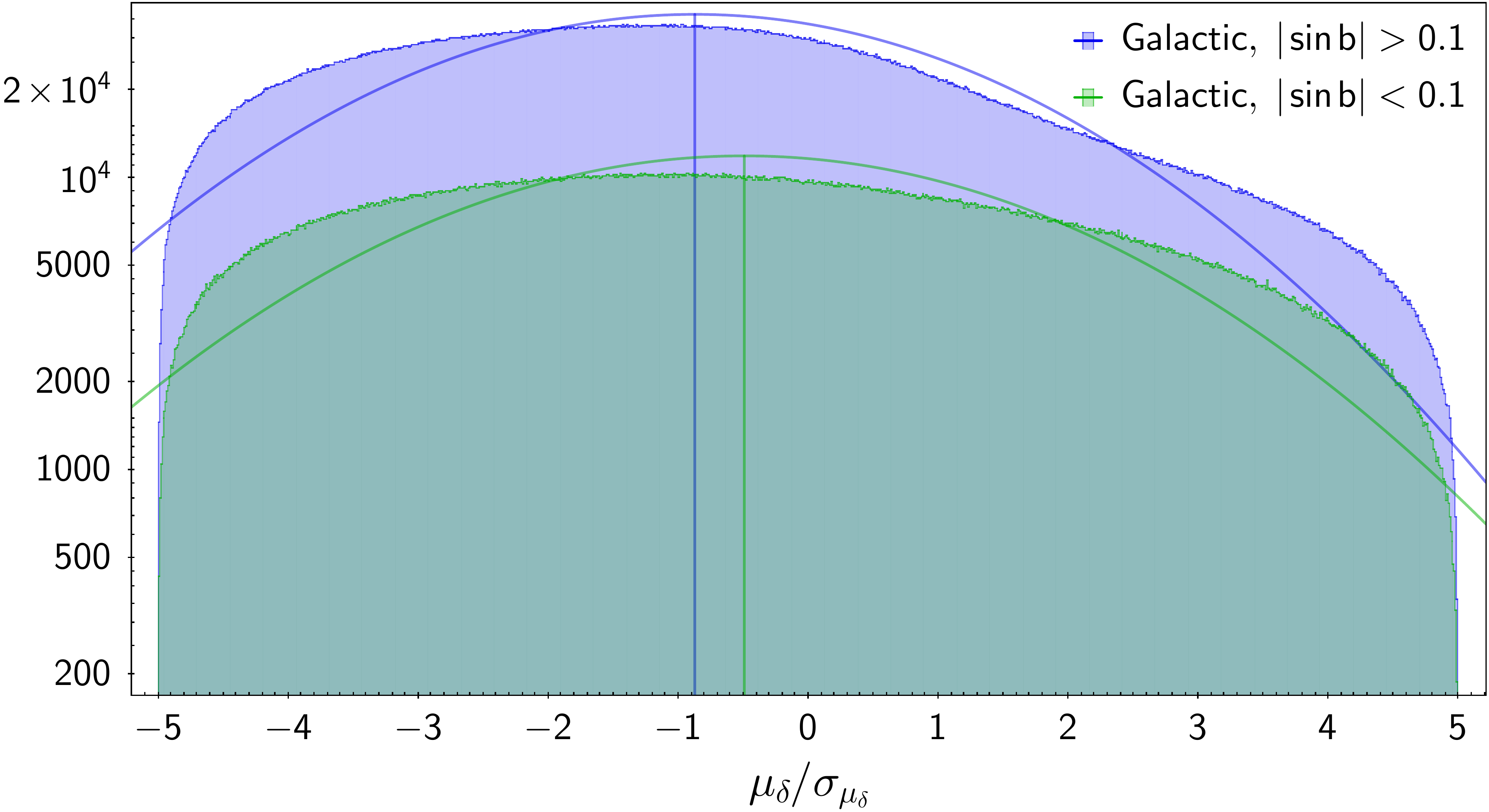}
  \includegraphics[width=\scalethree\hsize]{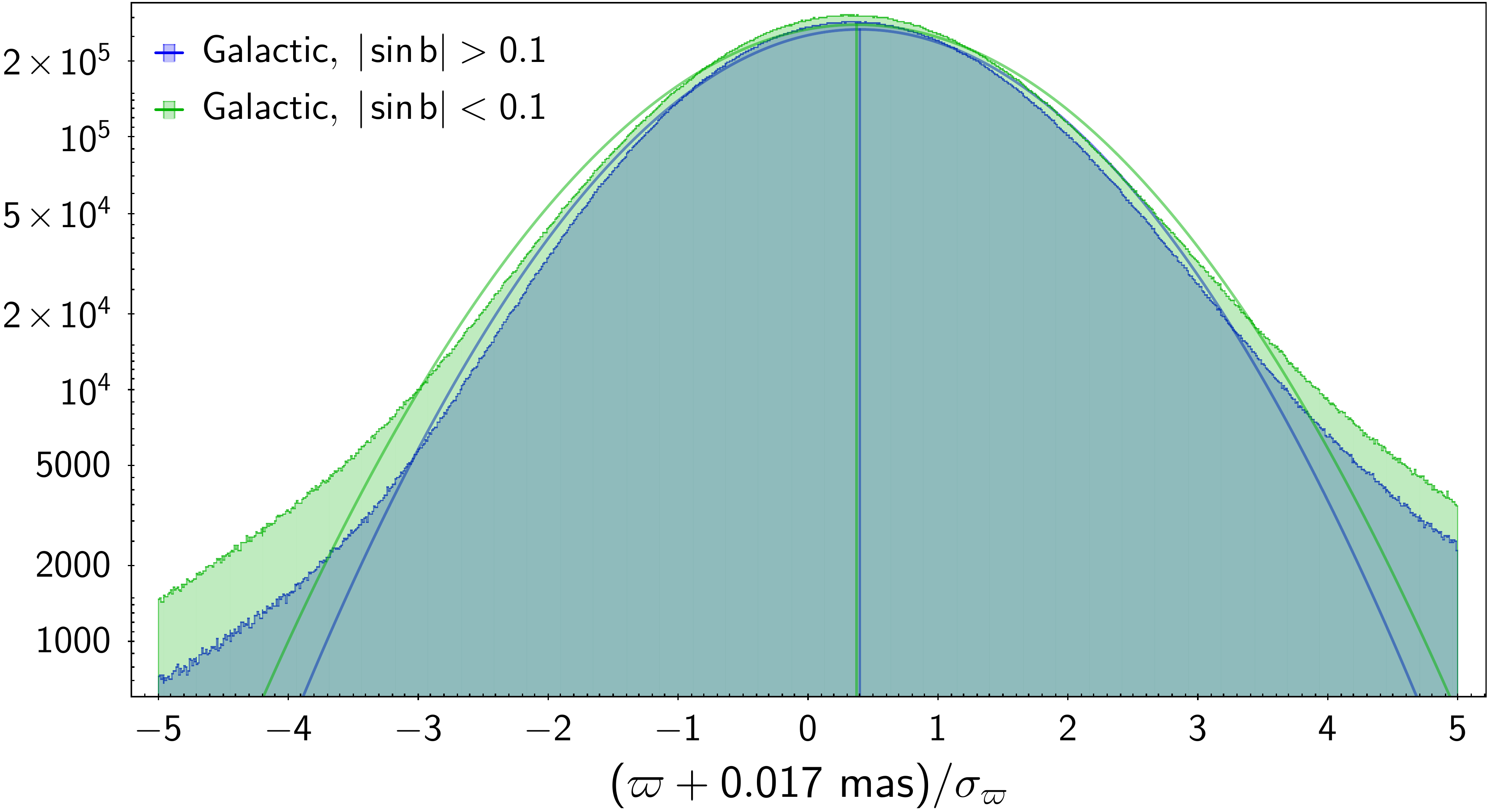}
  \includegraphics[width=\scalethree\hsize]{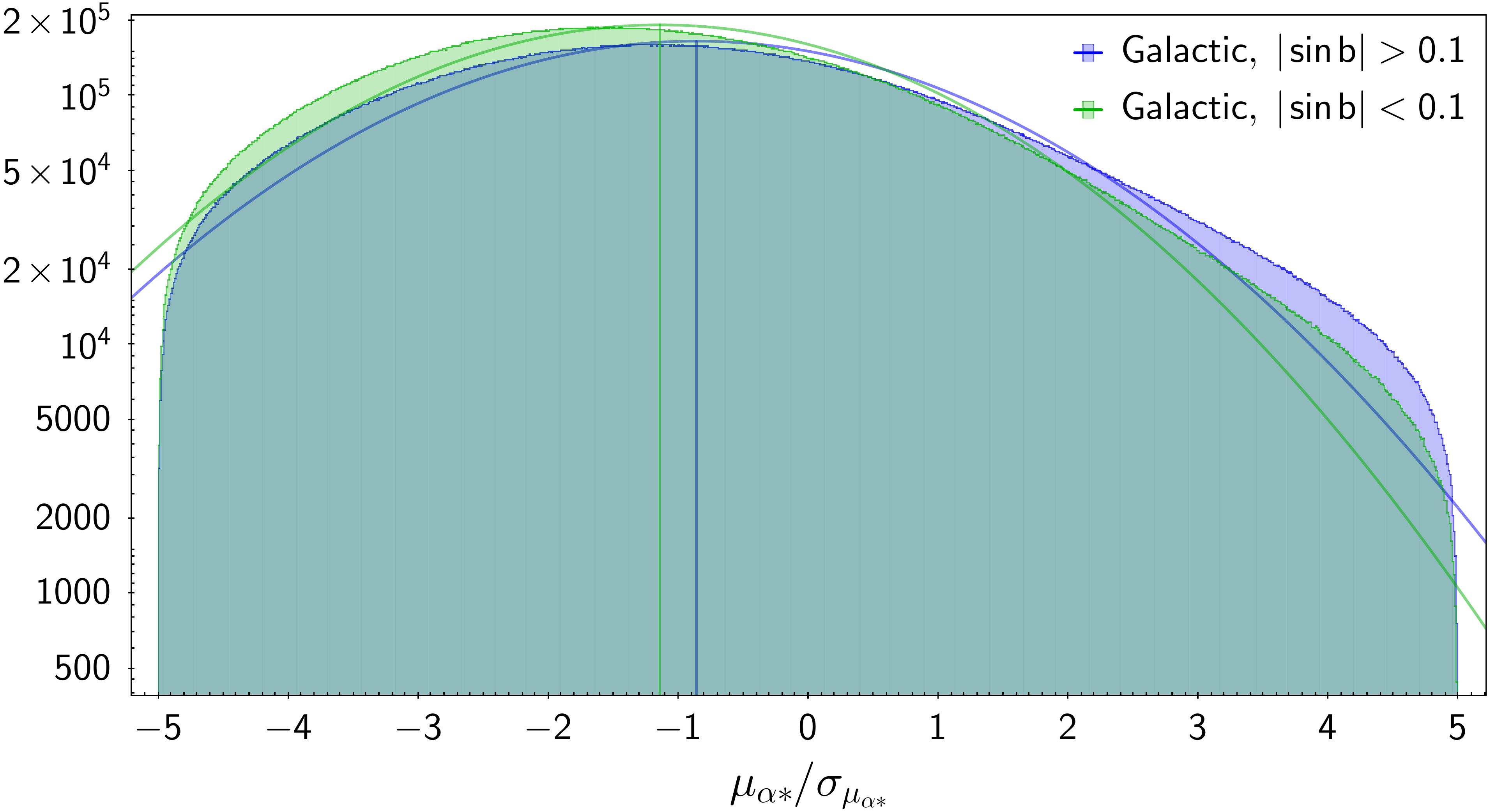}
  \includegraphics[width=\scalethree\hsize]{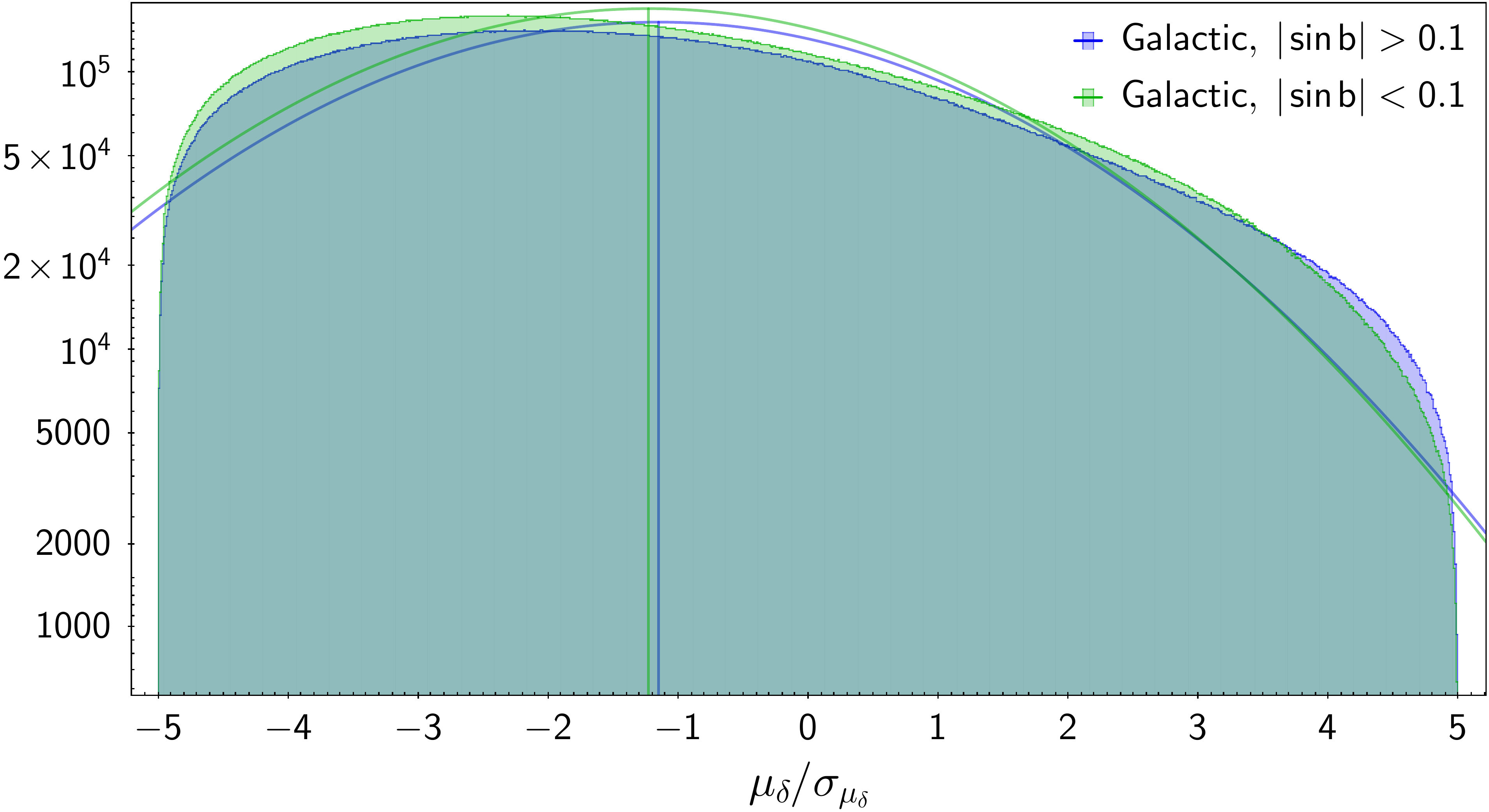}
  \caption{Histograms of the normalised parallaxes and proper motion components for the Galactic
  confusion sources.  
    \textit{Top:} sources with five-parameter solutions. \textit{Bottom:} sources with six-parameter solutions.
  The different colours show the selections indicated in the legend.
  }
\label{fig:histograms-normalized-Galaxy}
\end{center}
\end{figure*}

\begin{figure*}
\begin{center}
  \includegraphics[width=\scalethree\hsize]{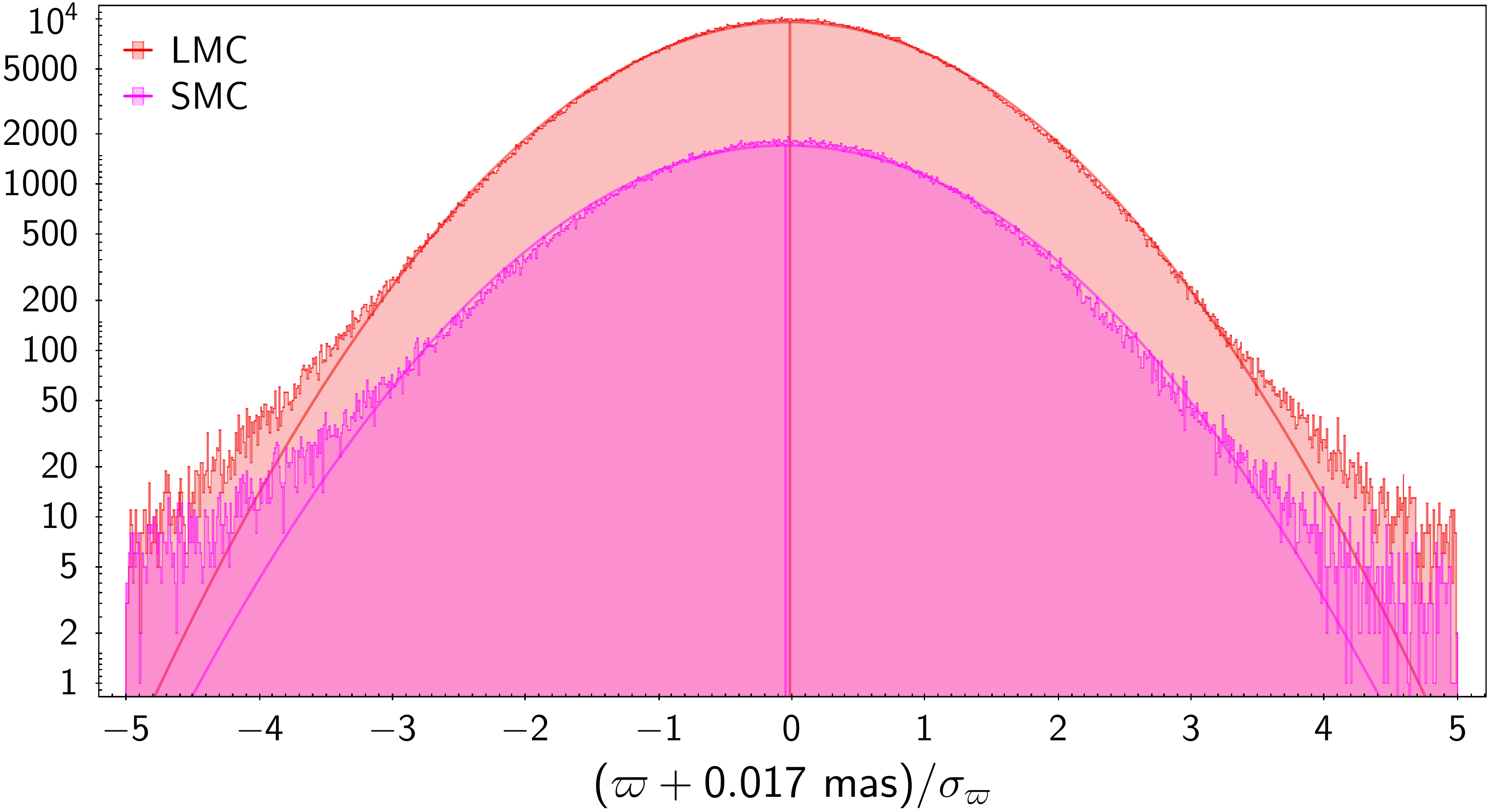}
  \includegraphics[width=\scalethree\hsize]{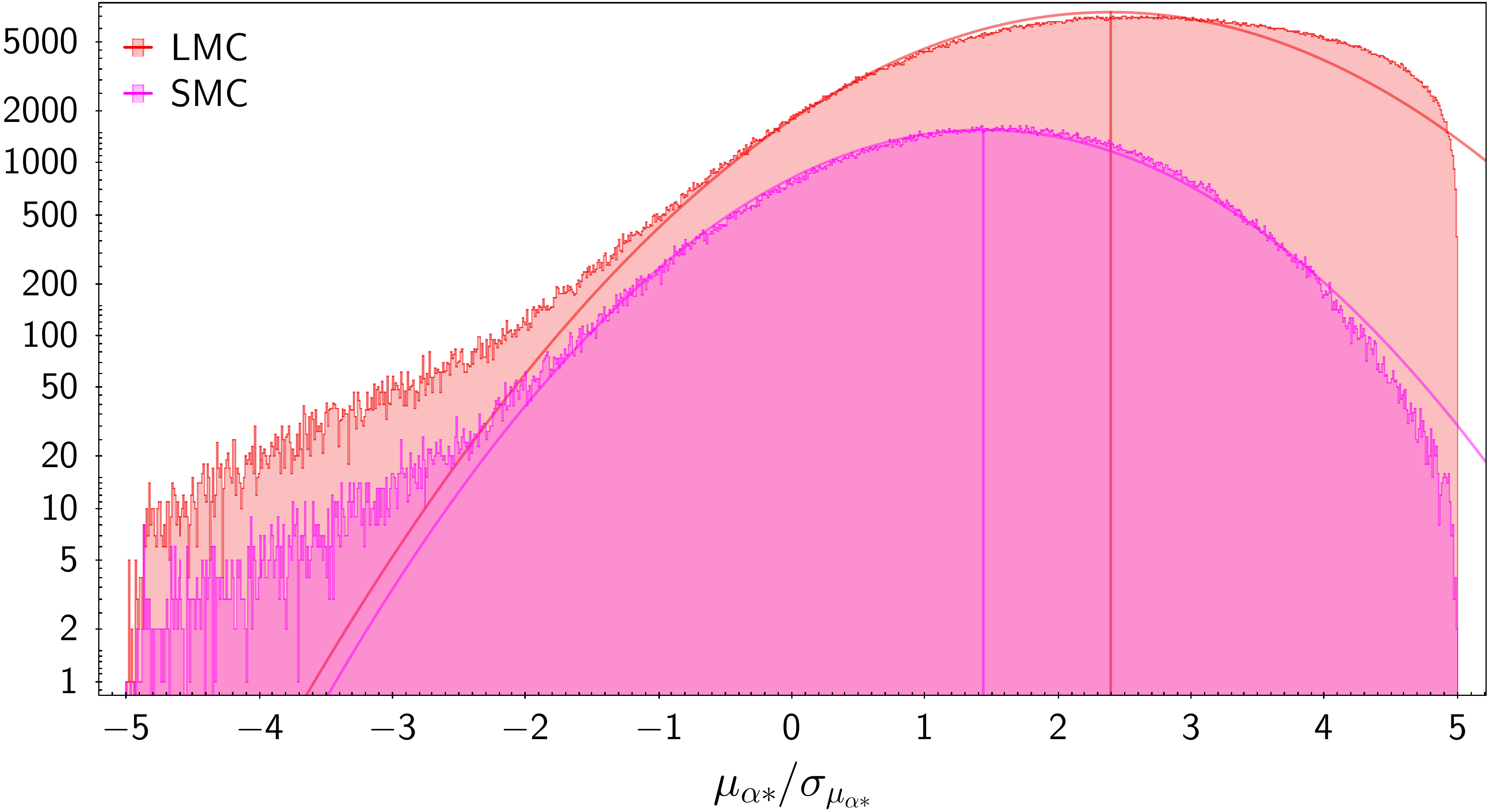}
  \includegraphics[width=\scalethree\hsize]{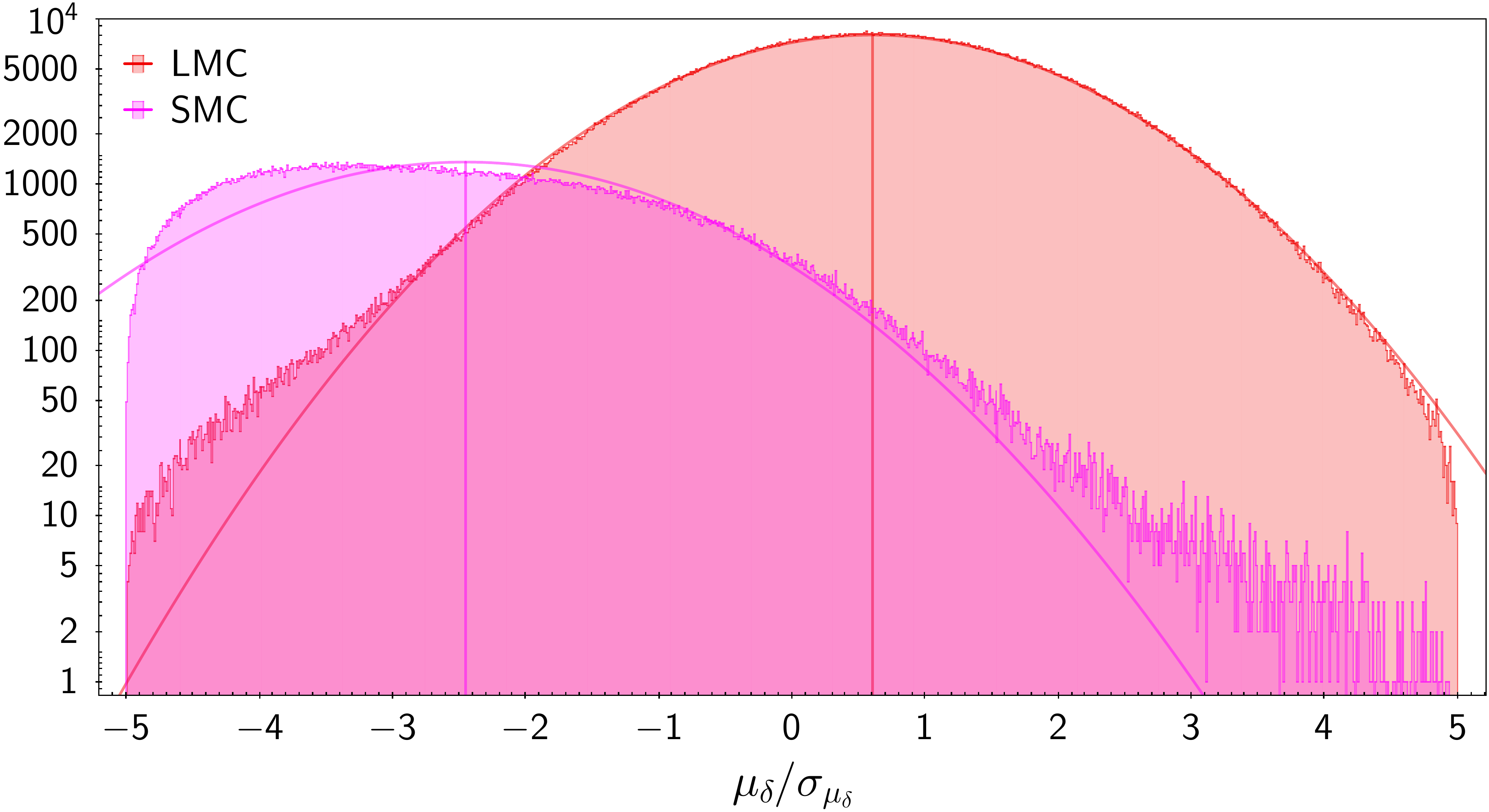}
  \includegraphics[width=\scalethree\hsize]{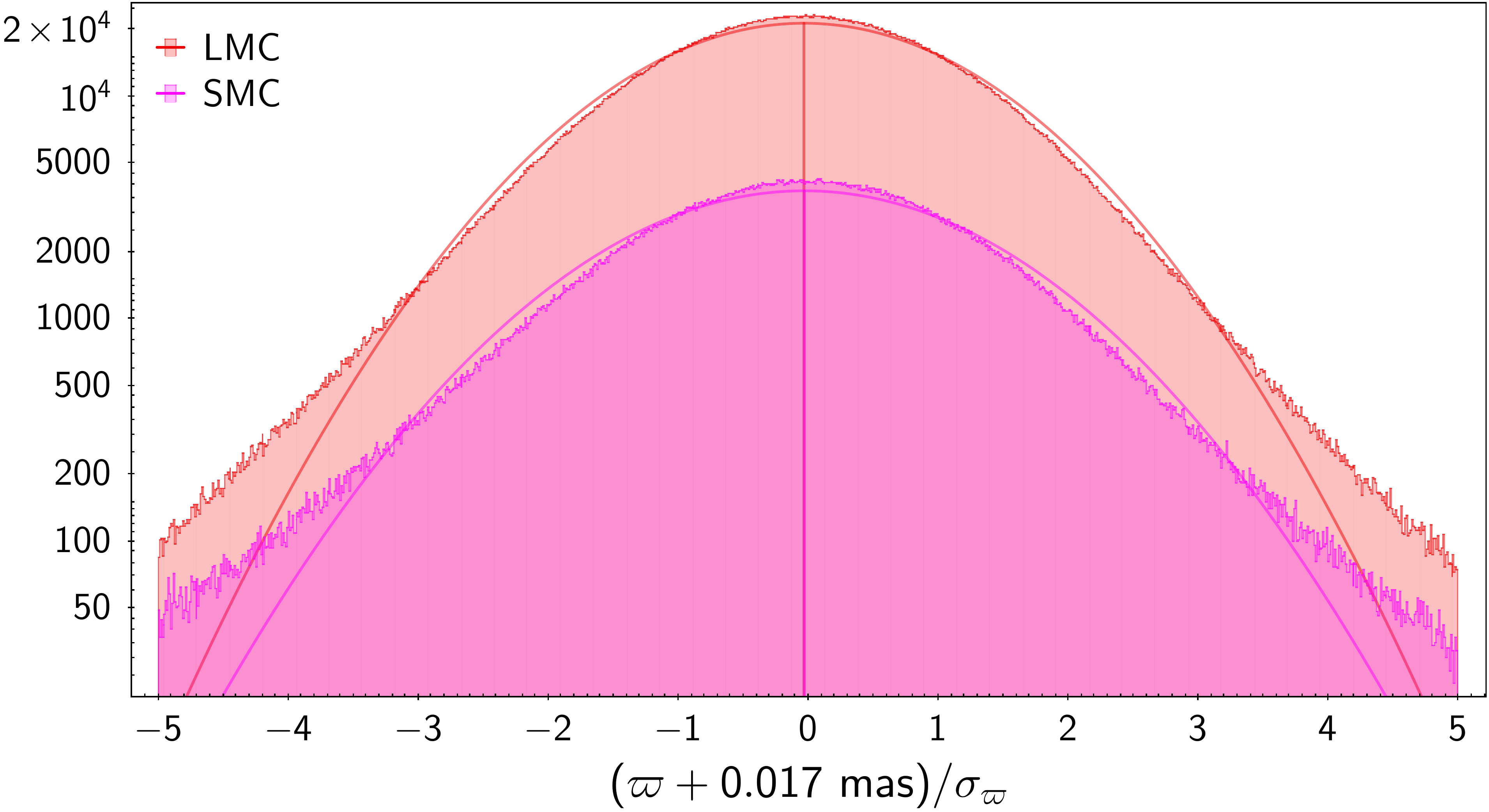}
  \includegraphics[width=\scalethree\hsize]{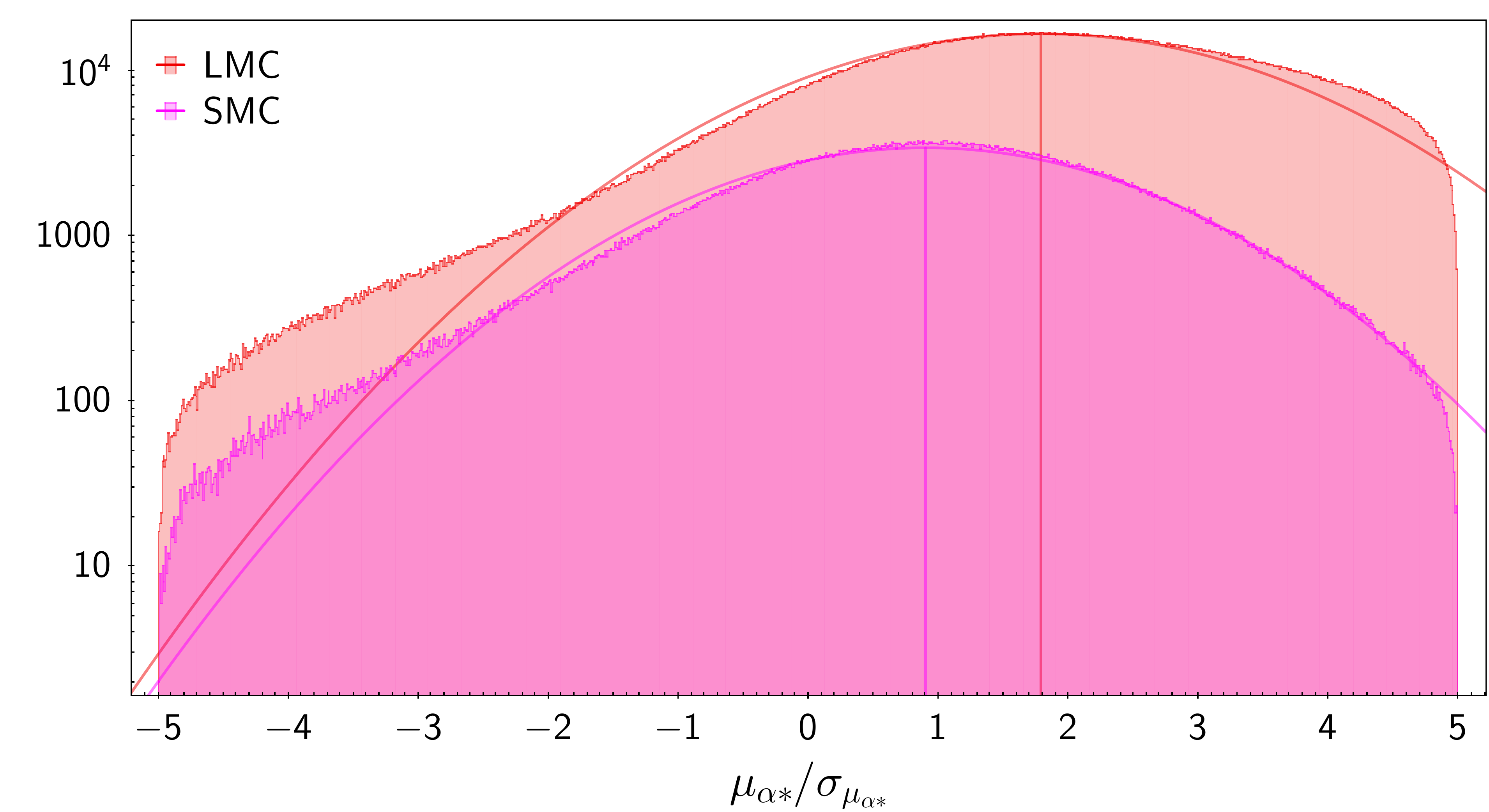}
  \includegraphics[width=\scalethree\hsize]{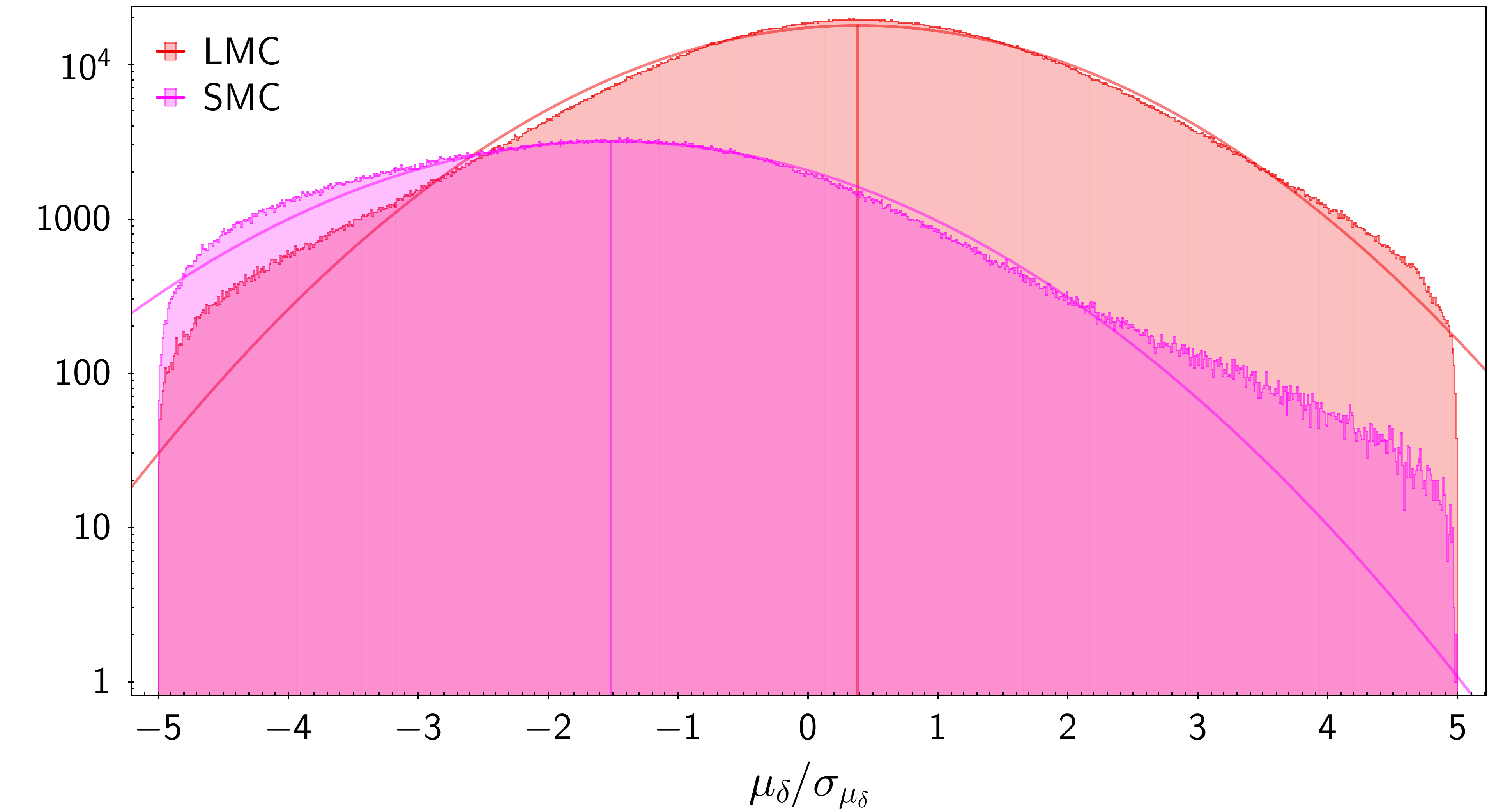}
  \caption{Histograms of the normalised parallaxes and proper motion components for the 
  confusion sources in the LMC and SMC areas.  
    \textit{Top:} sources with five-parameter solutions. \textit{Bottom:} sources with six-parameter solutions.
  }
\label{fig:histograms-normalized-LMCSMC}
\end{center}
\end{figure*}

Figures~\ref{fig:histograms-normalized-Galaxy} and \ref{fig:histograms-normalized-LMCSMC}
show the distributions of the normalised parallaxes and proper motions of the confusion sources 
separately for the Galactic sources (split by galactic latitude) and the LMC and SMC areas.
The distributions are strongly non-Gaussian (with the possible exception of parallaxes
in the LMC/SMC areas) and very different from those of the \gcrf{3}\ sources in 
Fig.~\ref{fig:pm-histograms}. In Sect.~\ref{sec:selection}, this difference was used to check 
the purity of the filtered sources in the external QSO catalogues and in the final \gcrf{3}.
The proper motions give the strongest indication of impurity. Interestingly, for the
Galactic confusion sources there is no clear dependence according to
$\left|\,\sin b\,\right|\gtrless 0.1$. On the other hand, 
the confusion sources in the LMC and SMC areas give rather different and distinct
distributions. The distributions for the five- and six-parameter solutions are not drastically different 
and mainly reflect the generally higher uncertainties of the six-parameter sources, which makes their 
distributions slightly less non-Gaussian.  

Selection of QSO-like objects in the \gaia\ catalogue, based exclusively on parallaxes 
and proper motions, is an interesting prospect already discussed in the literature 
(\citeads{2015A&A...578A..91H}; \citeyearads{2018A&A...615L...8H}).  
Among the \gedr{3} sources with five-parameter solutions, there are
23.6~million satisfying Eqs.~(\ref{eq:filterA})--(\ref{eq:filterB}), of which 1.2~million
(5.1\%) are also in \gcrf{3}. Allowing that the actual number of QSOs among the 
five-parameter sources in \gedr{3} could be a factor two higher (cf.\ Sect.~\ref{sec:properties}), 
we conclude that the purity of such a selection from \gedr{3} would be at most about 10\%  
-- too low to be of practical use, but still interesting and promising. In certain areas on 
the sky the percentage of true QSO-like sources in the sample is much higher. For example, 
excluding $\left|\,b\,\right|<20^\circ$ and the LMC/SMC areas, there are 5\,767\,163 confusion 
sources, of which 1\,410\,314 (24.5\%) are \gcrf{3} sources. The purity of such selections 
in future \gaia\ data releases will improve drastically thanks to the higher accuracy of 
parallaxes and (especially) proper motions (cf.\ Sect.~\ref{sect:selection_discussion}). 

\clearpage
\clearpage

\section{Estimating the spin and orientation corrections} 
\label{sec:estimating}

As described in Sect.~6.1 of \citet{2012A&A...538A..78L}, the frame rotator is an integral 
part of AGIS with three main functions: (i) to estimate the spin and orientation corrections 
needed to bring the current astrometric solution onto the ICRS; (ii) to apply the estimated
corrections to the astrometric parameters of the primary sources; and (iii) to apply the
corresponding corrections to the attitude parameters. This appendix describes the 
algorithm used for the first function in the construction of \textit{Gaia} EDR3.

The 2012 paper gave a stringent mathematical framework for all three functions and 
outlined how they could be implemented in AGIS. The frame rotator used for the first 
two releases of \textit{Gaia} data followed these prescriptions quite rigorously, which 
however resulted in code that was unnecessarily complex and not very transparent. For example, 
it permitted the use of different reference epochs for the AGIS solution, comparison catalogue, 
and orientation parameters, while in practice this was never needed. It also foresaw the use 
of radio stars observed by VLBI to help align the reference frame, which is no longer considered 
necessary or even desirable. On the other hand, the details of the numerical algorithm used
for the robust least-squares estimation (LSE) of the rotation parameters were not at all 
described in \citet{2012A&A...538A..78L}, making it very difficult for outside users to 
reproduce how the frame rotator worked in those releases.

In 2018 it was decided to implement a new version of the frame rotator to be used for 
the third and subsequent data releases. The main changes with respect to the earlier 
version are as follows:
only two kinds of sources are used for estimating the rotation parameters, namely 
QSO-like sources with high-accuracy external positions (the VLBI positions of ICRF3 S/X
in \gedr{3}), used for both orientation and spin, and 
QSO-like sources without accurate external positions, used only for the spin. 
The same reference epoch is used for rotation parameters as for the astrometric
data (2016.0 for (E)DR3), which permits to separate completely the solution for 
the orientation from that of the spin.
Furthermore, the observation equations are written directly in terms of the proper
motions and position differences (Sect.~\ref{sec:estimationObs}), rather than using
the rigorous but less transparent formalism of \citet{2012A&A...538A..78L}. 
The solution takes into account the correlation between the components 
in right ascension and declination of both \textit{Gaia} and external data
(Sect.~\ref{sec:estimationDecorr}), and has a transparent scheme for the detection 
and treatment of outliers (Sect.~\ref{sec:estimationOut}). 
Finally, the solution provides reasonable estimates of the statistical uncertainties of 
the rotation parameters also when the uncertainties of the input data are underestimated
(Sect.~\ref{sec:estimationUnc}), and may optionally include additional terms, such as a 
glide in the proper motions or more general expansions of the proper motions and 
position differences (Sect.~\ref{sec:estimationOpt}).

All these changes concern the estimation algorithm, that is the first function of 
the frame rotator. The remaining two functions, that is the application of the 
corrections to the source and attitude data, are still done as described in
Sects.~6.1.2 and 6.1.3 of the 2012 paper and are not further discussed here.

We denote by ${\cal Q}_0$ the set of QSO-like sources considered for the 
determination of the spin correction, of which the subset ${\cal Q}$ was actually used 
after elimination of outliers; similarly, ${\cal R}_0 $ and ${\cal R}$ are the set of
sources considered and used for the determination of the orientation. Although 
not required by the algorithm, we normally have ${\cal R}_0\subseteq{\cal Q}_0$.
The number of sources in each set is denoted $n({\cal Q}_0)$, etc. We use
$\sigma(x)$, $\rho(x,y)$, and $\text{Cov}(x,y)$ for the uncertainty, correlation 
coefficient, and covariance of the arbitrary parameters $x$ and $y$. The asterisk
in $\alpha^*$ indicates that the $\cos\delta$ factor is implicit, as in 
$\mu_{\alpha^*}=\dot{\alpha}\cos\delta$. To distinguish between data in the
current astrometric solution and the corresponding data in the external catalogue, we
put an overline ($\bar{\phantom{x}}$) on the latter. The prime $'$ denotes the 
matrix transpose.
 
\subsection{Observation equations}
\label{sec:estimationObs}

Let $\vec{\omega}=[\omega_X,~\omega_Y,~\omega_Z]'$ be the spin correction 
of the current solution, expressed by its components along the principal 
axes of the ICRS. (As indicated by the prime, $\vec{\omega}$ is a column matrix.) 
In matrix form, the observation equations in proper motion for source $i$ are
\begin{equation}\label{eq:obseqPm}
\vec{A}_i\vec{\omega} = \vec{\mu}_i \, ,
\end{equation}
with coefficient matrix
\begin{equation}\label{eq:obseqA}
\vec{A} = \begin{bmatrix}
-\cos\alpha_i\sin\delta_i &\quad & +\sin\alpha_i\sin\delta_i &\quad & +\cos\delta_i \\
+\sin\alpha_i && -\cos\alpha_i && 0
\end{bmatrix} \, 
\end{equation}
and right-hand side
\begin{equation}\label{eq:obseqDeltaPm}
\vec{\mu}_i = \begin{bmatrix} \mu_{\alpha^*i} \\
\mu_{\delta{}i} \end{bmatrix} \, .
\end{equation}
Here, $\alpha_i$, $\delta_i$, $\mu_{\alpha^*i}$ and $\mu_{\delta{}i}$ are the position 
and proper motion components of the source in the current solution, with reference epoch
$t_\text{ref}$ ($=2016.0$ for \textit{Gaia} EDR3).
For the weight normalisation and decorrelation (Sect.~\ref{sec:estimationDecorr}),
we also need the covariance of the right-hand side in Eq.~(\ref{eq:obseqDeltaPm}), 
\begin{multline}\label{eq:covPm}
\text{Cov}(\vec{\mu}_i) =\\
\begin{bmatrix} \sigma(\mu_{\alpha^*i})^2 
& \rho(\mu_{\alpha^*i},\mu_{\delta{}i})\,\sigma(\mu_{\alpha^*i})\,\sigma(\mu_{\delta{}i})~\\[6pt]
~\rho(\mu_{\alpha^*i},\mu_{\delta{}i})\,\sigma(\mu_{\alpha^*i})\,\sigma(\mu_{\delta{}i}) 
& \sigma(\mu_{\delta{}i})^2
\end{bmatrix} \, .
\end{multline}

Similarly, let $\vec{\varepsilon}=[\varepsilon_X,~\varepsilon_Y,~\varepsilon_Z]'$ be the
orientation correction at the reference epoch $t_\text{ref}$, expressed by its components
in the ICRS. The linearised observation equations in position for source $i$ are
\begin{equation}\label{eq:obseqPos}
\vec{A}_i\vec{\varepsilon} = \vec{\Delta}_i\, ,
\end{equation}
with the same coefficient matrix $\vec{A}_i$ as before. The right-hand side, 
\begin{equation}\label{eq:obseqDeltaPos}
\vec{\Delta}_i \equiv
\begin{bmatrix} \Delta\alpha^*_i \\ \Delta\delta_i \end{bmatrix}
= \begin{bmatrix} \left(\alpha_i-\bar{\alpha}_i\right)\cos\delta_i \\
  \delta_i-\bar{\delta}_i \end{bmatrix}
\, ,
\end{equation}
is the difference between the position $(\alpha_i,~\delta_i)$ in the current solution and 
the position $(\bar{\alpha}_i,~\bar{\delta}_i)$ according to the external catalogue.
We note that the external positions at $t_\text{ref}$ (= 2016.0 for EDR3) should be used here.
In particular, for \gedr{3} the galactic acceleration should be taken into account with the ICRF3 S/X positions
as prescribed in Sect.~5.3 of \citetads{2020A&A...644A.159C}.
The covariance of $\vec{\Delta}_i$ in Eq.~(\ref{eq:obseqDeltaPos}) is computed on the 
assumption that the positional errors of the current solution and of the external catalogue are independent:
\begin{equation}\label{eq:covPos}
\text{Cov}(\vec{\Delta}_i) = \text{Cov}(\alpha^*_i,~\delta_i) 
+ \text{Cov}(\bar{\alpha}^*_i,~\bar{\delta}_i) \, ,
\end{equation}
where  
\begin{multline}\label{eq:covPosG}
\text{Cov}(\alpha^*_i,~\delta_i) =\\ 
\begin{bmatrix}
\sigma(\alpha^*_i)^2 &\quad& \rho(\alpha^*_i,\delta_i)\,\sigma(\alpha^*_i)\,\sigma(\delta_i)~\\[6pt]
~\rho(\alpha^*_i,\delta_i)\,\sigma(\alpha^*_i)\,\sigma(\delta_i) &\quad& \sigma(\delta_i)^2
\end{bmatrix} \, ,
\end{multline}
and correspondingly for the external data. We note that
\begin{align}\label{eq:sigmaDeltaAlpha}
\sigma(\Delta\alpha^*_i) &= \sqrt{\sigma(\alpha^*_i)^2+\sigma(\bar{\alpha}^*_i)^2}\, ,\\
\label{eq:sigmaDeltaDelta}
\sigma(\Delta\delta_i) &= \sqrt{\sigma(\delta_i)^2+\sigma(\bar{\delta}_i)^2}\, ,\\
\label{eq:sigmaRho}
\rho(\Delta\alpha^*_i,\Delta\delta_i) &= \frac{\rho(\alpha^*_i,\delta_i)\,
\sigma(\alpha^*_i)\,\sigma(\delta_i)
+\rho(\bar{\alpha}^*_i,\bar{\delta}_i)\,\sigma(\bar{\alpha}^*_i)\,\sigma(\bar{\delta}_i)}%
{\sigma(\Delta\alpha^*_i)\,\sigma(\Delta\delta_i)}\, .
\end{align}

The total number of observation equations considered for $\vec{\omega}$ is 
$2n({\cal Q}_0)$, and for $\vec{\varepsilon}$ it is $2n({\cal R}_0)$. A feature of the 
algorithm described here is that the observation equations always come in pairs, 
as in Eqs.~(\ref{eq:obseqPm}) and (\ref{eq:obseqPos}), and that the two components 
in a pair are always treated together. This simplifies the decorrelation of data (see
below), and affects the way outliers are detected and removed. The resulting 
solution and statistics are independent of the coordinate system in which they are 
calculated.

\subsection{Weight normalisation and decorrelation}
\label{sec:estimationDecorr}

Because the observation equations have different uncertainties (heteroscedasticity) 
and are pairwise correlated, they should be solved by the generalised least-squares 
method for optimality. Equivalently, the ordinary (unweighted) least-squares method 
can be applied to the normalised equations,
\begin{align}\label{eq:obseqPmK}
\vec{K}_i\vec{A}_i\vec{\omega} &= \vec{K}_i\vec{\mu}_i\, , \\[6pt]
\label{eq:obseqPosK}
\vec{L}_i\vec{A}_i\vec{\varepsilon} &= \vec{L}_i\vec{\Delta}_i\, ,
\end{align}
where $\vec{K}_i$ and $\vec{L}_i$ are $2\times 2$ matrices chosen in such a way
that the transformed right-hand sides are uncorrelated and of unit variance.
This condition does not define $\vec{K}_i$ and $\vec{L}_i$ uniquely, since any 
orthogonal transformation of them will give the same result; the form used here 
is the inverse lower Cholesky factor of the covariance matrix, or
\begin{equation}\label{eq:obseqK}
\vec{K}_i = \begin{bmatrix} 
\frac{1}{\sigma(\mu_{\alpha^*i})}  &\quad& 0 \\[12pt] 
~\frac{-\rho(\mu_{\alpha^*i},\,\mu_{\delta{}i})}%
{\sigma(\mu_{\alpha^*i})\sqrt{1-\rho(\mu_{\alpha^*i},\,\mu_{\delta{}i})^2}} &\quad& 
\frac{1}{\sigma(\mu_{\delta{}i})\sqrt{1-\rho(\mu_{\alpha^*i},\,\mu_{\delta{}i})^2}}~
\end{bmatrix}\, 
\end{equation}
and
\begin{equation}\label{eq:obseqL}
\vec{L}_i = \begin{bmatrix} 
\frac{1}{\sigma(\Delta\alpha^*_i)}  &\quad& 0 \\[12pt] 
~\frac{-\rho(\Delta\alpha^*_i,\,\Delta\delta_i)}%
{\sigma(\Delta\alpha^*_i)\sqrt{1-\rho(\Delta\alpha^*_i,\,\Delta\delta_i)^2}} &\quad& 
\frac{1}{\sigma(\Delta\delta_i)\sqrt{1-\rho(\Delta\alpha^*_i,\,\Delta\delta_i)^2}}~
\end{bmatrix}\, .
\end{equation}
It is readily verified that, as required, $\vec{K}_i\text{Cov}(\vec{\mu}_i)\vec{K}_i{'}=\vec{I}$
and $\vec{L}_i\text{Cov}(\vec{\Delta}_i)\vec{L}_i{'}=\vec{I}$, where $\vec{I}$ is the identity matrix.

Given the subsets ${\cal Q}$, ${\cal R}$ of the accepted sources, the generalised least-squares 
solutions are obtained by minimising the sum of squared normalised residuals,
\begin{align}
\label{eq:obseqXpm}
X^2(\vec{\omega},{\cal Q}) &= \sum_{i\,\in\,{\cal Q}}\,
\bigl|\vec{K}_i\vec{\mu}_i-\vec{K}_i\vec{A}_i\vec{\omega}\,\bigr|^2\, , \\[6pt]
\label{eq:obseqXpos}
X^2(\vec{\varepsilon},{\cal R}) &= \sum_{i\,\in\,{\cal R}}\,
\bigl|\vec{L}_i\vec{\Delta}_i-\vec{L}_i\vec{A}_i\vec{\varepsilon}\,\bigr|^2\, .
\end{align}
This can be done by a range of standard numerical techniques, including orthogonal
transformations and the use of normal equations (e.g.\ \citeads{lawson1995solving},
\citeads{bjorck1996numerical}). For the frame rotator, the normalised observation 
equations in Eqs.~(\ref{eq:obseqPmK}) and (\ref{eq:obseqPosK}) are solved by QR 
decomposition, using Householder transformations.

\subsection{Residual statistics and treatment of outliers}
\label{sec:estimationOut}

Standard least-squares fitting is sensitive to outliers, that is data points that do not 
reasonably fit the observation equations, considering the uncertainties and the typical
scatter of residuals. Outliers must be expected among the external counterparts, as well as
among the other QSO-like sources, for a variety of reasons. It is therefore essential that 
the fit is robust against outliers. The method adopted for the frame rotator is to 
identify the outliers that must be removed from the originally considered samples 
(${\cal Q}_0$, ${\cal R}_0$) to create the subsets ${\cal R}$ and ${\cal Q}$ used in 
Eqs.~(\ref{eq:obseqXpos}) and (\ref{eq:obseqXpm}). Two things should be noted
concerning this process. Firstly, outliers are identified at source level, that is, a given 
source is either rejected in both coordinates ($\alpha$ and $\delta$, or $\mu_{\alpha^*}$
and $\mu_\delta$), or accepted in both coordinates. This makes sense from a physical 
viewpoint, because a ‘bad’ source is probably bad (at some level) in both coordinates. 
Secondly, because $\vec{\omega}$ and $\vec{\varepsilon}$ are treated separately, 
it happens that a source rejected in ${\cal R}$ (for the orientation) is accepted in 
${\cal Q}$ (for the spin), or vice versa. Again, this could be motivated on physical grounds.

Details of the subsequent algorithm are exactly the same for Eqs.~(\ref{eq:obseqXpm})
and (\ref{eq:obseqXpos}), and are therefore described in terms of the generic function
\begin{equation}\label{eq:obseqGen}
X^2(\vec{x},{\cal S}) = \sum_{i\,\in\,{\cal S}}\,X^2_i(\vec{x})\, ,
\end{equation}
where 
\begin{equation}\label{eq:obseqXi}
X_i(\vec{x}) = \bigl|\,\vec{d}_i-\vec{D}_i\vec{x}\,\bigr| \, .
\end{equation}
Here, $\vec{x}$ could be either $\vec{\omega}$ or $\vec{\varepsilon}$, with
${\cal S}$, $\vec{d}$, and $\vec{D}$ correspondingly defined according to 
Eq.~(\ref{eq:obseqXpm}) or (\ref{eq:obseqXpos}).

We note that the functions $X_i(\vec{x})$, which can be computed for all the sources 
in ${\cal S}_0$, are a natural measure of the discrepancy 
of source $i$ from the model for given parameter vector $\vec{x}$. A useful 
overall statistic is the median discrepancy
\begin{equation}\label{eq:obseqXmed}
X_{0.5}(\vec{x}) = \operatorname*{med\,}_{i\,\in\,{\cal S}_0}
\bigl(X_i(\vec{x})\bigr) \, ,
\end{equation}
and a robust estimator could be to simply minimise this
quantity. However, if the errors are approximately Gaussian, with a moderate
fraction of outliers, robust least-squares estimation is more accurate
and therefore preferred for the frame rotator.  If $\vec{x}$ is the true parameter 
vector, and all the errors are Gaussian with nominal covariances, $X_{0.5}(\vec{x})$
is expected to be around $\sqrt{\ln 4}\simeq 1.1774$ (which is the median of the
chi-distribution with two degrees of freedom).

The adopted solution $\vec{\hat{x}}$ and accepted subset ${\cal S}$ 
simultaneously satisfy the three conditions
\begin{equation}\label{eq:obseqHatEps}
\left. \begin{aligned}
\vec{\hat{x}} = \operatorname*{argmin}_{\vec{x}}
\sum_{i\,\in\,{\cal S}} X^2_i(\vec{x})\, , \\
\forall i\in {\cal S} : 
X_i(\vec{\hat{x}}) \le \kappa\,X_{0.5}(\vec{\hat{x}})\, , \\[6pt] 
\forall i\in {\cal S}_0\backslash {\cal S} : 
X_i(\vec{\hat{x}}) > \kappa\,X_{0.5}(\vec{\hat{x}})\,  
\end{aligned} \quad\right\}
\end{equation}
for the chosen clip limit $\kappa \ge 1$ (see below). It is not
obvious that these conditions lead to a unique solution; indeed, it is easy to construct
examples where this is not the case, for example if ${\cal S}_0$ consists of disjoint 
subsets with distinctly different rotation parameters. However, these are somewhat
contrived examples and in practice the non-uniqueness is hardly an issue. More relevant
is the question whether a solution always exists -- to which we do not know the answer.
We take the pragmatic view that a unique solution `close enough' to satisfying 
Eq.~(\ref{eq:obseqHatEps}) can always be found by the following algorithm. 

Start by provisionally accepting all the sources in ${\cal S}_0$, compute the 
least-squares estimate $\vec{\hat{x}}_0$ by minimising 
$X^2(\vec{x},{\cal S}_0)$, and hence obtain the statistic
$X_{0.5}(\vec{\hat{x}}_0)$. In the next iteration, let ${\cal S}_1$ 
be the subset of sources satisfying 
$X_i(\vec{\hat{x}}_0) \le \kappa\,X_{0.5}(\vec{\hat{x}}_0)$,
leading to the estimate $\vec{\hat{x}}_1$ minimising 
$X^2(\vec{x},{\cal S}_1)$ and the new statistic 
$X_{0.5}(\vec{\hat{x}}_1)$, from which the subset ${\cal S}_2$ is
constructed, and so on. This gives a sequence of subsets 
${\cal S}_k$, $k=0,\,1,\,2,\,\dots$ that may converge to a stable subset 
(i.e.\ ${\cal S}_{k+1}={\cal S}_k$ for some $k$). If so, the stable subset and 
the corresponding estimate $\vec{\hat{x}}_k$ is the desired solution. 
But even if the sequence does not converge, it will sooner or later be found 
that the same subsets reappear cyclically in the sequence, that is 
${\cal S}_{k+p}={\cal S}_k$, where $p>1$ is the cycle period. In such
a case any of the solutions in the cycle might be accepted. However, by 
adopting the rule that the sequence is stopped at the lowest $k$ where  
${\cal S}_{k}={\cal S}_{k-p}$ for some $p>0$, we arrive at a unique result
both for a converging sequence ($p=1$) and in the cyclic case ($p>1$).

To avoid having to save and compare the (potentially many and large) subsets 
${\cal S}_k$, we assign a fixed random number $r_i$ to each source in 
${\cal S}_0$, and save from previous iterations only the outlier `checksum'
$o_k=\sum_{i\,\in\,{\cal S}_0\backslash {\cal S}_k}r_i$. The iteration stops 
as soon as $o_k=o_{k-p}$ for some $p>0$, at which point we adopt 
${\cal S}={\cal S}_k$ and $\vec{\hat{x}}=\vec{\hat{x}}_k$ as the solution.

The clip limit $\kappa$ is a dimensionless number, typically in the range 2--5, to
be chosen by the researcher. A value $\kappa\ge 1$ ensures that at most 50\%
of the sources in ${\cal S}_0$ are rejected as outliers, but in practice a higher
value should always be used. In principle, an optimum clip level that minimises 
the expected uncertainty of $\vec{\hat{x}}$ can be estimated by careful 
modelling of the actual error distribution in ${\cal S}_0$, but for the frame rotator 
we always use a fixed $\kappa = 3$. In the nominal case of purely Gaussian errors 
and no actually bad data, this will (erroneously) reject about 1\% of the sources.
This appears to be an acceptable price to pay for an estimation algorithm that is
both efficient and very robust.

\subsection{Uncertainty of the estimates}
\label{sec:estimationUnc}

The covariance of $\vec{\hat{x}}$ is estimated as
\begin{equation}\label{eq:obseqCov}
\text{Cov}\left(\vec{\hat{x}}\right) = 
\left[ \sum_{i\,\in\,{\cal S}} \vec{D}_i'\vec{D}_i\right]^{-1}\times f \, ,
\end{equation}
where the first factor (the inverse of the weighted normal matrix) is
the nominal covariance obtained on the assumption that the data
uncertainties in $\text{Cov}(\vec{\Delta}_i)$ and
$\text{Cov}(\vec{\mu}_i)$ are correctly estimated, and that no outlier
rejection is applied.  An empirical correction factor $f>0$ takes into
account that the data uncertainties may be systematically wrong
(usually underestimated), and that the nominal covariance estimate is
biased by the outlier rejection process even if the data uncertainties
are correctly estimated.

In practice, the correction factor $f$ must be derived from the residuals of the least-squares
solution, or in our case from the statistics of $X_i(\vec{\hat{x}})$. An obvious possibility 
is to use the reduced chi-square (the square of the unit weight error $u$),
\begin{equation}\label{eq:obseqU2}
u^2 = \frac{X^2(\vec{\hat{x}},{\cal S})}{2n({\cal S})-m} \, ,
\end{equation}
as an estimate of $f$, where $n({\cal S})$ is the number of accepted sources and $m$ 
is the dimension of $\vec{x}$ (here, $m=3$). 
This is accurate if $n({\cal S})$ is large (as it usually is in our applications), the fraction of
outliers is very small, and the data uncertainties are biased by an approximately constant 
factor. However, when a significant fraction of the sources are rejected as outliers, the use 
of $f=u^2$ tends to severely underestimate the uncertainties of $\vec{\hat{x}}$. 
A more reasonable correction factor should therefore consider the error distribution in
the full sample ${\cal S}_0$, not just in the accepted subset ${\cal S}$. Discrepancy 
quantiles such as $X_{0.5}$ in Eq.~(\ref{eq:obseqXmed}) provide robust characterisation 
of the full error distribution.

Based on numerical simulations, we have adopted the following recipe for
the correction factor to be used in Eq.~(\ref{eq:obseqCov}):
\begin{equation}\label{eq:obseqF}
f = \max\left[ u^2,~X^2_{0.5}/(\ln 4)\right] \, .
\end{equation} 
In the ideal case of Gaussian errors, correctly estimated data uncertainties, and a negligible
fraction of outliers, we expect $u^2\simeq 1$ and $X^2_{0.5}\simeq\ln 4$, and hence $f\simeq 1$.
If the data uncertainties are biased by a certain factor, both $u$ and $X_{0.5}$ scale by the same
factor, and the recipe still works. For samples yielding a non-negligible fraction of outliers we
normally have $u^2<X^2_{0.5}/(\ln 4)$, and the higher factor then tends to give a more accurate 
estimate of the uncertainties, hence Eq.~(\ref{eq:obseqF}). 

\subsection{Optional parameters}
\label{sec:estimationOpt}

The estimation as described above has only three unknowns, that is the components of
$\vec{\omega}$ or $\vec{\varepsilon}$. The algorithm, including weight 
normalisation, decorrelation, and outlier treatment, is readily adapted to more general linear 
models, for example using VSH to represent higher-order distortions of the
proper motions and position differences \citepads{2012A&A...547A..59M}, or including 
terms depending on the colour and magnitude of the source. However, while such more
complex models are useful for investigating the properties of the final catalogue, they 
should normally not be used in the frame rotator.

One possible exception is the glide in proper motion, which is expected to be present on physical
grounds. Already in \textit{Gaia} EDR3 this effect is very clear, and found to be consistent 
with the expected Galactic acceleration of the solar system barycentre
\citepads{2021A&A...649A...9G}. The proper motion model implicit in Eq.~(\ref{eq:obseqPm}) 
neglects this effect, which was deemed acceptable for the frame rotator in the first 
three releases of \textit{Gaia} astrometry. For future releases this may not be accurate
enough (since the effects of rotation and glide can be correlated and mutually disturb each other).
One possible remedy is then to correct the observed proper motions in 
Eq.~(\ref{eq:obseqDeltaPm}) for the assumed Galactic glide.
However, a better option may be to estimate the glide $\vec{g}$ 
together with $\vec{\omega}$, using the augmented observation equations
\begin{equation}\label{eq:obseqPmG}
\vec{A}_i\vec{\omega} + \vec{B}_i\vec{g} = \vec{\mu}_i \, .
\end{equation}
The estimation follows the generic scheme outlined above, minimising
$X^2=\sum_{i\,\in\,{\cal Q}}\,|\,\vec{d}_i-\vec{D}_i\vec{x}\,|^2$ with
\begin{equation}\label{eq:obseqPmG1}
\vec{x}=\begin{bmatrix} \vec{\omega}\\ \vec{g}\end{bmatrix}\, , \quad
\vec{d}_i=\vec{L}_i\vec{\mu}_i\, , \quad\text{and}\quad
\vec{D}_i=\vec{L}_i\,\Bigl[\,\vec{A}_i~~\vec{B}_i\,\Bigr]\, .
\end{equation}
It should be noted that the application of the spin and orientation corrections to the
source and attitude data, that is functions (ii) and (iii) of the frame rotator, is in any case 
restricted to the pure rotations given by $\vec{\omega}$ and $\vec{\varepsilon}$. 
In particular, the published source parameters are never corrected for the (assumed or 
estimated) glide in proper motion.

\end{appendix}

\end{document}